%% file: main.tex
\documentclass[12pt]{iopart}

\usepackage{bm}
\usepackage{graphicx}
\usepackage{outlines}
\usepackage{enumitem}
\usepackage{cite,ulem}
\usepackage{xcolor}
\usepackage{amssymb}
\usepackage{hyperref}
\usepackage{subcaption}

\newcommand{\mcal}{\mathcal}
\newcommand{\as}{\alpha_s}
\newcommand{\bas}{{\bar\alpha}_s}

\def\Phperp{P_{hT}}
\newcommand{\zh}{\ensuremath{z_{h}}}

\newcommand{\df}{\mathrm{d}}

\newcommand{\xbj}{\ensuremath{x}}
\begin{document}

\vspace*{-2cm}
\hspace*{11.9cm}
INT-PUB-26-011

\vspace*{3mm}
\title[Precision QCD with the Electron-Ion Collider]{Precision QCD with the Electron-Ion Collider}

\vspace*{2mm}
\author{C. Alexandrou}
\address{Department of Physics, University of Cyprus, P.O. Box 20537, 1678 Nicosia, Cyprus and Computation-based Science and Technology Research Center, The Cyprus Institute, 20 Kavafi Str., Nicosia 2121, Cyprus}
\ead{alexand@ucy.ac.cy}

\author{M. Arratia}
\address{Department of Physics and Astronomy, University of California, Riverside, CA 92521, USA}
\ead{miguela@ucr.edu}

\author{E.C. Aschenauer}
\address{Physics Department, Brookhaven National Laboratory, Upton, New York 11973, USA}
\ead{elke@bnl.gov}

\author{A. Avkhadiev}
\address{Physics Division, Argonne National Laboratory, Lemont, IL 60439, USA}
\ead{aavkhadi@anl.gov}

\author{P.V. Balachandran}
\address{Department of Materials Science and Engineering, University of Virginia, Charlottesville VA 22903, USA}
\ead{pvb5e@virginia.edu}

\author{V. Bertone}
\address{IRFU, CEA, Universit\'e Paris-Saclay, F-91191 Gif-sur-Yvette, France}
\ead{valerio.bertone@cea.fr}

\author{I. Borsa}
\address{Institute for Theoretical Physics, University of Tübingen,
Auf der Morgenstelle 14, 72076 Tübingen, Germany}
\ead{ignacio.borsa@itp.uni-tuebingen.de}

\author{M. Cerutti}
\address{IRFU, CEA, Universit\'e Paris-Saclay, F-91191 Gif-sur-Yvette, France}
\ead{matteo.cerutti@cea.fr}

\author{X. Chu}
\address{Physics Department, Brookhaven National Laboratory, Upton, New York 11973, USA}
\ead{xchu@bnl.gov}

\author{W. Cosyn}
\address{Department of Physics, Florida International University, Miami,
Florida 33199, USA}
\ead{wcosyn@fiu.edu}

\author{ D. de Florian}
\address{ICIFI-Instituto de Ciencias Físicas, Universidad Nacional de San Martín, 25 de Mayo y Francia, 1650 San Martín, Pcia. de Buenos Aires, Argentina}
\ead{deflo@unsam.edu.ar}

\author{A. Dumitru}
\address{Department of Natural Sciences, Baruch College, CUNY,
17 Lexington Avenue, New York, NY 10010, USA}
\address{The Graduate School and University Center, The City University of New York, 365 Fifth Avenue, New York, NY 10016, USA}
\ead{adrian.dumitru@baruch.cuny.edu}

\author{M. Engelhardt}
\address{Department of Physics, New Mexico State University, Box 30001, MSC 3D, Las Cruces, NM 88003, USA}
\ead{engel@nmsu.edu}

\author{R. Fatemi}
\address{Department of Physics and Astronomy, University of Kentucky, Lexington, KY 40506, U.S.A.}
\ead{renee.fatemi@uky.edu}

\author{S. Forte}
\address{Dipartimento di Fisica, Universit\`a di Milano and INFN, Sezione di Milano, via Celoria 16, 20133 Milano, Italy}
\ead{forte@mi.infn.it}

\author{Y. Fu}
\address{Center for Theoretical Physics – a Leinweber Institute, Massachusetts Institute of Technology, Cambridge, MA 02139, USA}
\ead{yangfu@mit.edu}

\newpage
\author{L. Gamberg}
\address{Division of Science-Physics, Penn State University Berks, 2100 Stoudt Road, Wyomissing, PA 19610}
\ead{lpg10@psu.edu}

\author{H. Gao}
\address{Department of physics, Duke University, and the Triangle Universities Nuclear Laboratory, Durham, NC 27708, U.S.A.}
\ead{Haiyan.gao@duke.edu}

\author{T. Gehrmann}
\address{Physik-Institut, Universität Zürich, Winterthurerstrasse 190, 8057 Zürich, Switzerland}
\ead{thomas.gehrmann@uzh.ch}

\author{A. Gehrmann--De Ridder}
\address{Institute for Theoretical Physics, ETH, CH-8093 Z\"urich, Switzerland}
\ead{gehra@phys.ethz.ch}

\author{Y. Go}
\address{Physics Department, Brookhaven National Laboratory, Upton, New York 11973, USA}
\ead{ygo@bnl.gov}

\author{Y. Guo}
\address{Nuclear Science Division, Lawrence Berkeley National Laboratory, Berkeley, CA 94720, USA}
\ead{yuxunguo@lbl.gov}

\author{Y. Hatta}
\address{Physics Department, Brookhaven National Laboratory, Upton, NY 11973, USA}
\address{RIKEN BNL Research Center, Brookhaven National Laboratory, Upton, NY 11973, USA}
\ead{yhatta@bnl.gov}

\author{J. Haug}
\address{Institute for Theoretical Physics, University of Tübingen, Auf der Morgenstelle 14, 72076 Tübingen, Germany}
\ead{juliane-clara-celine.haug@uni-tuebingen.de}

\newpage
\author{T.J. Hobbs}
\address{High Energy Physics Division, Argonne National Laboratory, Lemont, IL 60439, USA}
\ead{tim@anl.gov}
   
\author{T. Horn}
\address{The Catholic University of America, Physics Department, 620 Michigan Ave NE, 20064 Washington DC, USA}
\ead{hornt@cua.edu} 

\author{E. Iancu}
\address{Universit\'{e} Paris-Saclay, CNRS, CEA, Institut de physique th\'{e}orique, F-91191, Gif-sur-Yvette, France}
 \ead{edmond.iancu@ipht.fr}

\author{J. Jalilian-Marian}
\address{Baruch College, CUNY, 17 Lexington Ave., 10010, New York, NY, USA}
\address{The Graduate Center, CUNY, 365 Fifth Ave., 10016, New York, NY, USA}
\ead{jamal.jalilian-marian@baruch.cuny.edu}

\author{Z.B. Kang}
\address{Department of Physics and Astronomy, University of California, Los Angeles, CA 90095, USA}
\ead{zkang@physics.ucla.edu}

\author{M. Klasen}
\address{Institute for Theoretical Physics, University of Münster, Wilhelm-Klemm-Str. 9, 48149 Münster, Germany}
\ead{michael.klasen@uni-muenster.de}

\author{Y.V. Kovchegov}
\address{Department of Physics, The Ohio State University, Columbus, OH 43210, USA}
\ead{kovchegov.1@osu.edu}

\author{B. Kriesten}
\address{High Energy Physics Division, Argonne National Laboratory, Lemont, IL 60439, USA}
\ead{bkriesten@anl.gov}

\newpage
\author{J.G. Lajoie}
\address{Oak Ridge National Laboratory, Oak Ridge, Tennessee 37831, USA}
\ead{lajoiejg@ornl.gov}

\author{W. Li}
\address{Department of Physics and Astronomy, Rice University, 6100 Main Stree, Houston, TX 77005, USA}
\ead{wl33@rice.edu}

\author{X. Li}
\address{Physics Division, Los Alamos National Laboratory, Los Alamos, New Mexico, 87545, USA}
\ead{xuanli@lanl.gov}

\author{Y. Li}
\address{4700 Elkhorn Ave, Suite 3300, Department of Computer Science, Old Dominion University, Norfolk, VA 23529-0162, USA}
\ead{yaohang@cs.odu.edu}

\author{H.-W. Lin}
\address{Department of Physics and Astronomy, Michigan State University, East Lansing, Michigan 48824, USA}
\ead{hwlin@pa.msu.edu}

\author{M.X. Liu}
\address{P-3 MS H846, Physics Division, Los Alamos National Laboratory, Los Alamos, NM 87545, USA}
\ead{mliu@lanl.gov}

\author{S. Liuti}
\address{Physics Department, University of Virginia, 382 McCormick Rd., Charlottesville, VA 22904}
\ead{sl4y@virginia.edu}

\author{C. Marquet}
\address{CPHT, CNRS, \'Ecole polytechnique,  Institut Polytechnique de Paris, 91120 Palaiseau, France}
\ead{cyrille.marquet@polytechnique.edu}

\author{P. Meinzinger}
\address{Physik-Institut, Universit\"at Z\"urich, Winterthurerstrasse 190, CH-8057 Z\"urich, Switzerland}
\ead{peter.meinzinger@uzh.ch}

\author{W. Melnitchouk}
\address{Jefferson Lab, 12000 Jefferson Avenue, Newport News, Virginia 23606, USA}
\ead{wmelnitc@jlab.org}

\author{S. Moch}
\address{II. Institute for Theoretical Physics, University of Hamburg, Luruper Chaussee 149, 22761 Hamburg, Germany}
\ead{sven-olaf.moch@desy.de}

\author{P. Nadel-Turonski}
\address{University of South Carolina, Columbia, SC 29208}
\ead{turonski@sc.edu} 

\author{P. Nadolsky}
\address{Department of Physics and Astronomy, Michigan State University, East Lansing, MI 48824, USA}
\ead{nadolsky@msu.edu}

\author{M. Neubert}
\address{Mainz Institute of Theoretical Physics,
Johannes Gutenberg University Mainz,
55099 Mainz, Germany}
\ead{matthias.neubert@uni-mainz.de}

\author{A. NieMiera }
\address{Department of Physics and Astronomy, Michigan State University, 567 Wilson Road, East Lansing, MI 48824, USA}
\ead{niemiera@msu.edu}

\author{E.R. Nocera}
\address{Dipartimento di Fisica, Università degli Studi di Torino and INFN, Sezione di Torino, Via Pietro Giuria 1, 10125 Torino, Italy}
\ead{emanueleroberto.nocera@unito.it}

\author{C. Pecar}
\address{Duke University, Durham, North Carolina 27708-0305}
\ead{connor.pecar@duke.edu}

\author{J. Penttala}
\address{Department of Physics and Astronomy, University of California, Los Angeles, CA 90095, USA}
\address{Mani L. Bhaumik Institute for Theoretical Physics, University of California, Los Angeles, CA 90095, USA}
\ead{janipenttala@physics.ucla.edu}

\author{C. Pisano}
\address{Dipartimento di Fisica, Universit\`a di Cagliari, Cittadella Universitaria, I-09042 Monserrato (CA), Italy}
\address{INFN, Sezione di Cagliari, Cittadella Universitaria, I-09042 Monserrato (CA), Italy}
\ead{cristian.pisano@unica.it}

\author{A. Prokudin}
\address{Penn State University Berks, 2100 Stoudt Road, Wyomissing, PA 19610}
\address{Theory Center, Jefferson Lab, Newport News, VA 23606, USA}
\ead{prokudin@jlab.org}

\author{J.-W. Qiu}
\address{Theory Center, Jefferson Lab, Newport News, VA 23606, USA}
\ead{jqiu@jlab.org}

\author{F. Ringer}
\address{Department of Physics and Astronomy, Stony Brook University, New York 11794, USA}
\ead{felix.ringer@stonybrook.edu}

\author{F. Salazar}
\address{Department of Physics, Temple University, Philadelphia, Pennsylvania 19122, USA}
\address{RIKEN-BNL Research Center, Brookhaven National Laboratory, Upton, New York 11973, USA}
\address{Physics Department, Brookhaven National Laboratory, Upton, New York 11973, USA}
\ead{farid.salazar@temple.edu}

\author{R. Sassot}
\address{Universidad de Buenos Aires, Facultad de Ciencias Exactas y Naturales,
Departamento de Fisica and IFIBA-CONICET, Ciudad Universitaria (1428) Buenos Aires,
Argentina}
\ead{sassot@df.uba.ar}

\author{J. Schoenleber}
\address{Physik-Department T31,
Technische Universität München,
James-Franck-Straße 1, D-85748 Garching, Germany}
\ead{jakob.schoenleber@tum.de}

\vspace*{-0.5mm}
\author{R. Seidl}
\address{Tokyo University, Quark Nuclear Science Institute, 113-0033 Tokyo-to, Bunkyo-ku, Hongo7-3-1, Japan}
\ead{ralf.seidl@qnsi.s.u-tokyo.ac.jp}

\vspace*{-0.5mm}
\author{V. Skokov}
\address{Department of Physics and Astronomy, North Carolina State University, Raleigh, NC 27695, USA}
\ead{vskokov@ncsu.edu}

\vspace*{-0.5mm}
\author{A.M. Sta\'sto}
\address{Department of Physics, The Pennsylvania State University, 104 Davey Lab, University Park, PA 16802, U.S.A.}
\ead{ams52@psu.edu}

\vspace*{-0.5mm}
\author{R. Sufian}
\address{Department of Physics, New Mexico State University, Las Cruces, NM 88003, USA}
\address{RIKEN-BNL Research Center, Brookhaven National Laboratory, Upton, NY 11973, USA}
\ead{rsufian1@nmsu.edu}

\vspace*{-0.5mm}
\author{S. Tiwari}
\address{Brookhaven National Lab, Upton, 11973 New York, United States}
\address{North Carolina State University, Raleigh, 11973 North Carolina, United States}
\ead{stiwari1@bnl.gov}

\vspace*{-0.5mm}
\author{M. Ubiali}
\address{DAMTP, University of Cambridge, Wilberforce Road, Cambridge, CB3 0WA, United Kingdom}
\ead{M.Ubiali@damtp.cam.ac.uk}

\vspace*{-0.5mm}
\author{R. Venugopalan}
\address{Brookhaven National Laboratory, Upton, NY 11973, USA}
\address{Center for Frontiers in Nuclear Science (CFNS) at Stony Brook University,
Stony Brook, NY 11794, USA}
\address{Higgs Centre for Theoretical Physics, The University of Edinburgh, 
 Edinburgh, EH9 3FD, UK}
\ead{raju.venugopalan@gmail.com}

\newpage
\author{W. Vogelsang}
\address{Institute for Theoretical Physics, University of Tübingen, Auf der Morgenstelle 14, 72076 Tübingen, Germany}
\ead{werner.vogelsang@uni-tuebingen.de}

\author{F. Wunder}
\address{Institute for Theoretical Physics, University of Tübingen, Auf der Morgenstelle 14, 72076 Tübingen, Germany}
\address{Deutsches Elektronen-Synchrotron DESY, Notkestrasse 85, 22607 Hamburg, Germany}
\ead{fabian.wunder@desy.de}

\author{F. Yuan}
\address{Nuclear Science Division, Lawrence Berkeley National
Laboratory, Berkeley, CA 94720, USA}
\ead{fyuan@lbl.gov}

\author{Y. Zhao}
\address{Physics Division, Argonne National Laboratory, Lemont, IL 60439, USA}
\ead{yong.zhao@anl.gov}

\author{W. Zhao}
\address{Key Laboratory of Quark and Lepton Physics (MOE) \& Institute of Particle Physics, Central China Normal University, Wuhan 430079, China}
\ead{WenbinZhao@ccnu.edu.cn}

\vspace{10pt}

\begin{abstract}
This document summarizes the discussions at the program ``Precision QCD with the Electron Ion Collider'',
held from May to June 2025 at the Institute for Nuclear Theory (INT) at the University of Washington.
The program was co-sponsored by the INT and by the Center for Frontiers in Nuclear Science (CFNS, Stony Brook
University). Over its five-week duration it brought together about 70 theorists, experimentalists and computer
scientists all interested in the physics program at the future Electron Ion Collider in preparation at 
Brookhaven National Laboratory. Key topics at the program were: higher-order perturbative-QCD calculations and techniques; nuclear structure and tomography; comparisons of phenomenological and lattice determinations of parton distribution functions; identification of signature observables for saturated gluons; assessment of the importance of AI techniques for EIC studies and detector development.
\end{abstract}

%
%
%
%
%


\tableofcontents

\newpage

\title[Precision QCD with the Electron-Ion Collider]{}

\vspace*{-2.5cm}
\section*{Introduction}
\input{intro.tex}

\clearpage
\section{Precision theory for hard scattering at the EIC, factorization and resummation}
{\it Editors: Daniel de Florian, Thomas Gehrmann\\  Contributors: 
Ignacio Borsa, Stefano Forte, Yang Fu, Aude Gehrmann-De Ridder, Juliane Haug, Huey-Wen Lin, Peter Meinzinger, Sven Moch, Matthias Neubert, Jianwei Qiu, Rodolfo Sassot, Werner Vogelsang, Fabian Wunder, Feng Yuan}
\input{week1.tex}

\clearpage
\section{Parton distributions and the interplay of EIC and LHC: Lattice QCD meets phenomenology} 
{\it Editors: Renee Fatemi, Huey-Wen Lin, Werner Vogelsang\\ Contributors: Constantia Alexandrou, Artur Avkhadiev, Valerio Bertone, Ignacio Borsa, Haiyan Gao, Thomas Gehrmann, Aude Gehrmann-De Ridder, Juliane Haug, Michael Klasen, Peter Meinzinger, Wally Melnitchouk, Pavel Nadolsky, Emanuele Nocera, Rodolfo Sassot, 
Vladimir Skokov, Raza Sufian, Fabian Wunder,  Maria Ubiali, Yong Zhao}
\input{week2.tex}

\clearpage
\section{Small-$x$ physics in the EIC era} 
{\it Editors: Zhong-Bo Kang, Werner Vogelsang\\ Contributors: Elke Aschenauer, Xiaoxuan Chu, Adrian Dumitru, Thomas Gehrmann, Edmond Iancu, Jamal Jalilian-Marian, Yuri Kovchegov, Wei Li, Cyrille Marquet, Peter Meinzinger, Jani Penttala, Cristian Pisano, Farid Salazar, 
Vladimir Skokov, Anna Stasto, Shaswat Tiwari, Raju Venugopalan, Wenbin Zhao}

\input{week3.tex}

\clearpage
\section{AI/ML for EIC physics} 
{\it Editors: Miguel Arratia, Huey-Wen Lin \\ Contributors: Prasanna Balachandran, Leonard Gamberg, Thomas Gehrmann, Yeonju Go, Tim Hobbs, Tanja Horn, Brandon Kriesten, Xuan Li, Ming Liu, Yaohang Li, Alex NieMiera, Connor Pecar, Felix Ringer, Werner Vogelsang}

\input{week4.tex}

\clearpage
\section{Jets and semi-inclusive reactions: Nucleon and nuclear tomography} 
{\it Editors: Miguel Arratia, Renee Fatemi, Zhong-Bo Kang\\ Contributors: Valerio Bertone, Matteo Cerutti, Wim Cosyn, Michael Engelhardt, Leonard Gamberg, Thomas Gehrmann, Yoshitaka Hatta, Yuxun Guo, John Lajoie, Ming Liu, Simonetta Liuti, Alexei Prokudin, Jakob Schoenleber, Pawel Nadel-Turonski, Ralf Seidl, Werner Vogelsang}
\label{ch:tomography}
\input{week5.tex}

\clearpage
\section*{Concluding remarks}
\input{outlook.tex}

\section*{Acknowledgments}

The program was supported in equal parts by the Institute for Nuclear Theory (INT) and by the Center for Frontiers in Nuclear Science (CFNS).
We extend our thanks to the directors of the two institutes, Sanjay Reddy and Abhay Deshpande, for providing us with the
opportunity to run this wonderful program. 

The administrative support offered by the INT has been magnificent. Our sincere gratitude goes to our local coordinator, Paris Nguyen, 
for his outstanding support throughout the program, from its early stages after approval to its completion. \\

C.A. acknowledges partial financial support by 
the European Joint Doctorate project 
AQTIVATE that received funding from the European
Union’s research and innovation program under the
Marie Sklodowska-Curie Doctoral Networks action,
Grant Agreement No 101072344 and the project  IMAGE-N (EXCEL-
LENCE/0524/0459), HyperON (VISION ERC-PATH
2/0524/0001),  co-financed
by the European Regional Development Fund and the
Republic of Cyprus through the Research and Innovation Foundation.
A.A.'s contributions are based upon work supported by Laboratory Directed Research and Development (LDRD) funding from Argonne National Laboratory, provided by the Director, Office of Science, of the U.S. DOE under Contract No. DE-AC02-06CH11357; and by the U.S. Department of Energy, Office of Science, Office of Nuclear Physics through Contract No.~DE-AC02-06CH11357.
The work of Valerio Bertone and Matteo Cerutti has been supported by l’Agence Nationale de la Recherche (ANR), project ANR-24-CE31-7061-01.
W.C. is supported by the U.S.~National Science Foundation
under award PHY-2239274.
The work of X. Chu and E.C. Aschenauer is supported by the U.S. Department of Energy under Contract No. DE-SC0012704.
X. Chu is also supported by the Laboratory Directed Research and Development (LDRD) project 25-029.
A.D. is supported by DOE Office of Nuclear Physics, Grant DE-SC0002307.
M.E. is supported by the U.S. DOE Office of Science, Office of Nuclear Physics, through grant DE-FG02-96ER40965.
The work of R. Fatemi is supported by the National Science Foundation under contract No. 2110293 and 2412373. 
The work of S.F. has received funding from the European Union NextGeneration EU program – NRP Mission 4 Component 2 Investment 1.1 – MUR PRIN 2022 – CUP G53D23001100006 through the Italian Ministry of University and Research (MUR).
Y.F. is 
supported by the U.S. Department of Energy, Office of Science, Office of Nuclear Physics under grant Contract Number DE-SC0011090.
This work is supported by the U.S. Department of Energy, Office of Science-Nuclear Physics,  under contract No. DE-SC0026320 (L.G.).
H. Gao acknowledges the support of the U.S. Department of Energy, Office of Science, Office
of Nuclear Physics under contract DE-FG02-03ER41231.
T.G. is supported by the Swiss National Science Foundation (SNSF) under contract 240015 and by the European Research Council (ERC) under the European Union's Horizon 2020 research and innovation programme grant agreement 101019620 (ERC Advanced Grant TOPUP).
A.G. is supported by the Swiss National Science Foundation (SNF) under contract 200021-197130. 
Supported in part by National Science Foundation grant PHY2309976 and US Department of Energy grant SC-0024691.
J. Jalilian-Marian is supported by the US DOE Office of Nuclear Physics through Grant No. DE-SC0002307 and the framework of the Saturated Glue (SURGE) Topical Theory Collaboration.
Yeonju Go is supported by the LDRD Program at Brookhaven National Laboratory and the U.S. Department of Energy under Contract No. DE-SC0012704.
The work of T.J.H. at Argonne National Laboratory was supported by the U.S. Department of Energy under contract DE-AC02-06CH11357.
The work of YK is supported by the U.S. Department of Energy, Office of Science, Office of Nuclear Physics
under Award Number DE-SC0004286 and within the framework of the Saturated Glue
(SURGE) Topical Theory Collaboration.
B. Kriesten was supported at Argonne National Laboratory by the U.S. Department of Energy under contract DE-AC02-06CH11357.
This work is supported by the DOE Office of Science, 
Office of Nuclear Physics.
Wei Li's work is supported by the Department of Energy grant number DE-SC0005131.
Xuan Li's research is funded by LANL's Laboratory Directed Research and Development (LDRD) program under project number 20230072ER.
Yaohang Li's work is supported by the U.S. Department of Energy, Office of Science, Office of Nuclear Physics, Office of Advanced Scientific Computing Research through the Scientific Discovery through Advanced Computing (SciDAC) program, under contracts DE-AC02-06CH11357, DE-AC05-06OR23177, and DE-SC0023472 for the award Femtoscale Imaging of Nuclei using Exascale Platforms, by 
the EXCLAIM collaboration under the DOE grant DE-SC0024644,
and by the Center for Nuclear Femtography (CNF), administrated by the Southeastern Universities Research Association under an appropriation from the Commonwealth of Virginia under contract No. C2024-FEMT-011-02.
The work of H.L. is partially supported by the US National Science Foundation under grant PHY 2209424,
by the U.S.~Department of Energy under contract DE-SC0024582, and by the Research Corporation for Science Advancement through the Cottrell Scholar Award. 
P.M. has been supported by the Swiss National Science Foundation (SNSF) under contract 240015 and by the European Research Council (ERC) under the European Union's Horizon 2020 research and innovation programme grant agreement 101019620 (ERC Advanced Grant TOPUP).
W.M. was supported by the DOE contract No. DE-AC05-06OR23177, under which Jefferson Science Associates, LLC operates Jefferson Lab.
The work by Pavel Nadolsky was supported by the Wu-Ki Tung Endowed Chair in particle physics.
This work was supported by the European Research Council (ERC) under the European Union’s Horizon 2022 Research and Innovation Program (ERC Advanced Grant agreement No. 101097780, EFT4jets), and the Cluster of Excellence PRISMA$^{++}$ (Precision Physics, Fundamental Interactions, and Structure of Matter, EXC 2118/2) funded by the German Research Foundation (DFG) under Germany’s Excellence Strategy (Project ID 390831469).
Alex NieMiera's work is supported in terms of a {\it High Energy Physics Computing Traineeship for Lattice Gauge Theory} at Michigan State University under contract No. DE-SC0024053.
E.R.N. was supported by the Italian Ministry of University and Research (MUR) through the “Rita Levi-Montalcini” Program.
Z.K. and J.P. are supported by the National Science Foundation under grant No.~PHY-2515057, and by the U.S. Department of Energy, Office of Science, Office of Nuclear Physics, within the framework of the Saturated Glue (SURGE) Topical Theory Collaboration.
The work of C.P. is supported by the European Union ``Next Generation EU'' program through the Italian PRIN 2022 grant n. 20225ZHA7W.
This work is supported by the U.S. Department of Energy, Office of Science, Office of Nuclear Physics through Contract No. DE-SC0020081.
This work was supported by the U.S. Department of Energy contract No.~DE-AC05-06OR23177, under which Jefferson Science Associates, LLC operates Jefferson Lab (A.P.),  and by the National Science Foundation under Grants No.~PHY-2310031, No.~PHY-2335114 (A.P.).
This work was supported in part by the U.S.\ Department of Energy (DOE) Contract No.\ DE-AC05-06OR23177, under which Jefferson Science Associates, LLC operates Jefferson Lab (J.Q.). 
F.R. is by the U.S. Department of Energy under grant number
DE-SC0025881.
F.S. is supported by the Laboratory Directed Research and Development of Brookhaven National Laboratory and RIKEN-BNL Research Center. F.S. also acknowledges support from the Saturated Glue (SURGE) and the Quark-Gluon Tomography (QGT) Topical Theory Collaborations, funded by the U.S. Department of Energy, Office of Science, Office of Nuclear Physics.
Supported in part by CONICET and UBACyT. 
J.S. was supported by the U.S. Department of Energy under Contract No. DE-SC0012704 and the Laboratory Directed Research and Development (LDRD) funds from Brookhaven Science Associates.
A.M.S. is  supported by the U.S. Department of Energy grant No. DE-SC-0002145 and within the framework of the of the Saturated Glue (SURGE) Topical Theory Collaboration.
M. Ubiali is supported by the European Research Council under the European Union’s Horizon 2020 research and innovation Programme (grant agreement n.950246) and also partially supported by the STFC grant ST/T000694/1.
Y.H. and R.V. are supported by the U.S. Department of Energy, Office of Science under contract DE-SC0012704 and within the framework of the Saturated Glue (SURGE) Topical Collaboration in Nuclear Theory. R.V. is also supported at Stony Brook by the Simons Foundation as a co-PI under Award number 994318 (Simons Collaboration on Confinement and QCD Strings). He acknowledges partial support from DOE Grant Ref. DE-SC0025732, Novel Holographic Approaches to the Non-perturbative Dynamics of Proton Spin.
R.V. thanks the UK Royal Society and the Wolfson Foundation for a Visiting Fellowship and the Higgs Center at the University of Edinburgh for their kind hospitality. 
I.B., J.H., W.V. and F.W. acknowledge support by Deutsche Forschungsgemeinschaft (DFG) through the Research Unit FOR 2926 (Project No. 409651613). 
The work of Y.Z. is based upon work supported by the U.S. Department of Energy, Office of Science, Office of Nuclear Physics through Contract No.~DE-AC02-06CH11357, and the Early Career Award through Contract No.~DE-SCL0000017. Y.Z. also acknowledges partial support by the U.S. Department of Energy, Office of Science, Office of Nuclear Physics under the umbrella of the \textit{Quark-Gluon Tomography (QGT) Topical Collaboration} with Award DE-SC0023646.

\newpage
\section*{References}
\bibliographystyle{unsrt}
\bibliography{refs, references}

\end{document}

%% file: intro.tex
The Electron-Ion Collider (EIC) will deliver unprecedented insight into the internal structure of nucleons and nuclei
and will probe QCD with a level of precision and versatility never before achieved. It  may well become
transformative for nuclear physics. As the collider and its flagship detector, ePIC, move further into detailed design and construction, 
the theoretical and computational communities face the urgent and timely challenge to ensure that the field is fully prepared to interpret, analyze, and maximize the scientific return of the forthcoming data. This set the motivation for the five-week international program 
``Precision QCD with the Electron-Ion Collider,'' hosted by the Institute for Nuclear Theory (INT) at the University of Washington in Seattle. The program was co-sponsored by the INT and the Center for Frontiers in Nuclear Science (CFNS, Stony Brook University). It convened about 70 leading experts in theory, experiment, phenomenology, lattice QCD, and artificial intelligence. 
The program built on the outstanding work done over the past
25 years on developing and promoting the EIC science case~\cite{Deshpande:2005wd,Boer:2011fh,Accardi:2012qut,AbdulKhalek:2021gbh,AbdulKhalek:2022hcn,Abir:2023fpo}.

Each week of the program addressed a central pillar of the EIC scientific mission. The opening week (conveners Daniel de Florian, Thomas Gehrmann) focused on precision perturbative QCD and factorization, emphasizing next-to-next-to-leading order techniques, resummation methods, and high-accuracy predictions for semi-inclusive and exclusive processes and building on lessons from the Large Hadron Collider. The second week (conveners Renee Fatemi, Huey-Wen Lin, Werner Vogelsang) concentrated on parton distribution functions (PDFs), lattice QCD, and global analyses, highlighting the growing synergy between first-principles calculations and phenomenological extractions. The third week (conveners Zhong-Bo Kang, Werner Vogelsang) turned to the small-$x$ regime and saturation physics, where nonlinear QCD dynamics and gluon density effects become dominant. Emphasis was placed on advancing the precision of theoretical predictions, quantifying uncertainties, and connecting formal developments, such as solutions to small-$x$ evolution equations and sub-eikonal corrections, to experimentally accessible observables. Recognizing the transformative role of modern computational methods, the fourth week (conveners Miguel Arratia, Huey-Wen Lin) focused on artificial intelligence and enhanced detector design. By integrating machine learning directly into simulation and analysis workflows, the community aims to accelerate detector optimization, improve the extraction of physical observables, and enable more sophisticated analyses of complex final states. These developments are particularly timely as the ePIC detector design continues to evolve. Finally, week five (conveners Miguel Arratia, Renee Fatemi, Zhong-Bo Kang)
synthesized many of these themes through a focus on jets, transverse momentum–dependent distributions (TMDs), generalized parton distributions (GPDs), and nucleon and nuclear tomography. Central topics were theoretical modeling, lattice calculations, global fits, and experimental strategies for accessing TMDs and GPDs, alongside advances in jet substructure and flavor tagging. The discussions underscored the EIC’s unique potential to deliver a three-dimensional picture of hadronic matter in momentum and coordinate space. 

The program was conceived as an intensive, coordinated effort to identify the most pressing theoretical challenges 
for the EIC era and to forge durable collaborations across subfields. 
It fostered a highly collaborative environment, with an emphasis on open discussion, early-career participation, 
and dialogue across the full community. The interactive structure of the program reflected the recognition that progress 
toward EIC precision physics demands both technical depth and cross-disciplinary exchange.

The present document collects the discussions at the program. It summarizes the status of each thematic area, identifies key open questions, 
and proposes coordinated strategies for future research. It lays out a coordinated vision for precision QCD at the Electron-Ion Collider,
based on advanced perturbative methods, lattice QCD, global analysis, explorations of small-$x$ dynamics, AI-driven innovation, and 
multidimensional study of hadron structure. We hope that the document will be a significant step toward ensuring theoretical and computational 
readiness for the EIC era. We also anticipate that it will serve as a useful guide as the EIC advances toward operation,
and be a valuable resource for unlocking the full scientific potential of the machine. 

That said, we stress that this document is not intended as a review of EIC science or, let alone, of the full field of QCD and hadronic structure. 
First and foremost, the document reflects the discussions at the program, by its participants.
Without question, not all topics relevant at the EIC were discussed in depth -- this would have required a much longer program.
Clearly, choices had to be made in identifying the core topics for the program, and likewise for
the various subtopics and contributions in each program week. Furthermore, EIC physics being a vibrant and advancing field, numerous developments have taken place even in the short time since our program, which may not all be reflected in this write-up. 
Consequently, even with over 950 references in this document, there will undoubtedly be
researchers of the field who will feel that their work is not properly addressed or acknowledged in this document. We apologize 
to these colleagues, stressing again the interactive and informal nature of the program.

%% file: week1.tex
\subsection{Higher-order calculations for the EIC 
}
\label{sec:higher_orders}

With its anticipated luminosity, the EIC will allow measurements of a multitude of final states to high statistical precision, often multi-differential in their kinematical variables. To fully exploit these data in precision probes of strong interaction dynamics in the underlying hard scattering process, in the proton structure or in the parton-hadron transition requires an equally high level of precision in the theoretical predictions. For processes that are characterized by a sufficiently large momentum transfer scale, this precision is attained through the computation of higher orders in perturbation theory. Depending on the process considered, the perturbative expansion is in powers of the QCD coupling constant $\alpha_s$ (LO, NLO, NNLO, $\ldots$) or in powers of logarithms that are resummed to all orders in the coupling (LL, NLL, NNLL, $\ldots$). A substantial body of work on higher-order corrections for hard processes in electron-proton collisions is already available, in the past largely motivated by the HERA collider physics program.

For fully inclusive or semi-inclusive processes, these results are expressed in terms of coefficient functions, usually known in closed analytical form. For unpolarized inclusive DIS, the massless coefficient functions have been known to N3LO~\cite{Vermaseren:2005qc} for quite some time; they were recently complemented by mass corrections~\cite{Ablinger:2025awb}, thereby enabling a fully consistent description of inclusive structure functions
to N3LO. These will be a key ingredient in N3LO-accurate determinations of collinear parton distributions, which are being enabled through the recent derivation of four-loop corrections to the Altarelli-Parisi splitting functions~\cite{Falcioni:2023luc,Falcioni:2023vqq,Falcioni:2024xyt,Falcioni:2024qpd,Gehrmann:2023cqm,Gehrmann:2023iah}. Polarized DIS is consistently known to NNLO with coefficient~\cite{Zijlstra:1993sh} and splitting functions~\cite{Moch:2014sna,Blumlein:2021enk}, supplemented by recent progress towards N3LO~\cite{Blumlein:2022gpp}. 
These form the basis of precision PDF fits, see Sections~\ref{sec:helicity_PDFs}--\ref{sec:unpol_PDFs} below. Coefficient functions for semi-inclusive hadron production in unpolarized and polarized deep inelastic scattering (SIDIS) were derived most recently to NNLO~\cite{Goyal:2023zdi,Bonino:2024qbh,Goyal:2024tmo,Bonino:2024wgg}, also accounting for weak interaction effects at large virtualities~\cite{Bonino:2025qta,Bonino:2025bqa}, as well as NNLO QED and QCD$\otimes$QED corrections \cite{Goyal:2025qyu}. They will be important ingredients in particular to NNLO-accurate determinations of hadron fragmentation functions (see Section~\ref{sec:FFs}) and of polarized parton distributions. Such determinations of parton distributions and fragmentation functions require an efficient code base, preferably publicly available. First steps towards speeding up the numerical evaluation of the SIDIS coefficient functions were initiated by discussions during this workshop~\cite{Haug:2025ava}.

Due to the presence of large logarithms, fixed-order corrections deteriorate in threshold regions.
Threshold resummation allows for taking these large logarithms into account.
For SIDIS, the resummation at NLL accuracy was formulated in \cite{Cacciari:2001cw, Anderle:2012rq, Anderle:2013lka}.
This was then extended to NNLL in \cite{Abele:2021nyo}, with the same authors providing N3LL and approximate N3LO corrections in \cite{Abele:2022wuy}.
The approximate NNLO coefficients given in \cite{Abele:2021nyo} were used for state-of-the-art fits in \cite{Borsa:2024mss, Bertone:2024taw}.
Recently, the N4LL resummation was performed and used to obtain approximate N4LO corrections to the SIDIS coefficients in \cite{Goyal:2025bzf}.

Higher-order predictions for more complex final states, such as jets or heavy quarks, but also for transverse-momentum dependent SIDIS are typically not expressed in the form of coefficient functions, but are obtained numerically through parton-level event generation. In these calculations, all real and virtual subprocess corrections to a given order are integrated over their respective phase spaces, while being subjected to the definition of the observable under consideration. This form of implementation allows to take full account of the final-state definition used in the experimental measurement, including the jet algorithm, and the fiducial cuts on object identification and isolation or on event selection. The parton-level event generator implementation requires an infrared subtraction scheme to extract and recombine infrared singular configurations between real and virtual corrections, such that only infrared-finite remainders are evaluated numerically. A variety of infrared subtraction schemes were developed at NLO~\cite{Catani:1996vz,Frixione:1995ms} and NNLO~\cite{Anastasiou:2003gr,Binoth:2004jv,Gehrmann-DeRidder:2005btv,Catani:2007vq,Czakon:2011ve,Caola:2017dug,DelDuca:2016ily,Bertolotti:2022aih}. NLO QCD calculations were accomplished for single-inclusive and di-jet production in photoproduction~\cite{Klasen:1996it,Frixione:1997ks}, and for the production of up to three jets in DIS~\cite{Nagy:2001xb}. These results are available in the form of open-source codes (DISENT~\cite{Catani:1996vz}, NLOJET~\cite{Nagy:2001xb}). Likewise, fully differential NLO QCD results are available for heavy quark production in DIS~\cite{Laenen:1992zk} and photoproduction~\cite{Frixione:1995qc,Cacciari:2001td}.

The LHC precision physics program has motivated a high degree of automation in NLO QCD calculations, leading to the development of multi-purpose event generator programs that incorporate NLO QCD corrections, which augment their predictions by matching to parton shower simulations. The leading
 codes used in LHC physics studies are
mg5$\_$aMCatNLO~\cite{Alwall:2014hca}, SHERPA~\cite{Sherpa:2024mfk}, POWHEG~\cite{Frixione:2007vw}, PYTHIA~\cite{Bierlich:2022pfr}
 and HERWIG~\cite{Bewick:2023tfi}. To adapt these programs to unpolarized electron-proton collisions requires some level of technical modification~\cite{Helenius:2024rth} which has been accomplished for several event generator codes
\cite{Hoeche:2023gme,Meinzinger:2023xuf,Helenius:2019gbd,Banfi:2023mhz,Meinzinger:2025pam}.
An extension towards polarized electron-proton collisions will be more challenging, thereby requiring the adaptation of the infrared subtraction to account for initial-state polarization~\cite{Borsa:2020yxh}.

Parton-level event generator calculations at NNLO QCD have been accomplished for a range of LHC benchmark processes~\cite{Heinrich:2020ybq}, working on a process-by-process basis. Results for processes in DIS or photoproduction can in principle be obtained with similar techniques; up to now, only di-jet production in DIS has been derived to this order~\cite{Currie:2017tpe}, and is available in the NNLOJET open-source code~\cite{NNLOJET:2025rno}. An extension to polarized collisions will require the development of an infrared subtraction technique~\cite{Gehrmann:2025xab}
for spin-polarized initial states at NNLO.

Precision high order calculations are greatly in need to solidify the theory framework to explore the nucleon tomography in terms of the transverse momentum dependent (TMD) parton distributions from the SIDIS observables~\cite{Boussarie:2023izj}. For this purpose, we need to go beyond the current state-of-the-art computations~\cite{Goyal:2023zdi,Bonino:2024qbh,Goyal:2024tmo,Bonino:2024wgg} to more differential with respect to the final state hadron's transverse momentum. Such  development is of crucial importance to understand the matching between the TMD factorization and collinear factorization in the intermediate transverse momentum region, where the current NLO calculation does not yield a satisfactory solution yet~\cite{Sun:2014dqm}, calling for the potentially important contributions from higher order perturbative QCD corrections. In addition, high transverse momentum differential cross sections for SIDIS itself would also benefit from those calculations to benchmark the precision for EIC physics in the future, see, for example, the tension between the theory and experiments on this in~\cite{Gonzalez-Hernandez:2018ipj}. Similar tension also exists for the associated Drell-Yan lepton pair production in fixed target kinematics~\cite{Bacchetta:2019tcu,Gauld:2021pkr}. It is interesting to notice that such developments have recently been carried out for finite transverse momentum SIDIS~\cite{Dong:2026eas} and low transverse momentum subtraction in SIDIS~\cite{Gao:2026tnd} both at NNLO. These advances will pave a way to consolidate the theory ground to explore nucleon TMDs through SIDIS. 

The perturbative precision predictions for inclusive cross sections, jet rates and specific hadronic final states that are 
available so far have all been derived for lepton-proton collisions. Their applicability to lepton-ion collisions must be investigated carefully in future studies. 
For the leading power contributions to cross sections, all existing perturbative precision predictions for lepton-proton collisions are valid for lepton-ion collisions when initial-state PDFs are replaced by nuclear PDFs, while extra care is needed for the size of power corrections and predictions at small-$x$ when the hard probe is no longer localized in the longitudinal direction and can interact with partons from different nucleons at the same impact parameter coherently~\cite{Eskola:2002yc,Qiu:2002mh,Qiu:2003vd}, see also Section~\ref{sec:FFs} below. The presence of the nuclear medium could have crucial implications on the formation of final state jets or on the hadronic fragmentation process 
due to coherent and incoherent multiple scatterings~\cite{Qiu:2001hj,Accardi:2012qut}.  
Early theoretical and experimental studies show rich and complex nuclear dependence for jet and hadron production in lepton-ion collisions~\cite{Luo:1994np,Guo:2000nz,Wang:2001ifa,HERMES:2007plz,HERMES:2005mar}, providing new opportunities to study QCD multiple scattering, multi-parton correlations, and mechanisms for the emergence of hadrons and jets~\cite{Accardi:2012qut}, clearly warranting further investigations.

\subsection{Resummation and improved threshold approximations 
}

The occurrence of 
threshold logarithms is a typical phenomenon when hard-scattering reactions are probed at
the boundary of phase space. They arise, for instance, when the energy available in a partonic reaction
is just sufficiently large for the observed final state to be produced. In this case, the phase space available 
for gluon bremsstrahlung vanishes, resulting in large logarithmic corrections. 
A classic example is the Drell-Yan process in the regime where $z\equiv Q^2/\hat{s}\sim 1$, with $Q$ being
the dimuon invariant mass and $\hat{s}$ the squared partonic center-of-momentum energy. Over the past
few decades, the resummation of threshold logarithms to all orders has developed into a cornerstone
of research in perturbative QCD. The importance of such resummations is two-fold.
On the one hand, because of the fact that the threshold logarithms become large in many kinematical situations 
and dominate the perturbative corrections, their resummation is often relevant for phenomenology,
all the way from fixed-target scattering to the LHC. At the same time, resummation offers insights
into the structure of perturbative corrections at higher orders, which among other things may 
yield benchmarks for explicit full fixed-order calculations in QCD or, alternatively,
provide approximations for higher-order corrections in cases where full calculations are not yet available. 

Taking the Drell-Yan cross section as an example, 
the leading large contributions near threshold arise as $\alpha_s^k\left[ \ln^{2k-1}(1-z)/
(1-z)\right]_+$ at the $k$th order in perturbation theory, where $\alpha_s$ is the strong 
coupling. There is a double-logarithmic structure, with two powers of the logarithm arising
for every new order in the coupling. Subleading terms have fewer logarithms, so that
the threshold logarithms in the perturbative series take the general form 
\begin{equation}\label{series1}
\hspace*{2.4cm}\sum_{k=0}^\infty\sum_{\ell=1}^{2k} \alpha_s^k \,{\cal A}_{k,\ell} \, \left(\frac{ \ln^{2k-\ell}
(1-z)}{1-z}\right)_+\,,
\end{equation}
with perturbative coefficients ${\cal A}_{k,\ell}$. One often refers to the all-order set of logarithms 
with a fixed $\ell$ as the $\ell$th {\it tower} of logarithms. Threshold logarithms exponentiate after taking an
integral transform conjugate to the relevant kinematical variable ($z$ in the above
example). Under this transform the threshold logarithms translate into logarithms of the 
transform variable $N$. The exponent may itself be written as a perturbative series and 
is only {\it single-logarithmic} in the transform variable. Thanks to this, knowledge of the two leading 
towers $\alpha_s^k \ln^{k+1}(N)$ and $\alpha_s^k \ln^k(N)$ in the exponent, along with the full virtual corrections
at order $\alpha_s^k$, is sufficient to predict the three
leading towers in the perturbative series~(\ref{series1}) for the cross section in $z$-space. 
This is termed ``next-to-leading logarithmic'' (NLL) resummation. At full next-to-next-to-leading logarithmic 
(NNLL) accuracy, one needs three towers in the exponent and the two-loop hard coefficients,
already providing control of five towers in the partonic cross section. 
 
While NLL resummation was the state of the art for many years, much progress has been made
on extending the framework to NNLL accuracy, or even beyond. The most advanced results have been
obtained for color-singlet processes, with the Higgs production cross section being a particularly 
prominent example~\cite{Catani:2003zt,Ahrens:2009cxz,Bonvini:2014joa,Catani:2014uta}.
There have also been applications to processes that are not characterized
by a color-singlet LO hard scattering reaction. Here the resummation framework becomes more complex because the interference
between soft emissions by the various external partons in the hard scattering process becomes sensitive to 
the color structure of the hard scattering itself, requiring a color basis for the partonic 
scattering process that leads to a matrix structure of the soft 
emission~\cite{Kidonakis:1997gm,Bonciani:2003nt}. An extensive list of 
color-non-singlet reactions of this type along with corresponding references to NLL studies may be 
found in~\cite{Catani:2013vaa}. Resummation studies beyond NLL have been presented in the context 
of top quark (pair) production~\cite{Beneke:2011mq,Czakon:2009zw,Yang:2014hya,Kidonakis:2010tc}, 
for single-inclusive photon or hadron production~\cite{Becher:2009th,Ferroglia:2013awa,Catani:2013vaa,Hinderer:2018nkb}, 
and for squark and gluino production~\cite{Beneke:2010da,Beenakker:2013mva}.

In terms of resummation for EIC observables, the main focus in recent years has been on SIDIS, primarily with the goal of obtaining
approximate fixed-order (NNLO and beyond) results, by expanding resummed results. Threshold resummation for SIDIS can be derived from that of the Drell-Yan rapidity distribution, which is related to it by crossing~\cite{Sterman:2006hu}, with the Bjorken variable $x$ and fragmentation variable $z$ of SIDIS mapped onto the momentum fractions $x_1$ and $x_2$ of the two incoming partons of the Drell-Yan process. The resummation can be performed in the absolute threshold (or double-soft) limit, in which both $z\to1$ and $x\to1$, and all extra radiation is soft, or in the single soft limit, in which either $z\to1$ at fixed $x$ or $x\to1$ at fixed $z$, and extra radiation is respectively collinear to either the incoming or the outgoing parton. The next-to-leading power correction to double-soft resummation can be viewed as the truncation to first order of the  power expansion of single-soft resummation  in powers of the non-soft variable,  either $1-x$ or $1-z$. 

Resummation to NLL for Drell-Yan in the double-soft limit was performed in~\cite{Catani:1989ne}, related to SIDIS in~\cite{Sterman:2006hu}, and explicitly carried out for SIDIS
in~\cite{Cacciari:2001cw,Anderle:2012rq,Anderle:2013lka}, with later extensions to NNLL~\cite{Abele:2021nyo}, N$^3$LL and next-to-leading power~\cite{Abele:2022wuy},
and even N$^4$LL~\cite{Goyal:2025bzf}. Resummation in the single soft limit was performed for Drell-Yan in~\cite{Lustermans:2019cau,Mistlberger:2025lee,DeRos:2026bcv}, and for SIDIS in~\cite{Forte:2026tva}, thereby generalizing the next-to-leading power result of~\cite{Abele:2022wuy}.

\begin{figure}
    \includegraphics[angle=270,width=0.49\linewidth]{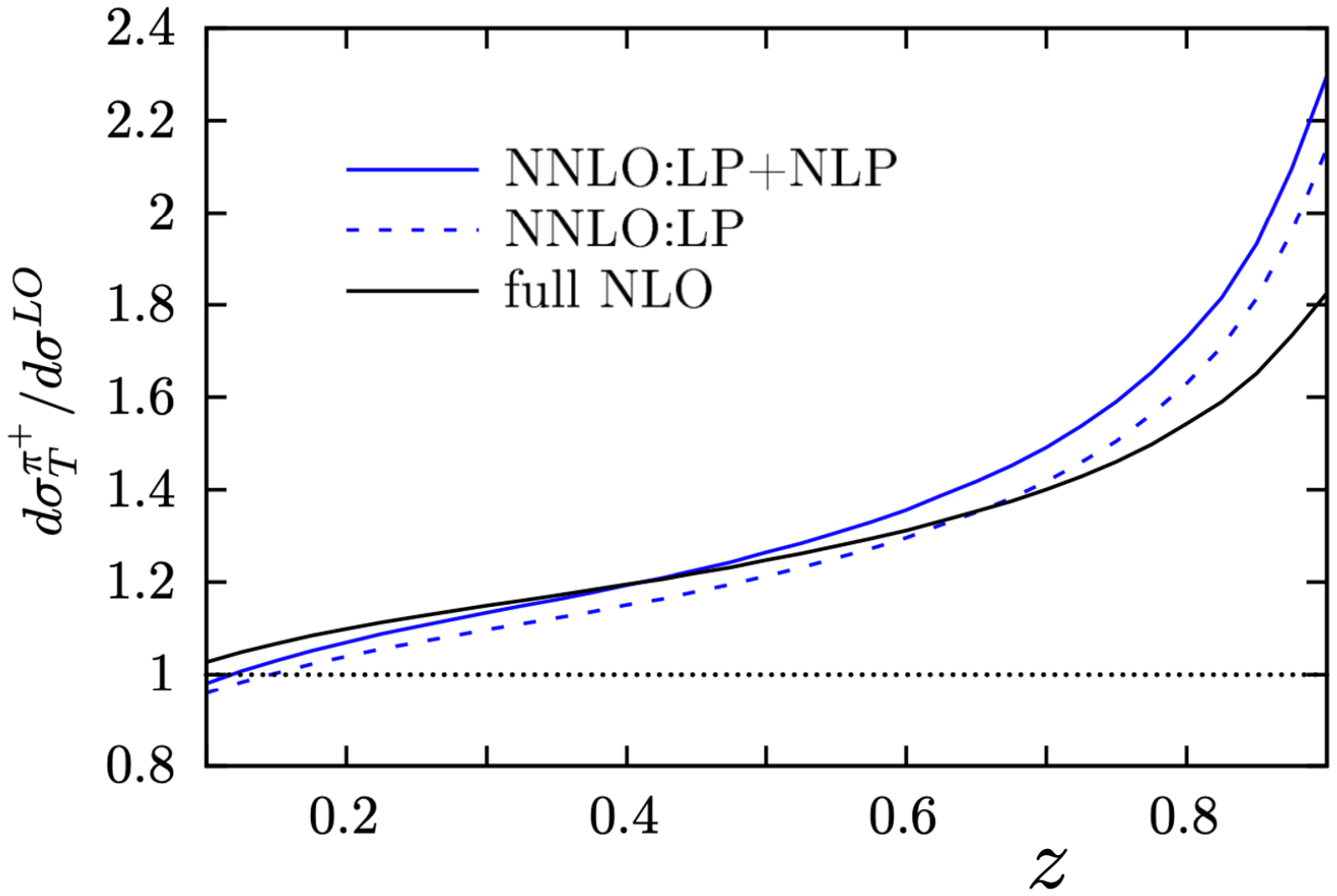}\\[-7.5cm]
    
    \hspace*{8cm}
    \includegraphics[width=0.49\linewidth]{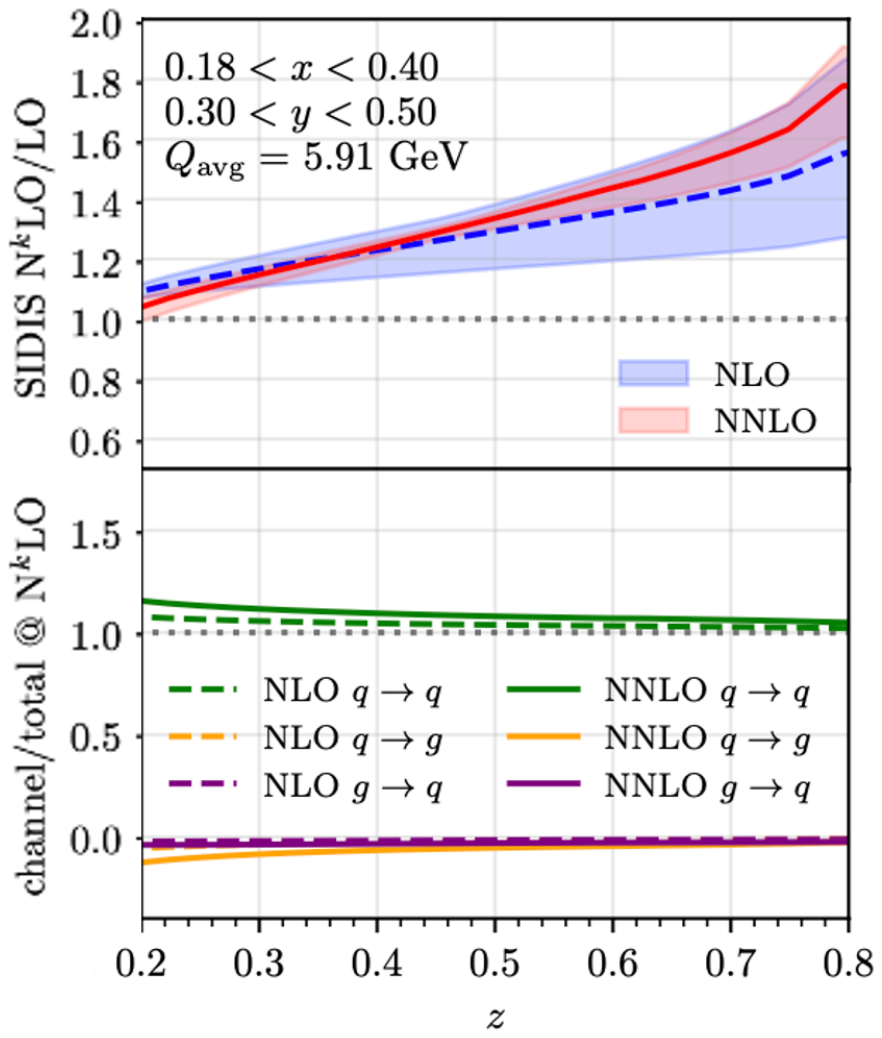}\\[-2.5cm]
    \caption{{\it Left}: Approximate NLO,NNLO $K$-factors for SIDIS of~\cite{Abele:2021nyo} for typical COMPASS kinematics. 
    {\it Right:} Full NNLO results of Ref.~\cite{Bonino:2024qbh}.}
    \label{fig:sidis}
\end{figure}

The NNLL double-soft results were  used in~\cite{Abele:2021nyo} to derive
approximations to the full NNLO corrections, subsequently used in determinations of NNLO fragmentation functions~\cite{Borsa:2022vvp}
and helicity parton distributions~\cite{Borsa:2024mss,Bertone:2024taw}. The approximations were
superseded by the full NNLO derivations for SIDIS in Refs.~\cite{Goyal:2023zdi,Bonino:2024qbh,Goyal:2024tmo,Bonino:2024wgg,Bonino:2025bqa,Goyal:2024emo,Bonino:2025qta}.
The left plot in Fig.~\ref{fig:sidis} shows the NLO and approximate NNLO ``$K$-factors'' for unpolarized SIDIS (using kinematics
typical for the COMPASS experiment) obtained in~\cite{Abele:2021nyo}, as functions of the fragmentation variable $z$
and for fixed bins in Bjorken-$x$ $(0.2<x<0.8)$ and $0.2<y<0.9$. The picture on the right shows the full NLO and NNLO results of~\cite{Bonino:2024qbh}. 
Although the two plots slightly differ in the kinematics chosen, one can see that in this case the approximate NNLO traces the full result
quite well. That said, the approximations to fixed-order results obtained from threshold resummation deteriorate the further one
goes away from the partonic threshold. This is true even for SIDIS when $x$ or $z$ are not both close to unity. 
Quite generally, it is important to validate the approximations by assessing their quality
at least via comparisons at NLO level and, even better, at NNLO if available. Furthermore, the study of subleading-power corrections 
near threshold becomes important~\cite{Bonocore:2015esa,vanBijleveld:2023vck,Ebert:2017uel}. 

In a similar context, as mentioned earlier, NNLO threshold approximations are also used for spin-dependent hard scattering reactions, in order
to extract helicity parton distributions~\cite{Borsa:2024mss,Bertone:2024taw}. Here the focus is on polarized
$pp\to \textrm{jet}+X$ or $pp\to h+X$ at RHIC, for which full NNLO corrections are not yet available.
Due to the fact that helicity PDFs and polarized partonic cross sections can change signs in the regimes
where they are probed, one finds that the threshold approximation typically is less accurate in the polarized
case. The study of corrections beyond the standard threshold logarithms is again useful here. We
foresee many interesting and phenomenologically relevant applications for the EIC here, 
among them in particular the inclusion of threshold resummation in predictions for high-$p_T$ cross sections~\cite{deFlorian:2013taa,Uebler:2015ria} and in TMD observables.

\subsection{The strong coupling constant 
}
The accurate determination of the strong coupling constant $\alpha_s$ is a major bottleneck in precision physics at present and future colliders. For instance, the yellow report 4 of the Higgs cross section working group~\cite{LHCHiggsCrossSectionWorkingGroup:2016ypw}(now Higgs working group) cited as a dominant uncertainty on the cross-section for  Higgs production in gluon fusion the PDF+$\alpha_s$ uncertainty.  This cross section starts at order $\alpha_s^2$, with the next order correction as large as the lowest order, and the subsequent order only half as large, so a $1\%$ uncertainty on $\alpha_s$ results in an uncertainty on the cross section as large as $3\%$. Indeed, in Ref.~\cite{LHCHiggsCrossSectionWorkingGroup:2016ypw}, the final uncertainty on the cross section was quoted as  $3.9\%$, and the uncertainty on $\alpha_s(M_Z)$ quoted by the then-current (2016) edition of the PDG~\cite{ParticleDataGroup:2016lqr} was  $\pm0.0011$ corresponding to $0.9\%$. 
Now, almost ten years later, all  other sources of uncertainty on this cross section have been either completely removed (such as the top--bottom interference, now known exactly) or reduced, often to the point of becoming completely negligible (such as higher order QCD corrections, now computed to NNLO with full top mass dependence). On the other hand,
in the current (2025) edition of the PDG~\cite{ParticleDataGroup:2024cfk} the uncertainty on $\alpha_s(M_Z)$ is quoted as $\pm0.0009$ corresponding to $0.8\%$, so essentially unchanged.

This example illustrates the increasing importance of the determination of $\alpha_s$ for precision physics. Here we will first briefly review the global determinations of $\alpha_s$, then turn to the current status and prospects at the EIC, and finally discuss several subtleties that arise when attempting the determination of the strong coupling to sub-percent accuracy.

\subsubsection{Global averages of $\alpha_s$.}

\begin{figure}
    \centering\vspace*{-4cm}
    \includegraphics[width=\linewidth]{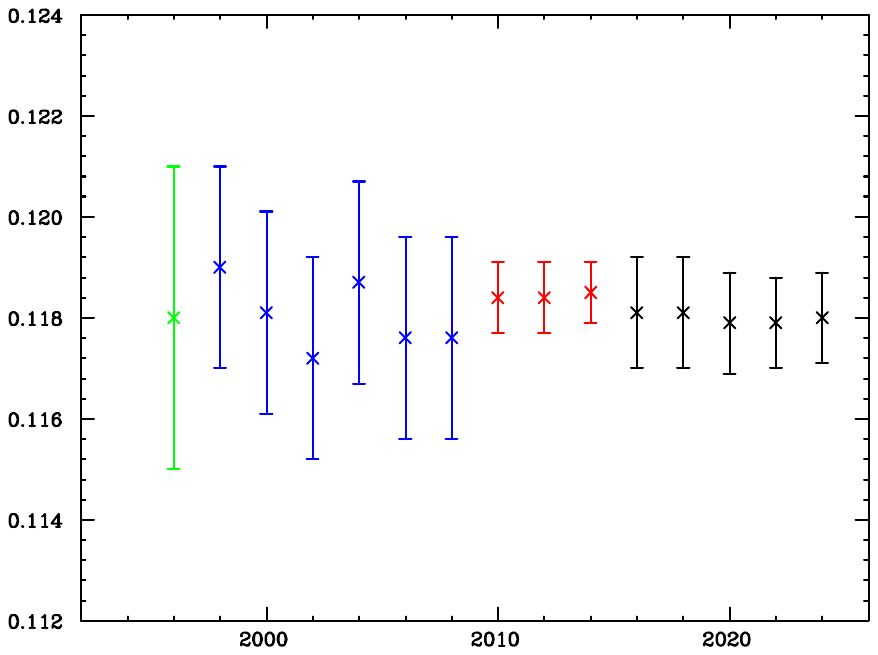} \vspace*{-6cm}
    \caption{The global determination of the value of the strong coupling $\alpha_s(M_Z)$ published by the PDG, as a function of time. Bands correspond to one-sigma uncertainties.}\label{fig:alphas_PDG}
\end{figure}

The strong coupling can be determined from a variety of different physics processes, as well as on the lattice, and accordingly a number of combined global determinations have been published over the years. The combined value and uncertainty on $\alpha_s(M_Z)$ determined by the PDG over the last 30 years are collected and displayed in Fig.~\ref{fig:alphas_PDG}. Other combined determinations have been performed and widely cited, see in particular Refs.~\cite{Bethke:2000ai,Bethke:2006ac,Bethke:2009jm}.

These combinations involve a certain degree of arbitrariness. For instance, the widely cited 2006 combination~\cite{Bethke:2006ac} quoted a value $\alpha_s(M_Z)=0.1189\pm0.0010$, while the contemporary 2006 edition of the PDG~\cite{ParticleDataGroup:2006fqo} reported $\alpha_s(M_Z)=0.1176\pm0.002$: the two uncertainties differ by a factor two, and the central values by about one sigma. These differences depend on subjective choices, such as in choosing  how to combine incompatible individual determinations. Figure~\ref{fig:alphas_PDG} provides an example of this: there is a clear discontinuity between the the 2008 and the 2010 edition, with specifically a very significant change in uncertainty. It turns out that the authors of the PDG combination had changed between these two editions, and that the 2010 edition~\cite{ParticleDataGroup:2010dbb} adopted the value published in Ref.~\cite{Bethke:2009jm}. On the other hand, a further change in uncertainty is apparent between the 2014 and the 2016 determinations.  In this case,  the authors are unchanged, and they trace the change in value mostly to a different estimate of the uncertainty on the lattice determination: it  is somewhat troubling that the uncertainty increases as a function of time.

\begin{figure}
    \centering
    \includegraphics[width=\linewidth]{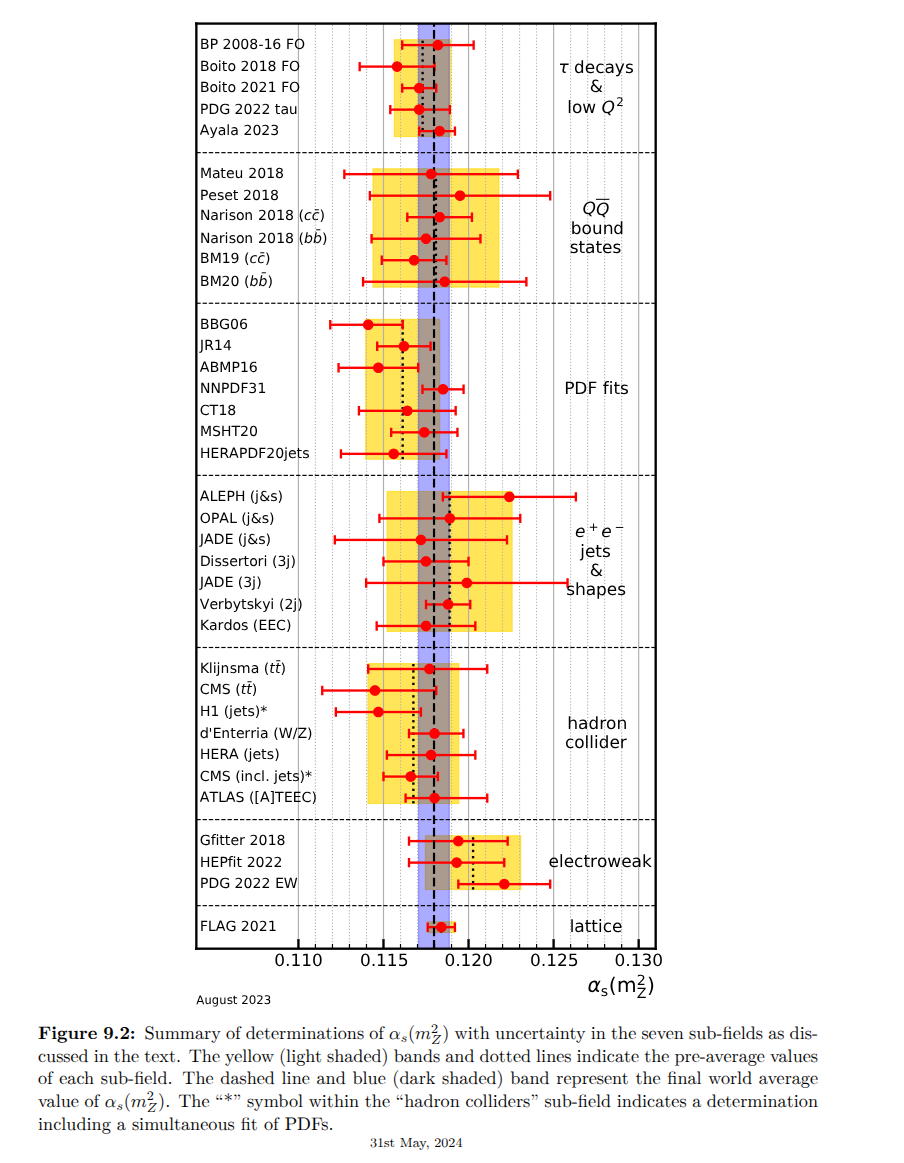}
    \caption{Summary of  the determinations of $\alpha_s$: taken from Ref.~\cite{ParticleDataGroup:2024cfk}.}\label{fig:alphas_PDG24}
\end{figure}
A significant issue in performing a combination can be understood by inspecting the current PDG determination: see Fig.~\ref{fig:alphas_PDG24}. The final value is obtained as an average of sub-averages, obtained by combining  determinations from similar processes. This procedure leads to a final value that may depend  on the way the categorization is performed. This involves some arbitrariness not only in choosing the different categories, but even in deciding which category each determination ought to be assigned to. For example, the "hadron collider" category includes several determinations (denoted by an asterisk) in which parton distributions are simultaneously determined. Clearly, these could also be included in the "PDF fits"  category, and indeed this is done for the determination from HERA jets. Other relevant issues in performing a global determination are, as mentioned, the way to combine incompatible results, and also how to treat situations in which there are reasons to believe that uncertainties may have been underestimated. In fact, it has been argued~\cite{Altarelli:2013bpa} that a reliable global average should only include a small number of very solid determinations, with other determinations only used as cross-checks.

Achieving the stated goal of arriving in the future at high-precision measurements with an uncertainty as low as 0.1\%~\cite{dEnterria:2022hzv} will only make issues of mutual compatibility and independence of results more serious. Section~\ref{sec:AlphasPitfalls} below addresses several issues that may affect this progress towards the increased precision.

\subsubsection{Current status and future developments at the EIC.} 

At the EIC, the strong coupling will be most accurately measured using deep-inelastic scattering (DIS) data. We briefly review the determination of $\alpha_s$ from DIS, then summarize recent determinations together with PDFs and on the lattice, and finally discuss future prospects at the EIC.

\paragraph{Deep-inelastic scattering.}
Deep‑inelastic scattering has historically been one of the most important sources of information on the strong coupling $\alpha_s$, because it directly probes the scaling violations predicted by perturbative QCD. In DIS, a lepton scatters off a nucleon, and the resulting structure functions depend on the resolution scale $Q^2$. QCD predicts how these structure functions evolve with $Q^2$ through the standard collinear evolution equations, and the rate of this $Q^2$ evolution is sensitive to $\alpha_s$.
The basic strategy for determining $\alpha_s$ is to compare high‑precision measurements of structure functions — mainly $F_2$, and in some cases $F_L$
and $xF_3$ — over a wide range in Bjorken $x$ and $Q^2$ with perturbative QCD calculations at NNLO or even approximate N$^3$LO, combined with global parton distribution function (PDF) fits.
In practice, the structure functions enter via the reduced cross section, which is what the fits actually use.

The key experimental inputs come from fixed‑target experiments (notably SLAC~\cite{Whitlow:1990gk}, BCDMS~\cite{BCDMS:1989eab} and NMC~\cite{NewMuon:1996fwh}), from the HERA collider experiments H1 and ZEUS~\cite{H1:2015ubc}, and from neutrino DIS measurements. HERA data are especially valuable because they cover a very large kinematic range with excellent precision, allowing a more reliable disentangling of perturbative scaling violations from non‑perturbative effects.

The dominant sources of uncertainty in $\alpha_s$ extracted from DIS include: 
\begin{itemize}
\item Experimental uncertainties in the structure function measurements  
\item Uncertainties in the PDFs, which are fitted simultaneously with $\alpha_s$ 
\item Theoretical choices, such as the heavy‑quark scheme and perturbative order  
\item Possible higher‑twist contributions at low $Q^2$ and large $x$  
\item Uncertainties from missing higher‑order QCD corrections, which are estimated through scale variations and comparisons between NLO, NNLO, and approximate N$^3$LO results.
\end{itemize}

Modern analyses address these issues by emphasizing high‑$Q^2$ data, using global PDF fits at consistent perturbative orders, and applying careful treatments of correlated systematic uncertainties. The impact of missing higher‑order contributions is reduced at NNLO and further at N$^3$LO, but remains one of the relevant theoretical systematics, especially at low $x$ and in regions where higher‑order threshold logarithms or small $x$ resummation effects may become important.

DIS determinations of $\alpha_s$ are generally consistent with the world average, though some datasets show mild tensions. Historically, certain DIS data have preferred slightly lower values than the PDG average and this trend persists in recent determinations~\cite{H1:2017bml,Alekhin:2024bhs,Alekhin:2025qdj,Ablat:2025gbp}, though as global PDF analyses have improved, the spread among different determinations has decreased\cite{H1:2017bml,Alekhin:2024bhs,Alekhin:2025qdj,Ablat:2025gbp}. Nevertheless, DIS remains central because it is sensitive to both the shape and the $Q^2$ dependence of the structure functions.

\begin{table}[h]
\centering
\begin{tabular}{l l}
\hline
PDF set & strong coupling  \\
\hline
ABMP16 NLO~\cite{Alekhin:2018pai}     & $\alpha_s(M_Z) =  0.1191 \pm 0.0011$ \\
ABMP16 NNLO~\cite{Alekhin:2017kpj}    & $\alpha_s(M_Z) =  0.1147 \pm 0.0008$ \\
ABMPtt NNLO~\cite{Alekhin:2024bhs}    & $\alpha_s(M_Z) =  0.1150 \pm 0.0009$ \\
ABMZ   NNLO~\cite{Alekhin:2025qdj}    & $\alpha_s(M_Z) =  0.1152 \pm 0.0008$ \\
CT25   NNLO~\cite{Ablat:2025gbp}     & $\alpha_s(M_Z) =  0.1183 + 0.0023 - 0.0020$ \\
MSHT20 NLO~\cite{Cridge:2021qfd}      & $\alpha_s(M_Z) =  0.1203 \pm 0.0015$ \\
MSHT20 NNLO~\cite{Cridge:2021qfd}     & $\alpha_s(M_Z) =  0.1174 \pm 0.0013$ \\
MSHT24 NNLO~\cite{Cridge:2021qfd}     & $\alpha_s(M_Z) =  0.1171 \pm 0.0014$ \\
MSHT24 aN$^3$LO~\cite{Cridge:2021qfd} & $\alpha_s(M_Z) =  0.1170 \pm 0.0016$ \\
NNPDF4.0 aN$^3$LO~\cite{Ball:2025xgq}  & $\alpha_s(M_Z) =  0.1194 + 0.0007 - 0.0014$ \\
\hline
\end{tabular}
\caption{Recent values of $\alpha_s(M_Z)$ in the $\overline{\textrm{MS}}$ scheme for $n_f=5$ quark flavors extracted in fits with various PDF sets.}
\label{tab:alphas_PDFs}
\end{table}

\paragraph{Determinations jointly with PDFs.} 

In several recent determinations, DIS data are combined with datasets of varying size in order to arrive at  a simultaneous determination together with the PDFs. These include the ABMP16 extraction of $\alpha_s(M_Z)$ at NLO~\cite{Alekhin:2018pai} and NNLO~\cite{Alekhin:2017kpj}, with an updated NNLO determination from the ABMPtt analysis~\cite{Alekhin:2024bhs}. 
Other recent studies examine $\alpha_s(M_Z)$ and its dependence on heavy‑quark masses in MSHT20~\cite{Cridge:2021qfd}, as well as approximate N$^3$LO extractions by MSHT~\cite{Cridge:2024exf} and NNPDF~\cite{Ball:2025xgq}, the latter also incorporating NLO QED effects. Increased cuts on $Q^2$ and the hadronic invariant mass $W^2$, together with setting higher‑twist terms to zero, have recently been used in a determination of $\alpha_s$ from high‑energy data~\cite{Alekhin:2025qdj}, showing consistency with fits that explicitly include higher‑twist contributions. A new  determination has also been published by the CTEQ‑TEA group~\cite{Ablat:2025gbp}. Some of these results show noticeable tensions with one another, which motivates continued detailed study.
According to the 2025 PDG~\cite{ParticleDataGroup:2024cfk}, the world average value of the strong coupling
(excluding lattice determinations) is
\begin{eqnarray}
\hspace*{2.8cm}
\label{eq:as-pdg2025}
  \alpha_s(M_Z)&=&0.1175 \pm 0.0010\, .
\end{eqnarray}
This result, like all values of $\alpha_s(M_Z)$ quoted in this section,
is in the $\overline{\textrm{MS}}$ scheme with $n_f=5$ quark flavors.

\paragraph{Determinations on the lattice.}

Lattice QCD determines the strong coupling constant $\alpha_s(\mu) = \bar{g}_s^2(\mu)/(4\pi)$ by computing a short-distance, Euclidean quantity $Q(\mu)$ on the lattice and matching it to its perturbative expansion in the $\overline{\textrm{MS}}$ scheme:
\begin{equation}
\hspace*{1cm}
Q(\mu) = c_1\,\alpha_{\overline{\mathrm{MS}}}(\mu)
       + c_2\,\alpha_{\overline{\mathrm{MS}}}^2(\mu)
       + c_3\,\alpha_{\overline{\mathrm{MS}}}^3(\mu) + \cdots
\end{equation}
The overall energy scale and quark masses are fixed from hadronic observables such as meson masses or decay constants. 
The coupling is then evolved to high scales using the nonperturbative $\beta$-function and matched perturbatively across flavor thresholds to obtain the conventional quantity $\alpha_s(M_Z)$ in the  $\overline{\textrm{MS}}$ scheme with $n_f=5$ quark flavors.
Several complementary lattice schemes are used to extract the running of the strong coupling.
Step-scaling techniques, such as the Schrödinger Functional or Gradient-Flow schemes, trace the scale dependence of the coupling in finite volume with $\mu = 1/L$.
Other approaches rely on short-distance observables, like the static-quark potential, current-current correlators, or Wilson loops, directly matched to continuum perturbation theory.  
In addition, decoupling strategies connect theories with different numbers of active quark flavors ($N_f$) through nonperturbative matching relations.

The uncertainties in lattice determinations of the strong coupling arise from several sources. 
Truncation of perturbative expansions and the limited knowledge of higher-order terms in the $\overline{\textrm{MS}}$ matching introduce the largest systematic errors, usually estimated through scale variations. 
Discretization effects, proportional to powers of $a\mu$, require controlled continuum extrapolations using multiple lattice spacings and $\mathcal{O}(a)$ improvement. 
Finite-volume corrections can appear when the lattice extent $L$ is too small, while topology freezing at very fine lattice spacings ($a \lesssim 0.05\textrm{fm}$) may bias short-distance observables.
The conversion from lattice to physical units through scales such as $r_0$, $t_0$, or $w_0$ adds only a minor ($\sim0.5\%$) uncertainty. 
Finally, at lower energy scales nonperturbative contributions, suppressed as $\mathrm{e}^{-\gamma/\alpha_s}$ or $1/\mu^2$, can distort the perturbative running when $\alpha_s$ becomes large.

Modern lattice determinations with $N_f=2+1$ or $2+1+1$ flavors yield
\begin{eqnarray}
\label{eq:as-flag}
\hspace*{2.8cm}
  \alpha_s(M_Z)&=&0.1185 \pm 0.0008\, ,
\end{eqnarray}
in excellent agreement with phenomenological averages, cf. Eq.~(\ref{eq:as-pdg2025}). 
The lattice precision is primarily limited by perturbative truncation and continuum-extrapolation uncertainties.
For a more detailed summary and references, we refer readers to~\cite{FlavourLatticeAveragingGroupFLAG:2024oxs}. 

\paragraph{Future prospects and the EIC.}

Lattice QCD now provides a very precise value of $\alpha_s$ but cannot simply replace phenomenological extractions: fixing $\alpha_s(M_Z)$ to the lattice value would still require refitting the PDFs, and independent determinations from DIS and other processes remain important consistency checks of QCD.
The EIC will greatly sharpen $\alpha_s$ extractions by providing high‑precision DIS data over a wide kinematic range, especially at moderate to large $x$ where current uncertainties are significant. Control of its systematics beyond the current state of the art, precise measurements of structure-function scaling violations over a theoretically clean $(x,Q)$ region, and cross checks with polarized beams will reduce both experimental and PDF‑related uncertainties, enabling a more accurate and internally consistent determination of $\alpha_s(M_Z)$ that can help resolve existing tensions among $\alpha_s$+PDF fits.  On the theory side, a more detailed understanding of electroweak corrections is likely to be required. Currently, DIS data  are published after subtracting the leading electromagnetic corrections~\cite{Mo:1968cg,Bardin:1988by,Bardin:1989vz,Kripfganz:1987yu}. However a  more accurate treatment of  QED$\times$QCD corrections is likely to be necessary at the EIC \cite{Cammarota:2025jyr}, as discussed in more detail in Section~\ref{subsec:qed} below. 

  \subsubsection{Pitfalls and subtleties.}
  \label{sec:AlphasPitfalls}
In view of the  precision on the determinations of $\alpha_s$ that the EIC aims to achieve, several subtleties become relevant: these are sometimes overlooked, but they become relevant once a determination with sub-percent accuracy is attempted. We briefly discuss  a few of them, some of which have been known for a long time, while others have been identified more recently.
\begin{figure}
    \includegraphics[width=\linewidth]{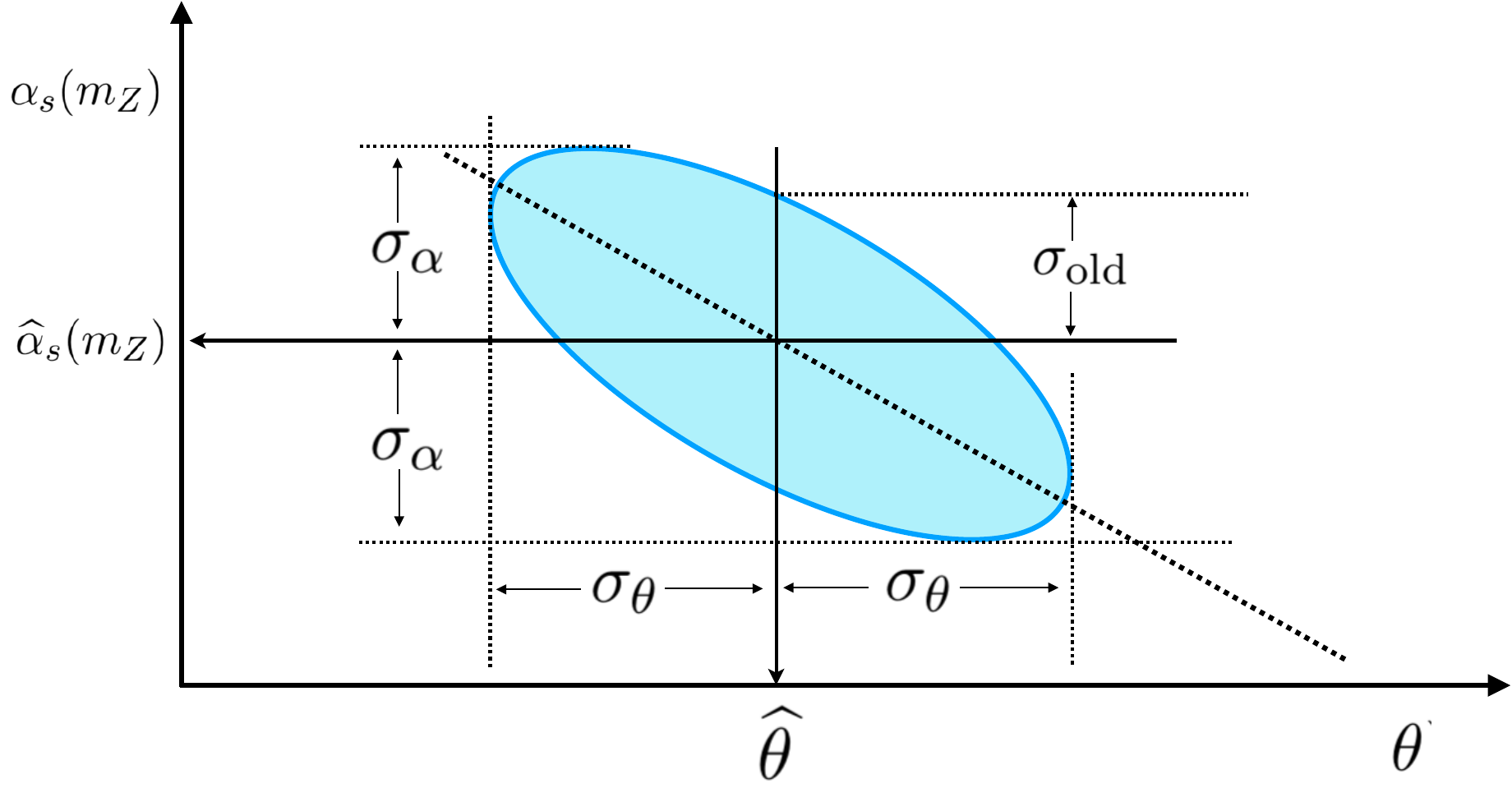} 
    \caption{A typical likelihood profile in $\alpha_s$-PDF space. The multidimensional PDF space is schematically represented as a single parameter $\hat\theta$; $\sigma_\alpha$ and $\sigma_\theta$ are  the one-sigma uncertainties on the PDF and $\alpha_s$ respectively, while $\sigma_{\rm old}$ denotes the value of the $\alpha_s$ uncertainty that is obtained by not determining simultaneously $\alpha_s$ and the PDFs.  (From Ref.~\cite{Ball:2018iqk })} \label{fig:alphas_PDF_ellipse}
\end{figure}

\paragraph{Correlation of $\alpha_s$ and PDFs.}
A first point of attention is related to the determination of $\alpha_s$ from processes that involve hadrons in the initial state and thus require the use of parton distribution functions (PDFs) as input. Several existing determinations of this kind are performed using pre-existing PDF sets, as opposed to simultaneously determining the PDFs and the strong coupling: indeed all the determinations shown in Fig.~\ref{fig:alphas_PDG24} in the "hadron collider" subcategory without an asterisk are performed in this way. In these determinations, a likelihood profile is extracted from the data as a function of $\alpha_s$, using as input for each value of $\alpha_s$ a PDF set previously determined as a best fit corresponding to that value. This is problematic for two independent potential sources of bias, one of which generally leads to a underestimate of the uncertainty and the other of which generally leads to a shift of the central value.  

The first source of bias is related to the fact that the PDFs and the strong coupling are correlated. As a consequence, a likelihood profile in $\alpha_s$-PDF space is generally an ellipsoid, whose axes are not aligned with the $\alpha_s$ direction (see Fig.~\ref{fig:alphas_PDF_ellipse}). If PDFs are not determined together with $\alpha_s$, the value of $\alpha_s$ and its uncertainty are determined from the likelihood profile that is found moving along the best-fit PDF line, which corresponds to the major axis of the ellipse in the schematic depiction of Fig.~\ref{fig:alphas_PDF_ellipse}, in which the (generally high-dimensional) PDF space is schematically represented as one-dimensional. This uncertainty is denoted as $\sigma_{\rm old}$ in the Figure, and it is generally smaller than the uncertainty determined instead from the full profile in $\alpha_s$-PDF space. For instance, the simultaneous determination of $\alpha_s$ and the PDFs in~\cite{Ball:2025xgq} explicitly verified that the uncertainty on $\alpha_s(M_Z)$ from the full likelihood profile is $\pm0.007$, but it becomes
$\pm0.005$ if moving along the best-fit line.

\begin{figure}
    \centering
    \includegraphics[width=\linewidth]{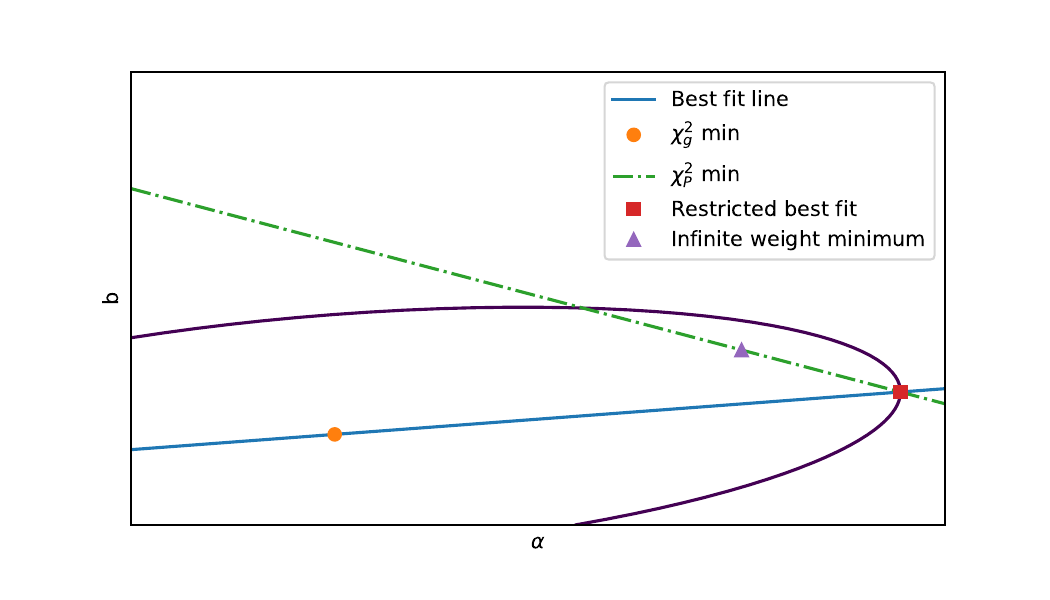} 
    \caption{Schematic representation of likelihood profiles in  $\alpha_s$-PDF space. The multidimensional PDF space is schematically represented as a single parameter $b$. The ellipse is the likelihood profile for a joint PDF-$\alpha_s$ determination, and the dot-dashed line is the maximum-likelihood curve for an experiment (approximated by a straight line in a small enough neighborhood) that does not fully determine the PDF (see text). From Ref.~\cite{Forte:2020pyp}}\label{fig:alphas_PDF_runaway}
\end{figure}
The second source of bias has a more subtle origin. It becomes relevant when a determination of $\alpha_s$ from an individual dataset is performed, using pre-existing PDFs. It is due to the fact~\cite{Forte:2020pyp} that in a combined PDF-$\alpha_s$ determination the value of $\alpha_s$ corresponding to the maximum likelihood along the likelihood profile for a single dataset using the global maximum-likelihood PDFs does not correspond to the maximum likelihood   $\alpha_s$ value for either that dataset or the global dataset. In order to understand this somewhat disconcerting fact, consider the situation depicted in Fig.~\ref{fig:alphas_PDF_runaway}, in which again schematically the PDF space is represented as one-dimensional. The ellipse shown in the figure corresponds to a fixed likelihood profile in joint $\alpha_s$-PDF space,  computed using some global dataset, and one axis of the ellipse  corresponds to the best-fit existing PDF for each $\alpha_s$ value. The dot-dashed line corresponds to a maximum-likelihood curve for an individual experiment, which may either be part of the global dataset, or provide a new dataset. It is a curve, rather than a point, because typically an individual experiment does not fully determine $\alpha_s$ and the PDFs, 
and it is approximated as a straight line in a small enough neighborhood. 

The maximum likelihood value of $\alpha_s$ from the new experiment found using a pre-existing best-fit PDFs must be along the best-fit curve for the new experiment, and also along the best-fit existing PDF line, hence it is the intersection of the solid and dot-dashed curves. It is obvious from inspection of the figure that all points along the dot-dashed line that are inside the ellipse provide a better fit to the PDFs, while providing an equally good fit to the new dataset. In fact one can reasonably argue that the value of $\alpha_s$ selected by the new dataset is the point along the dot-dashed line (all of which are equally good fits to the new datsets) that is maximum likelihood for the PDFs. This is denoted as a triangle in the figure, and it is in fact closer to the value of $\alpha_s$ that is maximum likelihood for the PDF-$\alpha_s$ fit, namely the center of the ellipse. This means that if the joint PDF-$\alpha_s$ determination is repeated with the new dataset included with a large weight, then the value of $\alpha_s$ denoted by the triangle will be found, highlighting the fact that this is the value actually preferred by the new dataset.

This situation is entirely generic, and it also holds when the new dataset does determine fully the PDFs~\cite{Forte:2020pyp}: it will happen whenever the value of $\alpha_s$ preferred by the new dataset does not coincide with that of the global fit. Clearly a consequence is that values of $\alpha_s$ determined from a new dataset using existing PDFs will lead to values that show an inflated deviation from the global best-fit, which in turn leads to instabilities when performing a global combination. This was recently~\cite{Forte:2025pvf} demonstrated explicitly in a closure test.

\paragraph{The treatment of multiplicative uncertainties.}
Two more sources of bias are related to the treatment of multiplicative systematic uncertainties, namely uncertainties that are proportional to the value of the quantity that is being measured, such as the luminosity uncertainty. First of all, it should be noted that it is a common practice for experiments to publish all systematic uncertainties in a  multiplicative form.
Hence it is important to distinguish uncertainties that are genuinely multiplicative, 
from those whose value is independent of it and that are simply presented as a percentage.

Genuinely multiplicative uncertainties may lead to  the so-called D'Agostini bias, due to the fact~\cite{DAgostini:1993arp}  that the maximum-likelihood determination of any parameter leads to a biased result if the likelihood is computed using the experimental covariance matrix, in which multiplicative uncertainties are normalized by the central value of the  measured quantity (experimental normalization, henceforth). A way out is given by the so called $t_0$ method, in which the value of multiplicative uncertainties is instead normalized by a fixed theory prediction~\cite{Ball:2009qv}, or the related $t$ method~\cite{Ball:2012wy,Gao:2013xoa}. 

\begin{figure}[t]
    \centering
    \includegraphics[width=.45\linewidth]{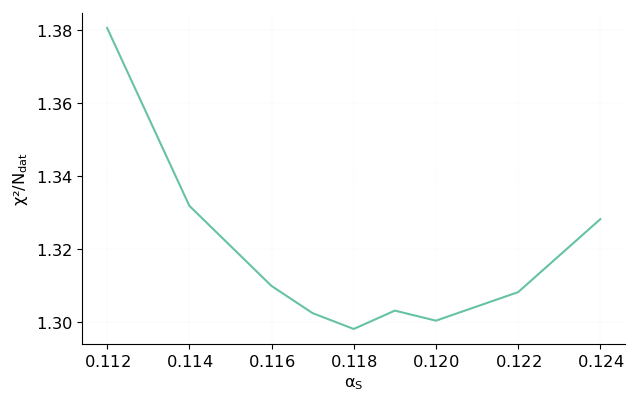} 
     \includegraphics[width=.45\linewidth]{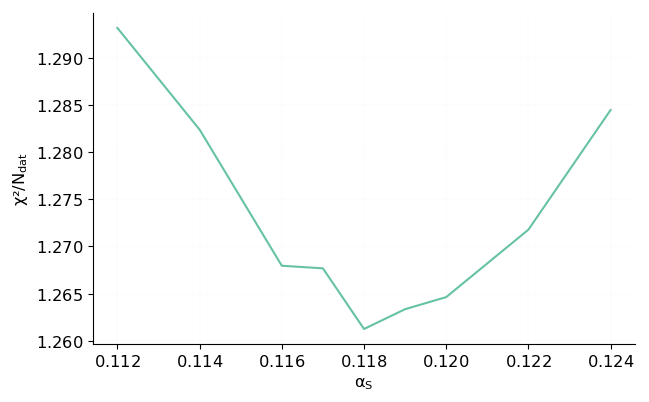}\\ 
     \includegraphics[width=.45\linewidth]{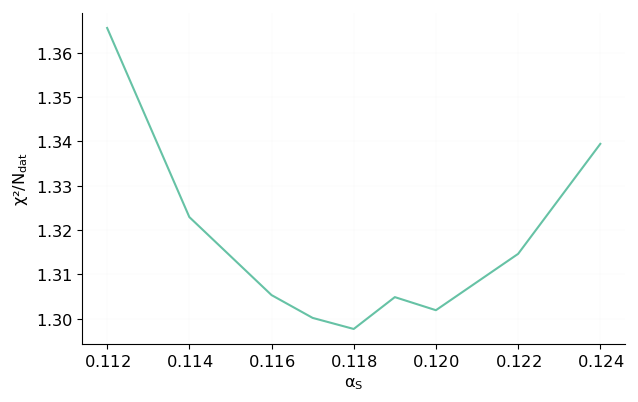} 
     \includegraphics[width=.45\linewidth]{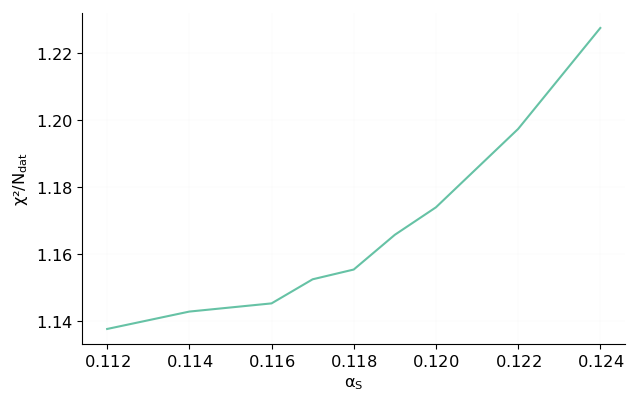} 
 \caption{The $\chi^2$ profile versus $\alpha_s$ for data from NMC (left), affected by additive uncertainties, and HERA (right), with multiplicative uncertainties, computed using the $t_0$ covariance matrix (top) or the experimental covariance matrix (bottom).}\label{fig:alphas_t0}
\end{figure}
The D'Agostini  bias is illustrated in Fig.~\ref{fig:alphas_t0}, which  shows results obtained in the joint PDF-$\alpha_s$ determination of~\cite{Ball:2011us}. In the top row, the $\chi^2$ computed using the $t_0$ method is shown on the left for a dataset (the NMC fixed-target DIS data) in which there are no significant multiplicative uncertainties, and on the right for a dataset (the combined HERA DIS data) whose uncertainties are all multiplicative. In the bottom row, the $\chi^2$ is shown again, now computed using the experimental covariance matrix. While for NMC the maximum likelihood value is unchanged, for the HERA dataset the profile becomes visibly biased.
Very similar findings were obtained also in the recent CT25 PDF+$\alpha_s$ determination~\cite{Ablat:2025gbp}. Here, progressively assuming experimental normalizations for systematic errors for the DIS, DY, and jet+$t\bar t$ production datasets  reduces the preferred $\alpha_s(M_Z)\approx 0.1183$ with the default extended $t$ method to a much lower value.

While all groups take steps to eliminate the well-understood D'Agostini's bias,  the $t_0$ or equivalent methods may lead to another potential, arguably smaller, source of a bias due to the fact that in  these methods the covariance matrix is computed using some theory prediction, usually with a fixed value of $\alpha_s$. 
Whether this procedure is viable for a given data set
can only be determined through a 
closure test.
This was done in~\cite{Ball:2025xgq}  where it was found that, somewhat counterintuitively,  the covariance matrix must be kept fixed. In a closure test with assumed underlying 
$\alpha_s(M_Z)=0.118$ it was found that the average over several tests led to the correct $\alpha_s(M_Z)=0.1182\pm0.0004$ with the $t_0$ kept fixed, but to a value off by more than three sigma, namely  $\alpha_s(M_Z)=0.1195\pm0.0004$, when the $t_0$ prediction is recomputed as a function of $\alpha_s$ for each value.

\paragraph{Positivity.} 
In PDF determinations the space of acceptable PDFs is usually restricted to PDFs that give non-negative predictions for physical cross sections. In~\cite{Ball:2025xgq}  it was found that the value of $\alpha_s$ obtained in a joint PDF-$\alpha_s$ determination depends on whether this positivity requirement is enforced or not: imposing positivity led to an upward bias in the value of $\alpha_s$. This bias was small enough that it could only be detected by repeating the closure test a large number of times, and verifying that the distribution of values was skewed and thus biased, even though the average deviation of $\alpha_s$ from the true value was consistent with its uncertainty. In the simultaneous determination  of PDFs and $\alpha_s$ of~\cite{Ball:2025xgq} the impact of imposing positivity was found to be about one standard deviation. 

\paragraph{The photon PDF.} Reference~\cite{Ball:2025xgq} found that the inclusion of a photon PDF in a joint PDF-$\alpha_s$ determination could lead to a shift of the value of $\alpha_s(M_Z)$ of order of $0.0003$. This may seem surprising, but it can be understood noting that the inclusion of a photon PDF may through the momentum sum rule affect the gluon PDF, which in turn is strongly correlated to the value of the strong coupling.  

\paragraph{Data compatibility and tolerance.}
The uncertainties on the values of $\alpha_s$ listed in  Table~\ref{tab:alphas_PDFs}  differ by more than a factor of three (between 0.0007 and 0.0023), reflecting in part different methodologies even when comparable datasets are used. Some of the relevant issues were recently investigated in~\cite{Ablat:2025gbp} in the context of the forthcoming CT25 PDF fit. Specifically, it was found that a  textbook determination of the one-sigma uncertainty based on using a $\Delta \chi^2 =1 $ within a Hessian fit would  yield $\sigma_{\alpha_s(M_Z)} \sim 0.0005$, which is likely too small given some tensions inside the fitted datasets as well as the subtleties discussed above. On the other hand, it was also found that the dependence of $\alpha_s(M_Z)$ on the PDF parametrization choice is weak in the CT25 PDF-$\alpha_s$ fit, yielding an error due to the choice of the PDF functional form of no more than 0.0005, compatible with $\Delta \chi^2 =1 $.

A larger source of uncertainty was found to be due to the inconsistency of the $\alpha_s$ preferences among the datasets. In reflection of this, the common error estimate based on the dynamical tolerance used by CT and MSHT appreciably depends on how the groups of experimental datasets are clustered when estimating the allowed limits on $\alpha_s$. Accounting for these experimental tensions using an average from four independent methods (based on the global and dynamic tolerance, and two Bayesian models) produces a larger uncertainty in the CT25 analysis, $\alpha_s(M_Z) =  0.1183^{+ 0.0023}_{- 0.0020}$, than the counterpart uncertainties from the other groups in Table~\ref{tab:alphas_PDFs}. The main message from~\cite{Ablat:2025gbp} is that the chosen methodology affects the $\alpha_s(M_Z)$ uncertainty estimate, with implications for predictions for Higgs and other precision physics. 

\subsubsection{Outlook on $\alpha_s$ at the EIC.}
A determination of $\alpha_s$ with sub-percent accuracy is likely to be necessary for future phenomenology at the high-luminosity LHC (HL-LHC). The EIC will offer the opportunity of performing joint measurements of $\alpha_s$ and the PDFs and, and of  studying their correlations, which will lead to a better understanding of the methdological issues involved, though it will also require paying attention to a number of issues and subtleties, most likely including all of those discussed above.  The determination of $\alpha_s$ at the EIC  is thus  likely to be an important synergy between the EIC and the HL-LHC.

\subsection{Fragmentation functions 
}
\label{sec:FFs}

\noindent Fragmentation Functions (FFs) are a key ingredient in the perturbative QCD
description of hard processes involving an identified final-state hadron, codifying how the color-carrying quarks and gluons give rise to color-neutral hadrons  \cite{Feynman:356451}.
Even though they are as fundamental as their initial-state counterparts, parton distribution functions (PDFs), their determination by means of global QCD analyses has always lagged behind those for PDFs both because of the scarcity of precise one-particle-inclusive data
in comparison to more inclusive processes, and also the lesser availability
of higher-order QCD corrections for these processes, whose calculation
is typically more involved \cite{Metz:2016swz}.

The EIC will provide unprecedented opportunities to advance our knowledge of FFs. Semi-inclusive DIS measurements, in both polarized and unpolarized configurations, will directly probe hadronization with a precision and kinematic reach that surpass that of previous fixed-target experiments \cite{AbdulKhalek:2021gbh}. In addition to extending the kinematic reach of SIDIS, the EIC will also enable studies of processes mediated by electroweak neutral and charged currents, offering complementary sensitivity to the flavor dependence of FFs. The combination of these data sets will allow the simultaneous refinement of FFs for different hadron species while also constraining the flavor and helicity dependence of PDFs \cite{Aschenauer:2019kzf,Borsa:2020lsz}. Realizing this potential requires that existing FFs, obtained from global fits to world data, be accurate and consistent enough to serve as a solid baseline.

\begin{figure}\label{fig:FFs_NNLO_comp}
    \centering
    \includegraphics[width=1\linewidth]{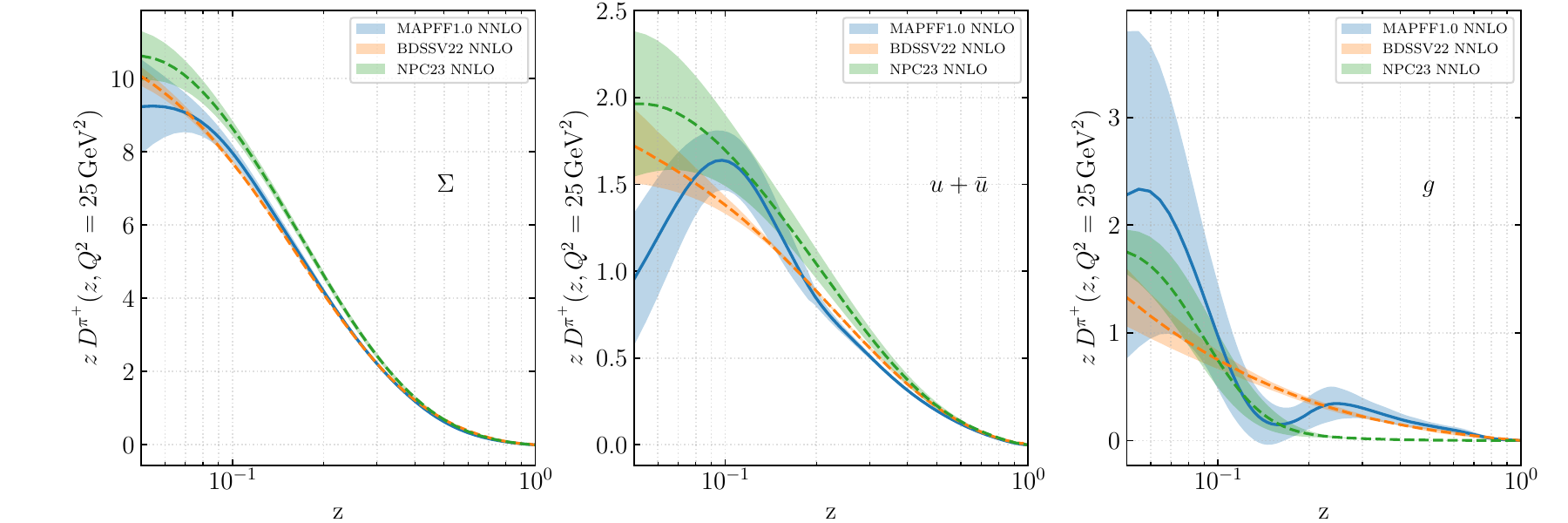}
    \caption{Comparison of NNLO parton-to-pion FFs for the singlet combination $\Sigma$, $u+\bar{u}$ and gluon at $Q^2=25\,\textrm{GeV}^2$, computed with {\tt MAPFF1.0}~\cite{AbdulKhalek:2022laj}, {\tt BDSSV22}~\cite{Borsa:2022vvp} and {\tt NPC23}~\cite{Gao:2025hlm}. Bands correspond to one-sigma uncertainties.}
    \label{fig:ffs_comp}
\end{figure}

At present, several global sets of NLO FFs for different hadrons are available, differing not only in the choice of fitted data, but also in the functional forms assumed at the initial scale and in the methodologies used to estimate uncertainties \cite{Metz:2016swz}.
First attempts at incorporating NNLO corrections for selected observables have also appeared recently \cite{Anderle:2015lqa, Borsa:2022vvp, AbdulKhalek:2022laj, Gao:2025hlm}. In this respect, the recent computations of the complete NNLO corrections to SIDIS \cite{Bonino:2024qbh,Bonino:2024wgg,Goyal:2023zdi,Goyal:2024tmo,Bonino:2025qta,Bonino:2025bqa} and to hadroproduction in proton-proton collisions \cite{Czakon:2025yti} represent major advances. Nevertheless, no full NNLO global extractions of FFs exist yet, and the comparison between different sets and their quoted uncertainties still lacks the refinement and sophistication of their PDF counterparts. This is highlighted in Fig.~\ref{fig:ffs_comp}, which presents a comparison of the recent NNLO extractions of parton-to-pion FFs from \cite{AbdulKhalek:2022laj, Borsa:2022vvp, Gao:2025hlm}. While the FF for the flavor singlet combination $\Sigma$, mainly constrained by single-inclusive annihilation data, shows reasonable agreement across the different extractions, the FFs for individual flavors, as well as that for the gluons, are less constrained, presenting substantial differences both in their central values and reported uncertainties.  Clearly, it is indispensable to better understand
the origin of the differences between FFs sets, which can be sizable, and to critically assess how
the different methodologies to estimate uncertainties reflect the
shortcomings of the perturbative description of one-particle-inclusive observables.

Exclusive observables related to the production of hadrons accompanied by a jet and/or a gauge boson in the final state
provide a unique opportunity to directly constrain the non-perturbative parton-to-hadron fragmentation functions. Among those, hadron-in-jet observables for which QCD corrections can be computed perturbatively are of particular relevance. In particular, distributions related to the momentum fraction of a charged hadron inside a jet have been computed at NNLO accuracy in  \cite{Czakon:2025yti} for single hadro-production and in association with a $Z$-boson (decaying leptonically) at NLO accuracy in \cite{Caletti:2024xaw}. In both cases, results were obtained using various NLO sets of charged hadron FFs and compared with ATLAS and LHCb data, respectively. It is found that the use of different FFs sets lead to strikingly different predictions including shape changes when higher order QCD corrections are included, partially incompatible with each other and reflecting that none of the NLO available FF sets used is able to describe the data over the whole kinematical range considered.
These comparisons highlighted the importance of including LHC data related to these hadron-in-jet observables to best constrain the fragmentation functions in global analysis of FFs. This has recently been done in \cite{Gao:2024nkz} showing  that these hadron-in-jet LHCb data related to associated production of a hadron and a $Z$-boson can be useful to better determine the gluon-to-hadron fragmentation functions, which other datasets leave largely unconstrained. 
Hadron-in-jet production in multi-jet final states in 
$e^+e^-$ annihilation, which is 
also known to NNLO~\cite{Bonino:2026dvr}, can provide equally tight constraints 
on gluon-to-hadron fragmentation. 
Hadron-in-jet observables may also be important for the tagging of heavy-flavor jets through 
identified heavy hadrons~\cite{Caletti:2024xaw,Generet:2025bqx}. 

Of special interest is furthermore the notorious inability of NLO FFs to reproduce
hadroproduction data sets obtained in large center-of-mass
energy proton-proton collisions from LHC simultaneously with measurements
produced at RHIC at much smaller energies, unless a very large
factorization scale uncertainty is taken into account \cite{dEnterria:2013sgr, Borsa:2021ran}. This may point to large QCD corrections beyond NLO or even factorization breaking effects.
The inclusion of observables including complete NNLO corrections for
hadroproduction processes in proton-proton collisions \cite{Czakon:2025yti} in global
analyses of FFs is a pending task in preparation for the use of FFs at EIC.

\subsection{QED and electroweak corrections for EIC physics }\label{subsec:qed}

Precision measurements in lepton-hadron scattering rely on the accurate determination
of the collision kinematics by means of measuring the outgoing lepton.
However, the lepton momentum might be subject to sizable changes through  {collision-induced} real photon emission,
which modify the underlying partonic kinematics with respect to the
assumed and measured momentum of the in- and outgoing lepton, respectively.

For HERA, different analytical calculations had been performed to correct
the cross section for QED~\cite{Kripfganz:1987yu,Kripfganz:1990vm} or EW effects~\cite{Bardin:1988by,Bardin:1989vz,Bohm:1986na}.
These corrections achieved full $\mathcal{O}(\alpha^3)$ accuracy: 
radiative corrections from the lepton and quark legs as well as virtual corrections,
and were typically expressed in terms of a correction factor, $\delta$,
to the Born cross section, differential in two variables,
\begin{equation}
\hspace*{2cm}
    \frac{\mathrm{d}^2}{\mathrm{d}x \mathrm{d}y} \sigma(ep \to eX) = \frac{\mathrm{d}^2 \sigma_\mathrm{Born}}{\mathrm{d}x \mathrm{d}y} \ (1 + \delta(x, y)) \ .
\label{eq:radiation_corrections}
\end{equation}
It had been observed that these types of corrections 
 $\delta(x,y)$  strongly depend on the experimental cuts, hence 
requiring their 
implementation in a fully differential form in an event generator.
This had been accomplished in the HERACLES~\cite{Kwiatkowski:1990es} and DJANGO6~\cite{Charchula:1994kf} codes for HERA physics,
which have however since been discontinued. For EIC physics, clearly novel approaches and implementations will be required. 

A recent development is the identification of a 
 new pinch-singular region in the DIS cross section 
in which the exchanged virtual photon goes on-shell~\cite{Cammarota:2025jyr}. This leads to additional 
cut requirements on the final state kinematics to ensure the 
deeply inelastic nature of the underlying parton-level process, 
consequently affecting the determination of the radiative correction factor $\delta(x,y)$.

A new joint QCD and QED factorization approach was proposed to treat the collision-induced QED radiation 
equally with QCD radiation as a part of production in lepton-hadron collisions rather than as a correction, 
allowing a systematic and consistent approach to account for the impact of collision-induced QED radiations 
from leptons as well as quarks~\cite{Liu:2020rvc,Liu:2021jfp}.  
For the inclusive lepton-hadron DIS, all leading-power collinear QCD and QED radiations off the scattered lepton 
and the colliding lepton and hadron are factorized into universal lepton fragmentation functions (LFFs), lepton distribution functions (LDFs) 
and parton distribution functions (PDFs), respectively.  
The newly identified pinch-singular and non-perturbative region in the phase-space integration can be naturally and consistently 
absorbed into the universal photon distribution of the colliding hadron in this joint QCD and QED factorization, 
leaving the process-dependent higher-order contributions in both QCD and QED infrared-safe. They can thus 
be calculated perturbatively in a joint expansion of powers of $\alpha_s$ and $\alpha_{em}$ 
without depending on any experimental cuts (or parameters) other than the standard factorization scale~\cite{Cammarota:2025jyr}.
In this joint factorization approach, there is no need to make the one-photon approximation for the DIS cross section, and 
contributions of two-photon exchange and its infrared sensitivity are naturally taken care of 
as parts of factorized higher-order QED corrections~\cite{Cammarota:2025jyr}.
The evolution equations and their kernels for lepton and hadron distribution functions
are calculated in both QCD and QED, while 
their solutions depend on non-perturbative boundary conditions that need to be determined from experimental data in 
close analogy to the established methodology of 
fitting PDFs and FFs in QCD.  
The predictive power of this joint QCD and QED factorization approach to lepton-hadron scatterings at the EIC relies on 
the universality of lepton and hadron distribution functions and our ability to calculate the high order perturbative correlations.  

The lepton-hadron semi-inclusive DIS (SIDIS) is another very important part of the EIC  physics 
program, providing new opportunities 
to study transverse momentum dependent (TMD) PDFs  and the 3D hadron structure in momentum space~\cite{Accardi:2012qut,Boussarie:2023izj}, see 
Section~\ref{ch:tomography} below.
The angular distributions between the leptonic plane (defined by the momenta of colliding and scattered lepton) and 
hadronic plane (defined by the momenta of colliding and scattered hadron) is sensitive to contributions from different TMDs, 
and consequently, corresponding azimuthal angular modulations could provide a powerful tool to isolate them~\cite{Boussarie:2023izj}.
However, the collision-induced QED radiation can change both the value and
direction of the exchanged virtual photon 
momentum, which alters the relation between the TMDs
and observed angular distributions~\cite{Liu:2020rvc}.  
While both collision-induced QCD and QED radiation can impact the TMDs of leptons and hadrons, 
a hybrid-factorization -- collinear factorization for observed leptons and TMD factorization for the observed hadron 
-- was proposed for studying TMD physics in SIDIS since transverse momentum broadening from QED radiation at the EIC energies 
is so much smaller than the $\Lambda_{\rm QCD}$ and typical parton transverse momentum in hadrons~\cite{Liu:2021jfp}.  

For precision studies at the EIC in this approach, knowledge of the universal LDFs and LFFs will be necessary.  
However, almost all published data from lepton-hadron and lepton-lepton collisions have had some kind of radiative corrections implemented.  
Therefore, it is crucial to analyze the future data, such as those from Jefferson Lab, or from recent measurement by the
ZEUS collaboration~\cite{ZEUS:2023zie} which, unlike previous studies, did not correct to
Born-level QED by means of truth-level Monte Carlo corrections~\cite{Lorkowski:2024hwx}.
By comparing these QED-uncorrected data 
with theoretical calculations performed in a joint QCD and QED factorization approach,
reliable LDFs and LFFs could be derived for the EIC era, as well as for future high energy lepton-lepton and lepton-hadron facilities.

While many of the event generators used for HERA predictions have been discontinued, 
tremendous progress has been achieved in the perturbative accuracy in other generators in the last decades (see Section~\ref{sec:higher_orders} above).
Virtually all analyses at the LHC make use of event generators, with NLO being the standard for perturbative SM predictions and
including resummation via parton-shower algorithms or dedicated interfaces to analytical resummation codes.

Going beyond this, electroweak corrections have seen increased interest in the last decade. 
NLO EW has been automated~\cite{Schonherr:2017qcj,Frederix:2018nkq} and
mixed NLO QCD$\oplus$EW~\cite{Zhang:2014gcy,Kallweit:2015dum,Biedermann:2017yoi,Kallweit:2017khh} and
even NNLO~QCD$\oplus$NLO~EW~\cite{Czakon:2017wor,Grazzini:2019jkl,Lindert:2022qdd} accuracies for certain processes have been achieved.
It can be foreseen that in the long-term, the latter should become available
for DIS as well in a parton-level event generator.

In terms of resummed predictions, the parton showers typically include QED splittings and can hence be used to resum photon emissions.
At higher orders, however, double counting has to be avoided and an appropriate matching
between the fixed-order computation and the resummation must be ensured.
It can be anticipated that a matching in QED can be accomplished in the coming years~\cite{Flower:2026byh},
and matching to mixed NLO QCD$\oplus$QED in the long-term, reaching NLL accuracy in the resummation.
As a different alternative, it seems feasible to compare the QED corrections through resummation of soft photons in the Yennie-Frautschi-Suura formalism~\cite{Yennie:1961ad}.

With an implementation in an event generator, studies of photon sensitivity, lepton dressing and
detector acceptance can be undertaken.
It will be informative to critically examine the different approaches in a comparison. Such benchmarking 
exercises will require QED-uncorrected data from 
lepton-hadron collisions such as the JLab and 
ZEUS data sets discussed above.

%% file: week2.tex
\subsection{Status of lattice calculations of unpolarized and helicity PDFs
\label{sec:lattice_PDFs}}

\begin{figure}[h!]
    \begin{minipage}{0.33\linewidth}
      \includegraphics[width=1.03\linewidth,height=\linewidth]{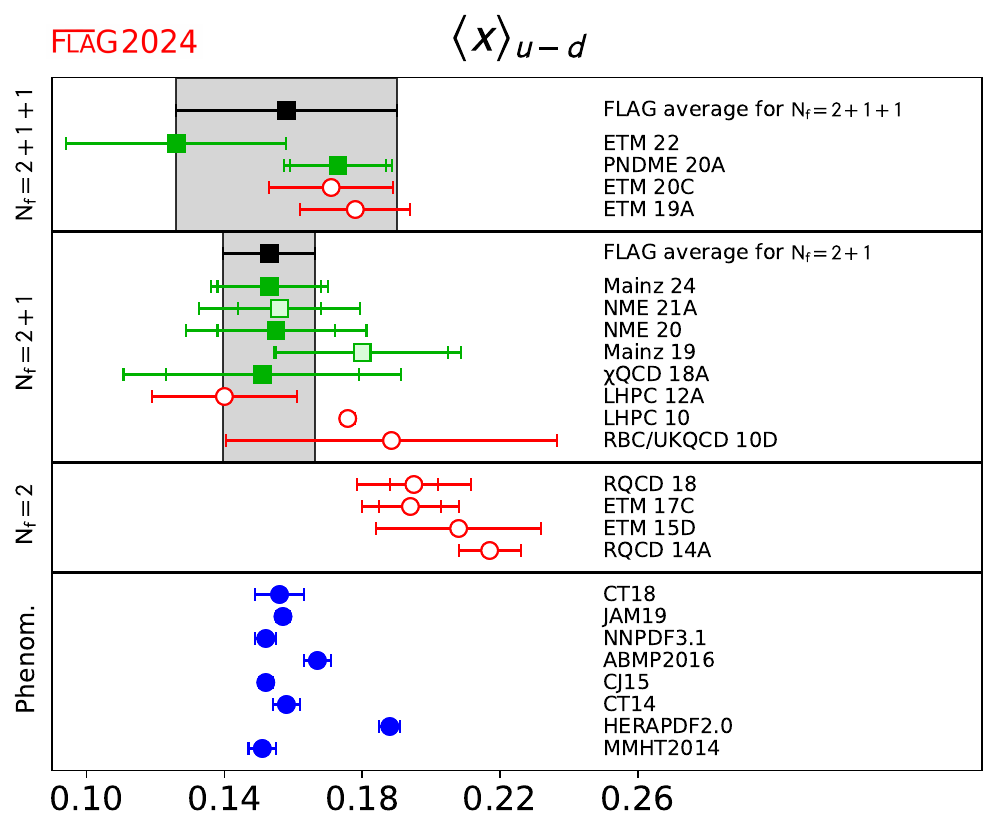}
    \end{minipage}\hfill
    \begin{minipage}{0.33\linewidth}
      \includegraphics[width=1.03\linewidth,height=\linewidth]{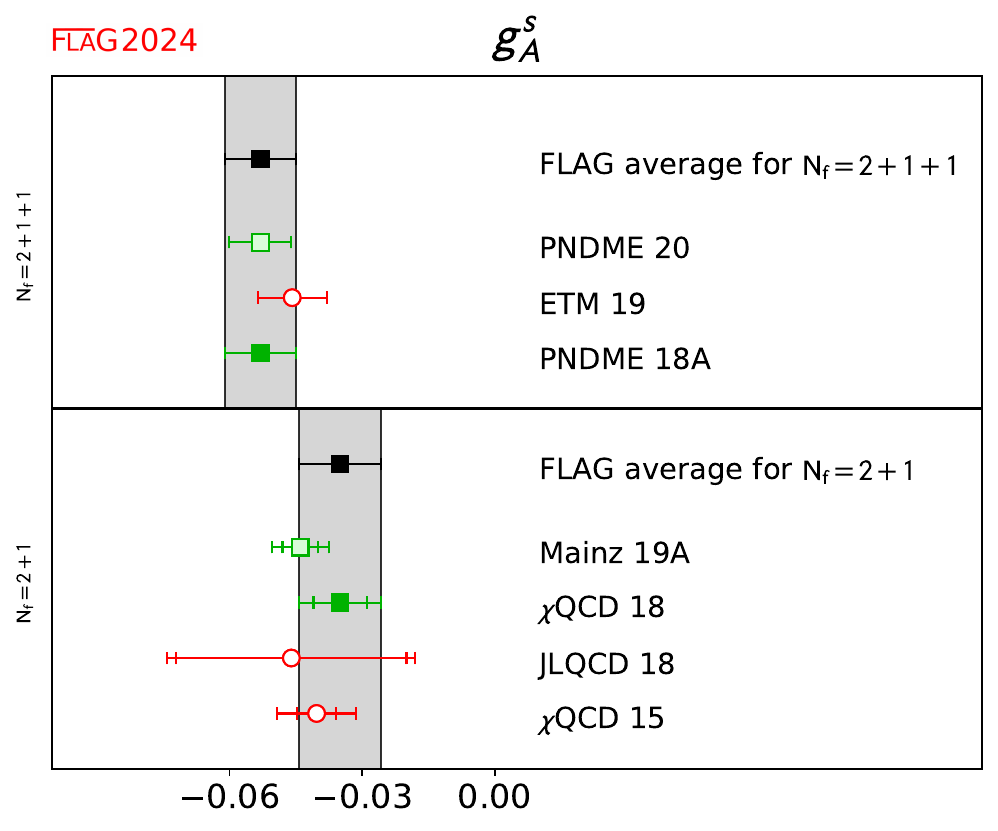}
    \end{minipage}\hfill
    \begin{minipage}{0.33\linewidth}
      \includegraphics[width=1.03\linewidth,height=\linewidth]{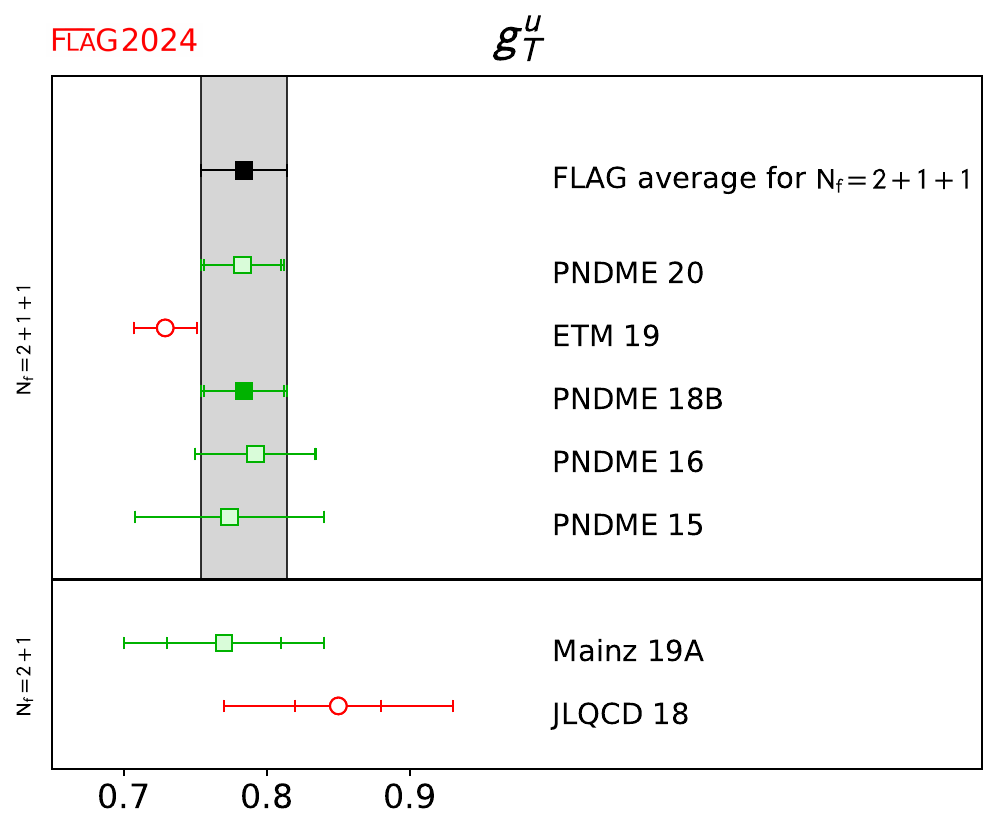}
     \end{minipage}
    \caption{Lattice QCD results on the nucleon isovector unpolarized (left), strange-quark contribution to axial charge, $g_{\mathrm{A}}^s$,  (middle) and up-quark contribution to tensor charge,  $g_{\mathrm{T}}^u$, (right) from the FLAG2024 report~\cite{FlavourLatticeAveragingGroupFLAG:2024oxs}.}
    \label{fig:FLAG-moments}
\end{figure}

Lattice QCD provides a first-principles, nonperturbative framework for studying hadron structure  by discretizing QCD on a spacetime lattice and performing a numerical simulation on supercomputers.mathrm
In lattice-QCD calculations, several sources of systematic uncertainty must be carefully controlled in order to obtain reliable physical results.
The most prominent of these are discretization effects, arising from the finite lattice spacing $a$, and requiring extrapolation to the continuum $a \to 0$ based on simulations at multiple lattice spacings.
Finite-volume effects, due to the limited spatial extent $L$ of the lattice, can modify long-range physics and are typically suppressed as $e^{-m_\pi L}$; these are mitigated through simulations at larger volumes or analytic finite-volume corrections.
Unphysical quark masses introduce additional systematics, as simulations often employ heavier-than-physical quarks to reduce computational cost; this necessitates a chiral extrapolation guided by chiral perturbation theory.
The determination of the lattice scale, through a reference observable such as the pion decay constant $f_\pi$, the nucleon ($M_N$) or the omega  ($M_\Omega$) mass or $w_0$, introduces scale-setting uncertainties that propagate to all dimensionful quantities.
Operator renormalization and matching to continuum schemes, whether perturbative or nonperturbative, contribute further systematic errors due to truncation or lattice artifacts.
Moreover, contamination from excited states in correlation functions can bias hadron observables, and must be controlled using multistate fits and varying source–sink separations.
Finally, many simulations neglect isospin breaking ($m_u = m_d$ and QED effects $\alpha_{\mathrm{em}} = 0$), which constitute additional sources of uncertainty, especially when the targeted precision is 1\% or less.
Overall, physical observables $O_{\mathrm{phys}}$ are extracted from lattice quantities $O_{\mathrm{latt}}(a, m_q, L)$ through controlled extrapolations to the continuum, infinite-volume, and physical-quark-mass limits, such that
$
O_{\mathrm{phys}} = O_{\mathrm{latt}}(a, m_q, L) + \textrm{stat. err.} + \textrm{syst. err.},
$
with a detailed error budget quantifying the remaining systematic effects.
For a detailed discussion of lattice systematics we refer the reader to Sec.~3.1.2 of the PDFLattice 2017 whitepaper~\cite{Lin:2017snn} and Sec.~2.1.1 of the Flavour Lattice Averaging Group (FLAG) report~\cite{FlavourLatticeAveragingGroupFLAG:2024oxs} on the rating of the specific work based on the parameters used in the lattice calculations.
For first Mellin moments of the nucleon,
such as nucleon charges, $g_{\mathrm{A},\mathrm{S},\mathrm{T}}^{u-d}$, $g_{\mathrm{A},\mathrm{S},\mathrm{T}}^{u,d,s}$~\cite{Alexandrou:2024ozj,Park:2025rxi} as well as electromagnetic~\cite{Djukanovic:2023beb,Djukanovic:2023jag,Alexandrou:2025vto} and axial form factors~\cite{Jang:2023zts,Alexandrou:2023qbg, Djukanovic:2022wru}, lattice computations have included or checked all aspects of the systematics, reaching in many cases close to 1\% accuracy. 
The  isovector second Mellin moments $\langle x \rangle_{u-d}$, $\langle x \rangle_{\Delta u-\Delta d}$ and $\langle x \rangle_{\delta u-\delta d}$, have also been computed by several lattice QCD groups. 
The latest summary and details on charges and isovector second Mellin moment of the nucleon can be found in Sec.~10 of Ref.~\cite{FlavourLatticeAveragingGroupFLAG:2024oxs}. 
Selected first and second moments summarized by the FLAG2024 report are shown in Fig.~\ref{fig:FLAG-moments}.
More difficult calculations on the lattice, such as the singlet quark flavor and gluon momentum fractions of the nucleon, have been done by fewer groups~\cite{Alexandrou:2020sml,Hackett:2023rif,Fan:2022qve} and have not yet been included in the FLAG summary; some summaries prior to 2020 can be found in Sec.~II.3 of PDFLattice2019 whitepaper~\cite{Lin:2020hdm}.

\begin{figure}[h!]
\begin{minipage}{0.27\linewidth}
\includegraphics[width=\linewidth]{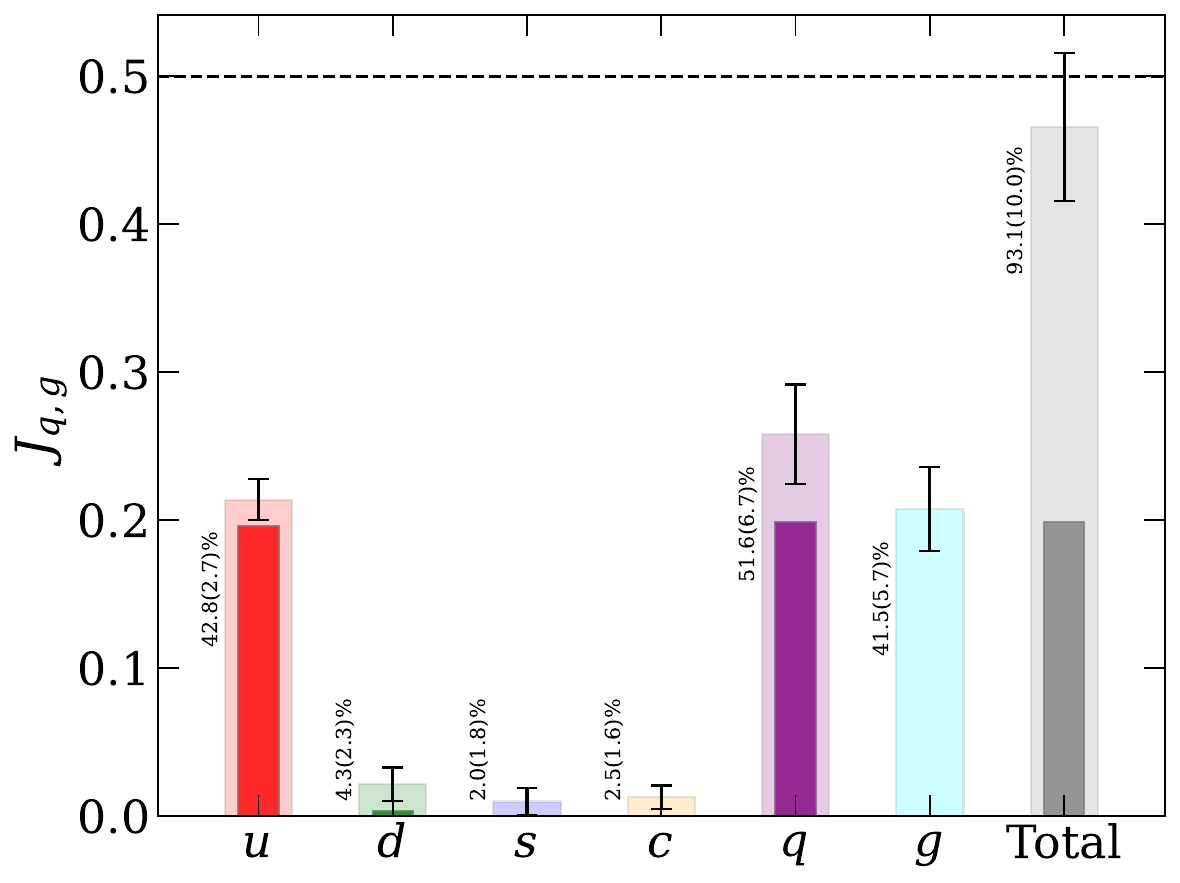}\\
\end{minipage}\hfill
\begin{minipage}{0.35\linewidth}
\includegraphics[width=\linewidth]{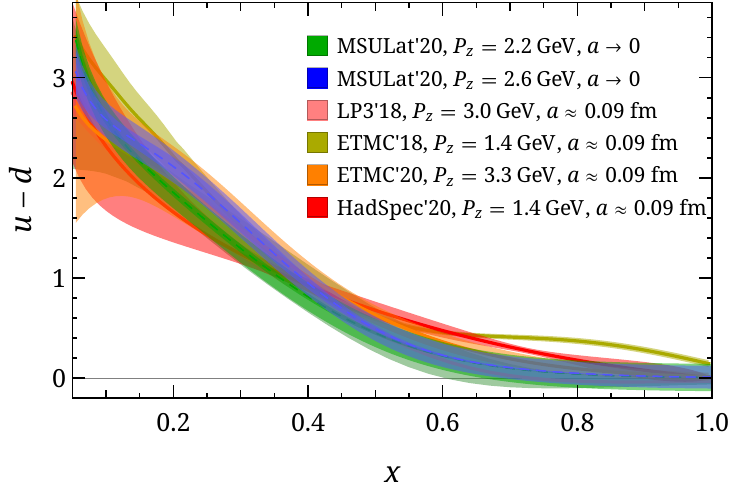}\\
\end{minipage}\hfill
\begin{minipage}{0.37\linewidth}
\includegraphics[width=\linewidth]{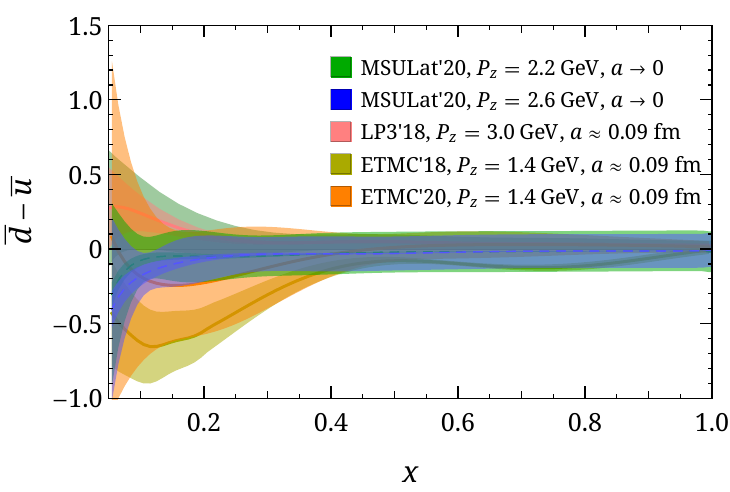}\\
\end{minipage}
\vspace*{-0.5cm}
\caption{
(left) Preliminary continuum extrapolated results at the physical pion mass of the contributions of quarks and gluons to the nucleon spin by ETMC in the $\overline{\rm MS}$ scheme at 2 GeV~\cite{Alexandrou2026}.
Unpolarized nucleon PDFs: the isovector quark (middle)  and antiquark (right)  PDFs from a lattice calculation ``MSULat'20''~\cite{Lin:2020fsj} in the physical-continuum limit, and from single-lattice-spacing calculations at (or extrapolated to) the physical pion mass  using the LaMET (``LP3'18''~\cite{Chen:2018xof} and ``ETMC'18''~\cite{Alexandrou:2018pbm}),  and pseudo-PDF  (``ETMC'20''~\cite{Bhat:2020ktg} and ``HadSpec'20''\cite{Joo:2020spy}) methods. 
}\vspace*{-0.4cm}\label{fig:selected-LQCD-Results}
\end{figure}

In addition, the second unpolarized moments of the nucleon for both quarks and gluons  have been computed for one gauge ensemble generated with physical quark masses, confirming the momentum and spin sums~\cite{Alexandrou:2020sml}. 
These results have recently been extended to include another two gauge ensembles allowing for taking the continuum limit. Preliminary results are shown by ETMC~\cite{Alexandrou2026} on the left-hand-side of Fig.~\ref{fig:selected-LQCD-Results}.

Progress has also been made on the pion and kaon structures. 
The quark flavor decomposition and gluon momentum fractions of the pion and kaon have been computed in the continuum limit and at (or extrapolated to) the physical pion mass~\cite{Good:2023ecp,ExtendedTwistedMass:2024kjf,NieMiera:2025inn}. 
ETMC computed the third and fourth Mellin moments of the pion and kaon  using one gauge ensemble at pion mass $m_\pi=260$~MeV~\cite{Alexandrou:2021mmi,Alexandrou:2020gxs}. 
Recently, two new methods, one based on Wilson flow~\cite{Shindler:2023xpd} and the other on the heavy quark operator product expansion~\cite{Detmold:2021uru}, presented a proof of concept computation of higher Mellin moments for the pion using one ensemble at $m_\pi=410$ MeV~\cite{Francis:2025rya,Francis:2025pgf} and 550~MeV~\cite{Detmold:2025lyb}, respectively. 

Beyond Mellin moments, within lattice QCD one can compute directly the $x$-dependence of PDFs and GPDs. Most computations, however, are still performed  using one gauge ensemble and/or heavier than physical pion mass.
All three types of PDFs (unpolarized, helicity, transversity) have been computed for the nucleon, as well as the unpolarized PDF for the pion and kaon. Most computations are done using  quasi- or pseudo-distributions, which can be regarded as complementary. There has been a lot of progress in improving the renormalization and using higher than NLO matching. A new decomposition of the nucleon matrix elements of the non-local operators in terms of gauge invariant amplitudes 
also allows for an efficient extraction of GPDs~\cite{Chu:2025kew,Bhattacharya:2025yba,Bhattacharya:2023jsc,Bhattacharya:2022aob}. 
Regarding the $x$-dependence, selected twist-2 unpolarized quark and antiquark PDFs are shown in the middle and right-hand panels of Fig.~\ref{fig:selected-LQCD-Results}, while the polarized PDFs are presented in Fig.~7 and Table~I of the PDFLattice2019 white paper~\cite{Lin:2020hdm}, which summarizes the state-of-the-art calculations including all results that are either computed directly at, or extrapolated to, the physical pion mass.
In addition to the usual lattice systematics, $x$-dependent PDFs can introduce additional systematics that are $O((\frac{\Lambda_\mathrm{QCD}}{xP_z})^n, (\frac{\Lambda_\mathrm{QCD}}{(1-x)P_z})^n)$ or $O({\Lambda_\mathrm{QCD}}{z})^n$, depending on the method, and with $n$ starting at 1 or 2 depending on the treatment, such as leading log and resummation~\cite{Gao:2021hxl,Su:2022fiu,Holligan:2023rex,Zhang:2023bxs,Ji:2023pba,Ji:2024hit,Holligan:2025baj}; some selected updated PDF results can also be found in Refs.~\cite{Lin:2025hka,Ji:2020ect,Burkert:2022hjz}.
The ensembles with smallest lattice spacing that have been used to calculate PDFs are at around 0.04~fm and 310 MeV pion mass, with boost momenta $P_z$ no more than 2~GeV, by the MSULat, BNL, Wuhan groups.

In contrast to recent progress in lattice QCD calculations of various quark GPDs using the quasi-PDF/LaMET~\cite{Ji:2013dva, Ji:2014gla} and pseudo-PDF~\cite{Radyushkin:2017cyf} formalisms, investigations of gluon PDFs have been relatively limited.  Compared to the quasi-PDF or quasi-GPD quark matrix elements, gluonic observables require significantly more statistics to achieve a reasonable signal-to-noise ratio. Meanwhile, extracting gluon PDFs from experimental data also remains challenging, and a precise determination of gluon PDFs in the mid-to-large~$x$ region provides a unique opportunity for lattice QCD to make a significant contribution to the field.
While there have been a few calculations of the unpolarized gluon PDFs in the nucleon, pion, and kaon~\cite{Fan:2020cpa,Fan:2021bcr,Fan:2022kcb,Salas-Chavira:2021wui,HadStruc:2021wmh,Delmar:2023agv,Good:2023ecp,Good:2024iur,NieMiera:2025inn}, mostly using the pseudo-PDF formalism, very recent progress has been made in calculating unpolarized gluon PDFs within the LaMET framework~\cite{NieMiera:2025mwj,Chen:2025xww,NieMiera:2025vcx}. These calculations have been performed in the continuum limit, representing a major step toward estimating various systematic uncertainties. Future progress is anticipated in performing these calculations near or at the physical pion mass, and in extracting gluon PDFs in the presence of mixing with quark flavor-singlet PDFs. To obtain reliable data that will have a significant impact on gluon PDF phenomenology, computational resources sufficient for roughly $O(10^6)$ measurements will be required. Notably, recent lattice QCD calculations of the pion gluon PDF have already shown a significant impact when included in global analyses~\cite{Good:2025nny}.

Compared to the unpolarized case, the calculation of polarized gluon PDFs remains challenging due to contamination terms in the Euclidean matrix elements that hinder extraction of PDFs from lattice data~\cite{Balitsky:2021cwr, HadStruc:2022yaw}. However, this problem has been shown to be bypassed by parametrizing lattice data at different momenta and removing the contamination term at leading-order approximation~\cite{Khan:2022vot, Chowdhury:2024ymm}. These lattice calculations provide numerical evidence disfavoring negative gluon polarization in the nucleon, consistent with earlier lattice QCD calculations of the gluon spin content in the nucleon~\cite{Yang:2016plb} using local matrix elements~\cite{Ji:2013fga}. The finding that the gluon helicity PDF cannot be negative in the moderate-to-large~$x$ region was further supported in~\cite{deFlorian:2024utd}, based on the fundamental requirement that physical cross sections must not be negative.

On the other hand, numerical calculations of gluon GPDs on the lattice within the quasi-PDF/LaMET formalism have not yet been performed, although a recent work~\cite{Schoenleber:2024auy} presented a method to achieve a Lorentz-covariant parametrization of matrix elements in terms of a linearly independent basis of tensor structures that is crucial for projecting lattice matrix elements onto light-cone GPDs. It has been found that the choice of lattice matrix elements for calculating various gluon GPDs is not unique, and some operator choices may exhibit smaller power corrections than others, providing an opportunity to study power corrections in lattice QCD. With numerical calculations of gluon GPDs expected in the near future, lattice QCD is anticipated to have a significant impact in constraining gluon GPDs, which remain poorly constrained by experimental data. 

The \textit{EIC Theory Alliance} whitepaper~\cite{Abir:2023fpo} outlines a coordinated roadmap and identifies the critical theoretical, computational, and human resources required to maximize the impact of lattice QCD in the EIC era. To conclude, we collect here a few of the salient points raised in that paper.
\paragraph{Theoretical and Computational Objectives:}
A principal aim of the LQCD program is to deliver precision calculations of the one- to three-dimensional structure of hadrons with full control of systematic uncertainties, including renormalization, finite-volume effects, and continuum extrapolations.
Theoretical developments are underway to establish robust connections between Euclidean lattice observables and light-cone parton distributions relevant for EIC measurements.
This includes the formulation of frame-independent approaches, improved matching procedures, and systematic inclusion of higher-order corrections.
\paragraph{Integration with Phenomenology.}
A central component of the roadmap is the integration of lattice QCD results with global QCD analyses and phenomenological models.
Collaboration between lattice practitioners, phenomenologists, and experimentalists will enable the use of LQCD inputs—such as Mellin moments and $x$-dependent distributions—in global fits of PDFs, GPDs, and TMDs. Such synergy will reduce model dependence and enhance the precision of hadron tomography at the EIC.
\paragraph{Infrastructure and Computational Resources.}
Achieving the required precision will depend on access to state-of-the-art high-performance computing facilities and software infrastructure.
The report emphasizes the need for sustained investment in algorithmic development, optimized lattice software, and utilization of emerging exascale resources, such as \textit{Frontier} and \textit{Aurora}, and \textit{Jupiter}.
Ongoing U.S. Department of Energy programs including SciDAC, INCITE, and ALCC are expected to provide critical support for large-scale LQCD simulations at fine lattice spacings and large hadron momenta. The EuroHPC Joint Undertaking is building an EU-wide infrastructure that includes, besides \textit{Jupiter}, a second exascale system as well as programs for funding code development. However, as AI  takes a central stage, EuroHPC is focusing on AI applications and there is ongoing effort by the lattice QCD community in Europe (e.g. EuroLFT) to emphasize the importance of large scale simulations and accompanied human support for the field. The Partnership of Advanced Computing in Europe (PRACE) through its \href{https://prace-ri.eu/prace-research-infrastructure/third-scientific-case/}{Scientific Case 2018-2026} and its forthcoming  \href{https://prace-ri.eu/prace-research-infrastructure/scientific-and-innovation-case/}{Scientific and Innovation Case} is also helping in this direction addressing also data storage needs.
\paragraph{Training and Workforce Development.}
The EIC Theory Alliance envisions a comprehensive training framework to prepare the next generation of lattice-QCD researchers.
Graduate students and postdoctoral fellows will receive multidisciplinary training in lattice techniques, theoretical QCD, and phenomenological analysis.
These efforts will ensure a strong and diverse workforce capable of advancing computational and theoretical frontiers aligned with the EIC program.
\paragraph{Precision Goals and Timeline.}
Within the next decade, the lattice-QCD community aims to achieve $10$--$20\%$ precision in quark TMD quantities and $20$--$40\%$ precision in full $(x, k_T)$-dependent distributions. Subsequent improvements in algorithms and computing power will extend these results to gluon observables, providing comprehensive nonperturbative input for EIC data analyses.
These advances will establish lattice QCD as a quantitative cornerstone of the EIC scientific mission.

\subsection{Status of helicity PDFs
\label{sec:helicity_PDFs}}

The accurate and precise determination of helicity PDFs~\cite{Ethier:2017zbq} is key to understanding the quark and gluon spin structure of the nucleon~\cite{Ji:2020ena}. In the last year or so,
three independent global QCD determinations of helicity PDFs, for the first time accurate to NNLO, have appeared: {\tt BDSSV24}~\cite{Borsa:2024mss}, {\tt MAPPDFpol1.0}~\cite{Bertone:2024taw} and {\tt NNDPFpol2.0}~\cite{Cruz-Martinez:2025ahf}. This increased theoretical accuracy, made possible by recent developments in perturbative computations (see Sec.~\ref{sec:higher_orders}), has been accompanied by a steady extension of the data set and by significant progress in fitting methodology. All of these improvements have contributed to solidifying our understanding of the contribution of partons' spin to the spin of the proton, checking their perturbative stability and independence from the details of each analysis.

\begin{figure}
    \centering
    \includegraphics[width=0.49\linewidth]{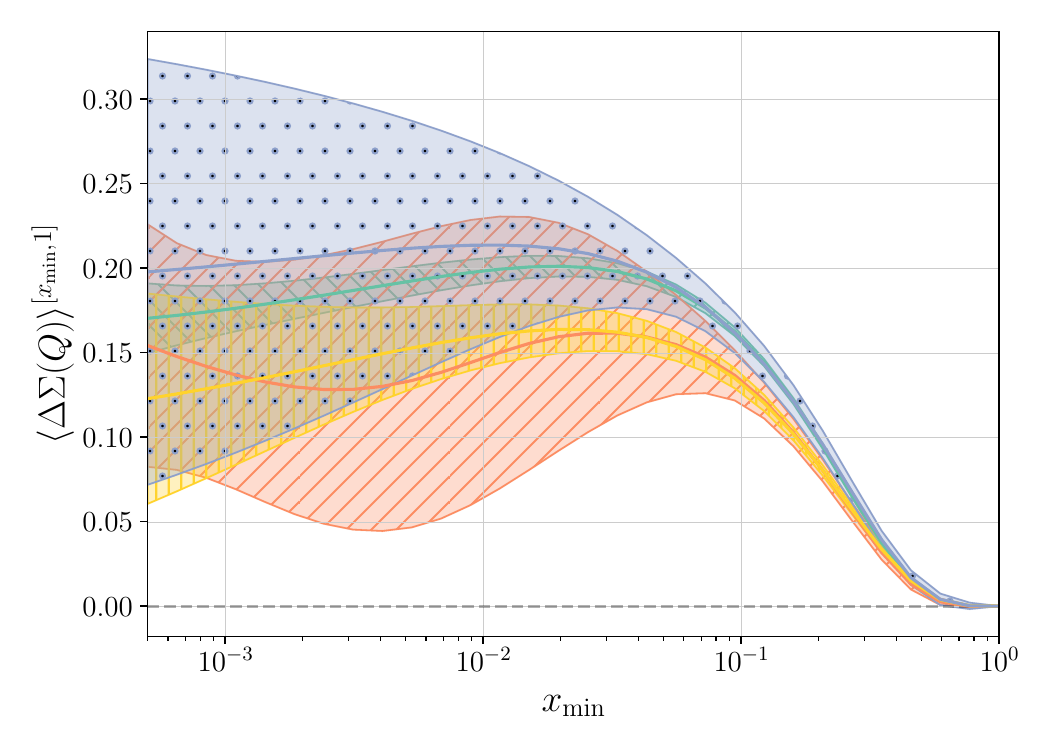}
    \includegraphics[width=0.49\linewidth]{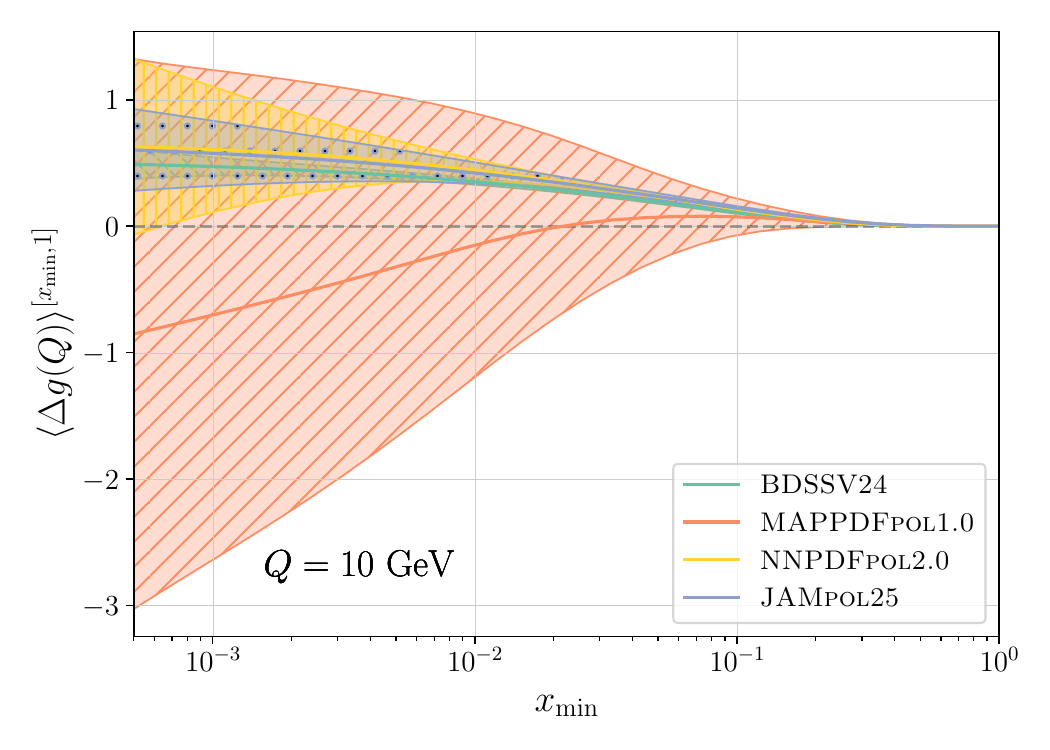}\\
    \caption{The singlet (left) and gluon (right) lowest truncated moments, Eq.~(\ref{eq:mom_frac}), as a function of $x_{\rm min}$ at $Q=10$~GeV, computed with the NNLO {\tt BDSSV24}~\cite{Borsa:2024mss}, {\tt MAPPDFpol1.0}~\cite{Bertone:2024taw} and {\tt NNDPFpol2.0}~\cite{Cruz-Martinez:2025ahf} helicity-dependent PDF sets. The result obtained with the NLO {\sc JAMpol25}~\cite{Cocuzza:2025qvf} PDF set is also displayed. Bands correspond to one-sigma uncertainties.}
    \label{fig:mom_frac}
\end{figure}

A snapshot of the current status is provided in Fig.~\ref{fig:mom_frac}, where we display the lowest truncated moments of the quark singlet and gluon helicity PDFs, defined as
\begin{equation}
    \langle \Delta f (Q) \rangle^{[x_{\rm min},1]}=\int_{x_{\rm min}}^1 dx\,\Delta f(x,Q)
    \qquad
    {\rm with\ } f =\Sigma,g\,,
    \label{eq:mom_frac}
\end{equation}
calculated as a function of $x_{\rm min}$ at $Q=10$~GeV. A similar comparison at the level of individual PDFs can be seen, {\it e.g.}, in Fig.~4.5 of Ref.~\cite{Cruz-Martinez:2025ahf}. For completeness, results obtained with the NLO {\sc JAMpol25} PDF set~\cite{Cocuzza:2025qvf} (see below for details) are also displayed. From inspection of this figure, we can make the following remarks:\\ 
{\it (i)} On average, the fraction of the proton spin carried by the spin of partons amounts to about 15\% for quarks and 70\% for gluons. Down to $x_{\rm min}\sim 0.02$, the former is constrained to a precision of about 10-15\% (thanks to inclusive polarized deep-inelastic scattering measurements), whereas the latter is constrained to a precision of about 5-10\% (thanks to single-inclusive jet and di-jet production measurements in polarized proton-proton collisions). \\
{\it (ii)} These numbers are consistent, within uncertainties, among different sets of helicity PDFs, although uncertainties vary widely depending on the extrapolation limit $x_{\rm min}$ and the set of PDFs. These differences are attributed to differences in experimental data and methodological details for each analysis. \\
{\it (iii)} As expected, uncertainties tend to increase as the extrapolation limit is lowered because of the lack of experimental data in that kinematic region. The extrapolation is generally agnostic in that no specific small-$x$ behavior is assumed. In principle, PDFs could be supplemented with a specific small-$x$ functional behavior, {\it e.g.} based on solutions of small-$x$ evolution equations (see Sec.~\ref{sec:smallx_helicity} and also Ref.~\cite{Adamiak:2021ppq,Adamiak:2023yhz} and references therein). However, such an extrapolation remains to be phenomenologically validated against new experimental measurements, as those planned at the future Electron Ion Collider in the US~\cite{AbdulKhalek:2021gbh} or in China~\cite{Anderle:2021wcy}.

The aforementioned recent NNLO global QCD analyses~\cite{Borsa:2024mss,Bertone:2024taw,Cruz-Martinez:2025ahf} are complemented by recent work on NNLO analyses based on inclusive deep-inelastic scattering measurements only~\cite{Taghavi-Shahri:2016idw} and by NLO analyses devoted to investigating specific phenomenological aspects of helicity PDFs. Examples are contributions of higher-twist corrections~\cite{Sato:2016tuz,Cocuzza:2025qvf}, the decomposition into quark and antiquark helicity distributions, including the interplay between PDFs and FFs in SIDIS~\cite{Ethier:2017zbq,Cocuzza:2022jye}, the role of positivity constraints~\cite{Zhou:2022wzm,Whitehill:2022mpq,Hunt-Smith:2024khs}, and the impact of lattice QCD data~\cite{Bringewatt:2020ixn,Karpie:2023nyg}. Concerning higher twist corrections, Ref.~\cite{Sato:2016tuz} revealed non-zero twist-3 quark helicity distributions, whereas it found twist-4 quark helicity distributions to be compatible with zero. Concerning the decomposition into quark and antiquark helicity distributions, Ref.~\cite{Ethier:2017zbq} confirmed SU(2) symmetry to 2\%, but highlighted the need for a 20\% breaking of SU(3) symmetry, as advocated in the literature long ago~\cite{Flores-Mendieta:1998tfv}. Good consistency between constraints on quark and antiquark helicity PDFs provided by measurements of $W$ boson production in polarized proton-proton collisions and by measurements of pion and kaon production in polarized SIDIS was also reported. The sensitivity to FFs is relatively weak, because SIDIS spin asymmetries are fit instead of cross sections, in which the dependence of the measurement on FFs largely cancels out between the numerator and the denominator. This contrasts with what happens with unpolarized SIDIS multiplicities used to determine FFs in the first place (see Sec.~\ref{sec:FFs}).  All of these findings are very much in line with Refs.~\cite{Borsa:2024mss,Bertone:2024taw,Cruz-Martinez:2025ahf}. Concerning positivity, Ref.~\cite{Zhou:2022wzm} observed that the gluon helicity PDF is largely unconstrained, and specifically that it can turn out to be negative if the common LO constraint~\cite{Altarelli:1998gn}, which bounds helicity PDFs to plus or minus their unpolarized counterparts, is relaxed. Their claim is that measurements of single-inclusive jet and di-jet spin asymmetries, which 
supposedly constrain the gluon helicity PDFs, are ineffective, because they receive their leading contribution from the gluon-gluon initiated channel, which depends quadratically on the gluon helicity PDF. Whereas this finding was not confirmed in Refs.~\cite{Bertone:2024taw,Cruz-Martinez:2025ahf}, which make use of a much more flexible PDF parametrization, Ref.~\cite{deFlorian:2024utd} demonstrated that a negative gluon helicity PDF violates the positivity of the polarized Higgs boson production cross section in a kinematic region accessible at RHIC. The same authors of Ref.~\cite{Zhou:2022wzm} have later acknowledged that a negative gluon helicity PDF is actually phenomenologically ruled out by asymmetry measurements of DIS structure functions at large $x$ or of SIDIS high-transverse momentum hadron production~\cite{Hunt-Smith:2024khs}. Finally, concerning the inclusion of lattice QCD data in PDF fits, Refs.~\cite{Karpie:2023nyg,Cocuzza:2025qvf} have explored constraints from nucleon isovector matrix elements and gluonic pseudo Ioffe-time distributions, respectively.

All of these developments testify to the active interest in helicity PDFs, but also expose some of the challenges and open issues that the community will have to face in preparation of the EIC. In the following, we summarize some of the compelling points that, in our opinion, should deserve some attention in the future, for intance as actions taken within the EIC theory alliance~\cite{Abir:2023fpo}.

\paragraph{PDF benchmarking.} As shown in Fig.~\ref{fig:mom_frac}, different helicity PDF sets differ in central values and uncertainties. These differences are commonly attributed to differences in the fit data set and in the methodological details of each analysis. However, their origin has never been systematically pinpointed, the reason being that all analyses give rather compatible results within their widely different uncertainties. In view of the precision demands of the future EIC, it now looks sensible to carefully benchmark helicity PDFs. In a spirit similar to what is regularly carried out by the PDF4LHC working group~\cite{Ball:2012wy,Rojo:2015acz,PDF4LHCWorkingGroup:2022cjn}, a common data set, in which the treatment of uncertainties is checked and standardized, and a common set of theoretical predictions, for which all theoretical settings are checked and standardized, should be defined and used as input to fits performed by the various groups. A comparison of these fits, in association with incremental, controlled variations of the experimental and theoretical input, will make it possible to understand whether the observed differences originate from differences in data, theory or methodology.
    \paragraph{Computing tools.} Development, benchmarking, documentation, deployment and maintenance of computing tools is also becoming increasingly crucial to fits of helicity PDFs. Many of the tools are publicly available, such as: {\tt APFEL++}~\cite{Bertone:2013vaa,Bertone:2017gds}, {\tt EKO}~\cite{Candido:2022tld} and {\tt yadism}~\cite{Candido:2024rkr,Hekhorn:2024tqm} for PDF evolution and computation of DIS and SIDIS cross sections; {\tt PineAPPL}~\cite{Carrazza:2020gss}, for the fast convolution of PDFs and partonic cross sections; {\tt LHAPDF}~\cite{Buckley:2014ana}, for the delivery of PDFs in a standardized format; the {\tt MAP}~\cite{valerio_bertone_2024_10933177} and {\tt NNPDF}~\cite{NNPDF:2021uiq} software, for fitting PDFs. Continuous and increasing availability of reliable pieces of software can only support progress and foster knowledge. As such, efforts in this direction should be encouraged, recognized, and rewarded.
    \paragraph{Interplay with FFs.} As demonstrated in Refs.~\cite{Borsa:2024mss,Bertone:2024taw,Ethier:2017zbq}, SIDIS is a crucial process for determining helicity quark and antiquark flavor separation. It requires knowledge of FFs, the dependence on which is however slight in spin asymmetries used to fit helicity PDFs. Nevertheless, in view of the increased precision to be reached by EIC measurements~\cite{Borsa:2020lsz,Zhou:2021llj} and of the possibility of measuring cross sections instead of asymmetries, this state of affairs may no longer hold true in the EIC era. An equally accurate determination of FFs would then be required (see Sec.~\ref{sec:FFs}), which may benefit from the same set of good practice measures outlined above.
    \paragraph{Input from lattice QCD.} As we have seen, helicity PDFs are affected by large extrapolation uncertainties at small $x$ that, in particular, hinder any definitive conclusion about the contribution of the gluon spin to the spin of the proton. It is therefore natural to wonder whether lattice QCD may help pin down this uncertainty. The potential impact of lattice computations of the lowest helicity quark moments was investigated in Ref.~\cite{Lin:2017snn}. There it was concluded that, in order to constrain the helicity up and down distributions better than what experimental data do, lattice uncertainties should be controlled down to 1-2\%; conversely, the strange helicity distribution, which is phenomenologically less known, may benefit from lattice constraints even if moments were controlled to a more realistic 15\% uncertainty. Inclusion of lattice QCD nucleon isovector matrix elements in coordinate space was performed in Ref.~\cite{Bringewatt:2020ixn}, whereas inclusion of lattice QCD Ioffe time pseudo-distributions was performed in Ref.~\cite{Karpie:2023nyg}. In both cases, good agreement with experimental data was found, however a noticeable reduction of PDF uncertainties was observed only in the former case. Therefore, inclusion of lattice moments and matrix elements remains the most promising way of further constraining helicity PDFs. All in all, inclusion of lattice QCD data in helicity PDF fits shall remain a possibility to be actively pursued, albeit with some caution associated with the necessity to control systematic uncertainties in lattice results (cf. Sec.~\ref{sec:lattice_PDFs}). In this respect, we recommend that results be always published without and with inclusion of these.

\subsection{Nucleon unpolarized PDFs and LHC $\leftrightarrow$ EIC interplay}
\label{sec:unpol_PDFs}

Large-scale phenomenological explorations of unpolarized PDFs in the nucleon are central to precision tests of QCD across a broad interval 
of scattering energies~\cite{Amoroso:2022eow}. Modern global PDF fits combine a wide range of experimental inputs, including fixed-target deep-inelastic 
scattering (DIS), HERA inclusive and jet data, fixed--target Drell--Yan measurements, and LHC observables such as vector-boson, jet, and top-quark production. 
State-of-the-art PDF sets are consistently determined at next-to-next-to-leading order (NNLO) in QCD~\cite{Alekhin:2017kpj,Hou:2019efy,Bailey:2020ooq,NNPDF:2021njg,Moffat:2021dji,ATLAS:2021vod,PDF4LHCWorkingGroup:2022cjn}, 
with progress towards N$^3$LO accuracy~\cite{McGowan:2022nag,NNPDF:2024nan,NNPDF:2024dpb,Cridge:2024icl} 
and the inclusion of electroweak and QED effects~\cite{Xie:2021ajm,Cridge:2021pxm,NNPDF:2024djq}.

\subsubsection{The place of EIC studies of unpolarized PDFs in future QCD phenomenology.}

The high luminosity of planned EIC experiments opens various opportunities for transforming determinations of unpolarized PDFs especially at 
large partonic momentum fractions $x$, in the region relevant for searches for physics beyond the Standard Model (BSM) at the LHC~\cite{AbdulKhalek:2021gbh,Ball:2022qtp}. 
Here, the systematic EIC program would complement and surpass past measurements of the nucleon structure in fixed-target DIS and Drell-Yan experiments. 
Currently, a combination of fixed-target measurements on diverse nucleon and nuclear targets is required to disentangle quark and antiquark PDFs of various flavors at 
$x>0.01$ and $Q$ up to a few tens of GeV, in order to constrain nucleon structure for precision LHC measurements and BSM searches at $Q$ above several hundred GeV.
Inclusive and semi-inclusive DIS at the EIC would provide similar information directly using proton beams, at sufficiently large $Q$ to minimize higher-order and 
higher-twist contributions, and without potential contamination from long-distance nuclear effects.

For the EIC to be successful in its role of an incisive QCD experiment, both the experimental measurements and theoretical analysis must evolve beyond 
the current standards toward higher levels of accuracy and precision while maintaining mutual agreement of various measurements, {\it i.e.}, guaranteeing their replicability. 
A useful goal in this case is to strive to suppress relative PDF uncertainties to be below 1-2 percent, which requires a coordinated 
community effort not only to consistently implement a variety of perturbative QCD contributions at NNLO and N$^3$LO in $\alpha_s$, but also 
to achieve commensurate control of multiple systematic and methodological factors across all stages of the measurement and analysis. 
Public PDF fitting codes and analysis frameworks are a crucial help in this direction~\cite{xFitter:2022zjb,NNPDF:2021uiq,Kotz:2025mcj,
Costantini:2025wxp,Costantini:2026mxm,Cruz-Martinez:2026rct}. In this section we identify the important measurements that the EIC can 
perform, as well as the challenges it must address to be successful. 

\subsubsection{Large-$x$ PDFs and the role of the EIC.} Experimental facilities probe PDFs in distinct and complementary kinematic regions. HERA $ep$ DIS data constrain valence (foremost, up-type) quarks, some sea quarks, and the low-$x$ gluon, while LHC measurements are primarily sensitive to sea-quark flavor decomposition and the gluon at medium and large $x$. Fixed-target data remain essential for flavor separation at intermediate $x$, although tensions with collider data are observed in the large-$x$ region. Even today, large data sets from BCDMS, NMC, and other fixed-target DIS experiments provide strong constraints on large-$x$ PDFs \cite{Wang:2018heo,Hou:2019efy,Jing:2023isu}, in particular, on the gluon PDF at $x>0.1$ via violation of Bjorken scaling. The growing assortment of LHC measurements in this region does not eliminate the need in high-precision DIS experiments, as those are not affected by complex systematic factors typical for the LHC and are free of the possible influence of contributions from non-SM physics that the LHC aims to detect. Furthermore, some PDF flavors are still insufficiently separated. At $x>0.1$ there is some degeneracy among the $d$, $s$, and $g$ PDFs, as well as among the $\bar u$, $\bar d$, and $\bar s$ ones. With the current data, the accuracy of such separation for nucleon PDFs is further obstructed by nuclear-medium modifications present in DIS data sets obtained on deuteron and nuclear targets.

The 2021 EIC Yellow Report \cite{AbdulKhalek:2021gbh} included ample examinations of the EIC's potential for advancing the accuracy of unpolarized PDF determinations. It emphasized the value of high-statistics DIS measurements in a clean experimental environment to achieve  substantial reductions in several PDF uncertainties at large $x$ and moderate $Q^2$. 
Valence-quark distributions will be particularly well-constrained, with additional sensitivity to the gluon achieved through scaling violations and longitudinal structure functions~\cite{Khalek:2021ulf}. 

\begin{figure}[tb]
    \centering
    \includegraphics[width=0.49\linewidth]{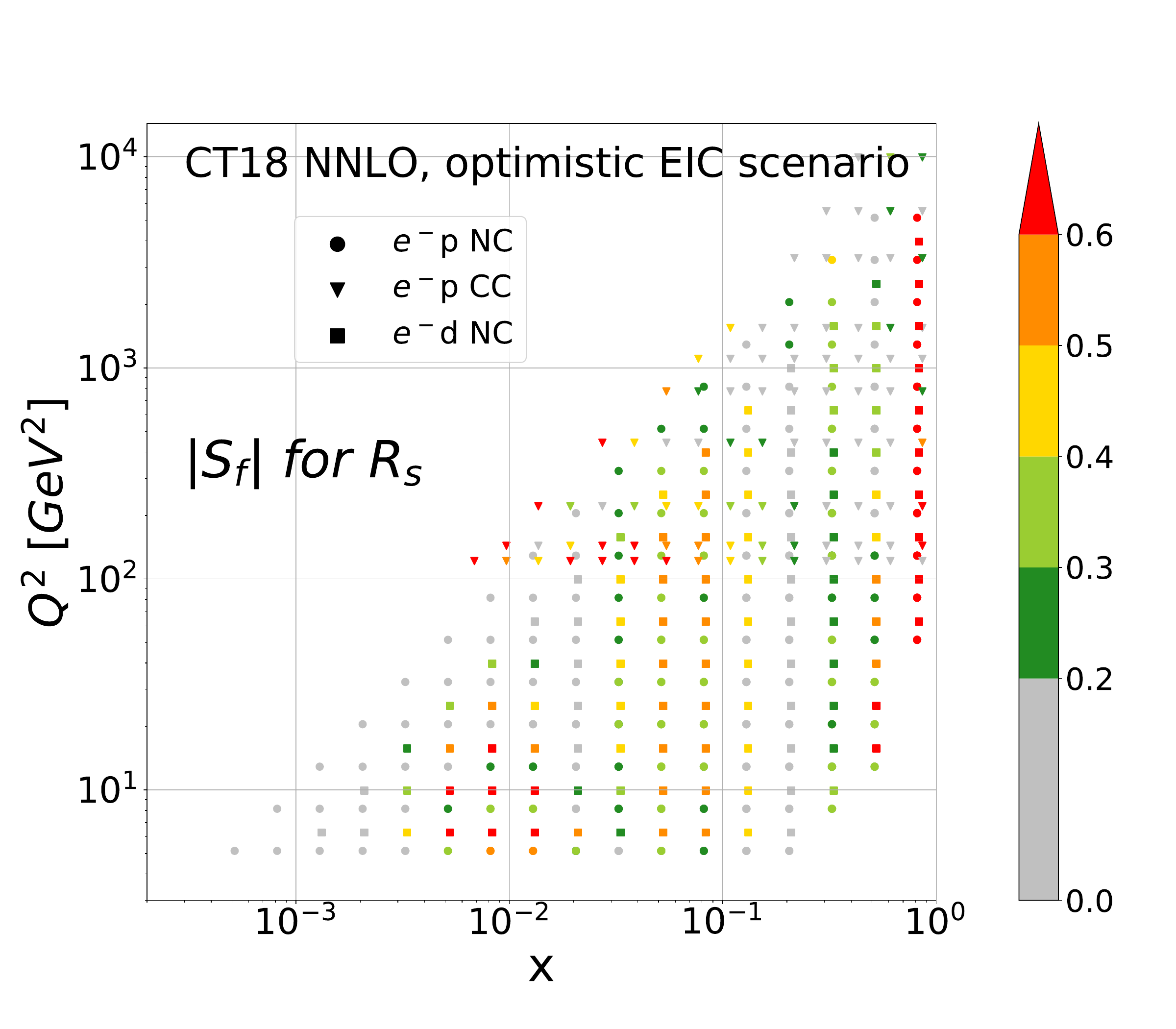}
    \includegraphics[width=0.49\linewidth]{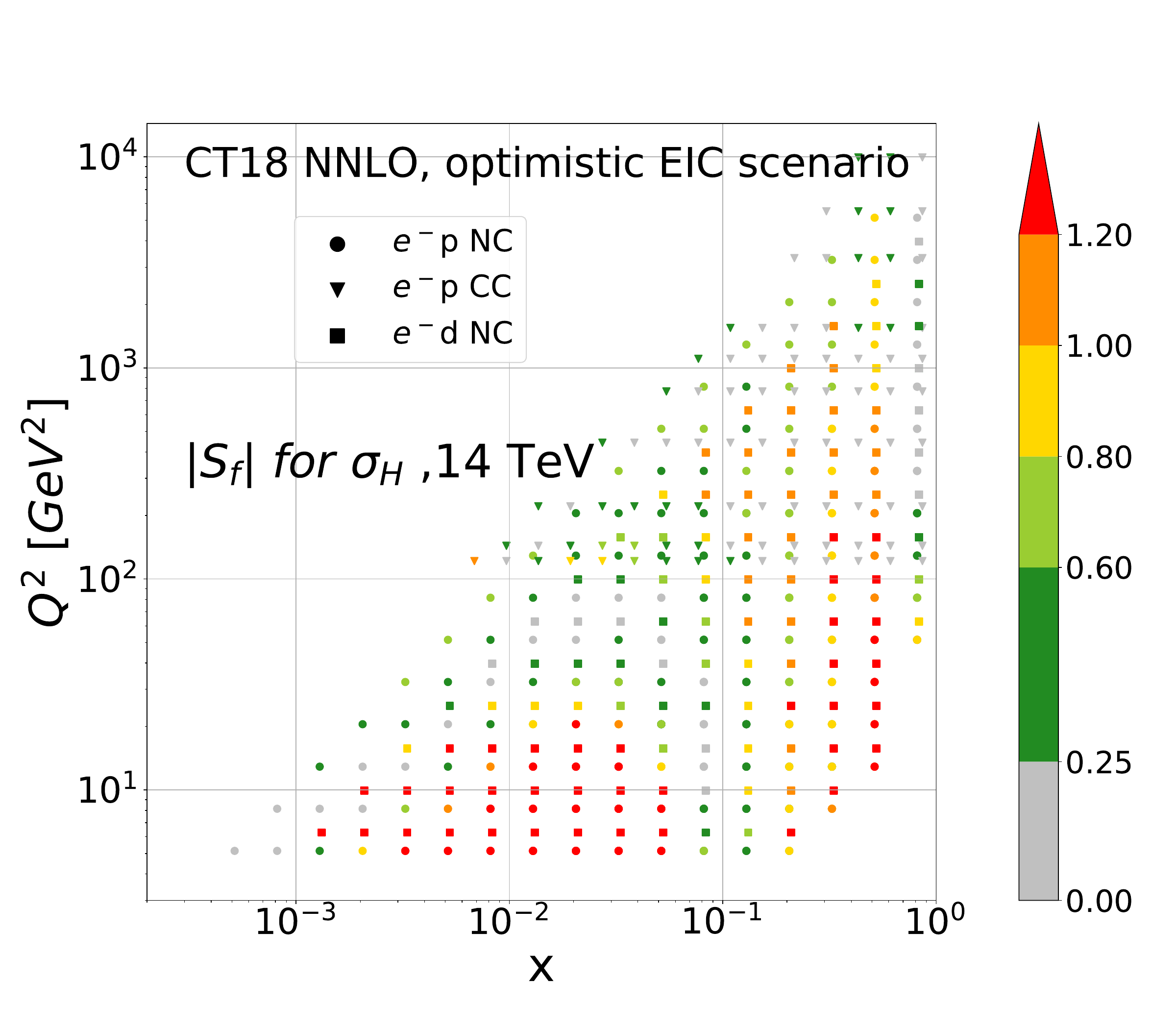}
    \caption{PDF-mediated sensitivities $|S_f|$ of the projected high-luminosity EIC data (left) to $R_s=(s(x,Q)+\bar s(x,Q))/(\bar u(x,Q) + \bar d(x,Q))$ at the $x$ and $Q^2$ values indicated on the axes; (right) to the PDF uncertainty of the Higgs boson production via gluon fusion at the LHC at 14 TeV. From \cite{AbdulKhalek:2021gbh}.}
    \label{fig:unpol_PDF_sensitivities}
\end{figure}

 Just to offer an example, Fig.~\ref{fig:unpol_PDF_sensitivities} applies the method of Hessian PDF sensitivities developed in Refs.~\cite{Wang:2018heo,Hobbs:2019gob} to quantify the EIC's potential for constraining various PDF combinations in neutral-current $ep$, $ed$ DIS, as well as charged-current $ep$ DIS. In the figures, the color of individual data points indicates their sensitivities according to the bar on the right. The sensitivities are normalized as in comparisons for the datasets in the CT14HERA2 and CT18 baselines in \cite{Wang:2018heo,Hou:2019efy}, also collected on the \texttt{PDFSense} website \cite{PDFSenseWebsite}, with the values above 0.4-0.5 identifying data points with substantial sensitivity. 

The left panel in Fig.~\ref{fig:unpol_PDF_sensitivities} demonstrates that, in addition to the already mentioned significant sensitivity to the valence $u$ quarks, a combination of $ed$ NC and $ep$ CC DIS measurements would offer an independent constraint on the strangeness-to-nonstrangeness ratio $R_s$ at $x>0.01$, where currently the primary relevant constraint arises from NuTeV dimuon DIS production on iron. The right panel shows that high-luminosity NC DIS at the EIC has a significant sensitivity to the combination of gluon PDFs that drives the main uncertainty in the SM Higgs boson production at the LHC. In this figure, sensitivities of individual colored datapoints over a large range of $x$ and $Q$ add up to the large total sensitivity of the EIC to gluon PDFs in the $x$ region relevant for Higgs physics. 
These sensitivity projections assume the optimistic scenario from the EIC Yellow Report with a high integrated luminosity, excellent control of systematic uncertainties, and reliable analysis techniques. In this case, the constraints of the EIC dataset would be comparable to those from the combination of fixed-target DIS datasets that currently constrain the PDFs. 
In the global fit, the net EIC impact would be determined by the relative strengths and mutual compatibility of constraints from the EIC and other experiments. 
An open question is how far in large $x$ the EIC can provide phenomenologically impactful constraints for the HL-LHC physics, which can be clarified by dedicated simultaneous impact studies for the projected EIC and HL-LHC measurements.

\subsubsection{BSM contamination and complementary Constraints.} 

\begin{figure}[tb!]
    \centering
    \includegraphics[width=0.7\textwidth]{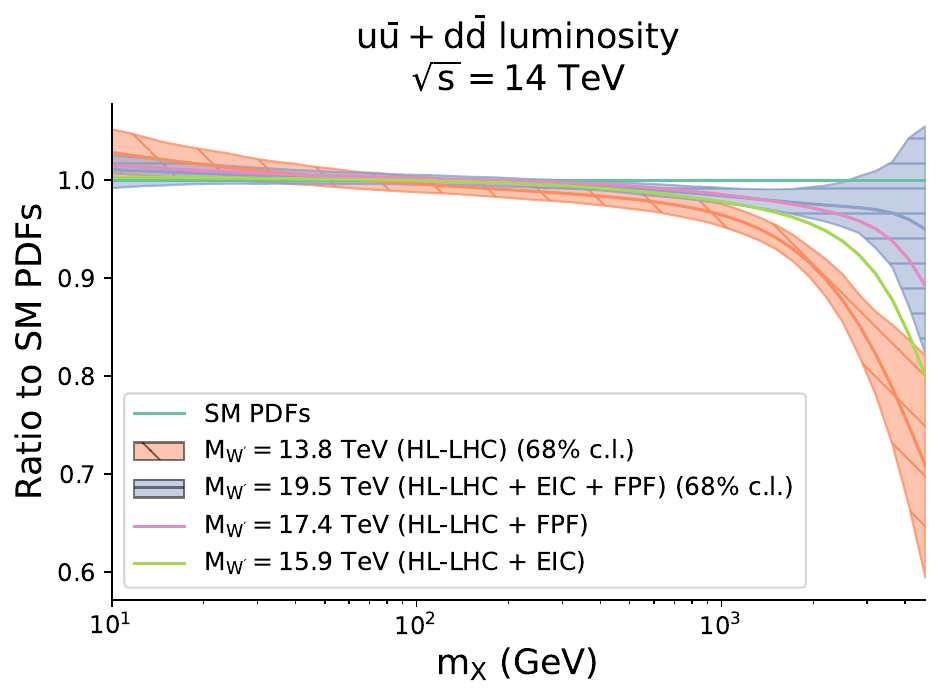}
    \caption{Impact of the BSM-induced bias on the $u\bar{u}+d\bar{d}$ luminosity for the different scenarios discussed in the main text, 
    all normalized to a consistent baseline PDF luminosity. 
    The orange bands correspond to the central value and PDF uncertainties of a global PDF fit including HL-LHC Drell-Yan high-mass projections. 
    The data have been generated according to a $W'$ model with $M_{W'}$ = 13.8 TeV and the fit is done assuming the SM.
The blue bands correspond to the same fit in which also the EIC and the CERN forward neutrino projected data are included
alongside the HL-LHC ones, thus moving the BSM-contamination threshold to $M_{W'}= 19.5$ TeV. The intermediate cases in which only EIC projections or only
FPF projections are included are shown with only their central values in green and pink respectively. From Ref.~\cite{Hammou:2024xuj}.}
    \label{fig:EICPDFsBSM}
\end{figure}

A critical issue for future precision phenomenology is the possibility that BSM effects may be inadvertently absorbed into PDFs~\cite{Carrazza:2019sec,Greljo:2021kvv,Iranipour:2022iak,Hammou:2023heg,Kassabov:2023hbm,Gao:2022srd,Costantini:2024xae,Hammou:2024xuj,Cole:2026eex}. 
In regions where PDF uncertainties are large, BSM contributions to high-energy observables can be compensated by distortions 
of PDFs without any significant deterioration of the quality of the global fit. This can lead to BSM-biased PDFs and potentially 
misleading conclusions in indirect BSM searches. An example of this mechanism is shown in Fig.~\ref{fig:EICPDFsBSM}, where 
PDF luminosities displaying different level of BSM-induced bias are compared.
The PDF luminosity plotted in the figure comes from a global fit of PDFs, in which -- on top of the full dataset included in an NNPDF4.0 analysis~\cite{NNPDF:2021njg} -- 
HL-LHC projections of high-mass NC and CC Drell-Yan invariant mass distributions are included in the fit. 
All data were generated by injecting a BSM model featuring a universally coupled $W'$~\cite{Farina:2016rws} with $M_{W'}$ set at different values, while the theory predictions 
in the PDF global fits are all produced assuming the SM. Each of the curves is compared to a consistent PDF luminosity obtained 
from a fit of data generated out of the SM using SM theory predictions. The orange band represents the $u\bar{u}+d\bar{d}$ luminosity fitted
injecting a $M_{W'} = 13.8$ TeV in the data. Here the luminosity shifts significantly as the 
PDFs have completely absorbed the BSM signal associated with $W'$. On the other hand, if the EIC NC and CC projections in the optimistic scenario discussed in Ref.~\cite{Khalek:2021ulf}   
and the CERN forward neutrino projected data discussed in Ref.~\cite{Cruz-Martinez:2023sdv} are included alongside the HL-LHC high-mass Drell-Yan projections, 
then the maximal BSM signal that can be absorbed in the PDFs is pushed to a larger value of $M_{W'}$ = 19.5 TeV. 
The corresponding PDF luminosities are displayed in blue. We observe a systematic reduction of the discrepancy with the consistent SM luminosities. 
The fit including EIC and CERN forward neutrino projections completely suppresses the discrepancy. The intermediate case
in which only EIC projections or only FPF projections are included are shown with only their central values in green
and pink respectively, according to the maximal BSM signal that can be absorbed in each of these cases. 

To conclude, EIC measurements are largely insensitive to high-scale BSM effects, making them a robust anchor for global PDF fits~\cite{Bissolotti:2023vdw}. 
The EIC's lower energy probing the large-$x$ region offers a mechanism to break the flat directions 
in PDF and BSM parameter space. By anchoring PDFs in kinematic regions that are insensitive to high-scale new physics, these data can expose inconsistencies that would otherwise 
remain hidden. 

\subsubsection{Prospects for testing lattice predictions for PDFs.} 
\begin{figure}[b]
    \centering
    \includegraphics[width=0.445\linewidth]{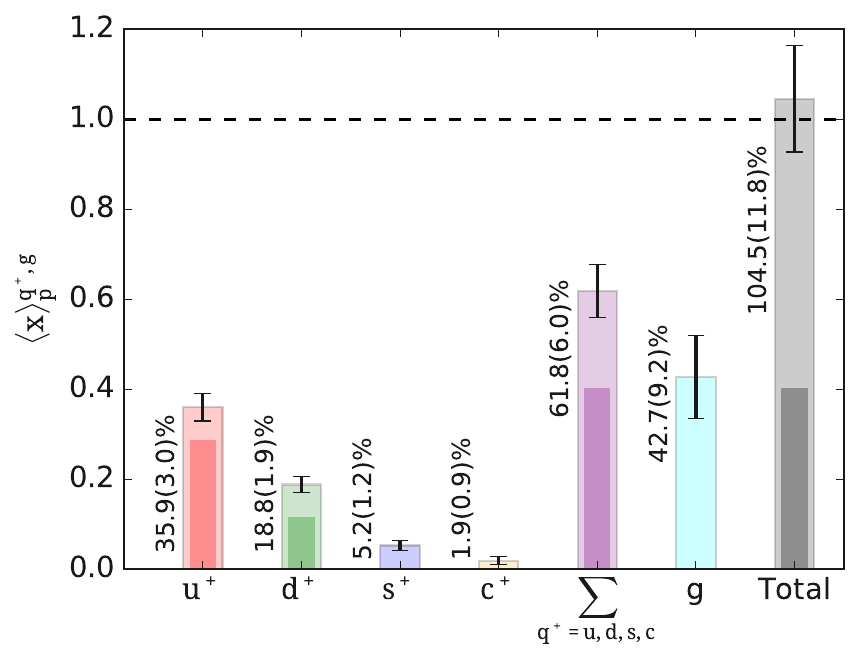}
    \includegraphics[width=0.535\linewidth]{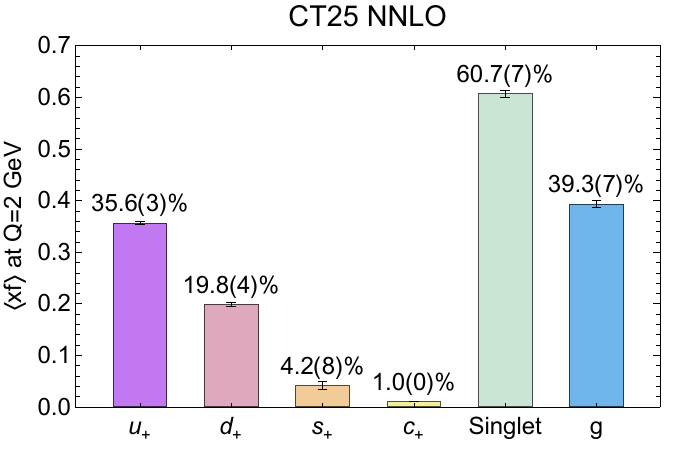}\\
    \caption{Partonic momentum fractions in a proton at $Q=2$ GeV obtained in (left) a lattice computation by the ETM Collaboration \cite{Alexandrou:2020sml} and (right) the CT25 NNLO PDF analysis. Numbers by the error bars indicate percentage momentum fractions and their respective uncertainties at 68\% CL (in parentheses) matching the rightmost significant figure of the central values.}
    \label{fig:unpol_mom_frac}
\end{figure}

Computations of PDFs and their Mellin moments is a rapidly developing domain of lattice QCD, with various insightful results expected by the time the EIC begins its operation~\cite{Lin:2017snn,Cichy:2018mum,Alexandrou:2020zbe}. In the last eight years, lattice QCD has progressed from first computations of the unpolarized nucleon PDFs to increasingly confident control of PDF systematics in the $x$ region between 0.2 and 0.8 that is also easily accessible at the EIC. Meaningful comparisons of lattice and phenomenological PDF models thus become increasingly possible. Unpolarized PDFs provide the best opportunity to calibrate various lattice methodologies via benchmark comparisons against accurate phenomenological PDF determinations from the experimental data. 
Higher Mellin moments, Ioffe-time distributions, and quasi-PDF matrix elements determined on the lattice may provide valuable information on the large-$x$ behavior, provided uncertainties from excited states, matching, and continuum extrapolation are reliably quantified. Establishing robust uncertainty budgets and performing lattice-specific closure tests will be essential steps toward integrating lattice results into global PDF analyses.

As an illustration of some of the opportunities, calculations of Mellin moments $\langle x^p \rangle_f$ of PDFs $f(x,Q)$ on the lattice mature quickly and can be already confronted against phenomenological determinations. An increasing number of such calculations is available for the Mellin moments in the nucleon \cite{Fan:2022qve,  Good:2023ecp, ExtendedTwistedMass:2024kjf, Alexandrou:2020sml}, pions and kaons \cite{Alexandrou:2020gxs,Alexandrou:2021mmi,ExtendedTwistedMass:2024kjf,   NieMiera:2025inn}.
Figure~\ref{fig:unpol_mom_frac} compares the second 
Mellin moments (partial momentum fractions in the proton) carried by specified PDFs in the ETMC lattice computation and CT25 NNLO global analysis. Not only do the central values of $\langle x\rangle_f$ agree within the current uncertainties, with further reduction of uncertainties can lattice QCD provide useful cross checks for the less well known momentum fractions, such as those for down and strange quarks. 

Given its excellent reach at large Bjorken $x$, the EIC is particularly suited for contributing constraints on even higher Mellin moments of $q_\pm = q\pm \bar q$ combinations of nucleon PDFs, including $\langle x^2\rangle$ for $u_- - d_-$ and $\langle x^3\rangle$ for $u_+ - d_+$. Hessian sensitivities computed in \cite{Hobbs:2019gob} indicate that the EIC would be very sensitive, compared to the current generation of hadronic experiments, to the lowest four moments of linear combinations of $u_\mp\equiv u\mp \bar u$, $d_\mp \equiv d \mp \bar d$ in neutral-current DIS with both electron and positron beams. 

\begin{figure}
    \centering
    \includegraphics[width=0.59\textwidth]{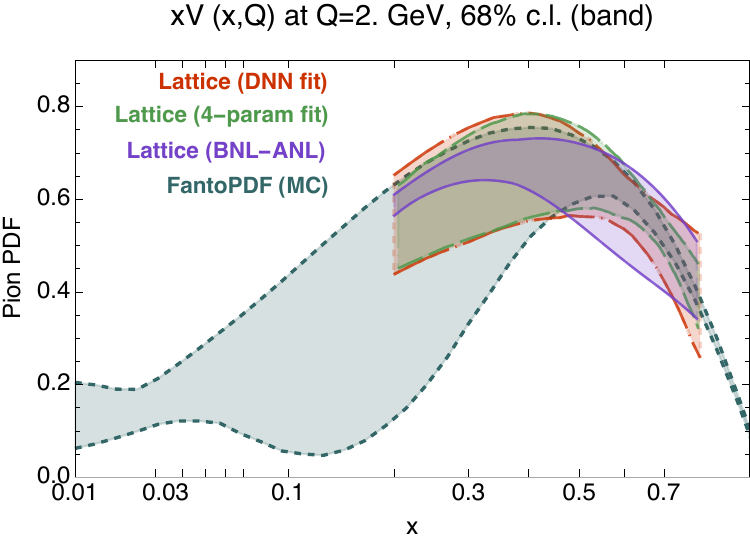}
    \includegraphics[width=0.39\textwidth]{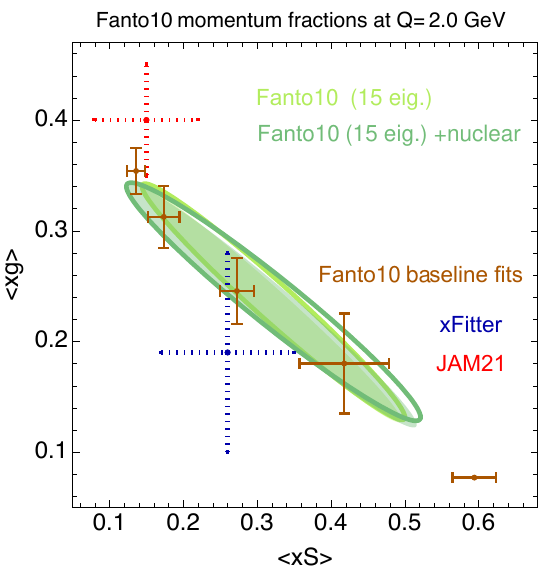}
    \caption{Left: Comparison of valence pion PDFs and uncertainties from the \texttt{Fanto10} NLO phenomenological analysis \cite{Kotz:2023pbu} and lattice QCD determinations \cite{Gao:2021dbh,Gao:2022iex}. Right: Pion total sea and gluon momentum fractions obtained with five \texttt{Fanto10} baseline fits (brown error bars) and combined ensembles at $1\sigma$ level \cite{Kotz:2023pbu,Kotz:2025lio}. We also show respective results from the JAM21 \cite{Barry:2021osv} (in red) and \texttt{xFitter} (in blue) studies \cite{Novikov:2020snp}.}
    \label{fig:FantoPI-nuclear_corr}
\end{figure}

Will the future comparisons of PDFs from global fits and lattice predictions be decisive? In the case of nucleon and meson PDFs, the answer will depend on the progress in quantifying and reducing uncertainties from a variety of independent factors both in the phenomenological and the lattice determinations.  This progress demands continued attention to advancements in PDF-fitting methodology, as discussed in the next subsubsection, as well as establishing standards for uncertainty quantification in the solution of relevant inverse problems arising in lattice studies.

As an example of the recent work along these lines, the pion PDFs are still relatively weakly constrained, and lattice QCD already provides valuable inputs to fits of pion PDFs \cite{JeffersonLabAngularMomentumJAM:2022aix}. However, the pion fits are affected by significant parametrization and other methodological uncertainties requiring dedicated methods that are still evolving. Thus, for instance, conclusions about the strength of constraints from the lattice compared to those arising from the experimental data depend on the estimated uncertainties of both, as illustrated by a comparison of error bands from the recent \texttt{FantoPDF} phenomenological NLO fit \cite{Kotz:2023pbu} and three lattice determinations in the left plot in Fig.~\ref{fig:FantoPI-nuclear_corr}. One might conclude from the figure that the BNL-ANL determination is already more constraining (has a smaller error) than the \texttt{FantoPDF} fit. In fact, this lattice determination is shown with only some of its total uncertainty, while the \texttt{FantoPDF} error band is much wider because it accounts for a large uncertainty due to the parametrization of pion PDFs that was not fully captured by the other analyses. The right plot in Fig.~\ref{fig:FantoPI-nuclear_corr} further drives the point that the parametrization uncertainty is important, this time in comparisons of Mellin moments. The figure compares the 
second moments of total sea and gluon PDFs in the pion, $\langle xS\rangle$ and $\langle xg\rangle$, obtained from recent analyses
by the JAM, \texttt{xFitter}, and CTEQ-\texttt{Fant\^omas} groups \cite{Novikov:2020snp, Barry:2021osv, Kotz:2025lio}. The cumulative uncertainty obtained with multiple functional forms of PDFs explored in the \texttt{FantoPDF 1.0} fit \cite{Kotz:2025lio}, here shown by the green ellipses, results in a wider span of the allowed  $\langle xS\rangle$ and $\langle xg\rangle$ combinations than in the JAM and \texttt{xFitter} pion fits to comparable datasets, but only using fixed parametrization forms.

\subsubsection{Methodological uncertainties and validation.} 
The demonstrated parametrization uncertainty in the case of pion Hessian PDFs is a part of the net methodological uncertainty present in all global fits, which also reflects the choices of priors, datasets, goodness-of-fit criteria, as well as algorithms for training and optimization of neural networks in the NN/MC fitting frameworks. Methodological uncertainties will be at the forefront of future EIC and HL-LHC studies involving PDFs. At present, despite the broad qualitative agreement among modern PDF sets~\cite{Chiefa:2025loi}, non-negligible quantitative differences persist in some phenomenologically important applications, particularly in parton luminosities relevant for high-mass processes. These differences may propagate into precision observables, such as extractions of $\alpha_s$~\cite{ATLAS:2023lsr}, the weak mixing angle~\cite{CMS:2024ony}, and the $W$-boson mass~\cite{ATLAS:2023fsi,CMS:2024lrd}, where PDF uncertainties constitute a 
substantial fraction of the total error budget, and where at times disagreement between the results obtained using different PDF sets is comparable 
to the total error of the measurement. As experimental uncertainties continue to shrink, understanding the origin, meaning, and robustness of PDF uncertainties has become the central challenge. PDF determination serves as a valuable testing ground for novel methodologies for precision data analysis and multivariate uncertainty quantification. The practical relevance of PDFs across many domains of high-energy physics motivates development and implementation of statistically sound methodologies in the PDF fits, as well as training of the HEP and nuclear physics communities in the proper usage of PDFs in diverse applications.

Several challenges in the PDF determination stem from the underlying inverse problem of inference of continuous 
functions for PDFs from a finite set of convolution integrals with heterogeneous experimental systematics. From the viewpoint of linear algebra, the general inference problem is 
undetermined unless additional assumptions are introduced about the PDFs to turn the problem into an (over)determined one. Consequently, PDF uncertainties reflect not only 
experimental errors but also theoretical assumptions and choices for the PDF parametrizations/neural network architecture, data selection, and fitting methodology.

Closure tests of several types are increasingly introduced by the PDF-fitting groups to check if the specific solution of the inverse problem is robust under 
such choices~\cite{DelDebbio:2021whr,Barontini:2025lnl,Harland-Lang:2024kvt}.  Closure tests provide a critical tool to validate PDF fitting frameworks, even though the statistical framework and the definition of the most effective test is still under development.
In a common "bias test", pseudo-data generated from a known underlying law are refitted, allowing one to assess whether the original PDFs are recovered within the quoted uncertainties. Closure tests can probe parametrization flexibility, fitting bias, and uncertainty propagation, and are essential for establishing the faithfulness of PDF error bands. Equally important are generalization tests against data not included in the fit~\cite{Cruz-Martinez:2021rgy,Chiefa:2025loi}. Thanks to modern tools enabling fast exact NNLO predictions~\cite{Carli:2010rw,Kluge:2006xs,Carrazza:2020gss,Devoto:2025cuf}, PDF sets can now be systematically confronted with unseen LHC and DIS measurements.

These comparisons also must account for experimental systematic, missing higher-order and power-suppressed QCD  uncertainties, together with their correlations. Existing studies indicate that present LHC data at $\sqrt{s}=13$~TeV only weakly discriminate among global PDF sets once all uncertainties are included~\cite{Chiefa:2025loi}, underscoring the need for complementary constraints.

An unresolved issue concerns the interpretation of uncertainties obtained with different statistical frameworks. The standard $\Delta \chi^2=1$ criterion could be reasonably applied to define the standard deviation on a parameter of the global fit \cite{Ablat:2025gbp} if this parameter is not strongly correlated with PDF shapes, and in the absence of hidden uncertainties. In practice, CTEQ, MSHT, and other groups use more complex definitions (tolerances) for PDF uncertainties. The relationship between Hessian determinations with non-trivial tolerances and Monte Carlo approaches is still under investigation \cite{Courtoy:2022ocu,Harland-Lang:2024kvt,Courtoy:2025ppd,Ablat:2025gbp}. Whether a tolerance choice close to unity can be regarded as statistically equivalent to Monte Carlo sampling in realistic global fits requires further systematic investigation.

\subsubsection{Parametrization dependence and tools.}
Parametrization dependence is another source of important epistemic uncertainty, especially in kinematic regions weakly constrained by data. While flexible functional forms and neural-network approaches have reduced overt bias, they may introduce new challenges related to overfitting, stability, and interpretability. Quantifying the impact of parametrization choices, and developing diagnostics to identify poorly constrained directions in PDF space, remains an active area of research.

Further progress will likely require a combination of improved statistical diagnostics, systematic closure testing across parametrization families, and enhanced tools for comparing and profiling PDF sets in a common framework, such as the recently published {\tt Fant\^omas} \cite{Kotz:2025mcj} and {\tt Colibri} platforms~\cite{Costantini:2026mxm}.

\subsubsection{Outlook and open questions.} 
Based on the discussion above, several open questions emerge for successful studies of unpolarized PDFs at the EIC. First, dedicated and realistic EIC impact studies should continue to quantify how far into the large-$x$ region EIC measurements can provide genuine discriminating power among modern PDF sets, and how this translates into improved sensitivity for BSM searches at the LHC. Second, further work is required to clarify the relationship between different statistical treatments of PDF uncertainties, including the interpretation of tolerances in Hessian fits and their connection to Monte Carlo approaches.

Third, the role of lattice-QCD inputs must be sharpened by identifying a set of lattice observables with maximal phenomenological impact, accompanied by robust and transparent uncertainty budgets. In particular, the potential of higher moments and related observables to constrain large-$x$ behavior deserves systematic investigation. Fourth, the genuine reduction of the PDF uncertainty requires advancements in control and representation of correlated systematic errors both in experiment and theory. Finally, further development of mathematical foundations of closure and generalization tests of global PDF analyses would enable reproducible assessments of methodological assumptions and reduce the risk of bias in precision and BSM interpretations.

PDFs remain a central limiting factor for precision SM measurements and BSM searches at current and future colliders. Continued progress requires advances in experimental control of systematics, methodological developments in global fits, and higher-order theoretical calculations with realistic uncertainty estimates.

The synergy between the LHC and the EIC is expected to play a crucial role in this effort. By providing precise, complementary constraints at large $x$, EIC measurements can both sharpen PDF determinations and prevent the absorption of potential BSM effects into PDFs. In this broader context, lattice-QCD inputs---while not yet competitive across the full kinematic range---may already provide valuable constraints in the large-$x$ region, further strengthening the interplay between theory and phenomenology.

\subsection{Nuclear PDFs}

Nuclear PDFs encode the modifications of quark and gluon distributions of protons and neutrons bound in nuclei with respect to those in free nucleons \cite{Klasen:2023uqj}. Clear physical interpretations of the observed shadowing (suppression) at low, antishadowing (enhancement) at intermediate, and the EMC effect (suppression, likely related to nuclear binding) at high $x$ still have to emerge, with nuclear and partonic interpretations currently competing. The same holds for clear evidence for saturation, e.g.\ at the EIC, arising from non-linear parton recombination and expected to be enhanced in nuclei.

Current global fits performed at NLO in QCD \cite{Duwentaster:2022kpv,Eskola:2021nhw,AbdulKhalek:2022fyi} differ not only in their proton baseline, but also in the assumed ansatz at the starting scale, the employed heavy-quark scheme, and the fitted experimental data. LHC data from weak boson to heavy (even top) quark production have significantly extended the kinematic range, but are mostly limited to lead nuclei. In some cases, their precision now requires going to NNLO \cite{Helenius:2021tof,Khanpour:2020zyu}, e.g.\ for CMS low-mass lepton pairs, but a few data sets (CMS $Z$-bosons at mid-rapdity, absolute ATLAS photon and CMS dijet cross sections) remain difficult to describe. A detailed comparison and benchmarking of the global nuclear PDF fits, as is customary for protons, would provide a combined estimate of the associated uncertainties. Also initial- and final-state higher-twist effects must eventually be disentangled \cite{Arleo:2025oos}.

Recently, xenon, oxygen, and neon have been successfully collided in the LHC. The EIC will for the first time provide clean DIS measurements of electrons on a large variety of nuclei in collider kinematics. It will thus significantly enhance the available information on the $A$-dependence of nuclear PDFs. Some sensitivity studies for the impact of these future EIC data exist \cite{Klasen:2017kwb, Khalek:2021ulf, Armesto:2023hnw}, but should be extended. New fixed-target data from JLab provide precise constraints at high $x$, but require the careful consideration e.g.\ of target-mass corrections \cite{Ruiz:2023ozv}. They allow for interesting model-dependent interpretations based, e.g.\ on short-range correlated nucleon pairs \cite{nCTEQ:2023cpo}. New precise neutrino data, e.g.\ from the Forward Physics Facility, would be very useful for flavor separation. Progress in lattice calculations on $x$-dependent PDFs for protons has been impressive, but is restricted to high $x$ and low $A$ (perhaps up to carbon in the future) for nuclei \cite{Winter:2017bfs, Detmold:2020snb}. However, even information on moments, e.g.\ for the strange quark, would be useful and could be combined with global fits.

\subsection{Pion PDFs}

As the lightest hadronic bound state in QCD, the pion displays unique physical characteristics at both low and high energies. On the one hand, it is the nearly massless pseudo–Nambu-Goldstone boson associated with chiral symmetry breaking, which is fundamental to understanding hadronic interactions at low energies. On the other hand, its internal quark and gluon structure can be resolved in high energy reactions, just like other hadrons. Historically, data from pion-induced Drell-Yan lepton-pair and prompt photon production on nuclear targets have provided constraints on the valence quark distributions at parton momentum fractions $x > 0.2$. To access the small-$x$ region, leading neutron electroproduction in DIS has been used to provide constraints on the sea quark and gluon PDFs via the Sullivan process~\cite{Sullivan:1971kd}, in which the proton splits into a $\pi^+$ and neutron, and detection of a forward neutron allows an interpretation in terms of DIS from a nearly on-shell pion.

Interest in pion PDFs has been generated recently with several analyses revisiting the determination of the valence quark PDFs in the high-$x$ region \cite{Aicher:2010cb, Barry:2018ort, Barry:2021osv}, including the effects of threshold resummation. The use of lattice data in global analyses of pion PDFs has also been explored by the JAM Collaboration. For example, reduced pseudo-ITD lattice data~\cite{Karpie:2018zaz,Joo:2019bzr} and matrix elements of current-current correlators \cite{Sufian:2020vzb} generated from the lattice by the HadStruc Collaboration were used to decrease the uncertainties on the PDFs and demonstrate consistency for the large-$x$ behavior found from phenomenology and from the lattice. Lattice data on gluonic pseudo-ITDs in the pion from the MSULat Collaboration~\cite{Good:2024iur} were also fitted simultaneously with experimental Drell-Yan and leading neutron electroproduction data by Good {\it et al.}~\cite{Good:2025nny} in order to significantly reduce the uncertainties on the pion's gluon PDF at $x > 0.2$. The reduced uncertainties revealed a significantly higher gluon density in the pion at large $x$ than in the proton.

Reduction of uncertainties on the gluon PDF at large $x$ can be obtained from the transverse momentum dependence of DY lepton pairs, although the impact of the existing data is relatively modest~\cite{Cao:2021aci}, and at present the behavior of the gluon PDF remains essentially unknown at $x > 0.1$. In fact, recently the Fant\^{o}mas group \cite{Kotz:2023pbu, Kotz:2025lio} noted that analysis of the DY data and a restricted set of LN data with a more general PDF parametrization based on B\'{e}zier curves may be consistent with a zero gluon distribution at the input scale.

Very recently, the first global QCD analysis of kaon PDFs was performed by the JAM Collaboration~\cite{Barry:2025wjx} using existing Drell-Yan kaon scattering data together with constraints from the lowest three lattice moments of $K$ (and $\pi$) PDFs computed by the ETM Collaboration \cite{ExtendedTwistedMass:2024kjf, Alexandrou:2021mmi}. 
The behavior of the valence and sea quark PDFs in the kaon will be further elucidated through measurements of the kaon-induced DY cross sections in the AMBER experiment~\cite{Adams:2018pwt} at CERN. Other experiments, such as at Jefferson Lab with 11~GeV or upgraded 22~GeV electron beams, as well as the Electron-Ion Collider (EIC), can access the kaon PDFs via the Sullivan process~\cite{Sullivan:1971kd} through detection of a $\Lambda$ hyperon in coincidence with the scattered electron, $e p \to e \Lambda X$. The Jefferson Lab experiments would kinematically access the valence region, having substantial overlap with AMBER, while the EIC would be sensitive to sea quark and gluon distributions at small $x$.

\subsection{Photon PDFs}

The photon is the fundamental gauge boson of QED. However, in high-energy reactions it can also exhibit a complex hadronic structure. There, it interacts not only directly with a hadronic target, but, similarly to vector mesons with the same quantum numbers, also through its partonic constituents. Similarly to hadron PDFs, the $Q^2$-dependence of photon PDFs is governed by perturbative evolution equations, but with additional inhomogeneous terms originating from the QED photon splitting to quark-antiquark pairs. Boundary conditions for the $x$-dependence at the starting scale can then be imposed by the Vector Meson Dominance model \cite{Klasen:2002xb}.

The photon structure function $F_2^\gamma$ has been measured at the $e^+e^-$ colliders PETRA, PEP and LEP~\cite{Nisius:1999cv}. At LO, it is directly related to the up-and down-type quark PDFs, weighted by their squared fractional charges. The coefficient and splitting functions for $F_1^\gamma$ and $F_2^\gamma$ are known up to NNLO~\cite{Moch:2001im}. Jet photoproduction is in addition sensitive to the gluon PDF. It has been measured in $ep$ collisions at HERA and $e^+e^-$ collisions at LEP, but this has so far not been exploited. Jet photoproduction has been calculated in approximate NNLO~\cite{Klasen:2013cba}, and the NLO corrections have been matched to parton showers in SHERPA~\cite{Meinzinger:2023xuf} and POWHEG~\cite{Feike:2025plq}.

\begin{figure}
    \centering
    \includegraphics[width=0.8\linewidth]{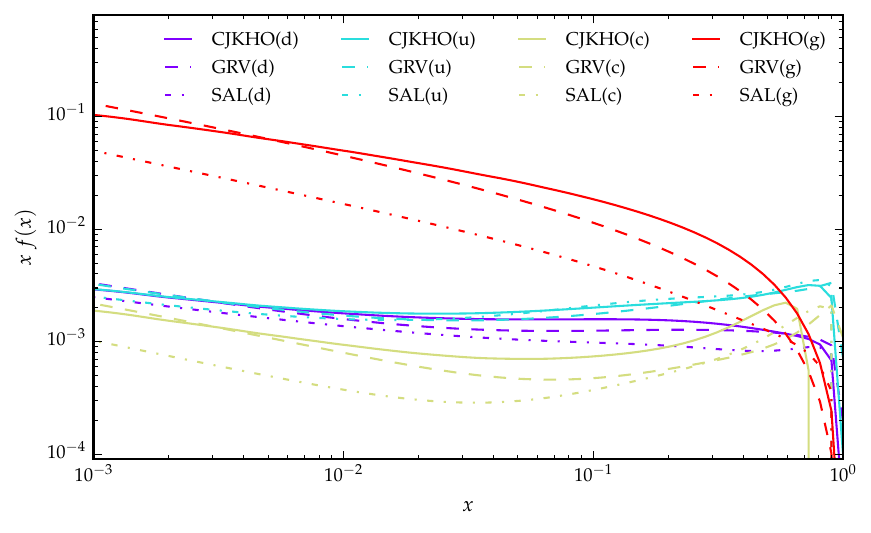}
    \caption{Comparison of up, down, charm and gluon PDFs in the photon at $\mu^2=20$ GeV$^2$ as determined in two of the most recent (2004-2005) analyses~\protect\cite{Cornet:2004nb,Slominski:2005bw} and a 1991 analysis~\protect\cite{Gluck:1991jc}. }
    \label{fig:photonpdfs}
\end{figure}

The up, down, charm and gluon PDFs in the photon are compared at $\mu^2=20$~GeV$^2$ in Fig.~\ref{fig:photonpdfs}. The 2004-2005 CJK \cite{Cornet:2004nb} and SAL \cite{Slominski:2005bw} NLO photon PDFs used all available data of $F_2^\gamma$ from the LEP experiments, whereas the older 1991 GRV \cite{Gluck:1991jc} parametrization could use only lower-energy measurements. The limited reach and precision of the experimental data is reflected in the spread of the distributions in Fig.\ \ref{fig:photonpdfs}, in particular at $x\leq0.1$ and for the gluon.

To prepare for photoproduction measurements at the EIC, a modern global photon PDF analysis including HERA and LEP jet data, based on the now available theoretical tools, would be highly desirable. At lower energy, BELLE measurements could contribute new structure function data, and the possible impact of lattice QCD calculations should be explored. A qualitative step forward can then be expected from high-precision EIC photoproduction measurements, not only for photon, but also proton and nuclear and even pion PDFs by exploiting the tagging of forward neutrons~\cite{Klasen:2001sg}.

\subsection{Experimental opportunities for PDFs} 

The precision extraction of PDFs requires input from experimental data that spans several decades of kinematic range, in both $x$ and $\sqrt{s}$. It is not currently possible for a single experiment or facility to provide all of the combinations of detector configuration and accelerator capabilities necessary to access the full kinematic landscape. Focusing on lepton-proton/ion scattering, there are several current and planned facilities that will collide beams at center-of-mass energies $1<\sqrt{s}<5\times10^3$ GeV$^2$ and probe momentum fractions down to $x\sim10^{-7}$.  

\begin{figure}
    \centering
     \begin{subfigure}[b]{0.45\textwidth}
         \centering \includegraphics[width=\textwidth]{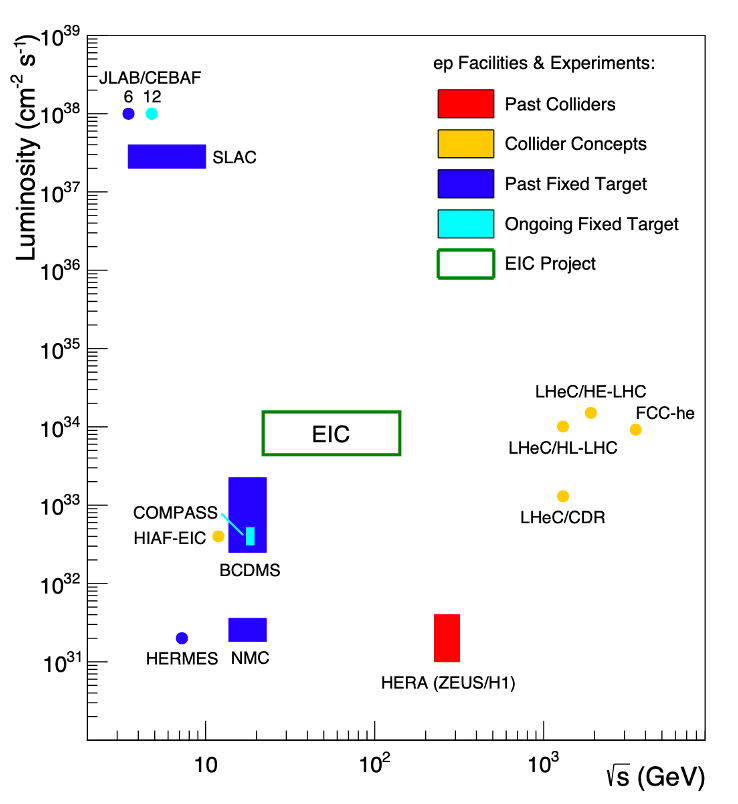}
         \caption{The landscape for selected $ep$ scattering facilities in terms of luminosity and center-of-mass scattering energies \cite{Ullrich:2020}.}
         \label{fig:FacilitySynergies}
     \end{subfigure}
     \hfill
     \begin{subfigure}[b]{0.45\textwidth}
         \centering
         \includegraphics[width=\textwidth]{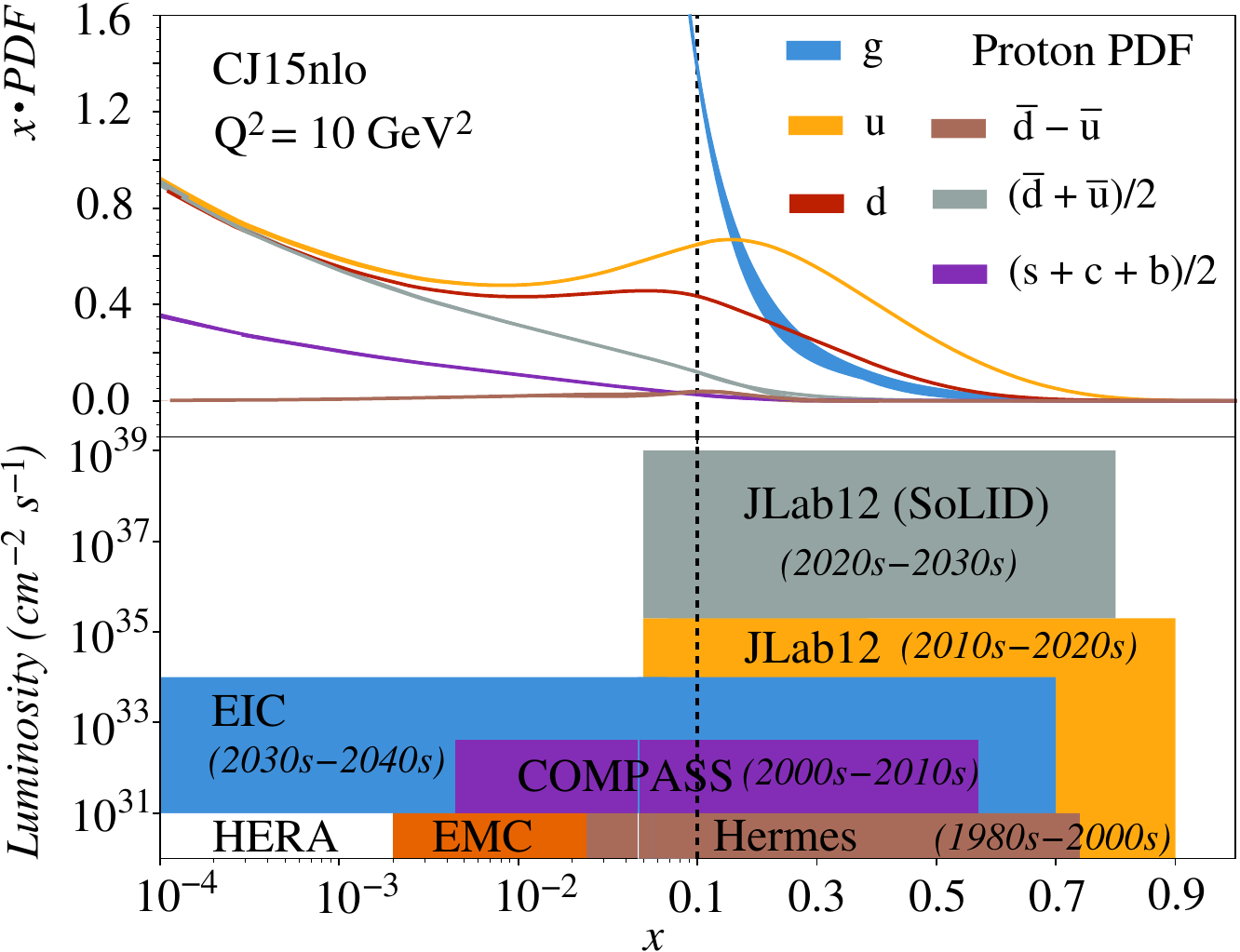}
         \caption{Kinematic regions of deep inelastic scattering and the comparative reach of EIC and 12-GeV CEBAF, as well as other facilities and experiments compared with parton
distributions from CJ15~\cite{PhysRevD.93.114017}. Taken from Ref.\ \cite{Arrington_2023} and adapted from~\cite{ARRINGTON2022103985}.}
         \label{fig:plotDIS}
     \end{subfigure}
        \caption{Synergies in luminosity and kinematic coverage for past, current and future electron-proton scattering experiments and facilities.}
        \label{fig:overall_figure}
\end{figure}

Figure \ref{fig:FacilitySynergies} compares the luminosity and scattering energy of facilities and experiments with current or planned deep-inelastic scattering programs. The fixed target program at the 12 GeV CEBAF at Jefferson lab (JLAB) provides some of the highest luminosities available.  Although the design luminosities planned for the EIC and the Large Hadron electron Collider (LHeC) are orders of magnitude below those at JLAB, they will run at higher luminosities than previous collider experiments and at higher $\sqrt{s}$ than fixed target programs, significantly expanding the range of $x$ and $Q^2$ that can be probed. Figure~\ref{fig:plotDIS} demonstrates the complementarity and synergistic aspect of the 12-GeV CEBAF and the future EIC. The program at JLAB will utilize high luminosity polarized electron beams, combined with a unique suite of detectors ~\cite{ARRINGTON2022103985}, to study hadron (nucleons, pions and kaons) and nuclear PDFs in the important valence quark region, while providing overlap with EIC measurements. The JLAB program includes studies of unpolarized and polarized PDFs and the three-dimensional tomography of the nucleon in momentum space (TMDs) and the hybrid momentum-coordinate space (GPDs) in Halls A, B and C. These measurements will also be pursued at the EIC\cite{Accardi:2012qut,AbdulKhalek:2021gbh}, but with a focus on the sea at low $x$. If realized, the LHeC\cite{Andre:2022xeh} and the LHC Forward Physics Facility (FPF)\cite{FPF:LOI,ANCHORDOQUI:20221} would usher DIS experiments into the TeV regime, potentially extending the $x$ coverage down to $~10^{-7}$. These facilities will be discussed in more detail in the following paragraphs.

 {\bf The Solenoidal Large Intensity Device (SoLID)}~\cite{Arrington_2023} is a next-generation instrument proposed for Hall A at JLab.
SoLID at the QCD intensity frontier is the capstone of the JLab 12-GeV science with exceptional potential to lead nuclear physics research worldwide. The three science
pillars of SoLID include imaging quarks in three-dimensional momentum space inside the nucleon with
SIDIS, extracting the nucleon mass and scalar profile densities and corresponding radii with threshold $J/\psi$ production, and searching for BSM physics and QCD/PDF studies with parity-violating DIS (PVDIS). The uniqueness of SoLID is its capacity to combine high luminosity with large acceptance handling
unprecedented luminosities ($10^{37}$ cm$^{-2}$ s$^{-1}$ for the SIDIS and J/$\psi$ and 10$^{39}$ for PVDIS configurations) and rates, orders of magnitudes higher than any existing or planned facility (see Figure~\ref{fig:plotDIS}). Additionally, the SoLID collaboration has developed a competitive program on GPDs, especially with the double DVCS (DDVCS) measurement.
Therefore, SoLID will provide data on all three pillars of its scientific program with unprecedented precision and/or sensitivity in the valence quark region.

The base equipment in {\bf Hall B} is the CEBAF Large Acceptance Spectrometer for operations at 12 GeV beam energy (CLAS12) to study electro-induced nuclear and hadronic reactions with efficient detection of charged and neutral particles over a large fraction of the full (4$\pi$) solid angle.  Fast triggering and high data-acquisition rates
allow operation at a luminosity of $10^{35}$ cm$^{-2}$ s$^{-1}$. These capabilities are being used in a broad program to study the
structure of the nucleons, nuclei, and mesons, using polarized and unpolarized electron beams and targets for beam energies up to 11 GeV, and especially for studies of TMDs through SIDIS, and GPDs via DVCS, DVMP, time-like Compton Scattering and DDVCS through potential future luminosity upgrade.

The base equipment in {\bf Hall C} consists of the high-momentum spectrometer (HMS), and the super high momentum spectrometer (SHMS). With routinely operating luminosities of $10^{38}$ to $10^{39}$ cm$^{-2}$ s$^{-1}$, they provide a unique ability to measure small cross sections which demand high luminosity and facilitate the careful study of systematic uncertainties. Important studies such as L-T separation of SIDIS can be performed in Hall C, providing unique tests of factorization and $Q^2$ evolution.  The neutral particle spectrometer (NPS), an addition to the base equipment, opens the door to study neutral pion production in both SIDIS and Deeply Virtual Meson Production processes for TMD and GPD studies and pion PDFs. 

The {\bf ePIC (Electron-Proton/Ion Collaboration) Detector}, the flagship experimental instrument being built at the EIC, will be located at one of two possible interaction regions. It is a general purpose detector designed to carry out the EIC’s broad physics program which requires precision reconstruction of the scattered electron and hadron momentum and energies for acceptances of $-3.5<\eta<3.5$ and $2\pi$ in azimuth. The ePIC detector package incorporates next-generation technologies and is designed to provide precision charged track vertex and momentum reconstruction and particle identification, as well as hadronic and  high-resolution electrogmagnetic calorimetery.  Far-forward and backward detectors are integrated into the interaction region, allowing for the detection and reconstruction of protons, neutrons, photons and low $Q^2$ electrons that are scattered along the beamlines. Data will be collected with a streaming data acquisition system and will use machine learning and AI techniques to classify events and perform real-time event reconstruction and analysis.

Global analyses~\cite{Borsa:2020lsz,Adamiak:2021ppq,AbdulKhalek:2021gbh} that incorporate simulated inclusive DIS scattering data with realistic EIC statistical and systematic errors show significant constraints on both the gluon and total quark helicity distributions, especially at lower $x$, but also extending into the valence region. The constraints on the flavor separated distributions are enhanced with the incorporation of SIDIS data as well as polarized $^3$He running. The impact is not limited to the polarized sector. A recent analysis of neutral current cross section measurements in the ePIC detector \cite{PhysRevD.109.054019} indicates that the strongest constraints relative to the HERAPDF2.0 and MSHT20 fits, come in the high $x$ regime and in particular for the up quark. For nuclear PDFs, ePIC will use heavy flavor channels to provide insights on the gluon densities in the anti-shadowing and EMC regions\cite{AbdulKhalek:2021gbh}. These measurements will be complementary to the constraints placed by measurements at the LHC in the shadowing region. Measurements of inclusive NC channels in $e+A$ will have significant impact on the $u/\bar{u}$ distributions\cite{PhysRevD.109.054019}.

In addition to the collinear 1D PDFs and nPDFs, the ePIC Collaboration is pursuing a rich program of TMD and GPD measurements. At the EIC TMDs will be accessed via jet and heavy flavor channels in addition to the traditional SIDIS channels. The higher $\sqrt{s}$ of the EIC will translate into smaller corrections, allowing the data to be more easily incorporated into global analyses. The GPD program will be one of the most challenging and luminosity hungry experimental programs at the EIC. Measurements of the DVCS channel require precision calorimetry to reconstruct both the electron and the photon and backward detectors, aligned closely to the beamline to tag the scattered proton. The ePIC detector is well equipped for these measurements and will be able to provide new insights on the sea quark and gluon TMDs and GPDs \cite{AbdulKhalek:2021gbh}.

\begin{figure}
    \centering
    \vspace{-1.5in}
    \includegraphics[width=0.80\linewidth]{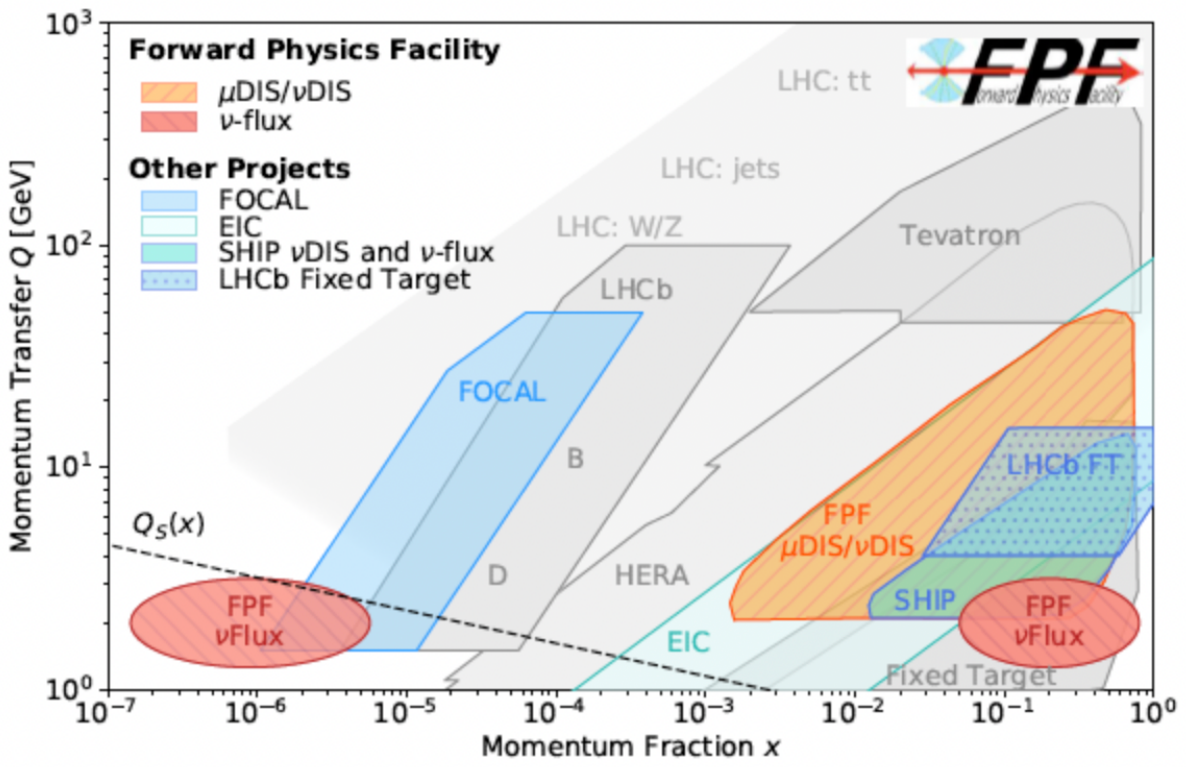}
    \vspace{-1.5in}
    \caption{Kinematic coverage of measurements at the proposed FPF including both
neutrino production (red) and neutrino/muon scattering (orange),
compared to past/ongoing experiments (gray), and other proposed
or planned experiments (blue) in the partonic momentum fraction $x$
versus momentum transfer $Q$ plane. Taken from Ref.\ \cite{FPF:LOI}.}
    \label{fig:FPF}
\end{figure}

{\bf The LHC Forward Physics Facility (FPF)}~\cite{FPF:LOI} is a proposed extension of the HL-LHC program to utilize the intense flux of high energy neutrinos and muons in the far-forward direction for a diverse set of physics programs including QCD physics. The FPF as a neutrino-ion collider has a center-of-mass energy range of 10-70 GeV, and the corresponding neutrino/muon DIS kinematics is shown in Figure~\ref{fig:FPF} as the yellow shaded area with a good overlap with the EIC and fixed target experiments. Additionally via forward neutrino production, the FPF has the capability to probe the gluon PDF down to $x \approx 10^{-7}$ and at very large-$x$, shown by the red ellipses in Figure~\ref{fig:FPF}. The FPF, complementary to the EIC and the 12-GeV CEBAF, provides opportunities to study many important QCD topics including quark and gluon PDFs of the nucleon and nuclei, the currently poorly studied $F_4$ and $F_5$ structure functions, the intrinsic charm content of the nucleon, gluon saturation, cold nuclear matter and more.

%% file: week3.tex
\subsection{Toward quantification of uncertainties 
}

\subsubsection{Introduction: Saturation/CGC framework, what progress do we need for the EIC?}
\label{sec:Intro}

Unlike QED and gravity where classical phenomena are ubiquitous and quantum features are, respectively, elusive or sub-dominant, QCD is an intrinsically quantum theory. Gluon saturation~\cite{Gribov:1983ivg,Mueller:1985wy} is the mechanism in QCD whereby classical dynamics can be realized, where the dynamics, in analogy to Maxwell's equations, and Einstein's equations, is described by the Yang-Mills equations~\cite{McLerran:1993ni,McLerran:1993ka,McLerran:1994vd}. In the gluon saturation regime, $F_{\mu\nu}^2\sim O(\frac{1}{\alpha_S})$, when $\alpha_S(Q_S^2)\ll 1$. Here 
$Q_S^2 \gg \Lambda_{\rm QCD}^2$ is emergent scale characterizing the saturation regime. It reflects the statement that a probe with a given resolution ($Q^2\gg \Lambda_{\rm QCD}^2$), will scatter off a hadron or nucleus with unit probability at sufficiently high boost (or small $x$). Thus classicalization and unitarization go hand-in-hand in the saturation regime. 

Gluon saturation thus corresponds to a non-perturbative strongly correlated regime of QCD whose dynamics is accessible through weak coupling. It is generated through copious bremsstrahlung with increasing boost, but the process of unitarization itself is not accessible in perturbation theory. (One can never generate field strengths of $1/\alpha_S$ in perturbation theory.) However, the kinematics of scattering in the Regge asymptotics of interest, and the corresponding generation of states of large color charge, allow one to construct a Born-Oppenheimer type EFT of quasi-static classical color sources at large $x$ and dynamical gauge fields at small $x$. This is the Color Glass Condensate (CGC)~\cite{Iancu:2003xm,Gelis:2010nm}, whose construction is robust for large nuclei with atomic number $A\gg 1$. The saturation scale in the CGC is a natural consequence of having large representations of color charge that are boosted to high energies. For a large nucleus, simple estimates (confirmed by more detailed comparisons with existing e+A data~\cite{Kowalski:2007rw})
give $Q_S^2 \propto A^{\alpha} (\frac{1}{x})^\lambda$, where simple estimates give $\alpha\approx 1/3$; the value of $\lambda$ can be computed in perturbation theory, and depends on the value of the running coupling.

The CGC effective action has a saddle point described by solutions to the classical Yang-Mills equations describing the dynamics of the small $x$ gauge fields in the presence of the highly boosted classical color currents that are static on the light cone. This saddle point classical configuration is sometimes called a shockwave~\cite{McLerran:1994vd,Balitsky:1995ub} because it is highly localized in light cone $x^-$, when boosted relative to a probe. Such a configuration breaks both Poincar\'{e} invariance and the global symmetry of large gauge configurations. One can construct shockwave quark and gluon propagators that match smoothly small fluctuations in the distinct pure gauge configurations across the shockwave boundary~\cite{Ayala:1995kg}. 

The problem of precision physics in the gluon saturation regime can then be understood as computing the 
scattering of a DIS probe off such a shockwave configuration. At small $x$, the probe is well-approximated as a quark-antiquark dipole that scatters off the shockwave. At next-to-leading order, the quark-antiquark pair emits a gluon with this $q\bar q g$ configuration now scattering off the shockwave. Though this process is naively suppressed by $\alpha_S$, there is a phase space enhancement of the scattering probability $\alpha_S \ln(1/x)$ which is $O(1)$ when $x\ll 1$. A systematic way to address the problem is analogous to the usual OPE picture of perturbative coefficient functions (often called "impact factors") and non-perturbative operators (expressed as correlators of lightlike Wilson lines), except now the scale separating the two is in a boost value (or rapidity or $x$),  as opposed to in the resolution (factorization) scale. 

The leading order description of the scattering of a quark-antiquark dipole can be expressed as a ``dipole" correlator of lightlike Wilson lines at the transverse spatial positions of the quark and the antiquark. The Wilson lines represent simply the color rotation experienced by the quark/antiquark upon their interaction with the shockwave background. The emission of a gluon from the dipole includes a contribution that results in  color rotation of the gluon off the background field, corresponding to an (adjoint) Wilson line. The less inclusive the final state is, the more sensitive it is at leading order to higher point Wilson line correlators; besides the aforementioned dipoles, they include quadrupoles, sextupoles, and so on. These  correlators encapsulate the rich non-perturbative many-body physics of the gluon saturation regime of QCD. Determining the energy evolution of these correlators is the principal motivation of the precision program of the EIC at small $x$. As we will discuss below, the renormalization group equations governing these constitute the Balitsky-JIMWLK hierarchy. 

\subsubsection{Higher order $\alpha_s$ calculations at small $x$.}
\label{sec:NLO}

While various signatures of gluon saturation dynamics have been observed at colliders, large theoretical uncertainties inherent to leading order computations, conceal the unambiguous discovery that gluons saturate inside hadronic matter. In recent years, tremendous efforts have been conducted to advance gluon saturation to a precision science, a timely endeavour as we approach the EIC era. These advances can be broadly classified as follows: i) the determination of evolution equations to next-to-leading logarithmic (NLL) accuracy $(\alpha_s^2 \ln 1/x )$ enhanced with (anti-)collinear resummation, ii) the analytic computation of process-dependent and perturbatively calculable impact factors at NLO, and iii) their proper numerical implementation and comparison to experimental data. The discussion here is mostly concerned with ii) and iii), with special emphasis on various two-particle correlation observables.\\

The leading logarithmic (LL) version of the BK-JIMWLK equations predicts too fast evolution with energy \cite{Albacete:2004gw}, suggesting the need to go beyond LL for controlled phenomenological predictions. An important subset of corrections beyond the LL evolution, pertaining to the running of the coupling, have been derived for the BK equation in \cite{Kovchegov:2006vj,Balitsky:2006wa} and for JIMWLK in \cite{Lappi:2012vw}. The former corrections have been extensively applied to phenomenology (see e.g. \cite{Albacete:2009fh,Albacete:2010sy,Lappi:2013zma,Mantysaari:2018zdd}). Presently, the next to leading logarithm (NLL) contributions to the BK equation have been derived in \cite{Balitsky:2008zza,Balitsky:2009xg,Balitsky:2009yp}. The first numerical results \cite{Lappi:2015fma}  showed that these corrections are unstable and contain potentially large negative terms. As in the case of BFKL \cite{Kwiecinski:1997ee,Salam:1998tj,Ciafaloni:1999yw,Altarelli:1999vw,Ciafaloni:2003rd,SabioVera:2005tiv}, the BK equation at this order requires resummation of large (anti-)collinear logarithms which have been implemented in its collinearly improved version \cite{Iancu:2015vea,Ducloue:2019ezk,Ducloue:2019jmy}.  The stability of the collinearly improved BK equation has been numerically confirmed in \cite{Lappi:2016fmu}, restoring predictive power. Likewise, the JIMWLK set of equations at NLL has been obtained in \cite{Balitsky:2013fea,Kovner:2013ona,Kovner:2014lca,Lublinsky:2016meo}, and the collinearly improved version in \cite{Hatta:2016ujq}.\\

For precision computations of physical processes, one also requires high order computations of the corresponding impact factors. The impact factors for DIS structure functions, back-bone of small $x$ observables, have been obtained at NLO for massless quarks in \cite{Balitsky:2010ze,Beuf:2017bpd,Hanninen:2017ddy}, with the first numerical results reported in \cite{Ducloue:2017ftk}. These efforts combined with the most dominant NLL contributions to the BK equation have resulted in the first fit of HERA data at NLO+NLL accuracy within the saturation framework \cite{Beuf:2020dxl}. The fit provides an excellent description of the reduced cross section across different values of Bjorken-$x$ and $Q^2$. The impact factors have been derived for exclusive light vector mesons in \cite{Boussarie:2016bkq}, exclusive dijets in \cite{Boussarie:2016ogo}, and exclusive dihadrons in \cite{Fucilla:2022wcg}, and diffractive structure functions in \cite{Beuf:2024msh}. The evaluation of heavy vector meson production has recently become available in \cite{Mantysaari:2021ryb} including numerical results in forward kinematics.  Recent developments at small $x$ have also been put forward for semi-inclusive processes, including single hadron production in proton-nucleus collisions \cite{Chirilli:2011km,Chirilli:2012jd}, single inclusive hadron/jet production \cite{Caucal:2024cdq,Altinoluk:2024vgg,Bergabo:2022zhe,Altinoluk:2025dwd}, dijet production \cite{Caucal:2021ent}, dihadron production \cite{Bergabo:2022tcu,Bergabo:2023wed} and dijet+photon production in DIS \cite{Roy:2019hwr}. The first numerical implementation of single hadron production provided good agreement with data at low transverse momentum \cite{Stasto:2013cha} but yielded negative results at high $p_T$. However, largely missing in the literature are the NLO computations for single inclusive and double-inclusive heavy-flavor in DIS, as well as jet+photon and dijet production in proton-nucleus collisions.\\

Beyond the CGC program, recent works \cite{Stewart:2023lwz,Neill:2023jcd} have shown that the physics of small $x$ DIS 
and the onset of saturation can be formulated systematically within 
soft-collinear effective theory (SCET) with Glauber operators. The authors in \cite{Stewart:2023lwz} demonstrate that saturation corresponds to the 
breakdown of a well-defined power-counting parameter governing successive  
Glauber exchanges, and identify new radiation modes that reconcile the 
appearance of $Q_s$ with the transition to nonlinear evolution. In \cite{Neill:2023jcd} the authors construct a small $x$ factorization 
theorem for the DIS hadronic tensor in terms of a universal collinear function 
and a gauge-invariant soft operator $S_{\mu\nu}$ that generalizes the dipole 
amplitude in the CGC. Rapidity renormalization in this framework reproduces 
BFKL evolution at NLL accuracy and establishes a clean operator definition for 
impact-factor like objects that are universal across processes such as DIS and 
Drell--Yan. Elevating existing CGC calculations to NLO might provide a bridge with extant SCET approaches to saturation. 

Recently, the CGC/TMD correspondence has been the subject of extensive studies, which can schematically be divided into three directions: i) this correspondence has been extended to additional processes~\cite{Altinoluk:2018byz,Altinoluk:2020qet,Tong:2022zwp,Tong:2023bus} including in particular particle production in diffractive processes~\cite{Iancu:2021rup,Iancu:2022lcw,Iancu:2023lel,Hatta:2022lzj,Hauksson:2024bvv} and in the target fragmentation region~\cite{Caucal:2025qjg}; ii) sub-eikonal~\cite{Altinoluk:2024tyx,Altinoluk:2024zom,Agostini:2024xqs,Altinoluk:2023qfr,Altinoluk:2022jkk} and kinematic power corrections~\cite{Altinoluk:2019fui,Altinoluk:2019wyu,Mantysaari:2019hkq,Boussarie:2021ybe} to tree-level TMD factorized cross sections have been investigated at small $x$; iii) the connection between CGC and TMDs has been generalized beyond leading order ~\cite{Mueller:2012uf,Mueller:2013wwa,Kovchegov:2015zha
,Taels:2022tza,Caucal:2022ulg,Caucal:2024bae,Caucal:2024vbv,Duan:2024nlr,Duan:2024qev,Mukherjee:2023snp} by consistently incorporating various quantum evolution effects that resum large logarithms; namely, small $x$ BK or JIMWLK equation~\cite{Balitsky:1995ub,Kovchegov:1999yj,JalilianMarian:1997jx,JalilianMarian:1997gr,Kovner:2000pt,Weigert:2000gi,Iancu:2000hn,Iancu:2001ad,Ferreiro:2001qy}, Collins-Soper-Sterman (CSS) evolution~\cite{Collins:1981uk,Collins:1981uw,Collins:1984kg,Collins:2011zzd}, and DGLAP evolution~\cite{Gribov:1972ri,Altarelli:1977zs,Dokshitzer:1977sg}.\\

Gluon production and correlations at {\it central} rapidity have been studied in considerable detail within the CGC/saturation formalism. However, these studies were performed at leading order. Recently~\cite{Ram}, progress has been made on NLO calculations of single inclusive gluon production—reproducing JIMWLK evolution in the target, BFKL evolution in the projectile, running coupling corrections, and DGLAP evolution for the fragmentation function; the results were obtained for quantum color charge densities in the projectile and thus can be applied in DIS—though the results have not yet been published.

\subsubsection{Sub-eikonal corrections.}
\label{sec:subeik}

The CGC/saturation framework and small-$x$ physics at leading order constitute a well-established and mature theoretical framework, supported by robust theoretical foundations and extensive phenomenological studies~\cite{Morreale:2021pnn}. Current developments in small-$x$ physics proceed along two complementary directions: the calculation of next-to-leading order (NLO) corrections and the calculation of sub-eikonal corrections. While both approaches aim to systematically improve upon the leading-order description of scattering processes, the nature of these corrections differs fundamentally.

The NLO corrections represent the natural next step in the systematic expansion of the LO small-$x$ formalism and, as such, preserve its fundamental assumptions. Specifically, these calculations are performed within the shock-wave approximation, which provides an effective description of scattering processes characterized by large rapidity separation. At NLO, the underlying kinematical assumptions remain unchanged: emission processes are characterized by strong ordering in longitudinal momentum fractions. Since this emission regime dominates the kinematical region where $x \propto 1/\sqrt{s}$ is small, it naturally defines the domain of applicability for NLO corrections. Formally, the calculation is performed in the limit $x\to0$; therefore, for observables relevant to this discussion, this approximation becomes exact at asymptotically high collision energies.

In contrast, sub-eikonal corrections are fundamentally different in character. These corrections aim to extend the applicability of the leading-order formalism to encompass effects associated with finite values of $x$. In particular, they account for deviations from strict longitudinal momentum ordering and the consequent breakdown of the shock-wave approximation in the kinematical region of moderate $x$.

The calculation of sub-eikonal corrections may carry significant phenomenological implications. In practice, the regime of very small $x$ (equivalently, very high energy) is experimentally challenging to access. Specifically, at the energies anticipated at the future Electron-Ion Collider ($\sqrt{s}_{\rm max} = 140$~GeV), terms suppressed by powers of energy or $x$ may substantially modify predictions based on the eikonal limit. To achieve quantitative predictive power in this experimentally accessible region, the leading-order small-$x$ formalism must be systematically extended to incorporate sub-eikonal corrections suppressed by the targets boost factor $\gamma$.  

A naive parametric estimate suggests that NLO corrections, controlled by $\alpha_s(\mu \approx 4~\mathrm{GeV}) \approx 0.15$, are more significant than sub-eikonal corrections, which are controlled by Lorentz boost factor of the proton/nucleus $\gamma \approx 0.01$. However, one cannot exclude the possibility of numerically large prefactors that could modify this naive estimate at EIC energies. Consequently, the precision physics program at the EIC necessitates both the analytical computation and numerical evaluation of sub-eikonal corrections.

Sub-eikonal corrections have now been computed for a range of quantities: quark and gluon propagators~\cite{Altinoluk:2014oxa,Altinoluk:2015gia,Altinoluk:2015xuy,Agostini:2019avp,Agostini:2019hkj,Altinoluk:2020oyd,Altinoluk:2021lvu,Agostini:2022ctk,Agostini:2022oge,Altinoluk:2022jkk,Agostini:2023cvc,Agostini:2024xqs,Altinoluk:2024zom,Altinoluk:2024dba}; $t$-channel quark exchanges between projectile partons and the dense target~\cite{Altinoluk:2023qfr,Altinoluk:2024tyx,Altinoluk:2025ang}; DIS structure functions~\cite{Altinoluk:2025ivn}; quark and gluon helicity evolution together with related observables such as single and double spin asymmetries~\cite{Kovchegov:2015pbl,Kovchegov:2016zex,Kovchegov:2016weo,Kovchegov:2017jxc,Kovchegov:2017lsr,Kovchegov:2018znm,Kovchegov:2018zeq,Kovchegov:2020kxg,Kovchegov:2020hgb,Adamiak:2021ppq,Kovchegov:2021lvz,Kovchegov:2021iyc,Cougoulic:2022gbk,Kovchegov:2022kyy,Borden:2023ugd,Kovchegov:2024aus,Borden:2024bxa}; rapidity evolution of gluon TMD distributions interpolating between moderate and small $x$ beyond eikonal accuracy~\cite{Balitsky:2015qba,Balitsky:2016dgz,Balitsky:2017flc}; and orbital angular momentum~\cite{Hatta:2016aoc,Kovchegov:2019rrz,Boussarie:2019icw,Kovchegov:2023yzd,Kovchegov:2024wjs}.

However, a realistic assessment of their numerical significance compared to the leading eikonal order (when it contributes) has not yet been performed, although simplified calculations have been carried out in Refs.~\cite{Agostini:2019avp,Agostini:2022ctk,Agostini:2019hkj,Adamiak:2021ppq,Agostini:2022oge,Agostini:2023cvc,Agostini:2024xqs}. The principal obstacle is that sub-eikonal corrections require explicit modeling of the transverse field components $F_{ij}$ and their dynamical evolution within the target, which are related to $F_{+-}$ contributions to the transverse momentum-dependent distributions (TMDs). This requirement extends beyond the scope of the commonly employed McLerran-Venugopalan (MV) model. While certain components of an extended MV framework have been discussed in the literature~\cite{Cougoulic:2020tbc,Li:2024fdb,Agostini:2025vvx}, a complete and self-consistent formulation has not yet been established.

To illustrate this point, we consider a specific process: dijet production in deep inelastic scattering (DIS) in the near back-to-back kinematic regime. In the small-$x$ limit, this process has been calculated to NLO precision both in $\alpha_s$ (see Sec.~\ref{sec:NLO}) and in the eikonal expansion (see Refs.~\cite{Altinoluk:2022jkk}). At moderate $x$, calculations are available to twist-three accuracy~\cite{Mukherjee:2026cte}. 
In what follows, we focus on the latter approach, which captures both the small-$x$ limit and sub-eikonal corrections while simultaneously providing access to higher-order effects that can be systematically estimated.

The cross section for dijet production via a longitudinally polarized virtual photon is given by 
\begin{eqnarray}\label{eq:lg_btb}
|i\mathcal{M}|^2_L
&\approx& e^2 \int dx^{-}\, d^2x_\perp\, dy^{-}\, d^2y_\perp \,
e^{i x P^+ (x^- - y^-)-i \Delta_\perp \cdot (x_\perp - y_\perp)}
\nonumber \\
&\times&
\frac{16\,z^2\bar{z}^2\,\epsilon_f^2}{(P_\perp^2+\epsilon_f^2)^4}
\,\mathrm{Tr}\Big[
 O_1 + O_2 + O_3 + O_4
\Big]
+ \mathcal{O}\left(\frac{\Delta_\perp^2}{P_\perp^2}\right),
\end{eqnarray}
where $P_\perp = \bar{z} k_{1\perp} - z k_{2\perp}$ denotes the total transverse momentum of the dijet system, $\Delta_\perp = k_{1\perp} + k_{2\perp}$ represents the transverse momentum imbalance, and $z$ ($\bar{z}$) is the longitudinal momentum fraction of the photon carried by the quark (antiquark), defined as $z = k_1^-/q^-$ ($\bar{z} = k_2^-/q^-$), and  $\epsilon_f^2 = z\bar{z} Q^2$. The longitudinal momentum transfer is given explicitly by $x P^+ = \frac{1}{2 q^- z \bar{z}} \left( P_\perp^2 + \epsilon_f^2 + z\bar{z} \Delta_\perp^2\right)$.

The operator structures are classified according to their field content; here we restrict ourselves to the two body operators (three body operators at twist three constitute  the coefficients $O_{3,4}$)
\begin{eqnarray}
&&\hspace*{-2.5cm}O_1=
\Big[
2P^iP^j
- (z-\bar{z})(P^i\Delta^j + P^j\Delta^i)
+ 8(z-\bar{z})\frac{(\Delta \cdot P)P^iP^j}{P_\perp^2+\epsilon_f^2}
\Big]
\overline{F}_{-i}(x^-,x_\perp)\,\overline{F}_{-j}(y^-,y_\perp),
\nonumber \\[4pt]
&&\hspace*{-2.5cm}O_2=
-\frac{(z-\bar{z})}{z\bar{z}\,q^-}\,
P^i P_\perp^2
\Big[
\overline{F}_{-i}(x^-,x_\perp)\,\overline{F}_{+-}(y^-,y_\perp)
+ \overline{F}_{+-}(x^-,x_\perp)\,\overline{F}_{-i}(y^-,y_\perp)
\Big]\,.
\end{eqnarray}
The amplitude can be expressed in terms of field-strength insertions dressed by fundamental Wilson lines along lightlike paths. For notational compactness, we used the shorthand $\overline{F}$ for the parallel-transported field-strength operator,
\begin{equation}
\hspace*{0.8cm}\overline{F}_{\mu\nu}(x^-,x_\perp)
\equiv
[\infty,x^-]_{x_\perp}\,F_{\mu\nu}(x^-,x_\perp)\,[x^-,\infty]_{x_\perp}\,.
\end{equation}
In the near back-to-back kinematic regime, the small-$x$ limit is obtained by setting $x=0$ in the exponential factor, such that $\exp(i x P^+ (x-y)^-) \to 1$, and neglecting contributions from $F_{+-}$ and $F_{ij}$. (Note that $F_{ij}$ does not appear at twist-three accuracy for longitudinal photon polarization but is present for transverse polarization.) This small-$x$ limit can be simulated using the conventional McLerran-Venugopalan (MV) model and evolved via the JIMWLK equation; see, e.g., Refs.~\cite{Dominguez:2011wm,Metz:2011wb,Dumitru:2015gaa,Marquet:2016cgx}.

Sub-eikonal corrections incorporate the subleading term in the expansion of the exponential, $\exp(i x P^+ (x-y)^-) \to 1 + i x P^+ (x-y)^-$, beyond the strict eikonal approximation, together with additional contributions involving $F_{+-}$ (and $F_{ij}$ for transverse photon polarization). While the magnitude of the exponential's argument can be estimated straightforwardly as discussed above, evaluating the $F_{+-}$ and $F_{ij}$ contributions requires extending the conventional MV model to incorporate sub-eikonal corrections in its formulation. In the dilute regime—characterized by $A_i \sim g^0$, with the $x^+$ dependence of $A_-$ at order $g^0$ and $A_- \sim 1/g$—this extension was accomplished in Ref.~\cite{Li:2024fdb} and awaits implementation for the dijet simulations. The general formulation in the dense regime, where $A_i \sim 1/g$, is currently under development~\cite{KovnerAndCo}.

\subsubsection{Simulating JIMWLK on a Quantum Computer.}

The JIMWLK evolution equation was originally formulated as the evolution of the weight functional $W[\rho(\vec{x})]$ with rapidity $Y$. For numerical implementations \cite{Lappi:2012vw} it was, however, advantageous to write JIMWLK as an evolution of the scattering amplitude $S[W[\rho(\vec{x})]]$ describing the scattering of the projectile on the target. These scattering amplitudes are constructed from infinite Wilson lines that arise from the eikonal scattering of color charges on the target, 
\begin{equation}
 \hspace*{1.5cm}  V_j({\vec{x}})\:=\:\mathcal{P}\exp\{i\int_{-\infty}^{\infty}\:dx^- \tau^a_j A_{-}^a(x^-,x_\perp)\}\,,
\end{equation}
where the $\tau_j$ are the generators in spin-$j$ representation. JIMWLK at leading logarithmic accuracy is then written as a Langevin equation on the Wilson lines ${V}_j({\vec{x}})$, representing a random walk in color space $SU(N_c)$:
\begin{eqnarray}
&&\hspace*{-1cm}{V}^{Y+dY}_{1/2}({\vec{x}})(Y + dY)\,=\,\exp\Big\{-i \frac{\sqrt{\alpha_s dY}}{\pi}\:\int_{\vec{z}} \vec{K}_{\vec{x} - \vec{z}}.\big[{V}^Y_{1/2}(z')\vec{\xi}_{\vec{z'}}{V}^{Y\dagger}_{1/2}(\vec{z'})\big]\Big\}\:{V}^Y_{1/2}({\vec{x}})\:\nonumber\\
&&\hspace*{-1cm}\times\,\exp\Big\{i\frac{\sqrt{\alpha_s dY}}{\pi}\:\int_{\vec{z}} \vec{K}_{\vec{x} - \vec{z}}.\vec{\xi}_{\vec{z'}}\Big\}\,,\label{eq: langevin}
\end{eqnarray}
where $\vec{K}_{\vec{x}}\:=\:{\vec{x} \over x^2}$ is the square root of BFKL kernel in two space dimensions and the noise $\xi$ is Gaussian.

Numerically, the Wilson line ${V}_{1/2}^Y(\vec{x})$ at the initial rapidity $Y_0$ is sampled from the color charge distribution $W[Y_0]$ and evolved to rapidity $Y$ using the Langevin equation~(\ref{eq: langevin}) for a given noise realization. Observables such as dipoles and quadrupoles are then computed from the evolved ${V}_{1/2}^Y(\vec{x})$. The expectation value at rapidity $Y$ is obtained by averaging over many initial Wilson lines and noise realizations, rendering the calculation computationally intensive and feasible only on high-performance computing facilities. Moreover, the NLL JIMWLK equation does not admit a Langevin formulation~\cite{Kovner:2014lca}, motivating the development of alternative solution methods. A key advance in this direction was made in~\cite{Armesto:2019mna,Li:2020bys}, where the JIMWLK equation was shown to admit a Lindblad form, as encountered in open quantum systems \cite{Lidar:2019qog}.
\par

In an open quantum system, when there is a separation of time-scales between the system-bath interactions and the bath-bath interactions (Born-Markov approximation), the evolution of the system with time is given by the Lindblad master equation \cite{Lidar:2019qog}, which has the following form
\begin{equation}
  \hspace*{1.cm}   
    \frac {\partial \rho} {\partial \tau} = -i [ \mathcal{H}, \rho] + \sum_r  {f}_r \big( \textbf{Q}_r \rho \textbf{Q}_r^\dagger - \frac{1}{2} \{ \textbf{Q}_r^\dagger \textbf{Q}_r , \rho\} \big),
\end{equation}
where ${f}_r$ is the interaction rate or decay rate, $\mathcal{H}$ is the system Hamiltonian and $\textbf{Q}_r$ are the Lindblad or jump operators which represent the interaction of the system with the bath.\par
In the CGC effective theory described above, there is a natural split in the QCD degrees of freedom based on their rapidity. Consequently in Refs.~\cite{Armesto:2019mna}, it was demonstrated that the JIMWLK evolution (in the limit of dilute/dense color charge densities) can be written as a Lindblad equation for rapidity evolution of the hadron density matrix $\rho$
\begin{equation}
 \hspace*{1.6cm}  {d \over dY}\hat{\rho}\:=\:\int\:{d^2 z_\perp \over 2\pi}\:\big[{Q^a}_i[z_\perp], \big[{Q^a}_i[z_\perp], \hat{\rho}\big]\big]\,,
\end{equation}
 where the system consists of the valence degrees of freedom and the environment are the soft degrees of freedom, with the operator ${Q^a}[z_\perp]$ representing a change of the system (valence) degrees of freedom due to interactions with the environment (soft vacuum). This is precisely the equation we would like to solve on a quantum computer \cite{Agrawal:2026kis}.\par
 Let us start by describing the Hilbert space of the system. It is the space of valence color charge density $|j\rangle$ or, equivalently of the gluon field $|\alpha\rangle$ at each transverse position $x_\perp$. It is infinite-dimensional and therefore requires truncation. As a simplification, we consider SU(2) gauge symmetry instead of SU(3) and reduce the two-dimensional space to a one-dimensional radial lattice by assuming azimuthal symmetry. As a truncation for the bosonic phase space at each lattice site, we use the so-called representation basis of lattice gauge theory (LGT) (see \cite{Raychowdhury:2019iki,Klco:2019evd,Ciavarella:2024fzw,Byrnes:2005qx}) at the cost of truncating the infinite Wilson lines, typically used in CGC, to a unit Wilson link in the $x^-$ direction,
 \begin{equation}
 \hspace*{1.6cm}  U^j(\vec{r},\mu,\alpha)\:=\:\exp\Big\{ig\:a_s\:\tau^a_j\:\alpha^{a}_\mu(\vec{r}) \Big\}\,.
\end{equation}
 A typical basis element has the form
\begin{equation}
\label{eq: rep basis hilbert space}
 \hspace*{1.5cm}  |q\rangle\:=\:\bigotimes_{y}\:|q^y\rangle\:=\:\bigotimes_{y}\:|j_q^y,n_{qL}^y,n_{qR}^y\rangle~.
\end{equation}
The evolution equation finally takes the form
\begin{eqnarray}
\frac{d}{dY}\rho_{pq}(Y) &=&a_\perp^2\:\sum_{r_\perp,s,t} 2\pi r_\perp\:\Big(\mathbf{Q}^a_{ps}(r_\perp) \mathbf{Q}^a_{st}(r_\perp) \rho_{tq}\:-2\:\mathbf{Q}^a_{ps}(r_\perp)\rho_{st}\mathbf{Q}^a_{tq}(r_\perp)\:\nonumber \\
&+&\rho_{ps}\:\mathbf{Q}^a_{st}(r_\perp)\:\mathbf{Q}^a_{tq}(r_\perp)\Big)~,
\label{eq: JIMWLK in rep basis}
\end{eqnarray}
where the jump operators $\mathbf{Q}^a_{ps}(r_\perp)$ are defined in the representation basis.\par
\begin{figure}[htbp]
  \centering
  \includegraphics[width=0.6\textwidth]{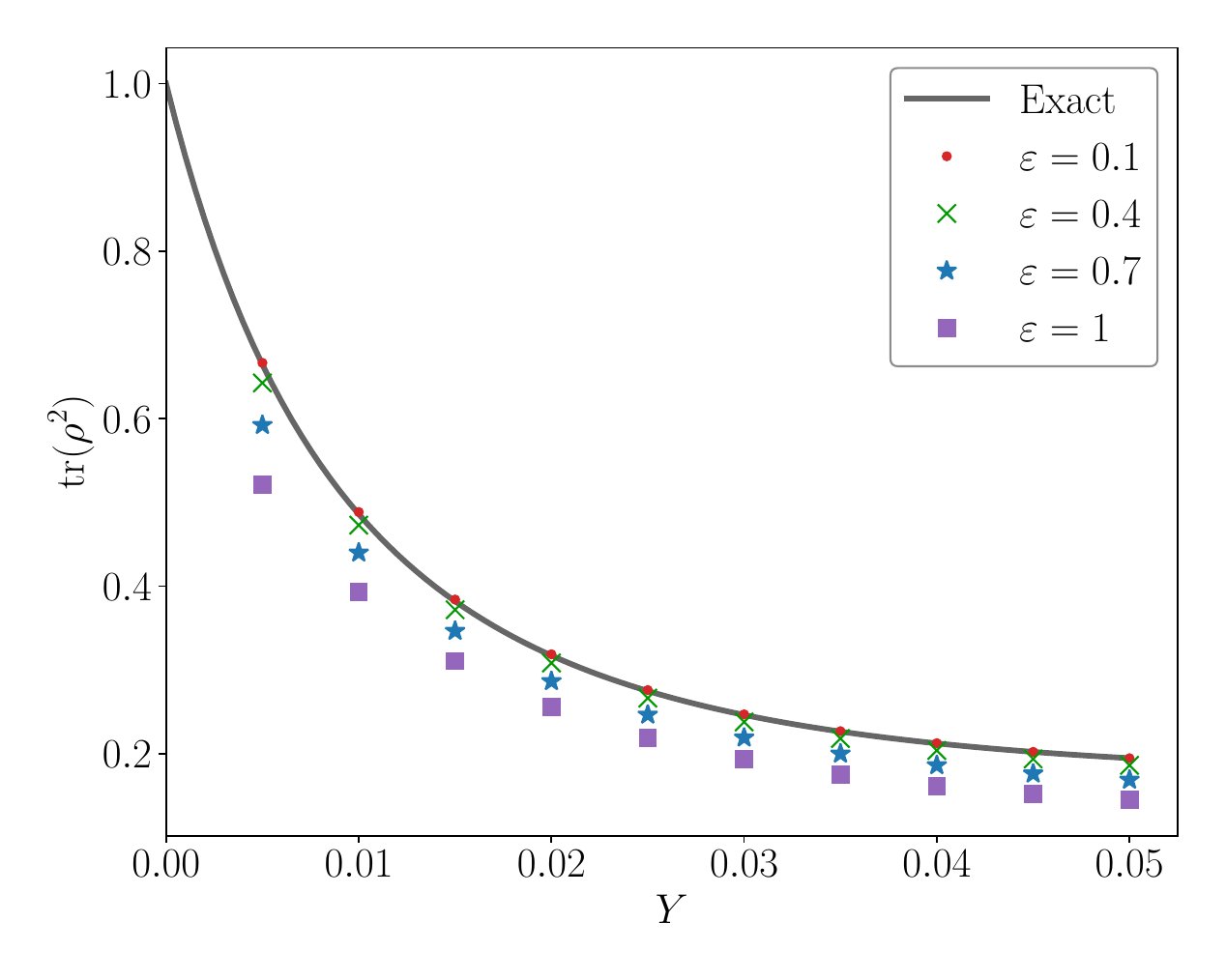}
  \caption{Purity versus rapidity $Y$ for different choices of expansion parameter $\epsilon$ (see \cite{Agrawal:2026kis,Schlimgen:2022aji,Schlimgen:2021hxs}).}
  \label{fig:purityplot}
\end{figure}

The solution of Eq.~\ref{eq: JIMWLK in rep basis} on a quantum computer has been done using the method described in \cite{Schlimgen:2022aji,Schlimgen:2021hxs}, which encodes the non-Hermitian jump operators on the quantum computer by increasing the size of the Hilbert space. The evolution of purity with rapidity $Y$ for a Gaussian initial condition and maximal spin $J_{\rm max} = 1/2$ is given in Figure \ref{fig:purityplot}. 

\subsection{Current hints at saturation and EIC prospects
\label{sec:currentcgc}} 

It has been found that the gluon density inside the proton grows rapidly at small momentum fractions. QCD predicts that this growth can be regulated by nonlinear effects, ultimately leading to gluon saturation.  
Within the color glass condensate (CGC) framework, nonlinear QCD effects are predicted to suppress and broaden back-to-back angular correlations in collisions involving heavy nuclei.  

Previous results from collisions between hadronic systems, i.e., $p+$A or $d+$A at RHIC and the LHC provide a window into extracting the parton distributions of nuclei at small $x$.  On the one hand, both RHIC \cite{STAR:2006dgg,PHENIX:2011puq,STAR:2021fgw} and the LHC \cite{LHCb:2015coe,ATLAS:2019jgo} observed this predicted suppression in nucleus-involved collisions compared to the baseline $p$+$p$ collisions at different collision energies, suggesting hints of gluon saturation. On the other hand, none of the experiments has claimed the observation of the broadening phenomenon predicted by CGC to date.
At PHENIX~\cite{Adare:2011sc}, the widths of the away-side peak in mid-forward rapidity correlations are similar between $p$+$p$ and central $d$+A. For forward-forward correlations, possible broadening in $d$+A is inconclusive due to the high pedestal in $d$+A compared to the baseline $p$+$p$ that possibly biases the width extraction.  
At STAR~\cite{STAR:2021fgw}, the pedestal in forward di-$\pi^0$ correlations is similar between $p$+$p$ and $p$+A, allowing a more reliable extraction of the width. However, no broadening is observed, the away-side widths remain unchanged between $p$+$p$ and $p+$A.  
At the LHC, LHCb~\cite{LHCb:2015coe} did not report the width for forward di-hadron 
correlation, and ATLAS~\cite{ATLAS:2019jgo} observed no difference in the di-jet correlation widths between $p$+$p$ and $p$+A.

\begin{figure*}[t]
\begin{center}
\includegraphics[width=0.47\textwidth]{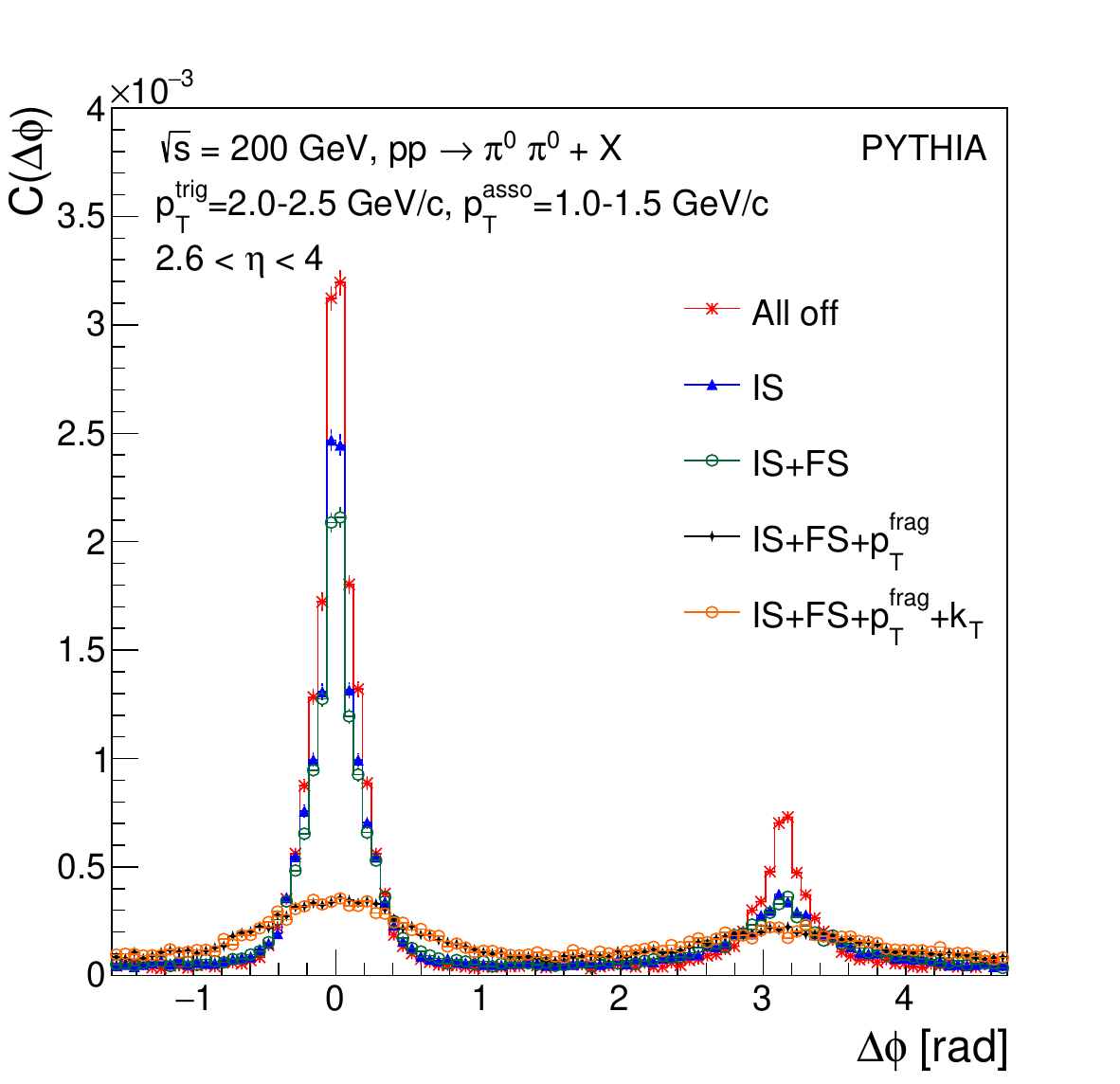}
\includegraphics[width=0.47\textwidth]{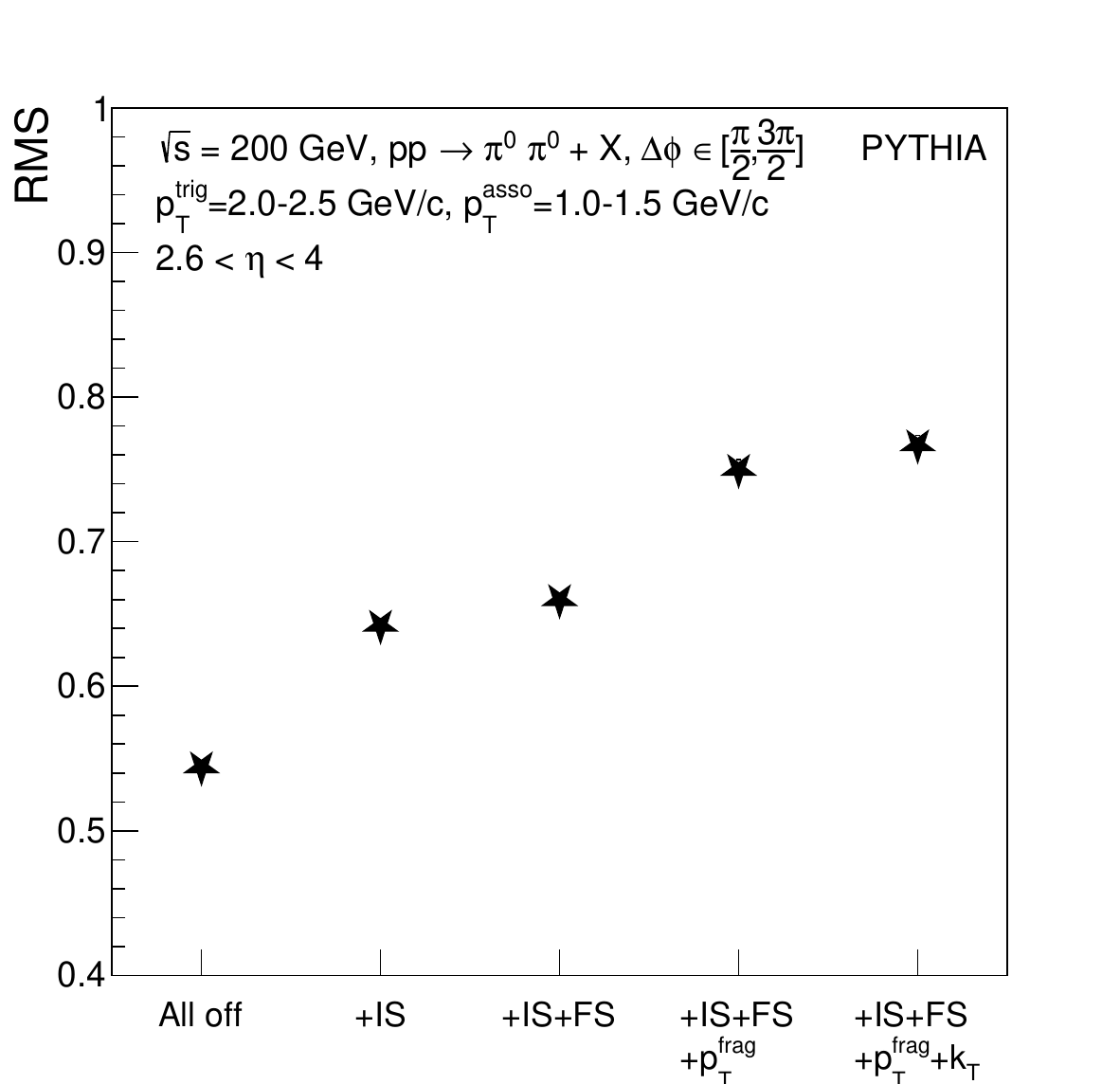}
\end{center} 
\caption{Left: The correlation function of forward di-$\pi^{0
}$ production for different intrinsic $k_{T}$ assumptions, parton showers, and $p_T^{\mathrm{frag}}$. Right: The RMS of the away-side peak for different assumptions. The Gaussian width of $k_{T}$ and $p_{T}^{\mathrm{frag}}$ are 0.5 and 0.4 GeV/$c$, respectively. Taken from \cite{Cassar:2025vdp}.}
\label{fig:correlation}
\end{figure*}

The study \cite{Cassar:2025vdp} investigates the contributions by
intrinsic transverse momentum ($k_{T}$), which is associated with saturation physics, as well as by parton showers (effectively included into the Sudakov factor)
and transverse motion from fragmentation ($p^{\mathrm{frag}}_{T}$), which are not saturation dependent, to the width of the
correlation function. Figure~\ref{fig:correlation} shows the correlation functions
from each source and their RMS widths of the
away-side peak. The largest contributions to the broadening come from initial-state
parton shower and fragmentation.
The conclusion is that the non-saturation dependent effects, especially the initial-state
parton shower and $p^{\mathrm{frag}}_{T}$, which occur independently of the collision system, smear the back-to-back correlation more than gluon saturation does, making the broadening phenomenon difficult to observe. The same conclusion was drawn from the prior simulation study \cite{Zheng:2014vka} for $e$$+$$p$ and $e$$+$A collisions at the EIC. It is also supported by the theoretical prediction \cite{Caucal:2023fsf} indicating large  
Sudakov effects.

It was proposed about twenty years ago that two-particle correlations provide one of the most sensitive channels for probing saturation physics. However, the definition and presentation of this observable have been rather nonuniform across both experimental and theoretical studies. 

First, the correlation function, $C(\Delta\phi)$, is usually defined as
\begin{equation}
 \hspace*{3cm}  C(\Delta\phi) = \frac{N_\mathrm{pair}(\Delta\phi)}{N \times \Delta\phi_\mathrm{bin}},
\end{equation}
where $N_\mathrm{pair}$ is the yield of correlated trigger-associated hadron pairs,
$\Delta\phi$ is the azimuthal angle difference between the trigger and associated hadrons,
and $\Delta\phi_\mathrm{bin}$ is the bin width of the $\Delta\phi$ distribution.
The normalization factor $N$ in the denominator, however, is defined differently in the literature.
Some publications use the trigger hadron yield $N_\mathrm{trig}$, while others use the total pair yield $N_\mathrm{pair}$, in which case the correlation function becomes effectively self-normalized.

\begin{table}[htbp]
\centering
\caption{Definitions of the correlation function in major publications.}
\label{tab:normalization}
\resizebox{\textwidth}{!}{
\begin{tabular}{|c|c|c|c|}
\hline
\textbf{Experiment/Theory} & \textbf{Normalization} & \textbf{System} & \textbf{Details} \\
\hline
STAR~\cite{STAR:2006dgg,STAR:2021fgw} & $N_\mathrm{trig}$ & $p$$+$$p$, $p$$+$Al, $p/d$$+$Au & Compare area ratio \\
PHENIX~\cite{PHENIX:2011puq} & $N_\mathrm{trig}$ & $p$$+$$p$, $d$$+$Au & Compare area ratio $\times$ $R_{d+Au}$ \\
LHCb, ATLAS~\cite{LHCb:2015coe,ATLAS:2019jgo} & $N_\mathrm{trig}$ & $p$$+$$p$, $p$$+$Pb & Compare area ratio \\
\hline
\cite{Kharzeev:2004bw} & $N_\mathrm{pair}$ & $p$$+$$p$, $d$$+$Au & $N_\mathrm{pair}$ from [$-\pi/2$, $3\pi/2$] \\
\cite{Stasto:2011ru} & $N_\mathrm{trig}$ & $d$$+$Au & Issue with $p$$+$$p$ \\
\cite{Stasto:2018rci} & $N_\mathrm{pair}$ & $p$$+$$p$, $p/d$$+$Au & $N_\mathrm{pair}$ from [$\pi/2$, $3\pi/2$] \\
\cite{Lappi:2012nh} & $N_\mathrm{trig}$ & $p$$+$$p$, $p/d$$+$Au &  \\
\cite{Lappi:2012nh} & $N_\mathrm{pair}$ & $p$$+$$p$, $p/d$$+$Au & $N_\mathrm{pair}$ from pedestal \\
\cite{Albacete:2010pg} & $N_\mathrm{trig}$ & $p$$+$$p$, $d$$+$Au &  \\
\cite{Albacete:2018ruq} & $N_\mathrm{trig}$ & $p$$+$$p$, $p$$+$Au & Compared with STAR data \\
\hline
\end{tabular}
}
\end{table}

Second, the quantification of suppression from $p/d$$+$A relative to the $p$$+$$p$ baseline also varies among experiments and theoretical calculations.
The STAR~\cite{STAR:2006dgg,STAR:2021fgw}, LHCb~\cite{LHCb:2015coe}, and ATLAS~\cite{ATLAS:2019jgo} collaborations compare the integrated area of the back-to-back correlation peak (after pedestal subtraction) between $p/d$$+$A and $p$$+$$p$ collisions.
The ratio of these areas is used to quantify the suppression.
In contrast, PHENIX~\cite{PHENIX:2011puq} additionally multiplies the area ratio by the nuclear modification factor $R_{d+Au}$, effectively applying an extra normalization to the suppression signal.
Theoretical works, on the other hand, adopt a variety of other conventions for presenting the suppression.

These inconsistencies in normalization and in the definition of the suppression observable lead to nonuniform results and may cause misinterpretations of the underlying physics.
A summary of major publications and their normalization schemes for the correlation function is presented in Table~\ref{tab:normalization}.
We recommend that the normalization of the correlation function should be performed with respect to $N_\mathrm{trig}$.
In this case, the function represents the probability of finding an associated particle given a trigger particle.
Self-normalization using $N_\mathrm{pair}$ fixes the total area of the correlation function to a constant value across all systems, thereby preventing the use of the area difference to quantify suppression, since all systems would yield identical areas.
We also argue that the additional normalization by $R_{d+Au}$ is not appropriate, as $N_\mathrm{trig}$ already accounts for the number of binary collisions.
Applying $R_{d+Au}$ again results in an over-normalization of the correlation function.

\begin{figure*}
    \centering
    \includegraphics[width=1\textwidth]{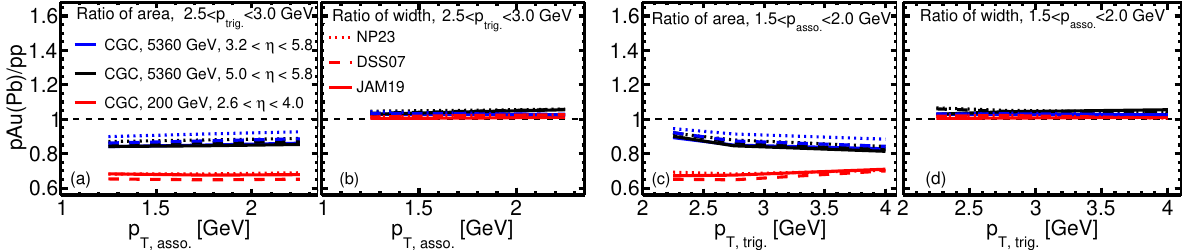}
    \caption{Relative area and width of back-to-back di-$\pi^0$ correlations at forward pseudorapidities calculated by the CGC theory with three different fragmentation functions  in p-Au  and p-Pb relative to p-p collisions, for $p_{T,\rm trig}$ = 2.5 - 3 GeV/c as functions of $p_{T,\rm asso}$ (left two panels) and  for $p_{T,\rm asso}$ = 1.5 - 2 GeV/c as functions of $p_{T,\rm trig}$ (right two panels). Taken from \cite{Caucal:2025zkl}.}
    \label{fig:ratio_ff_com}
\end{figure*}

Recently, the Color Glass Condensate calculations (using an improved TMD approximation) describe forward di-$\pi^0$ production around the away-side peak measured by STAR in both proton–proton and proton–gold collisions. While the coincidence probability, $CP(\Delta\Phi)$, is sensitive to hadron fragmentation, the ratios of the integrated area and the width of the correlation remain robust against changes in the fragmentation function as shown in Figure \ref{fig:ratio_ff_com}. This robustness suggests that these ratios reliably probe the initial-state dynamics and can be used to investigate gluon saturation effects \cite{Caucal:2025zkl}. Furthermore, Ref. \cite{Caucal:2025zkl} predicts that at the kinematics accessible to the ALICE FoCal experiment ($\sqrt{s_{NN}}=$ 5.36 TeV), di-hadron correlations will experience significantly less suppression than observed at STAR. Despite probing larger saturation scales at FoCal (due to accessing smaller-$x$ via higher center-of-mass energy and more forward rapidities), the corresponding smaller average fragmentation energy implies that partons have higher transverse momentum, making them more sensitive to the dilute regime.

\subsection{Helicity at small $x$  
\label{sec:smallx_helicity}}

Our understanding of the proton structure would be incomplete without understanding its spin structure. The latter has been a challenge for decades, exemplified in the proton spin puzzle \cite{EuropeanMuon:1987isl, Jaffe:1989jz, Ji:1996ek, Boer:2011fh, Aidala:2012mv, Accardi:2012qut, Leader:2013jra, Aschenauer:2013woa, Aschenauer:2015eha,  Proceedings:2020eah, Ji:2020ena, AbdulKhalek:2021gbh}. The proton spin puzzle is formulated in terms of the spin sum rules, like the Ji sum rule \cite{Ji:1996ek} and the Jaffe-Manohar sum rule \cite{Jaffe:1989jz} given here: 
\begin{equation}\label{JMsumrule}
  \hspace*{3cm}  S_q+L_q+S_G+L_G = \frac{1}{2}\, . 
\end{equation}
The components are the spin and orbital angular momentum (OAM) of quarks ($S_q$ and $L_q$) and gluons ($S_G$ and $L_G$). The spin components can be written in terms of integrals of helicity PDFs,
\begin{equation}\label{spins}
    S_q(Q^2) = \frac{1}{2}\int\limits_0^1\mathrm{d}x \, \Delta\Sigma(x,Q^2)\,, \quad\quad S_G(Q^2) = \int\limits_0^1 \mathrm{d}x \, \Delta G(x,Q^2) \, .
\end{equation}
As the integrals cover the region down to $x=0$, where no experimental measurement can be made, theoretical understanding of helicity PDFs at small $x$ is crucial for solving the proton spin puzzle.

Theoretical understanding of small-$x$ asymptotics of helicity PDFs began to emerge in the mid 1990s with the pioneering works of Bartels, Ermolaev, and Ryskin (BER) \cite{Bartels:1995iu,Bartels:1996wc} employing the infrared evolution equations (IREE) framework \cite{Gorshkov:1966ht,Kirschner:1983di,Kirschner:1994rq,Kirschner:1994vc,Blumlein:1996hb,Griffiths:1999dj}. These papers aimed to resum all orders of the double-logarithmic parameters $\as\ln^2(1/x)$ and $\as\ln(1/x)\ln(Q^2/\Lambda^2)$ for helicity PDFs  $\Delta G(x,Q^2)$ and $\Delta \Sigma(x,Q^2)$ and for the related $g_1$ structure function. The phenomenology based on these works allowed for the possibility that a large amount of the proton spin may be carried by small-$x$ partons \cite{Blumlein:1995jp,Blumlein:1996hb,Ermolaev:1999jx,Ermolaev:2000sg,Ermolaev:2003zx,Ermolaev:2009cq}. The question of the small-$x$ asymptotics of helicity PDFs has been revisited over the last decade in \cite{Kovchegov:2015pbl, Hatta:2016aoc, Kovchegov:2016zex, Kovchegov:2016weo, Kovchegov:2017jxc, Kovchegov:2017lsr, Kovchegov:2018znm, Kovchegov:2019rrz, Boussarie:2019icw, Cougoulic:2019aja, Kovchegov:2020hgb, Cougoulic:2020tbc, Chirilli:2021lif, Kovchegov:2021lvz, Cougoulic:2022gbk, Borden:2023ugd, Adamiak:2023okq, Borden:2024bxa, Borden:2025ehe}, employing the $s$-channel evolution/shock wave formalism \cite{Mueller:1994rr,Mueller:1994jq,Mueller:1995gb,Balitsky:1995ub,Balitsky:1998ya,Kovchegov:1999yj,Kovchegov:1999ua,Jalilian-Marian:1997dw,Jalilian-Marian:1997gr,Weigert:2000gi,Iancu:2001ad,Iancu:2000hn,Ferreiro:2001qy} at the sub-eikonal level \cite{Altinoluk:2014oxa,Balitsky:2015qba,Balitsky:2016dgz, Kovchegov:2017lsr, Kovchegov:2018znm, Chirilli:2018kkw, Jalilian-Marian:2018iui, Jalilian-Marian:2019kaf, Altinoluk:2020oyd, Boussarie:2020vzf, Boussarie:2020fpb, Kovchegov:2021iyc, Altinoluk:2021lvu, Kovchegov:2022kyy, Altinoluk:2022jkk, Altinoluk:2023qfr,Altinoluk:2023dww, Li:2023tlw, Altinoluk:2024dba, Altinoluk:2024tyx, Altinoluk:2025dwd, Altinoluk:2025ewj}. These works resulted in small-$x$ evolution equations \cite{Kovchegov:2015pbl, Kovchegov:2016zex, Kovchegov:2017lsr, Kovchegov:2018znm,  Cougoulic:2022gbk, Borden:2024bxa} for polarized dipole amplitudes, which also resum all powers of the double-logarithmic parameters $\as\ln^2(1/x)$ and $\as\ln(1/x)\ln(Q^2/\Lambda^2)$ for helicity PDFs. These shock-wave equations close in the large-$N_c$ (KPS-CTT) and large-$N_c\&N_f$ limits (KPS-CTT-BCL). Exact analytic solutions of these equations were constructed in both limits \cite{Borden:2023ugd, Borden:2025ehe}. 

Both the BER and KPS-CTT-BCL formalisms agree with the small-$x$ limit of the polarized DGLAP splitting functions and anomalous dimensions to all three known loops \cite{Altarelli:1977zs,Dokshitzer:1977sg,Zijlstra:1993sh,Mertig:1995ny,Moch:1999eb,vanNeerven:2000uj,Vermaseren:2005qc,Moch:2014sna,Blumlein:2021ryt,Blumlein:2021lmf,Davies:2022ofz,Blumlein:2022gpp}. (A scheme transformation is required to establish agreement at three loops.) However, the two formalisms differ slightly in the intercepts for helicity PDFs. Defining the intercept $\alpha_h$ by the leading small-$x$ asymptotics in the power-law form,
\begin{equation}\label{asymptotics_general}
    \Delta \Sigma (x \ll 1, Q^2) \sim \Delta G (x \ll 1, Q^2) \sim g_1 (x \ll 1, Q^2) \sim \left( \frac{1}{x} \right)^{\alpha_h} ,
\end{equation}
we observe that, at large $N_c$, the BER formalism gives \cite{Kovchegov:2016zex} (here $\bas = \as \, N_c /(2 \pi)$)
\begin{equation}\label{BER_intercept}
\hspace*{0.4cm}
    \alpha_h^{\mathrm{BER}} = \sqrt{\frac{17 + \sqrt{97}}{2}} \, \sqrt{\bas} \approx 3.66394 \, \sqrt{\bas} , 
\end{equation}
while the exact solution of the KPS-CTT equations leads to \cite{Borden:2023ugd}
\begin{equation}\label{analytic_intercept}
\hspace*{0.4cm}
   \alpha_h = \frac{4}{3^{1/3}} \, \sqrt{\textrm{Re} \left[ \left( - 9 + i \, \sqrt{111} \right)^{1/3} \right] } \,\sqrt{\bas} \approx 3.66074 \, \sqrt{\bas}\,.
\end{equation}
The two results are given by entirely different expressions, which evaluate to two very close, yet different, numbers. The discrepancy persists in the large-$N_c\&N_f$ limit. At the moment, the origin of the disagreement remains unclear, in large part due to significant differences between the two formalisms. One possible way to resolve the controversy may result from calculating the polarized DGLAP gluon-gluon ($GG$) anomalous dimensions at four loops. In the small-$x$ (small Mellin moment variable $\omega$) limit and at large $N_c$, the two formalisms appear to disagree on the order-$\as^4$ anomalous dimensions: 
\begin{equation}\label{intro_largeNcanomdims}
    \Delta\gamma_{GG}^{(\mathrm{KPS-CTT-BCL})\,(3)}(\omega) = \frac{496\, \bas^4}{\omega^7}\,, \quad\quad  \Delta\gamma_{GG}^{(\mathrm{BER})\,(3)}(\omega) = \frac{504 \, \bas^4}{\omega^7}\,.
\end{equation}
Other anomalous dimensions ($qq, qG, Gq$) also exhibit differences at four loops at large $N_c\&N_f$ \cite{Borden:2024bxa}.

\begin{figure}[t] 
\begin{centering}
\includegraphics[width=\textwidth]{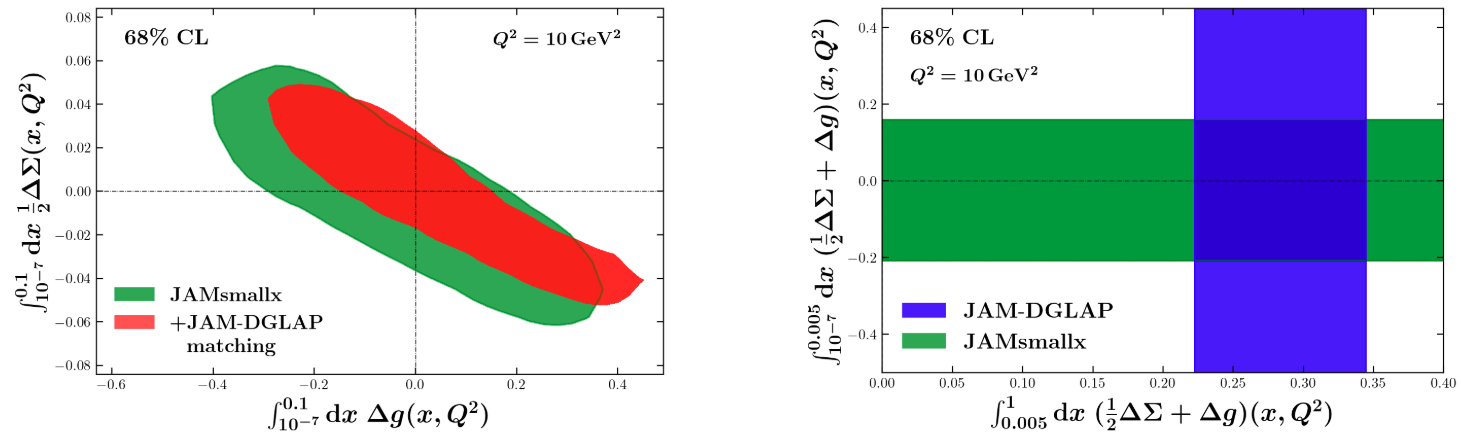}
\caption{(Left) Comparison of the $10^{-7}< x < 0.1$ truncated integrals of helicity PDFs obtained in the JAMsmallx analyses from \cite{Adamiak:2025dpw} with (red) and without (green) matching onto JAM-DGLAP extractions.
(Right) Exclusion plot of two truncated integrals:~the vertical axis shows the truncated integral of the total parton helicity contribution to the proton spin over the extrapolation/prediction range of $10^{-7}< x < 0.005$, while the horizontal axis shows the truncated integral over the data-driven range $0.005< x< 1$. The JAMsmallx analysis (green) does not fit data for $x > 0.1$, 
while the JAM-DGLAP analysis ~\cite{Cocuzza:2022jye} (blue) has uncontrolled extrapolation for $x < 0.005$.
The overlapping region is the best prediction for the total parton helicity that simultaneously satisfies both analyses. 
    \label{DGLAP_trunc}
}
\end{centering}
\end{figure}

The differences between the two calculations, while significant, appear to be numerically small. Therefore, one can attempt to perform data analyses based on either of the two existing formalisms. Phenomenology based on the KPS-CTT formalism with running coupling constant was performed in \cite{Adamiak:2021ppq, Adamiak:2023yhz} (JAMSmall$x$) fitting the world polarized DIS and SIDIS data for $x < 0.1$ and $Q^2 > m_c^2$. The predictions for EIC, obtained by extrapolating the results of the data analysis to smaller values of $x$ while using the KPS-CTT evolution, unfortunately exhibited a wide spread among different replicas, due to the inability of the current data to sufficiently constrain the initial conditions for the evolution. One way to better constrain the initial conditions for helicity evolution is by incorporating empirical knowledge of the proton at moderate $x$ \cite{Adamiak:2025mdy,Dumitru:2024pcv}. Another way to improve the accuracy of the EIC predictions is by including more data in the analysis. To that end, some of the longitudinally polarized ${\vec p} +{\vec p}$ data from RHIC were included in the analysis in \cite{Adamiak:2025dpw} based on the gluons-only small-$x$ result for the double-spin asymmetry $A_{LL}$ from \cite{Kovchegov:2024aus}. When the theoretical expression including quarks into $A_{LL}$ at small $x$ becomes available, further phenomenological improvements are anticipated, including a more comprehensive analysis of the polarized ${\vec p} +{\vec p}$ data. Note also that the existing phenomenology has been based on evolution from \cite{Cougoulic:2022gbk}, in which the treatment of quarks is incomplete: future phenomenology will be based on the complete KPS-CTT-BCL evolution equations (see \cite{Borden:2024bxa}). The results of the analysis in \cite{Adamiak:2025dpw} are presented in Fig.~\ref{DGLAP_trunc}.

Future theoretical challenges also include calculating higher-order corrections to the KPS-CTT-BCL, beyond just the running coupling included in phenomenological applications. The next correction comes from resumming single logarithms, that is, powers of $\as \, \ln (1/x)$. This was attempted in \cite{Kovchegov:2021lvz}. Single logarithmic corrections to helicity evolution may come from the longitudinal (SLA$_L$) and transverse (SLA$_T$) integral. The latter is in stark contrast to the unpolarized (BFKL/BK/JIMWLK) small-$x$ evolution, which has no such logarithms of energy arising from transverse integrals. The transverse contributions can be UV and IR. While the IR-divergent transverse integrals do not give a logarithm of energy, they give a logarithm of the dipole transverse size, which may result in a logarithm of energy after another step of evolution. Hence, they must be included. These IR contributions appear to come from the evolution steps inside the shock wave, breaking down the standard small-$x$ picture of only emissions outside the shock wave contributing to the evolution. This contribution is similar to how DGLAP evolution is derived in the shock-wave picture \cite{Balitsky:1987bk}. Incorporating these contributions into small-$x$ evolution is challenging, but, if successful, may lead to unifying DGLAP and small-$x$ evolution. Other searches of DGLAP and small-$x$ evolution matching are under way as well, with some initial attempts shown in Fig.~\ref{DGLAP_trunc}. However, we are yet to find a controlled systematic prescription for such matching. 
\\


Exclusive processes may also involve helicity flips, even in the {\it eikonal}
limit due to orbital angular momentum exchange, as is
well known from early work on ``$s$-channel helicity non-conservation''
in exclusive $J/\psi$ production~\cite{Kuraev:1998ht,Ivanov:2004ax}.
Improved theoretical precision for such processes requires these contributions.
The eikonal amplitude for the
production of a quarkonium state in
$\gamma(q,\lambda) \, p(P,\Lambda) \to Q(\Delta,\bar{\lambda})\,
p'(P',\Lambda')$, in dipole model factorization, is given by
$
{\cal M}_{\lambda\Lambda;\bar{\lambda}\Lambda'}({\vec{\Delta}_{\perp}}) = 2
N_c \int {\mathrm{d}}^2 {\vec{r}_{\perp}} {\cal Q}_{\Lambda\Lambda'}({\vec{r}_{\perp}},{\vec{\Delta}_{\perp}}) {\cal A}_{\lambda
  \bar{\lambda}}({\vec{r}_{\perp}},{\vec{\Delta}_{\perp}})\,.
$
Here, ${\cal A}_{\lambda\bar{\lambda}}({\vec{r}_{\perp}},{\vec{\Delta}_{\perp}})$ denotes
the photon-quarkonium LC wave function
overlap. ${\cal Q}_{\Lambda\Lambda'}({\vec{r}_{\perp}},{\vec{\Delta}_{\perp}})$ represents
$C$-even BFKL Pomeron or $C$-odd Odderon exchange,
\begin{eqnarray}
\hspace{-2cm}
   {\cal Q}_{\Lambda\Lambda'}({\vec{r}_{\perp}},{\vec{\Delta}_{\perp}}) &=&
   \int{\mathrm{d}}^2 {\vec{b}_{\perp}} \, e^{-i {\vec{\Delta}_{\perp}}\cdot{\vec{b}_{\perp}}}\nonumber\\ 
& \times& \left\{
1 - \frac{1}{2 N_c}\left[\frac{\langle P' \Lambda'|{\rm tr}
    \left[V({\vec{x}_{\perp}})V^\dag({\vec{y}_{\perp}}) \pm V({\vec{y}_{\perp}})V^\dag({\vec{x}_{\perp}})\right]
    |P\Lambda\rangle}{\langle P\Lambda|P\Lambda\rangle}\right]\right\}\,,
\label{eq:pom1}
\end{eqnarray}
with ${\vec{x}_{\perp}} = {\vec{b}_{\perp}} + {\vec{r}_{\perp}}/2$, ${\vec{y}_{\perp}} = {\vec{b}_{\perp}} - {\vec{r}_{\perp}}/2$, and $V({\vec{x}_{\perp}})$
are eikonal Wilson lines at transverse position ${\vec{x}_{\perp}}$.
We choose the $+$ sign
for ${\cal P}_{\Lambda\Lambda'}({\vec{x}_{\perp}},{\vec{y}_{\perp}})$ and the $-$ sign for $i{\cal
  O}_{\Lambda\Lambda'}({\vec{x}_{\perp}},{\vec{y}_{\perp}})$. Which of these actually appears in
${\cal M}_{\lambda\Lambda;\bar{\lambda}\Lambda'}$ is determined by the quantum numbers of the
quarkonium state: ${\cal P}_{\Lambda\Lambda'}({\vec{x}_{\perp}},{\vec{y}_{\perp}})$ is
relevant for vector quarkonium ($J/\psi$) production while ${\cal
  O}_{\Lambda\Lambda'}({\vec{x}_{\perp}},{\vec{y}_{\perp}})$ appears in the amplitudes for
pseudo-scalar $\eta_c$, scalar $\chi_{c0}$, axial-vector $\chi_{c1}$,
or tensor $\chi_{c2}$ production. The spin-dependent Pomerons and
Odderons can also be formulated in a GTMD
approach~\cite{Boussarie:2019vmk,Hagiwara:2020mqb}.

The standard BFKL Pomeron and Odderon correspond to no
helicity flip of the proton, $\Lambda = \Lambda'$. However, despite
the eikonal limit, the matrix elements of the dipole operator for
$\Lambda = - \Lambda'$ in general are not zero.
All of these matrix elements evolve with energy according to the same
BFKL (or BK) equations, regardless of $\Lambda, \Lambda'$.

The $J/\psi$ production amplitude with proton helicity flip
can be parametrized in terms of two ``spin dependent BFKL Pomerons''
associated with
$({\vec{r}_{\perp}}\cdot{\vec{\Delta}_{\perp}})
(\vec{\epsilon}_{\perp}^\Lambda\cdot{\vec{r}_{\perp}})$ 
and $(\vec{\epsilon}_{\perp}^\Lambda\cdot{\vec{\Delta}_{\perp}})$
correlations, respectively,
where ${\vec{r}_{\perp}}$ is the dipole vector and
$\vec{\epsilon}_{\perp}^\Lambda = -
(\Lambda,i)/\sqrt{2}$ the helicity vector for the proton:
$
{\cal P}_{\Lambda\Lambda'}({\vec{r}_{\perp}},{\vec{\Delta}_{\perp}}) = \delta_{\Lambda,\Lambda'}
{\cal P}({\vec{r}_{\perp}},{\vec{\Delta}_{\perp}}) +
\cos(\phi_{r\Delta})\Lambda e^{i \Lambda\phi_r}\delta_{\Lambda,-\Lambda'}{\cal P}_S({\vec{r}_{\perp}},{\vec{\Delta}_{\perp}}) +
\Lambda e^{i \Lambda\phi_\Delta}\delta_{\Lambda,-\Lambda'}{\cal P}^\perp_S({\vec{r}_{\perp}},{\vec{\Delta}_{\perp}})\,.
$
In the leading twist limit of small $r_\perp \Delta_\perp$, the
helicity flip amplitude involves only the combination ${\cal P}_S + 2
{\cal P}^\perp_S$ which is related to the GPD $E_g(x,t)$
\cite{Koempel:2011rc,Hagiwara:2020mqb,Hatta:2022bxn}.  Both these Pomerons
are non-zero only at finite momentum transfer, $\Delta_\perp \ne
0$.
For a more detailed discussion of vector quarkonium production
with the spin-dependent BFKL Pomeron, see~\cite{Benic:2025ral}.

Along similar lines, the helicity flip Odderon features in amplitudes
for exclusive production of $C$-even quarkonium (also in T-odd
nucleon energy correlators measuring the single-spin asymmetry of charged hadrons~\cite{Mantysaari:2025mht}).  Unlike the
Pomeron counterpart, it does
not vanish in the forward limit~\cite{Boussarie:2019vmk} (see, also,
\cite{Ma:2003py,Boer:2015pni,Benic:2024fbf}) because three gluon
exchange {\it can} transfer net orbital angular momentum even when
$\vec q_1 + \vec q_2 +\vec q_3 = 0$.
In fact, in the forward limit, three gluon exchange
necessarily involves a helicity flip of the proton as well as
in the $\gamma \to \chi_{c1}$ transition, so the
$\chi_{c1}$ is longitudinally polarized~\cite{Benic:2024fbf}.
Taking the collinear
limit on the proton side and the NR heavy $c$-quark limit, the
amplitude can be written in
a form involving the ``$d$-type''
trigluon correlator $O(x)$ at small $x$~\cite{Benic:2024fbf}.
The cross section
$\lim_{t\to 0}\, {\mathrm{d}} \sigma_{\rm Siv}/{\mathrm{d}} t \propto
|x f_{1T}^{\perp (1) g}(x)|^2$
is proportional to the square of the $\vec{k}_\perp^2$ moment of the
small-$x$ gluon Sivers function,
$x f_{1T}^{\perp (1) g}(x)$.
The angular distribution of the $\chi_{c1}\to J/\psi + \gamma$
decay,
$
W(\theta,\phi) \propto \big(1 + \lambda_\theta\cos^2\theta + \dots\big)/(3 + \lambda_\theta)
$
is characterized by the coefficient
$
\lambda_\theta = (2r - 1)/(2r + 3)\,,
$
where $r$ denotes the ratio of the Sivers to Primakoff cross sections
at $t\to 0$.  If photon exchange is dominant, $\lambda_\theta \to
-1/3$; if three gluon exchange dominates then $\lambda_\theta \to
1$.


\subsection{TMD and SIDIS physics at small $x$}

Despite its solid conceptual foundations and its compelling applications to phenomenology (as briefly summarized in Sec.~\ref{sec:currentcgc}; see also the review papers~\cite{Gelis:2010nm,Lappi:2010ek,Albacete:2014fwa,Morreale:2021pnn} and Refs. therein), gluon saturation remains an elusive phenomenon, difficult to measure in practice. This is so because this phenomenon occurs in a limited region of phase-space --- small values of $x\lesssim 10^{-2}$ and semi-hard transverse momenta (or virtualities), in the ballpark of 1~GeV--- which is experimentally difficult to access and where non-perturbative effects, like QCD confinement, can obscure the physical interpretation. 
The characteristic transverse momentum scale for small-$x$ gluons in a hadron wavefunction is
the saturation momentum $Q_s(x,A)$.  It  grows with decreasing longitudinal fraction $x$ and with increasing nuclear mass number $A$, roughly like  $Q_s^2(x,A)\propto A^\delta/x^\lambda$ with $\delta\simeq 1/3$ and $\lambda\simeq 0.25$. But for realistic experimental conditions, this scale is {\it semi-hard}. Even for a large nucleus like Pb  ($A=208$), the value of $Q_s$ will not exceed 1 to 2~GeV in the typical ``small-$x$'' kinematics at the EIC, i.e. for $x\sim 10^{-2} \div  10^{-3}$. Inclusive quantities, like the DIS structure function  $F_2(x, Q^2)$ at small $x\le 10^{-2}$, probe saturation only in a convoluted way, that is, by mixing partonic configurations over a wide range in $k_\perp$. Besides, when $Q^2\lesssim Q_s^2$ they are also influenced by non-perturbative, hadronic, configurations. The observation of {\it geometric scaling}~\cite{Stasto:2000er} in the small-$x$ data for electron-proton DIS at HERA strongly suggests that the effect of saturation may extend to much larger virtualities $Q^2\gg Q_s^2$. Yet, this interpretation is  ambiguous since, for such large values of $Q^2$, the data can also be described by the conventional pQCD approach, based on collinear factorization. More exclusive observables, like the production of hadrons with transverse momenta $P_\perp\lesssim Q_s$, are affected by ambiguities as well, due to non-perturbative effects like hadronization. And jet production (which would be insensitive to such effects) can hardly be measured for semi-hard $P_\perp$.

Yet, over the recent years, it has been progressively realized that signatures of saturation can also be found in semi-inclusive processes at small $x$ which are {\it relatively hard} --- in the sense that {\it some} of the external transverse scales (transverse momenta or virtualities) are much larger than $Q_s(x,A)$. This is interesting since hard  jets, or hadrons, with $P_\perp\gg Q_s$ are easier to measure in the final state and less sensitive to non-perturbative physics.  So long as $x\sim P_\perp^2/s\ll 1$, such processes can still be computed within the CGC effective theory~\cite{Gelis:2010nm,Iancu:2002xk,Iancu:2003xm}. Despite involving hard scales, they can still be sensitive to gluon saturation if the observables are chosen well.

The typical example is back-to-back production of a pair of jets, or hadrons, in ``dilute-dense''  --- electron-nucleus ($eA$) or proton-nucleus ($pA$) --- collisions.
The transverse momenta $\bm{k}_{1\perp}$ and $\bm{k}_{2\perp}$ of the measured jets (or hadrons) are assumed to be much larger than both the nuclear saturation momentum $Q_s(x,A)$ and the dijet  imbalance $K_\perp\equiv |\bm{k}_{1\perp}+\bm{k}_{2\perp}|$. The relevant hard scale  is the dijet relative momentum $P_\perp\equiv |\bm{k}_{1\perp}-\bm{k}_{2\perp}|/2$ and obeys $P_\perp\simeq k_{1\perp}\simeq k_{2\perp}\gg K_\perp, \, Q_s$. The imbalance $K_\perp$ 
is produced via scattering off the nuclear target and thus becomes sensitive to gluon saturation in the semi-hard regime at $K_\perp\lesssim Q_s$~\cite{Kharzeev:2004bw,Marquet:2007vb}. The effects of saturation can be measured from the distribution of the produced jets in the azimuthal angle $\Delta\Phi$ between $\bm{k}_{1\perp}$ and $\bm{k}_{2\perp}$: nuclear effects are expected to lead to the broadening and the suppression of the back-to-back peak at $\Delta\Phi=\pi$ in $eA$ collisions compared to $ep$ collisions with similar kinematics~\cite{Albacete:2010pg,Lappi:2012nh,Zheng:2014vka,Stasto:2018rci,Mantysaari:2019hkq}.
The {\it suppression} of the back-to-back peak has indeed been observed in d+Au collisions at RHIC~\cite{Adare:2011sc,STAR:2021fgw} and in p+Pb collisions at the LHC~\cite{ALICE:2015ppz,ATLAS:2019jgo}, but for its {\it broadening} the situation is less clear: the measured peak appears to be quite wide, but it does not look {\it wider} in $pA$ collisions (d+Au or p+Pb) than in proton-proton collisions ($pp$) with similar kinematics --- a process where there should be no  saturation. This absence of additional, nuclear broadening can be linked to soft gluon radiation -- often referred to as {\it Sudakov effects} -- to be discussed shortly.

For the $eA$ collisions to be measured at the EIC, there are additional hard processes that could be used to study saturation at small $x$. When the photon virtuality $Q^2$ is sufficiently large, $Q^2\gg Q_s^2(x,A)$, the semi-inclusive production of  a single jet, or hadron (SIDIS) is a hard process, irrespective of the value of the transverse momentum $K_\perp$ of the measured jet (or hadron). At small $x$ and for $Q^2\gg K_\perp^2$, this is a leading-twist process which measures the sea quark distribution in the nuclear target. When  $K_\perp\lesssim Q_s$, it probes {\it quark} saturation --- an induced phenomenon, which is the consequence of gluon saturation and of the fact that sea quarks are produced in pairs via the decay of small-$x$ gluons~\cite{McLerran:1998nk,Mueller:1999wm,Marquet:2009ca}. 

Still in the context of $eA$ collisions, the sensitivity to saturation is even stronger for {\it diffractive} processes \cite{Kowalski:2008sa}, in which the nucleus scatters elastically and thus emerges unbroken in the final state. In this case, there is a rapidity gap between the emergent nucleus and the produced particles (partonic fluctuations of the photon that are put on-shell by the collision). The elastic cross section is controlled by strong scattering and hence it is particularly sensitive to gluon saturation. Indeed, unlike the total (inclusive) cross section, which by the optical theorem is proportional to the imaginary part Im~$\mathcal{A}_{el}$ of the elastic amplitude, the elastic cross section is proportional to its modulus squared $|\mathcal{A}_{el}|^2$. At small-$x$, $\mcal{A}_{el}$ involves a color-singlet exchange, a.k.a. the hard Pomeron,  and is
proportional to the gluon occupation number. So, the elastic scattering is naturally biased towards the saturated gluons, and as a result hard diffraction is known to be a semi-hard process (sensitive to physics at the scale $Q_s$), even at large $Q^2$ \cite{Golec-Biernat:1998zce,Golec-Biernat:1999qor}. Similarly, in the case of diffractive SIDIS, the produced quark has a semi-hard momentum $K_\perp\lesssim Q_s$. (For much larger values $K_\perp\gg Q_s$, the cross section is power-suppressed.) Likewise, in the diffractive production of back-to-back dijets, the typical imbalance $K_\perp$ is of order $Q_s$~\cite{Iancu:2021rup,Hatta:2022lzj}.

A remarkable property of the hard processes at small $x$ is that they exhibit {\it TMD factorization}, like their counterparts at moderate values of $x$. Indeed, explicit calculations~\cite{Marquet:2009ca,Dominguez:2010xd,Dominguez:2011wm,Metz:2011wb,Dominguez:2011br,Xiao:2017yya,Marquet:2017xwy,Iancu:2021rup,Iancu:2022lcw,Iancu:2023lel,Hauksson:2024bvv,Hatta:2022lzj,Altinoluk:2024tyx,Caucal:2025xxh,Caucal:2025qjg} with the CGC effective theory show that, when computed  {\it  to leading power in the hard scale} ---  i.e. by keeping only the leading-order terms in the double expansion in powers of $Q_s/P_\perp$ and $K_\perp/P_\perp$ --- the respective cross sections can be written as  products of a  {\it partonic cross section} (or {\it ``hard factor''}), which encodes the dependence upon the hard scale $P_\perp$ (or $Q^2$), and a  {\it transverse-momentum dependent parton distribution function} (``TMD'' in short),  which refers to {\it small-$x$ partons (gluons or sea quarks) in the dense target}. 

The TMDs in general describe parton distributions in the three-dimensional space spanned by the parton longitudinal momentum fraction $x$ and its transverse momentum $\bm{K}_\perp$. They satisfy the two-parameter CSS evolution equations which describe evolution of the TMDs with UV and rapidity scales. As such they carry more detailed information about the hadronic structure than the standard (or ``collinear'') parton distribution functions (PDFs) which are one-dimensional and describe the distribution in $x$ alone~\cite{Sterman:1993hfp,Collins:2011zzd}. Roughly speaking, the PDFs are obtained by integrating the TMDs over $K_\perp$ up to the hard scale $P_\perp$, which acts as a transverse resolution scale.

There are some formal similarities between the traditional TMDs at moderate $x$~\cite{Collins:2011zzd,Boussarie:2023izj} and those emerging from the CGC at small $x$ --- they correspond to the same operator definitions in terms of non-local bilinear forms of quantum (quark and gluon) fields that are joint by gauge links to ensure local gauge invariance --- but also some important differences. The CGC TMDs are explicitly known (to leading order in $\alpha_s$ at least) and they apply at small $x$ only. They include the effects of gluon saturation via Wilson lines which describe multiple scattering in the eikonal approximation and which materialize the gauge-links from the operator definitions already at tree-level. Remarkably, the {\it sea} quark TMDs~\cite{Marquet:2009ca,Caucal:2025xxh} as well as the {\it diffractive} (quark and gluon) TMDs~\cite{Iancu:2021rup,Hatta:2022lzj,Iancu:2022lcw,Hauksson:2024bvv} have  for the first time been unveiled within the CGC approach. See also the discussions in Refs.~\cite{Bhattacharya:2025fnz,Kovchegov:2025yyl,Benic:2026idy}.
 
Besides the fact that it facilitates the applications to phenomenology, TMD factorization has another important virtue: it offers a framework for resumming large radiative corrections to all orders. Within the CGC framework, such corrections are most conveniently computed as additional parton emissions (real or virtual) within the wavefunction of the dilute projectile --- the proton for $pA$ collisions and the photon for $eA$ collisions. The full next-to-leading order (NLO) corrections are known for several among the aforementioned processes (notably those referring to $eA$ collisions)~\cite{Ayala:2016lhd,Ayala:2017rmh,Iancu:2020mos,Caucal:2021ent,Taels:2022tza,Bergabo:2022tcu,Iancu:2022gpw,Bergabo:2022zhe,Caucal:2024cdq,Altinoluk:2025dwd,Fucilla:2022wcg,Fucilla:2023mkl} and for arbitrary kinematics. 
Here however we are merely interested in the ``hard'' kinematics and especially in the fate of TMD factorization. To that aim, we need the NLO corrections only to leading power in the hard scale. It turns out that it is more efficient, and also more illuminating, to directly compute them within this approximation, rather than try and extract them from the full NLO results~\cite{Mueller:2012uf,Mueller:2013wwa,Taels:2022tza,Caucal:2022ulg,Caucal:2023fsf,Hauksson:2024bvv,Caucal:2024bae,Altinoluk:2024vgg,Caucal:2024vbv,Caucal:2025mth}.  
Specifically, the NLO corrections are obtained by integrating over the phase-space of unmeasured gluon emissions. Given the hierarchy of longitudinal and transverse momentum scales in the problem, and the soft and collinear divergences of the parton branching rates, these integrals can generate large logarithms:
 the  rapidity (or BFKL) logarithm $Y\equiv \ln(1/x)$, the Sudakov double logarithm $\ln^2(P_\perp^2/K_\perp^2)$, and the DGLAP logarithm $\ln(P_\perp^2/Q_s^2)$. Such large corrections can dilute, or even hide, the effects of gluon saturation in applications to phenomenology. 

Consider, for example, the effects of the Sudakov  double-logs on the cross section for back-to-back dijet production in $eA$ collisions. Within the CGC picture for this process, these large radiative corrections are generated via soft and collinear gluon emissions in the {\it final state} (after the hard partonic process)~\cite{Mueller:2012uf,Mueller:2013wwa}. The typical emissions have transverse momenta $k_{g\perp}$ within the range $P_\perp\gg k_{g\perp}\gg Q_s$. Via their recoil, they can strongly enhance the imbalance $K_\perp$ between the produced jets, thus obscuring the signatures of gluon saturation in observables like the dijet distribution in the azimuthal angle $\Delta\phi$~\cite{Zheng:2014vka,Albacete:2018ruq,Stasto:2018rci,vanHameren:2016ftb,vanHameren:2019ysa,Al-Mashad:2022zbq,vanHameren:2023oiq,Cassar:2025vdp,Caucal:2025zkl}. Moreover, these radiation effects act in the same way in $pA$ and $pp$ collisions: in both cases, they produce a wide back-to-back peak at $\Delta\phi=\pi$. This may explain why previous measurements of two-particle correlations at RHIC~\cite{Adare:2011sc,STAR:2021fgw}  and the LHC~\cite{ALICE:2015ppz,ATLAS:2019jgo} have not observed additional nuclear broadening in $pA$ collisions compared to $pp$.  

 Yet, the existence of large radiative corrections does not necessarily mean that the searches for saturation in hard processes are {\it a priori} hopeless. Corrections enhanced by kinematical logarithms occur to all orders, but their systematics is well understood, so they can be isolated and resummed by solving appropriate evolution equations. The rapidity logarithms pose no conceptual problem: the CGC effective theory was specially tailored to facilitate their resummation, by solving the {\it B-JIMWLK hierarchy of equations}~\cite{Balitsky:1995ub,JalilianMarian:1997jx,JalilianMarian:1997gr,Kovner:2000pt,Weigert:2000gi,Iancu:2000hn,Iancu:2001ad,Ferreiro:2001qy} (or the {\it BK} equation~\cite{Kovchegov:1999yj} when appropriate). On the other hand, the treatment of the collinear (Sudakov and DGLAP) logarithms within the CGC formalism is less obvious. The standard tools used for their resummation within the collinear factorization at moderate values of $x$~\cite{Sterman:1993hfp,Collins:2011zzd} --- the DGLAP equation~\cite{Gribov:1972ri,Altarelli:1977zs,Dokshitzer:1977sg} for the PDFs and the CSS equation~\cite{Collins:1981uk,Collins:1981uw,Collins:1984kg,Collins:2011zzd,Boussarie:2023izj} for the TMDs --- are not {\it a priori}  built-in within the CGC effective theory. Nevertheless, there is recent work deriving the CSS equations in the presence of  a background field which may offer a hint on how to unify the two formalisms in general~\cite{Altinoluk:2025ewj}.  
 
Interestingly enough, recent results demonstrate that the DGLAP and CSS equations also emerge from the CGC approach at small $x$, after computing the NLO corrections. This has been explicitly verified for several processes (so far only in the context of $eA$ collisions): diffractive~\cite{Hauksson:2024bvv} and inclusive SIDIS~\cite{Caucal:2024vbv} in the hard regime at $Q^2\gg K_\perp^2$, and 
inclusive~\cite{Caucal:2024bae,Caucal:2025mth} and diffractive~\cite{Iancu:2025jsu} di-jets in the back-to-back configuration ($P_\perp\gg K_\perp$). By solving these evolution equations, one can keep the potentially large radiative corrections under control and thus perform calculations that are accurate enough to be able to disentangle the effects of gluon saturation. In particular, the solutions to the CSS equation ensure the resummation of the Sudakov logarithms within the structure of the TMDs.

What is particularly subtle about the emergence of the DGLAP and CSS equations from the CGC is the fact that these equations refer to parton distributions (PDFs and TMDs) in the {\it nuclear target}, while in practice they have been obtained by computing gluon emissions within the wavefunction of the {\it photon projectile}. That is, the effects of the NLO corrections  have been effectively transferred from the wavefunction of the projectile to that of the target. Moreover, even though these equations preserve the same general structure as in the collinear factorization at moderate $x$, they also exhibit new features which are specific to this small-$x$ context, like the interplay with the high-energy evolution and the sensitivity to gluon saturation. These new features enter via the initial condition for the DGLAP equation and via the boundary condition for the CSS equation, and thus affect the respective solutions~\cite{Caucal:2024bae,Iancu:2025jsu}.

\begin{figure}[t]
    \centering \hspace*{-0.2cm}
    \includegraphics[width=0.51\textwidth]{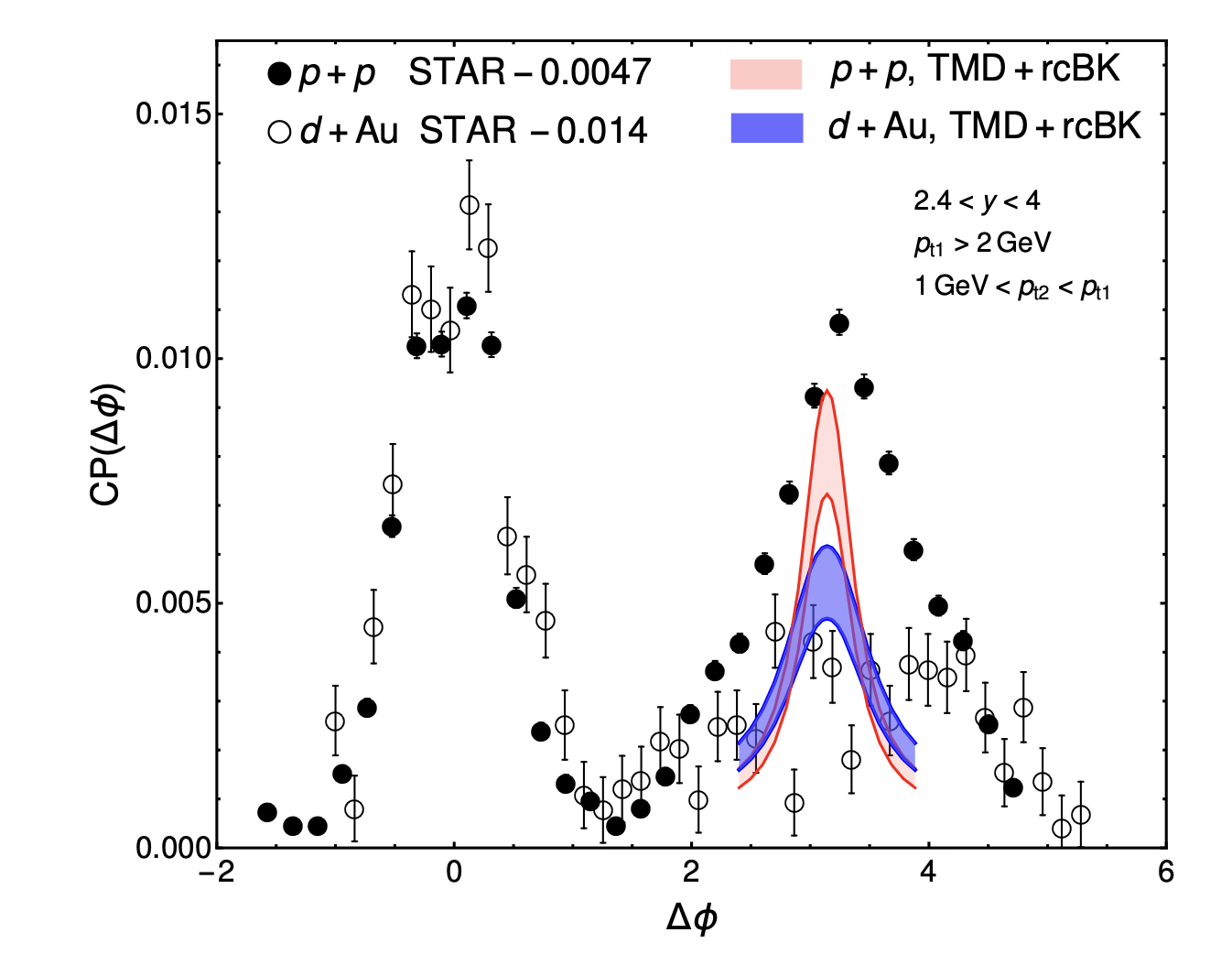}  \quad
       \includegraphics[width=0.42\textwidth]{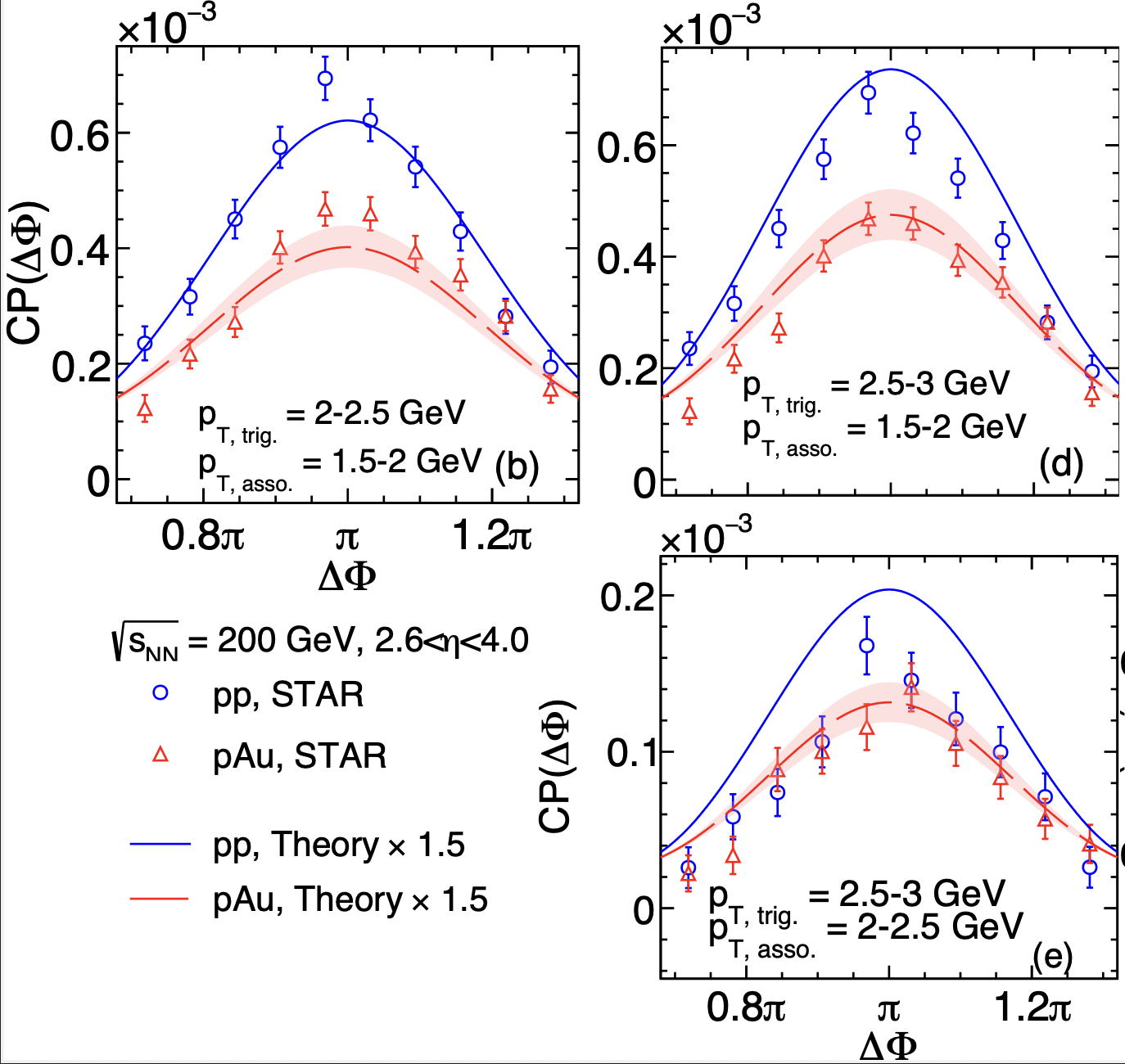}
      \caption{\small Di-hadron ($\pi^0$) distributions in the azimuthal angle $\Delta\phi$ in proton-proton and 
 deuteron-gold (or proton-gold) collisions: STAR data at $\sqrt{s}=200$~GeV vs. two theoretical calculations.
   {\it   Left:}  High-energy evolution (rcBK) but no Sudakov resummation \cite{Albacete:2018ruq}; the data are from \cite{Braidot:2010zh}. {\it Right:}  High-energy evolution (rcBK) plus  Sudakov resummation
      \cite{Caucal:2025zkl}; the data are from \cite{STAR:2021fgw}.
}
    \label{fig:DeltaPhi}
\end{figure}

To illustrate the effects of the Sudakov resummation, we show in Fig.~\ref{fig:DeltaPhi} the results of two theoretical calculations~\cite{Albacete:2018ruq,Caucal:2025zkl} for the di-hadron distribution in the azimuthal angle $\Delta\phi$, for both $pp$ and  $pA$ collisions. Both calculations use TMD factorization and include the high-energy evolution of the gluon TMD (within a mean field approximation based on the BK equation with running coupling). The early calculation in Ref.~\cite{Albacete:2018ruq} does not include the CSS evolution and thus
predicts a relatively narrow back-to-back peak at $\Delta\phi=\pi$ --- considerably narrower than the respective data at RHIC~\cite{Braidot:2010zh}. The recent calculation in Ref.~\cite{Caucal:2025zkl} also includes the resummation of the Sudakov logarithms via solutions to the CSS equation. The respective prediction for the away peak is significantly broader and in fairly good agreement with the RHIC data in \cite{STAR:2021fgw}
(especially for the p+Au collisions). Other applications of Sudakov resummation to the phenomenology of hard processes at small $x$ can be found e.g. in~\cite{Zheng:2014vka,Stasto:2018rci,vanHameren:2016ftb,
vanHameren:2019ysa,Al-Mashad:2022zbq,vanHameren:2023oiq,Cassar:2025vdp}.

\subsection{New observables at LHC and EIC}

Ultra-peripheral collisions (UPCs) at hadron colliders, or exclusive DIS/photoproduction at the EIC
may have the potential to discover the elusive hard Odderon of
perturbative QCD (see~\cite{Ewerz:2003xi} for a review of the early
literature).  This represents a $C$-conjugation odd $t$-channel
exchange in eikonal amplitudes, which in perturbation theory starts
out as three gluons.  The amplitude is proportional to the cubic
Casimir of color-$SU(N_c)$ whereas Pomeron exchange amplitudes
involve the quadratic Casimir.
Theory~\cite{Kovchegov:2003dm,Hatta:2005as} predicts a strikingly
different small-$x$ evolution of the $C$-odd versus
the $C$-even exchange: with increasing energy the target
proton becomes more
transparent~\cite{Lappi:2016gqe,Yao:2018vcg,Benic:2024pqe}.
Furthermore, processes involving Odderon exchange provide
valuable, complementary
insight into the structure of the proton since the Odderon couples to
$C$-odd color charge fluctuations. For example, some
model wave functions predict much weaker but substantially
``harder'' coupling of the Odderon to the proton than for the
Pomeron~\cite{Dumitru:2019qec,Dumitru:2022ooz,Benic:2024fbf}.

\begin{figure}
    \centering
    \includegraphics[width=0.49\linewidth]{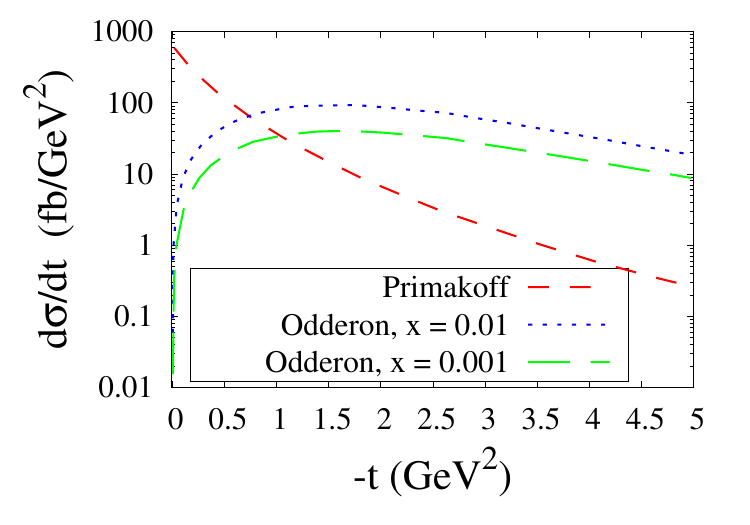}
    \vspace*{-0.5cm}
    \caption{Photoproduction cross section of $\chi_{c1}$ axial-vector
      charmonium in $\gamma-p$ scattering, as a function of momentum
      transfer~\cite{Benic:2024fbf}.}
    \label{fig:chi_c1-photoproduction}
\end{figure}
The exclusive production of  $\chi_{c1}$
axial-vector charmonium may represent a suitable process for the discovery
of the hard Odderon~\cite{Benic:2024fbf} as this meson has a
fairly large branching ratio of $34.3\%$ to $J/\psi +
\gamma$. Furthermore, this also provides access to the helicity
structure of the process via the angular distribution of the decay
products (see also Sec.~\ref{sec:smallx_helicity}).  For high momentum
transfer, $|t| > 1-2$~GeV$^2$, Odderon exchange is predicted to
dominate over photon exchange, see
Fig.~\ref{fig:chi_c1-photoproduction}, and may even exceed the
background from vector $\psi(2S)$ production (via Pomeron exchange) with
subsequent radiative decay to $\chi_{c1}+\gamma$~\cite{Siddikov:2025orq}.

\begin{figure}
    \centering
    \includegraphics[width=0.49\linewidth]{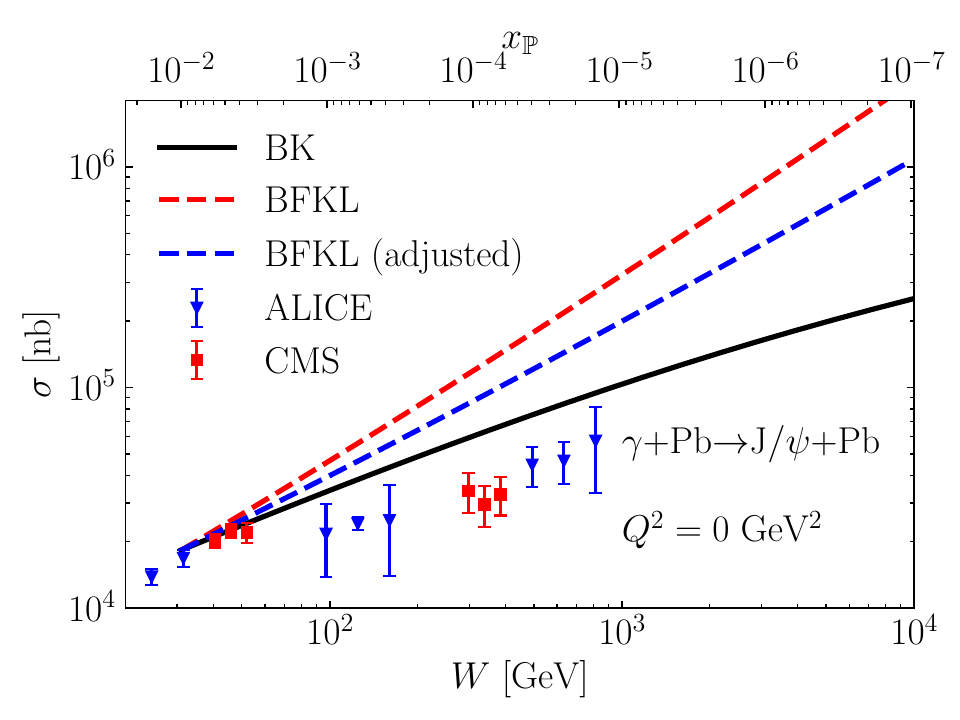}
    \vspace*{-0.5cm}
    \caption{
    Exclusive photoproduction of $J/\psi$ for lead target, calculated using BFKL and BK evolutions for the dipole amplitude.
    Taken from Ref.~\cite{Penttala:2024hvp}.
    }
    \label{fig:BFKL_vs_BK}
\end{figure}

A powerful process for studying gluon saturation is exclusive heavy vector meson production.
At leading order in perturbation theory, this process requires an exchange of two gluons with the target as opposed to one, making it very sensitive to the gluon density.
Additionally, the heavy quark mass provides a scale that makes this process perturbative even in the photoproduction limit that can be studied in the UPCs at the LHC.
For this reason, this process has been studied extensively in the small-$x$ limit~\cite{Bautista:2016xnp,ArroyoGarcia:2019cfl,Hentschinski:2020yfm,Peredo:2023oym,Mantysaari:2018nng,Penttala:2024hvp} for deviations between linear BFKL and nonlinear BK evolutions, with the energy dependence of the nuclear data showing better agreement when including saturation effects (see Fig.~\ref{fig:BFKL_vs_BK}).
However, the experimental data seems to indicate even more nuclear suppression than predicted by gluon saturation alone~\cite{Mantysaari:2022sux,Mantysaari:2023xcu,Mantysaari:2024zxq,Mantysaari:2025ltq}.
Resolving this discrepancy is crucial for a consistent description of exclusive heavy vector meson production for both proton and nuclear targets, and would serve as strong evidence of nonlinear QCD effects in heavy nuclei.

As these works have been carried out only at leading order, NLO corrections offer one possible explanation to the problem.
While exclusive heavy vector meson production has been computed at NLO in the nonrelativistic limit~\cite{Mantysaari:2021ryb,Mantysaari:2022kdm}, relativistic corrections for $J/\psi$ photoproduction have been shown to be large~\cite{Lappi:2020ufv},
making these results inapplicable for comparisons to the LHC data.
Including relativistic corrections at higher orders in perturbation theory is generally difficult due to the nonperturbative nature of the meson wave function.
Possible avenues for this includes calculating velocity corrections order by order using nonrelativstic QCD (NRQCD)~\cite{Bodwin:1994jh}, or a resummation of a subset of velocity corrections to all orders using the generalized Gremm--Kapustin relation~\cite{Gremm:1997dq,Bodwin:2006dn}.
Such resummations have been applied successfully for other processes, see e.g. Refs.~\cite{Bodwin:2007fz,Li:2019ncs,Babiarz:2025agk}.

Turning to inclusive heavy vector meson production, the process 
$e\,p \to e\, J/\psi\, X$ has attracted great interest as a probe of TMD gluon distributions. It has also been shown that the TMD expression of the cross section requires a so-called {\it shape functions}~\cite{Echevarria:2019ynx,Fleming:2019pzj} 
instead of the usual long distance matrix elements (LDMEs) of NRQCD, which combines soft gluon radiation and the formation of the bound state, in order to
describe quarkonium production at small transverse momentum. Its perturbative tail has been calculated by matching the cross sections valid at low and high transverse momenta and, more recently, within the SCET formalism~\cite{Boer:2023zit,Echevarria:2025oab}. It is expected that with the upcoming EIC and more data provided by $pp$ facilities, such as the LHC in fixed target mode, extractions of the TMD shape functions will become available in the future and new features of heavy quarkonium production will be uncovered. 

Due to the elusive nature of gluon saturation, it is important to study a wide range of different observables to precisely determine genuine saturation effects from the experimental data.
One class of observables that has recently garnered renewed attention are energy correlators~\cite{Basham:1978bw,Basham:1978zq}, where one studies the outgoing energy flow instead of individual particles (see Ref.~\cite{Moult:2025nhu} for a recent review).
These energy correlator observables have the advantage of being less sensitive to the nonperturbative final-state hadronization effects, allowing one to focus more on other parts of the collision such as the target structure.
In the context of small-$x$ physics, this has been studied for transverse energy--energy correlators~\cite{Kang:2023oqj,Kang:2025vjk,Bhattacharya:2025bqa,Ganguli:2025aqa,Kang:2026hig} and nucleon energy correlators~\cite{Liu:2023aqb,Mantysaari:2025mht}, indicating the potential as a probe for gluon saturation and Odderon interaction.
While experimental data of energy correlators in the small-$x$ region is still lacking, these theoretical predictions indicate that future studies of energy correlators can provide important information about the high-energy structure of nucleons and nuclei.

%% file: week4.tex
\subsection{AI for inverse problems} 

The determination of physics parameters from experimental observables, lattice QCD inputs, and theoretical priors is a ubiquitous inverse problem in particle physics phenomenology relevant to the EIC and other current and future collider facilities. 
The variational autoencoder inverse mapper (VAIM) \cite{Almaeen:2021,Almaeen:2024guo, Hossen:2024qwo} provides a convenient framework for cross-domain inference \cite{Kriesten:2021sqc,Kriesten:2019jep,Kriesten:2020wcx}, integrating heterogeneous physics constraints which can be benchmarked against global analyses \cite{Almaeen:2022imx,Hossen:2024qwo}. 
The latent representation, the embedding space produced by the encoder model, 
catches ``missing" physics-based information abstracted away in the forward mapping, allowing for physics interpretability.
%

There has been an increased exploration of lattice-QCD--aware ML-based approaches which bridge lattice calculated data and phenomenological extractions of quantum correlation functions, disentangling the PDFs and identifying many underlying PDF shape profiles which may be statistically compatible with the constraints~\cite{Karpie:2019eiq, DelDebbio:2020rgv, Dutrieux:2024rem, Chowdhury:2024ymm, Zhang:2020gaj, Boyda:2022nmh}. 
VAIM builds on these studies as an end-to-end framework, which makes it a particularly insightful tool to investigate open problems in physics-based analyses such as the parametrization dependence of PDF fits, as well as the lattice inverse problem. 
Recently, VAIM was deployed to interrogate the extraction of the gluon PDF from the reduced pseudo-Ioffe-time distributions (RpITDs)~\cite{Balitsky:2021bds,Kriesten:2025gti}.

\begin{figure}
    \centering
    \includegraphics[width=0.9\linewidth]{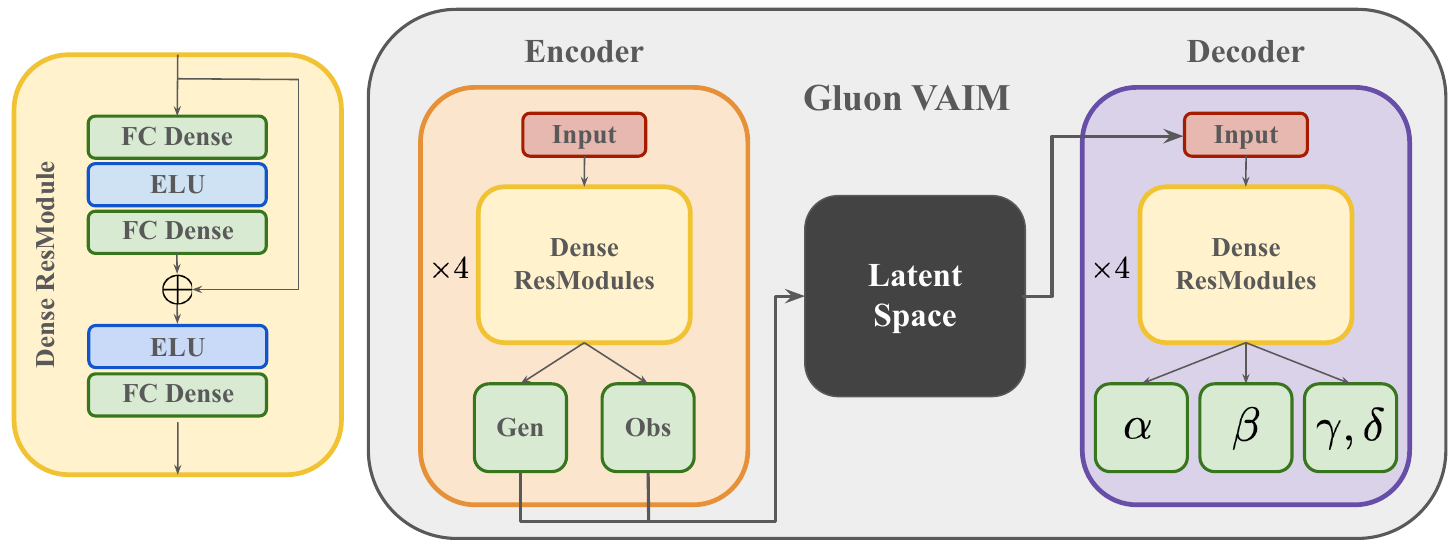}
    \caption{Overview of C-VAIM architecture.}
    \label{fig:placeholder}
\end{figure}
It was found in the study of Ref.~\cite{Kriesten:2025gti} that the VAIM could disentangle the gluon PDF from the lattice calculated RpITDs with realistic uncertainties that were comparable (within the lattice $x$ region of validity) with those in phenomenological global PDF fits to experimental data. 
The resulting error band is an ensemble of the many solutions which are statistically compatible with the training data uncertainties, arising from the ill-posed-ness of the inverse problem of inverting the convolution integral in the definition of the matching relation between the PDF and the RpITD.
The Pearson correlation was calculated over the decoder of the VAIM model to study correlated effects between the RpITDs and the subsequent gluon PDFs; it was found that many of the RpITDs have similar constraining power in specific $x$ regions of the gluon PDF.
%

In a pilot study~\cite{Kriesten:2023uoi}, 
VAIM was used to reconstruct the $x$-dependent information of PDFs from the integrated Mellin moments as calculable on the lattice.
Two scenarios were explored with varied sizes of the latent representation based on the number of input Mellin moments. 
Both scenarios produced comparable results in the extraction of the PDF, with the smaller latent representation performing slightly worse in reconstruction. 
However, when one inspects the learned physics content of the two models one finds that the larger latent representation allowed for the creation of spurious correlations. 
In the smaller model, the squeezing of information into a tighter latent representation forced the model to 'learn' the meaningful physics correlations. 
This spurred investigations into explainable AI as applied to PDF theory assumptions \cite{Kriesten:2024are}.
%

VAIM-based calculations like those presented above serve as an improvable and extensible framework to assess the impact of lattice information and its phenomenological implications in global QCD analyses for the forth-coming EIC.

\subsection{AI/ML in heavy-ion physics}
Accurate and efficient event simulation will be a critical component of the EIC physics program, where large-scale modeling of detector response to complex collision environments is essential. A survey for physics event generation is provided in ~\cite{Alanazi:ijcai2021p588}. Generative AI models can play a transformative role in this effort. Ref.~\cite{Torbunov:2024iki} has demonstrated that Denoising Diffusion Probabilistic Models (DDPM) can reproduce full-detector responses in heavy ion collisions with superior accuracy and stability compared to Generative Adversarial Networks (GANs), while achieving an effective speedup of up to 1,800 relative to conventional Geant4-based simulations. Although GANs can generate events at an even faster rate, the higher fidelity achieved by diffusion models is particularly valuable for physics applications that demand precision and reproducibility. Once trained on a relatively modest dataset (on the order of millions of events), these models can be used to generate statistically large samples (billions of events) at minimal computational cost, providing a scalable and high-fidelity framework for detector optimization and physics analyses at the EIC. In addition to fidelity, diffusion models also demonstrate better expressivity, which can conveniently represent particle multiplicity and learn physics constraints such as energy and momentum conservations~\cite{Devlin:2023jzp}.

Accurate extraction of physics signals from large and fluctuating backgrounds is essential for precision measurements in heavy-ion and EIC experiments. The Unsupervised Unpaired Image-to-Image Translation framework, UVCGAN-S~\cite{Go:2025exm}, has been developed to address this challenge by performing jet background subtraction directly from calorimeter-level data. Unlike supervised machine learning approaches that rely on labeled signal and background pairs, typically obtained from simulations and therefore prone to model-dependent biases, UVCGAN-S learns the underlying mapping between signal and background domains directly from unlabeled experimental data. By employing cycle-consistency principles, the framework preserves full event-level information and reconstructs background-free calorimeter images, enabling faithful jet signal extraction. When applied to simulated sPHENIX calorimeter images, UVCGAN-S achieves up to a factor-of-two improvement in jet momentum resolution and a substantial reduction in fake-jet rates relative to conventional subtraction methods. The approach further exhibits strong generalization capability, accurately reconstructing out-of-distribution (quenched) jets even though it is trained exclusively on unquenched samples. Although validated using simulated data, the method is inherently data-driven and can be applied directly to experimental datasets, providing a robust framework for realistic background subtraction in imaging-based detectors at the EIC.

The emergence of large-scale foundation models (FMs), such as Large Language Models (LLMs), has established a new paradigm in AI by enabling general representation learning and versatile task performance. Similar transformative potential is now being realized in nuclear and particle physics. Recent developments in experimental analysis have introduced foundation models trained directly on low-level detector signals~\cite{Park:2025ebs, Giroux:2025elr}, inspiring a new paradigm of scientific foundation models. The Foundation Model for Nuclear and Particle Physics (FM4NPP)~\cite{Park:2025ebs} employs self-supervised training on Time Projection Chamber (TPC) data from the sPHENIX experiment in proton–proton collisions and demonstrates neural scalability with architectures containing up to 188 million parameters.  Using frozen FM with lightweight task-specific adapters, the model consistently outperforms baseline networks across multiple downstream tasks and exhibits strong data-efficient adaptability. Analysis of the learned representations indicates that they are task-agnostic yet easily specialized through simple linear mappings for diverse applications.

The implications for the EIC are substantial. The FM4NPP framework aims to integrate multiple detector modalities, collision systems, and experimental environments, including the EIC, RHIC, and LHC, together with metadata and theoretical inputs within a unified training framework. Beyond advancing reconstruction, calibration, and classification, such models have the potential to enable autonomous reasoning and adaptive optimization in real-time data acquisition and analysis. Establishing a scalable, cross-experiment AI foundation will accelerate scientific discovery by linking simulation, theory, and experiment into a cohesive learning ecosystem, allowing researchers to focus on interpretation and physical insight while AI systems manage the complexity of large-scale data exploration.

\subsection{AI/ML applications at the EIC} 
In high-energy collision experiments with high event rates and large raw data volumes, hardware triggers have traditionally been unavoidable due to technological and bandwidth limitations. The future EIC detector is, however, expected to operate in a fully streaming (triggerless) readout mode. This capability will enable high-precision measurements, particularly in the low-$p_{\rm T}$ region, by eliminating data loss associated with trigger inefficiency. Achieving this requires the use of AI-driven data compression frameworks, such as the Bicephalous Convolutional Autoencoder (BCAE)~\cite{Huang:2021ymz,Huang:2023ype}, which performs real-time encoding of detector signals into compact latent representations suitable for permanent storage and later reconstruction. The BCAE framework has been demonstrated using TPC data from the sPHENIX experiment in heavy-ion collision environments, achieving higher compression efficiency and reconstruction fidelity than conventional lossy algorithms. Its variable-sparsity extension, BCAE-VS~\cite{Huang:2024tir}, is optimized for sparse data conditions such as proton–proton or electron–ion collisions, employing key-point identification and sparse convolution to dynamically adjust the compression ratio based on input complexity. Integration of such AI-based compression frameworks into streaming readout systems will be essential for managing the unprecedented data rates expected at the EIC while preserving the complete physics information necessary for precision QCD measurements.

\begin{figure}
    \centering
    \includegraphics[width=0.8\linewidth]{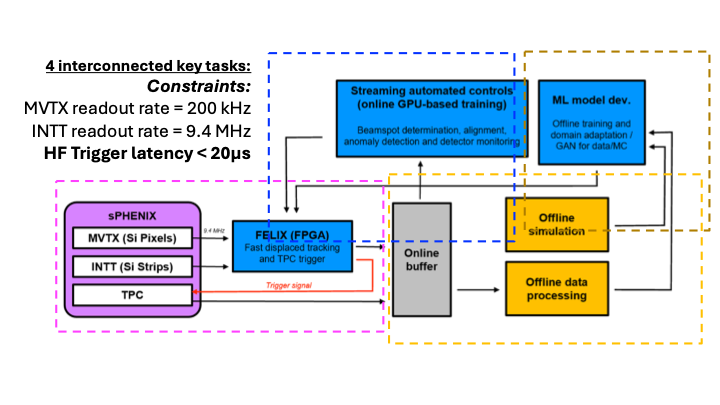}
    \vspace*{-0.5cm}
    \caption{
    Data flow chart for Fast-ML heavy flavor trigger system. 
    }
    \label{fig:Fast-ML-sPHENIX}
\end{figure}

Since 2022, a multi-institutional collaboration spanning Nuclear Physics, High Energy Physics, Computer Science, and Electronics Engineering,  researchers from LANL, MIT, FNAL, NJIT, ORNL, and GIT, has been advancing an AI-driven real-time processing demonstrator for the sPHENIX experiment at RHIC. Supported by the DOE Nuclear Physics AI/ML initiative, this effort targets the development of intelligent, low-latency data processing architectures that bridge current collider experiments with future capabilities at the Electron-Ion Collider. The demonstrator integrates streaming readout, intelligent control, and FPGA-accelerated inference to identify rare heavy-flavor events among collision rates on the order of 10 MHz, while operating within the O(15) kHz data acquisition limit, see Figure~\ref{fig:Fast-ML-sPHENIX} . By leveraging graph-based machine learning algorithms trained on detailed detector simulations and implemented in firmware through $hls4ml$ and FlowGNN frameworks, the system enables rapid signal identification, adaptive data reduction, and real-time detector feedback control.
A key milestone of this effort was achieved through the design and implementation of an AI/ML-based system utilizing selected streaming data exclusively from two silicon subsystems within sPHENIX, the MVTX (slow response, $\sim$5$\mu$s) and INTT (fast timing, $\sim $50$n$s) detectors. This proof-of-concept demonstrated that physics-driven trigger decisions can be made efficiently using only these critical subsystems, while remaining within the constraints of available hardware resources (LUT, FF, BRAM, etc.) on state-of-the-art FPGA boards (FELIX-712 and FELIX-182). The study validated the feasibility of using subsystem-limited data streams for effective heavy-flavor tagging in real time, underscoring the potential of machine learning–based trigger systems to optimize bandwidth and computational budgets without compromising physics performance. This initial demonstration not only showcases the viability of hardware-aware AI/ML and FPGA co-design for high-throughput discovery science but also establishes a clear pathway toward future AI/ML applications at the EIC. The techniques developed here, real-time inference, intelligent data selection, and adaptive background rejection, are poised to enhance rare signal tagging and suppress high-background events in next-generation collider experiments at EIC. Together, these advances mark a significant step toward realizing fully autonomous, AI-augmented data acquisition systems for future precision nuclear physics programs. 

\begin{figure}[bth]
\centering
\includegraphics[width=0.48\linewidth]{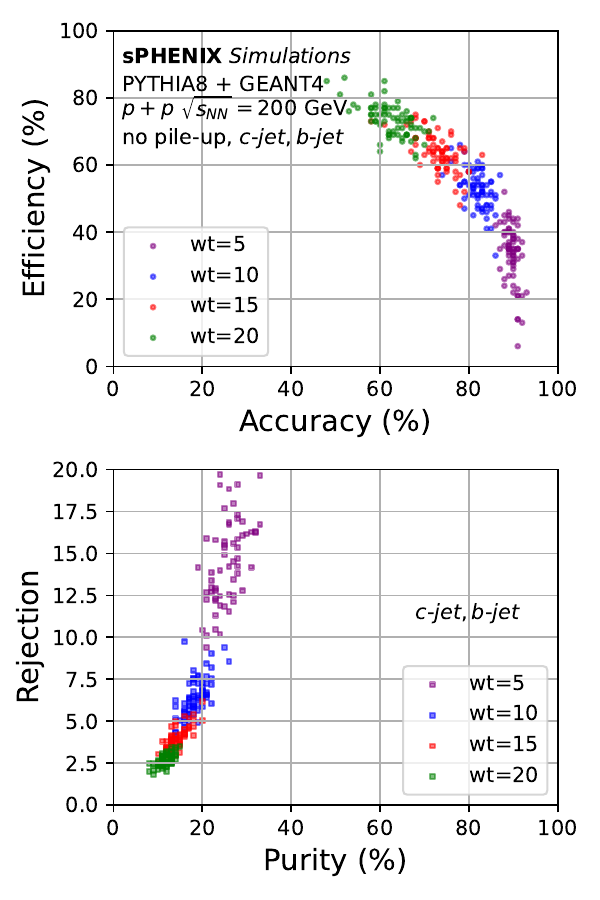}
\includegraphics[width=0.48\linewidth]{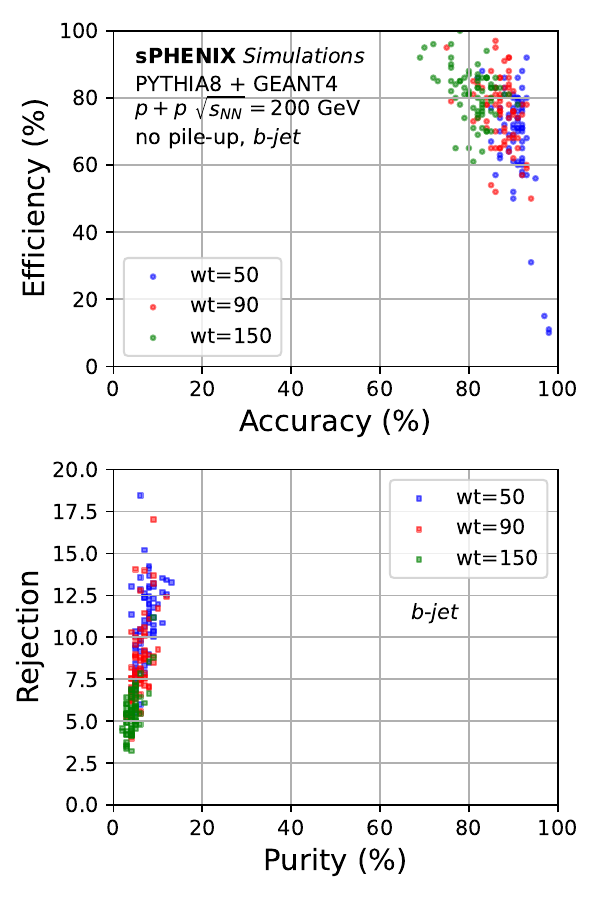}
\caption{Correlation metrics for heavy-flavor jet tagging (left column) and bottom jet tagging (right column) under LSTM model training using the sPHENIX 200 GeV $p+p$ simulation.}\label{fig:LSTM_jet_tag}
\end{figure}

In heavy ion experiments at the Relativistic Heavy Ion Collider (RHIC) and the Large Hadron Collider (LHC), jets are among the most important channels for testing perturbative Quantum Chromodynamics (pQCD) and probing the properties of the nuclear medium. Jets originating from charm and bottom quarks are of particular interest, as they are expected to undergo different parton energy loss mechanisms and hadronization processes compared to those from light quarks. Traditionally, heavy flavor jet tagging relies on precise measurements of the kinematic variables of jets and the particles within them. However, applying a series of selection on event-, jet-, and particle-level observables often limits the jet tagging purity and efficiency. In recent years, Machine Learning (ML) based approaches, including Convolutional Neural Networks (CNNs), Graph Neural Networks (GNNs), and transformer-based architectures, have been extensively employed in jet tagging analyses at the LHC, notably within the ALICE, ATLAS, and CMS experiments~\cite{Qu:2019gqs, Qu:2022mxj, Stein:2022nvf, CMS-DP-2024-066}. 

The sPHENIX experiment at RHIC is designed to reconstruct full jets and advance heavy flavor measurements using integrated tracking, as well as electromagnetic and hadronic calorimeters. A Neural Network based Long-Short-Term-Memory (LSTM) ML model \cite{6795963} has been utilized to study heavy flavor (charm+bottom) and bottom jet tagging at sPHENIX in 200 GeV $p+p$ simulation. The LSTM model uses cell states that retain memory of previous inputs, and at each time step gate control determines how information is updated, remembered, or discarded. The full 200 GeV $p+p$ simulation set was divided into 70 random subsets, and the model was trained sequentially on each subset for 400 epochs. For every subset, 50$\%$ of the events were used for training and 50$\%$ for testing. The correlations of efficiency, accuracy, background rejection, and purity for heavy flavor and bottom jet tagging are shown in Figure~\ref{fig:LSTM_jet_tag}. Jets in this simulation sample have a dominant transverse momentum $p_{T}$ around 10 GeV/c, and most of them originate from light quarks. Using unbiased inputs of track and jet kinematic variables such as track $p_{T}$ and Distance of Closest Approach (DCA) for LSTM training, a tagging performance comparable to that of the traditional cut based method can be achieved for heavy flavor jets. Bottom jets contain a larger number of tracks and exhibit longer decay length for particles from bottom hadron decays and within the jet, compared to jets of other flavors. The initial simulation studies indicate that the LSTM based ML model can successfully capture these features. Furthermore, the bottom jet tagging performance such as the purity versus efficiency achieved using the LSTM based ML model is better than that obtained with the traditional cut based method.

Heavy flavor jet and jet substructure measurements at the future EIC will play an essential role in mapping out the flavor dependent parton energy loss mechanisms in cold nuclear matter and the flavor dependent hadronization processes \cite{Li:2023koj, Li:2025lxr}. Jets reconstructed at the EIC are expected to require a large cone radius due to the significantly lower particle density in $e+p$ and $e+A$ collisions. Extending the ML based jet tagging approaches developed at sPHENIX to the EIC-related studies will enhance the precision and impact of the associated measurements.

\subsection{Using AI in search for new phenomena} 

The expanding landscape of beyond standard model (BSM) physics has recently spurred the development of systematic methods for making
quantitative determinations of model separability over a wide range of possible theories, particularly with controllable uncertainties and the identification of anomalies. 
This task naturally poses a classification problem with rigorous uncertainty quantification, separating total uncertainty into aleatoric (data overlap), epistemic (modeling and distributional), and distributional/OOD components.
It is convenient to project BSM scenarios with their numerous degrees of freedom onto a latent embedding space where we can then compare the theories on equal footing, and perform classification with uncertainty quantification and anomaly detection.

The separation of uncertainties into their components comes from the definition of the likelihood function.
In the context of classification, the negative log-likelihood can be rewritten as a sum of two separate contributions:
\begin{equation}
 \hspace*{-2.1cm}   \mathbb{E}_{p_{\mathrm{true}}(\mathbf{x},y)}\left[\mathcal{L}^\mathrm{NLL}(y,\mathbf{x},\theta) \right] \,=\, \mathbb{E}_{p_{\mathrm{true}}(\mathbf{x})}  \Bigg[ D_\mathrm{KL}\Big( p_{\mathrm{true}}(y| \mathbf{x}) \Big \| p( y| \mathbf{x}, \theta)\Big) 
    +\mathbb{H} \Big(p_{\mathrm{true}}(y| \mathbf{x})\Big) \Bigg]\ ,
\end{equation}
where $\mathbb{E}(\cdot)$ is the expectation value, $\mathbb{H}(p) = -p\ln p$ is the entropy, $D_\mathrm{KL}(p\| q) = p \ln (p/q)$ is the KL divergence between probability distributions $p$ and $q$, $p_{\mathrm{true}}(y|x)$ is the true underlying data distribution which is sampled to generate the data $x$ and labels $y$, and $p(y|x,\theta)$ is the modeled distribution over the labels given the data and model parameters $\theta$. Notice that this factorization has two contributions: (i) the KL divergence term which depends on how well the statistical model represents the data through parameters $\theta$ and choice of statistical distribution to represent that data --- the total epistemic contribution (which can be split into modeling uncertainty and distributional uncertainty), and (ii) the entropy term which is inherent to the underlying true dataset and is independent of the modeling methodology --- the total aleatoric contribution.

\begin{figure}
    \centering
    \includegraphics[width=0.8\linewidth]{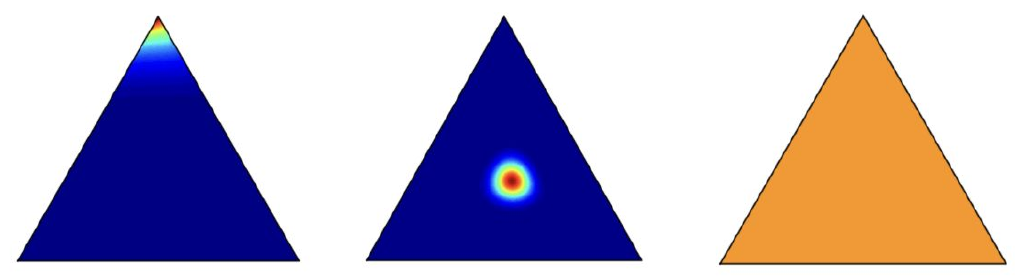}
    \caption{Example output of Dirichlet prior network demonstrating cases of ({\it left}) low aleatoric and low epistemic uncertainty, ({\it middle}) high aleatoric and low epistemic uncertainty, and ({\it right}) out of distribution samples.}
    \label{fig:dpn_example}
\end{figure}

Recent advances in evidential deep learning techniques (see Ref.~\cite{ulmer2023priorposteriornetworkssurvey} and references therein) allow for a machine learning algorithm to model these uncertainties and quantify them through information theoretic metrics.
Dirichlet prior networks (DPNs) model the prior probability distribution over the parameters of the categorical distribution, letting us read off evidence and uncertainties in analytic expressions of the Dirichlet concentration parameters; effectively implementing the above decomposition of the negative log-likelihood.
Figure \ref{fig:dpn_example} shows an example of the output of a DPN classification with three distinct classes represented as a categorical distribution with three parameters drawn as contours on a triangular 2-simplex with applicable use-cases related to combinations of epistemic and aleatoric uncertainties. 
In Ref.~\cite{Kriesten:2024ist}, using a typical multi-GeV $\nu$DIS experiment such as CDHSW
as an example, DPNs were deployed on an interpretable latent projection of various BSM anomalous electroweak interaction
scenarios which shift the CKM matrix elements within discovery ranges but result in sub-percent shifts in the calculated hadronic structure functions at leading order.
From the modeled concentration parameters of the DPN, various metrics were used to map contours of the uncertainty.
The aleatoric uncertainty can be written in terms of an expected value over the entropy of the posterior distribution,
and the distributional uncertainty (flagging OOD samples) can be written as the mutual information over modeled categorical distributions.

\begin{figure}[t!]
\centering
\includegraphics[width=0.8\columnwidth]{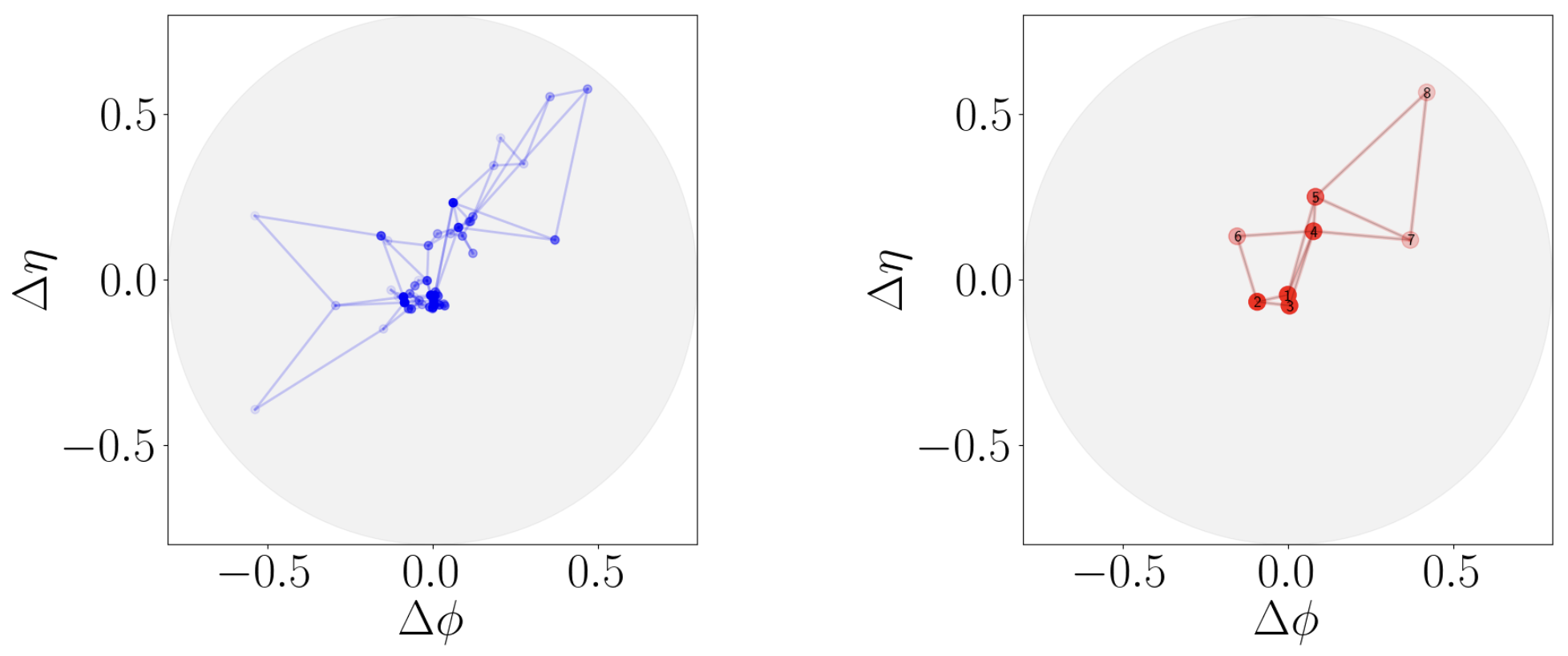}
\caption{Illustration of event representation as a sparse geometric graph, adapted from [59]. Left: Laman graph built from jet constituents. Right: graph constructed from reclustered subjets.~\label{fig:sparsegraph}}
\end{figure}

Machine-learning methods for anomaly detection using the full information content of collider events provide a complementary approach to traditional searches for new physics in high-energy collisions. These techniques offer sensitivity to deviations from expected event distributions without relying on an explicit signal hypothesis~\cite{Metodiev:2017vrx,Cerri:2018anq,Blance:2019ibf,Dillon:2021nxw,Hallin:2022eoq,Buhmann:2023acn}. Based on the idea that resonant or non-resonant new physics signals may manifest themselves more clearly using the full information instead of low-dimensional projections or observables, these methods can leverage the low-level particle information inside jets or the full collider event. Event-level representations based on geometric or graph structures are particularly suited for this purpose, as they encode the topology of hadronic final states and the correlations between their kinematic degrees of freedom. In recent work, an unsupervised graph autoencoder architecture was introduced inspired by graph theory~\cite{Araz:2025oax}. Events are represented as sparse graphs where the nodes correspond to hadrons or subjets and the edges encode their relative angular or momentum-space relationships in the rapidity-azimuth plane, see Fig.~\ref{fig:sparsegraph}. By enforcing geometric constraints based on Laman and uniquely rigid graphs, the model limits the accessible correlations and regularizes the learning process. The best performance, measured in terms of the Significance Improvement Characteristic curve, is obtained for graphs with an intermediate level of sparsity and certain unique graph constructions. The network is trained directly on background-rich data to identify outlier events, without assuming a specific signal model. Applied to collider benchmark datasets~\cite{Kasieczka:2021xcg}, the sparse graph representation provides a transparent and data-driven way to capture higher-order correlations that are otherwise difficult to parametrize in terms of standard jet observables. Complementary to these methods are weakly supervised methods where generative models are employed to create a signal-free background data set. Here, for example, diffusion models are trained on sideband regions where little to no signal is present, which is then interpolated into the signal region. Then an ML-based classifier can be trained to identify anomalous signals.

At the future Electron–Ion Collider, similar strategies can be employed to search for anomalous event topologies that could indicate new dynamics at the parton or hadron level, including possible manifestations of BSM physics. The flexible event-level representation provided by sparse graphs makes it possible to carry out model-independent searches that are sensitive to a wide range of unconventional final states, extending the reach of LHC-style anomaly detection to a new kinematic regime. In addition, different QCD-focused applications may be achieved. For example, the same or related frameworks can be used to identify systematic deviations from DGLAP-based expectations in inclusive and semi-inclusive observables, providing a new methodology to search for both new fundamental interactions and emergent collective behavior such as the color-glass-condensate (CGC). In this way, unsupervised anomaly detection can serve as a general tool to uncover previously unexplored structures in data at the EIC.

\subsection{Toward shared practices for data analysis in the EIC era}

As the Electron-Ion Collider (EIC) program develops, it presents an opportunity to
reconsider and refine how experimental measurements and theoretical predictions are
interfaced with one another. Building on experience from HERA, RHIC, JLab, and the LHC,
we outline here a set of heuristic practices intended to
strengthen the connection between theory and experiment. These points are not meant as
``prescriptive rules'', but rather as a starting framework for discussion and continued
evolution in the EIC era.

A key focus must be on maximizing the long-term impact and interpretability of
experimental data. Providing full covariance matrices, including both statistical
and systematic uncertainties, will significantly enhance the utility of
measurements for global analyses and inference. In cases where multiple
observables probing similar physics quantities 
are measured simultaneously, documenting correlations will
enable consistent combined fits and more robust theoretical interpretation.
This can be achieved by publishing the complete covariance matrices or, if this is not
practical, through detailed breakdown of systematic uncertainties.

Another guiding principle is the clear separation of instrumental and theoretical
effects. Experimental results should, where feasible, be unfolded to remove detector
effects, allowing direct comparison to present and future theoretical predictions.
We note that unfolding techniques entered the field only after the HERA era. Their
use at the EIC will be vital for full exploiting the measurements and
will likely contribute to analyses techniques at the EIC being able to outperform those at HERA. 
In a similar vein, it is advisable that corrections applied to data be restricted to those derived purely
from (GEANT~\cite{GEANT4:2002zbu}) detector simulations. Model-dependent adjustments, such as hadronization
effects, may be more transparently applied at the theory
level. When additional non-instrumental corrections are unavoidable, sufficient information should be
provided to allow them to be reversed.

Careful treatment of binning also warrants discussion. Measured
cross sections correspond to integrals over finite bin ranges, and interpreting them as values
at bin centers~\cite{Lafferty:1994cj} introduces model-dependent assumptions. A possible good practice is
to compare measurements with theoretical calculations integrated over identical bin
ranges, rather than applying bin-centering corrections to the data. If such
corrections are supplied, presenting their details as supplementary material alongside the
original bin definitions will preserve flexibility for future reinterpretation.
In a similar spirit, reporting fully inclusive DIS cross sections, rather than
reduced cross sections~\cite{H1:2009pze} evaluated at representative kinematic points in each bin, may help avoid
model dependence.

For jet measurements, alignment between experiment and theory in jet definitions
is essential for meaningful comparisons. Reporting measurements at hadronic level, within
clearly defined fiducial regions will be highly beneficial. Quite generally, 
on the theory side, transparent estimation of truncation uncertainties can help clarify
theoretical systematics. This may be achieved by comparing results at successive perturbative orders
and assessing the scale dependence of the predictions. Benchmarking global fits against common observables
and extracted functions may further illuminate the impact of forthcoming EIC data.

Analysis methodology itself can benefit from continued refinement.
Blinded analyses, potentially developed and validated using realistic Monte Carlo
pseudo-data, may help mitigate bias. Standardized output formats for
event generators (such as HEPMC3~\cite{Buckley:2019xhk}) could streamline integration into experimental
workflows.

Taken together, these ideas are meant to foster dialogue between the
experimental and theoretical communities. As the EIC program matures, the proposed 
practices will undoubtedly evolve, informed by new data, new techniques, and
shared experience.

%% file: week5.tex
\subsection{TMD factorization and nucleon structure}

The precise determination of the three dimensional (3D) structure of hadrons is one of most active fields of nuclear physics research and is a fundamental goal of future Electron-Ion Colliders (EICs)~\cite{Accardi:2012qut,AbdulKhalek:2021gbh,Anderle:2021wcy}. The information associated with the three dimensional momentum structure is encoded in the Transverse Momentum Distributions (TMDs)~\cite{Ji:2004wu,Ji:2004xq,Collins:2011zzd,Aybat:2011zv,Aybat:2011ge,Echevarria:2011epo} which enter as TMD Parton Distribution Functions (TMD PDFs),  TMD Fragmentation Functions (TMD FFs), and the TMD Jet Fragmentation Functions (TMD JFFs) \cite{Bain:2016rrv,Makris:2017arq,Kang:2017glf,Kang:2019ahe,Kang:2020xyq,Kang:2021ffh}; they are non-perturbative quantities which arise from QCD factorization theorems~\cite{Collins:1981uk,Collins:1981uw,Collins:1984kg,Ji:2004xq,Collins:2011zzd,Aybat:2011zv,Echevarria:2011epo}. In recent years there has been progress in calculating them from first principles in lattice QCD; see for instance~\cite{Musch:2011er,Engelhardt:2015xja,Yoon:2017qzo,Engelhardt:2023aem,Ji:2020ect,Ebert:2020gxr,Ebert:2022fmh}.

QCD factorization theorems are central to disentangling the perturbative and non-perturbative contributions to the cross section while evolution and resummation formalisms, such as the Collins-Soper-Sterman (CSS)  formalism~\cite{Collins:1984kg,Collins:2011zzd,Collins:2014jpa,Collins:2016hqq}, allow us to study how TMDs evolve between different energy scales. The evolution equations describe how transverse momentum is generated perturbatively by considering the emission of soft and collinear gluons which in turn allows us to study the intrinsic transverse momenta of the bound partons in hadrons.  The evolution equations have been extensively studied in the literature \cite{Balazs:1995nz,Balazs:1997xd,Ellis:1997sc,Bozzi:2010xn,Banfi:2012du,Echevarria:2015uaa,Catani:2015vma,Collins:2016hqq,Monni:2016ktx,Ebert:2016gcn,Kang:2017cjk,Chen:2016hgw,Chen:2018pzu,Becher:2019bnm,Scimemi:2019cmh,Bacchetta:2019sam,Ebert:2020dfc,Ebert:2022fmh}.
These studies have facilitated global extractions of the non-perturbative partonic structure of hadrons in, for instance, Refs.~\cite{Scimemi:2017etj,Scimemi:2019cmh,Bacchetta:2019sam,Bacchetta:2022awv,Echevarria:2020hpy,Bacchetta:2020gko,Bury:2020vhj,Bury:2021sue,Bury:2022czx,Moos:2023yfa,
Bacchetta:2024qre,Bacchetta:2025ara,Barry:2025glq}.

Transverse Momentum Dependent factorization is well developed at the leading power  (LP) in the hard scale. The factorization theorems at LP are applicable to processes in which $\Lambda_{\rm QCD}\lesssim q_\perp\ll Q$ where $\Lambda_{\rm QCD} \sim M$ represents a non-perturbative scale, and  where $q_\perp$, $M$, and $Q$ refer to the transverse momentum of the final state, the hadron mass, and the hard scale of the interactions, respectively~\cite{Collins:1981uk,Collins:1981va,Collins:1984kg,Collins:2011zzd,Collins:2016hqq,Boussarie:2023izj}.   They have been demonstrated in the literature at leading power for the benchmark processes, Semi-Inclusive DIS (SIDIS)~\cite{Meng:1991da,Meng:1995yn,Ji:2004wu,Collins:2011zzd,Aybat:2011zv}, Drell-Yan~\cite{Collins:1984kg,Idilbi:2004vb}, and back-to-back two hadron production in $e^+e^-$ collisions~\cite{Collins:1981uw,Collins:1981uk,Collins:1981va,Collins:2011zzd}. The factorization theorems~\cite{Collins:2011zzd,Boussarie:2023izj} for these processes each involve different TMD PDFs and TMD FFs: SIDIS 
processes involve a TMD PDF for the initial state proton and a TMD FF for the final state hadron, the Drell-Yan process involves two TMD PDFs for the initial state hadrons, and di-hadron production involves two TMD FFs, one for each final state hadron.

Hadron in jet measurements, such as  $pp\rightarrow {\rm (jet + \pi)}\, + X$ offer an interesting avenue for the study of the nucleon structure and in particular the TMD FFs as proposed in Ref.~\cite{Yuan:2007nd}. Recent developments include studies within the SCET formalism, see e.g. Ref.~\cite{Kang:2017glf}, that allowed to interpret LHC data in terms of TMDs.

For quarks, there are eight different TMD PDFs~\cite{Collins:2011zzd,Boussarie:2023izj} at leading power and one can study the distribution of unpolarized quarks in transversely polarized nucleons (Sivers function~\cite{Sivers:1989cc}), the distribution of transversely polarized quarks in unpolarized nucleons (Boer-Mulders function~\cite{Boer:1997nt}),  longitudinally  polarized quarks in  transversely polarized nucleons, and vice versa, (both known as Worm-gear functions~\cite{Tangerman:1994eh,Kotzinian:1995cz}) and transversely polarized quarks whose polarization is orthogonal to  the transverse polarization of the nucleon (Pretzelosity function~\cite{Mulders:1995dh}).

An important feature of the TMD framework concerns the universality of the
non-perturbative TMD distributions, which are defined through gauge-invariant
quark and gluon correlators containing Wilson lines that resum soft gluon
interactions. The
structure of the Wilson-line gauge links depends on the color flow of the
process~\cite{Boer:1999si,Belitsky:2002sm,Bomhof:2006dp}, leading to characteristic process dependence. A well-known
example is the predicted sign change of time-reversal--odd TMD PDFs,
such as the Sivers and Boer-Mulders function~\cite{Sivers:1989cc,Boer:1997mf,Goldstein:2002vv}, between SIDIS and the Drell--Yan process~\cite{Brodsky:2002cx,Brodsky:2002rv,Collins:2002kn}.

\paragraph{Next-to-Leading-Power Observables: Factorization and  NLP TMDs.}

In the previous subsection we briefly reviewed the leading-power (LP)
TMD parton distributions and
fragmentation functions that describe semi-inclusive processes in the
region $\Lambda_{\rm QCD}\lesssim q_\perp\ll Q$.
However, once contributions suppressed by powers of the hard scale are
considered, the LP TMD correlators are no longer sufficient to describe
the full structure of the cross section.
At next-to-leading power (NLP) in the $M/Q$ expansion, additional
azimuthal and spin-dependent observables arise that are encoded in a
new class of partonic correlation functions—commonly referred to as
subleading-power or NLP TMDs—which extend the TMD framework by
capturing power-suppressed transverse-momentum and quark--gluon
correlations inside hadrons.

At next-to-leading power in the hard scale, observables enter the
semi-inclusive cross sections through azimuthal and spin correlations
between the final-state hadrons and leptons.
In semi-inclusive deep-inelastic scattering (SIDIS), azimuthal
correlations were first studied in the perturbative regime at large
transverse momentum by Georgi and Politzer~\cite{Georgi:1977tv},
who argued that such angular correlations should be largely insensitive
to nonperturbative effects and could therefore serve as a clean test of
perturbative QCD.

In contrast, Cahn~\cite{Cahn:1978se,Cahn:1989yf}, based on a simple
kinematic analysis, pointed out that nonperturbative contributions
associated with intrinsic partonic transverse momentum can generate
azimuthal modulations already at zeroth order in the strong coupling.
Although this work predates the modern formulation of TMD factorization,
the so-called Cahn effect~\cite{Cahn:1978se} demonstrated that power corrections of
order $q_\perp/M$ and $M/Q$ contain valuable information about the
internal structure of hadrons.
Indeed, some of the earliest investigations of the transverse motion of
partons in the nucleon emerged from studies of power-suppressed
contributions in SIDIS~\cite{Ravndal:1973kt}.

The azimuthal $\cos\phi$ modulation identified by Cahn can be understood
as arising from the interplay between the intrinsic transverse momentum
of partons and the kinematics of the hard scattering process~\cite{Cahn:1989yf}.
Although originally derived within a simple parton-model framework,
this effect provided one of the earliest indications that
power-suppressed azimuthal observables require additional
transverse-momentum--dependent partonic correlators beyond those that
appear at leading power.
In modern terminology, the Cahn effect may therefore be interpreted as
an early phenomenological manifestation of next-to-leading-power (NLP)
TMD structure in semi-inclusive processes.
The importance of these NLP TMD contributions is further underscored by
the fact that, although they are formally suppressed by $M/Q$ relative
to leading-power observables, they are often numerically significant.  In particular, in the kinematic regime of fixed-target experiments
these contributions can be sizeable, as emphasized in the
phenomenological analysis of Ref.~\cite{Anselmino:2005nn}.

The theoretical foundations of subleading-power transverse-momentum
dependent distributions were established in a series of pioneering
studies by the Amsterdam group.
Early work in a parton-model framework already explored NLP azimuthal modulation contributions
to SIDIS~\cite{Kotzinian:1994dv}.
This program was subsequently developed in
Refs.~\cite{Levelt:1993ac,Mulders:1995dh,Boer:1997mf,Boer:2003cm,Bacchetta:2004zf}, which provided a systematic classification of TMD parton correlation
functions and their associated spin and momentum structures.

Within this framework, the SIDIS cross section was expressed at tree
level in terms of a complete set of leading-power (LP) and
next-to-leading-power (NLP) TMD parton distribution and fragmentation
functions~\cite{Mulders:1995dh,Bacchetta:2006tn}.
A key observation of these studies is that the tree-level factorization
contains four subleading-power contributions:
one kinematic contribution arising from the leptonic tensor and three
subleading-power distribution functions associated with the hadronic
tensor.
However, due to the QCD equations of motion (EOM), the number of
independent hadronic contributions can be reduced to two.
The same methodology was later applied to $e^+e^-$ annihilation in
Ref.~\cite{Boer:1997mf} and subsequently to the Drell--Yan process in
Ref.~\cite{Lu:2011th}. Within this formalism, the subleading-power TMDs were systematically
identified and it was established that their understanding is required
for a complete description of SIDIS and related semi-inclusive reactions.

Beyond their phenomenological importance, these functions are also of
interest in their own right, since they provide access to quark--gluon--quark
correlation functions that encode novel aspects of hadron structure.
Such correlations may be interpreted as quantum interference effects and
have been related, for example, to the average transverse forces acting
on partons inside polarized hadrons~\cite{Burkardt:2008ps}.
Future facilities such as the Electron--Ion Collider (EIC), with its
large kinematic reach and high precision, will provide ideal conditions
for exploring these effects in detail.

Experimentally, the unpolarized azimuthal dependence of the SIDIS
cross section has been measured by several collaborations
~\cite{EuropeanMuon:1983tsy,EuropeanMuon:1986ulc,Adams:1993hs,
Breitweg:2000qh,Chekanov:2002sz,Mkrtchyan:2007sr,CLAS:2008nzy,
Airapetian:2013bim,Adolph:2014zba}.
In addition, the first observed single-spin asymmetries (SSAs) in SIDIS
were sizeable power-suppressed longitudinal target asymmetries for
pion production measured by the HERMES Collaboration
~\cite{Airapetian:1999tv,Airapetian:2001eg}.
These measurements triggered some of the above theoretical studies, and 
in fact, preceded the first measurements of the leading-power Sivers
and Collins asymmetries.
They played an important role in the development of the modern program
of three-dimensional momentum imaging of hadrons and in theoretical
studies of TMD factorization at both leading and subleading power.

Going beyond the tree-level description, however, revealed potential
complications.
In particular, the appearance of uncanceled rapidity divergences in
time-reversal--odd subleading-power TMDs~\cite{Gamberg:2006ru}
suggested that TMD factorization at NLP might be more subtle than at
leading power.  As a consequence, progress in calculating perturbative corrections to
subleading-power TMD observables has historically been limited,
with most analyses performed at tree level.
 Nevertheless, achieving a quantitatively reliable description of the
azimuthal correlations observed in experiments—and ultimately at the
EIC—requires a consistent treatment of perturbative corrections,
factorization, and resummation beyond tree level.

An early study incorporating both resummation at NLP was carried out in the context of
Drell--Yan scattering in Ref.~\cite{Boer:2006eq}.
That work examined, in particular, the impact of resummation on the
angular structure of the cross section and on the Lam--Tung
relation~\cite{Lam:1978pu}. In SIDIS resummation has been studied in 
various combinations of polarizations of the initial lepton and nucleon and the produced hadron~\cite{Koike:2006fn}.  Subsequently, one-loop corrections to the subleading-power SIDIS cross section at both low and high transverse momentum 
 were studied in Ref.~\cite{Bacchetta:2008xw}, where the matching
between the TMD region ($\Lambda_{\rm QCD}\ll q_\perp\ll Q$) and the
large-transverse-momentum region
($\Lambda_{\rm QCD}\ll q_\perp\sim Q$) was investigated.

More recently, the theoretical understanding of NLP contributions has
advanced significantly.
A number of works have revisited the problem using modern theoretical
tools~\cite{Chen:2016hgw,Bacchetta:2008xw,Bacchetta:2019qkv,
Vladimirov:2021hdn,Rodini:2022wki,Ebert:2021jhy,Gamberg:2022lju,
Rodini:2023plb}.
In Ref.~\cite{Vladimirov:2021hdn}, the operator product expansion for
TMD factorization was analyzed using a background-field method.
The evolution of NLP TMD operators was subsequently studied in
Ref.~\cite{Rodini:2022wki}.
In Ref.~\cite{Ebert:2021jhy}, a factorization theorem for SIDIS at NLP
was derived within soft-collinear effective theory (SCET), while
Ref.~\cite{Gamberg:2022lju} explored a generalization of the
Collins--Soper--Sterman (CSS) approach to include NLP effects.

\paragraph{Present Status.}

Various sources for power suppressed terms have been identified and discussed in the literature. This includes corrections associated with kinematic prefactors involving contractions between the leptonic and hadronic tensors, which are sometimes referred to as kinematic power corrections.  Another type of contribution involve subleading terms in quark-quark correlators involving Dirac structures that differ from the leading power ones in which are sometimes called intrinsic power corrections.
Finally there are contributions from hadronic matrix elements of (interaction dependent) quark-gluon-quark operators~\cite{Mulders:1995dh}, referred to as  quark-gluon-quark correlators, or $qgq$ correlators for short. These  are sometimes also referred to as dynamic power corrections, and  it is the $qgq$ correlators that actually introduce new independent subleading power TMDs, while all other $\Lambda/Q$ suppressed power corrections can be expressed in terms of leading power TMDs~\cite{Bacchetta:2006tn,Ebert:2021jhy,Gamberg:2022lju}. At present  a prevailing view is that it is advantagous to express the factorization in terms of TMD PDFs that are composed of good light-cone components~\cite{Ebert:2021jhy,Vladimirov:2021hdn}. 

The fully differential SIDIS cross section --- assuming one-photon exchange between the lepton and the nucleon, and unpolarized produced hadrons in the final state --- can be decomposed into 18 structure functions~\cite{Diehl:2005pc, Bacchetta:2006tn}. 
For low transverse momenta of the final-state hadron, eight of those structure functions are leading in a $\Lambda/Q$ expansion. 
Another eight are suppressed by a factor $\Lambda/Q$, while the remaining two are suppressed by a factor $\Lambda^2/Q^2$. 

Focusing on the ten subleading contributions, in the notation of Refs.~\cite{Bacchetta:2006tn}, the SIDIS cross
section is given by
\begin{eqnarray}
\hspace*{-2.6cm}\frac{\df^6\sigma_{\rm subleading}}{\df\xbj \, \df y \, \df \zh \, \df \phi_S \, \df\phi_h \, \df\Phperp^2}
   & =&
   	\frac{\alpha_{em}^2}{x\,y\,Q^2}\biggl(1-y+\frac12y^2\biggr)
        \biggl\{p_1 F_{UU,L}+
          \cos(\phi_h)\,p_3 \, F_{UU}^{\cos(\phi_h)}
	\nonumber\\
   & +& \lambda\sin(\phi_h) \, p_4 \, F_{LU}^{\sin(\phi_h)}
	+ S_L\sin(\phi_h) \, p_3 \, F_{UL}^{\sin(\phi_h)}
	\nonumber\\ & +& \lambda\, S_L \cos(\phi_h) \, p_4 \, F_{LL}^{\cos(\phi_h)}\phantom{\frac11}
	+ S_T\sin(2\phi_h-\phi_S)\,p_3\,F_{UT}^{\sin(2\phi_h-\phi_S)}
   \nonumber\\
   &      +& S_T\sin(\phi_S)\,p_3\,F_{UT}^{\sin(\phi_S)} 
         + S_T \sin( \phi_h-\phi_S) \,p_1\, F_{UT,L}^{\sin( \phi_h-\phi_S)} \phantom{\frac11}\nonumber\\
   &  	+& \lambda\,S_T\cos(\phi_S)\,p_4\,F_{LT}^{\cos(\phi_S)}
        + \lambda\,S_T\cos(2\phi_h-\phi_S)\,p_4\,F_{LT}^{\cos(2\phi_h-\phi_S)}
	\biggr\} \,,\nonumber \\
	&& \label{e:SIDIS_subleading}
\end{eqnarray}
where the kinematic prefactors $p_i$ in can be found in Ref.~\cite{Boussarie:2023izj}. The structure functions $F_{UU,L}$ and $F_{UT,L}^{\sin( \phi_h-\phi_S)}$ are of ${\cal O}(\Lambda^2/Q^2)$ for small transverse momenta of the final-state hadron. 
These structure functions can be expressed in terms of combinations of
leading power and  $qgq$ correlators~\cite{Ebert:2021jhy,Boussarie:2023izj}.  

\paragraph{Challenges and Opportunities: TMD Physics at the EIC.}

A central challenge in extending TMD factorization beyond leading power
is the consistent treatment of rapidity divergences~\cite{Chiu:2012ir, Balitsky:2023hmh}
 and the associated
operator mixing among subleading-power TMD correlators.
In particular, the appearance of rapidity divergences in certain
time-reversal--odd NLP TMDs initially suggested a potential breakdown of
naive factorization beyond leading power~\cite{Gamberg:2006ru}.
However, once the QCD equations of motion (EOM) relations among the
subleading-power operators are properly implemented, nontrivial
consistency conditions emerge that relate intrinsic, kinematic, and
dynamical twist-three contributions.
These relations could ensure that physical observables depend only on a
reduced set of independent nonperturbative functions and play a crucial
role in organizing the cancellation of rapidity divergences between the
various operator structures entering the factorized cross section.

In this regard, one advantage to using the basis of leading power and $qgq$ TMDs is that the
latter correlators also contain the same soft function as at leading power~\cite{Ebert:2021jhy,Vladimirov:2021hdn}. This occurs because the soft gluons probe the $qg$ pair at the same transverse position, and only see the corresponding product of operators in its combined color triplet state. 
Therefore the bare soft function can be absorbed into these correlators just like at leading power~\cite{Collins:2011zzd,Aybat:2011zv,Boussarie:2023izj}.

A second essential ingredient for a phenomenologically viable NLP
formalism is the consistent matching between the small- and
large-transverse-momentum regimes.
At leading power this is achieved through the Collins--Soper--Sterman
(CSS) $W{+}Y$ construction~\cite{Collins:1984kg,Catani:1989ne,Collins:2016hqq}, which combines the resummed TMD
contribution dominating the region
$\Lambda_{\rm QCD} \lesssim q_\perp \ll Q$
with the fixed-order collinear result relevant for
$q_\perp \sim Q$.  Early phenomenological attempts to describe the transition between the
low  and high-transverse momentum regimes for NLP azimuthal modulations were carried out in
Refs.~\cite{Chay:1991jc,Oganesian:1997jq}, where the small- and
large-transverse-momentum contributions were combined using a sharp
momentum cutoff.

However, extending  this framework to NLP observables requires a systematic
treatment of the power-suppressed contributions, including recently delineated kinematic power corrections~\cite{Piloneta:2025jjb,Balitsky:2026nux}, 
in both regions and
their matching across the intermediate transverse-momentum domain~\cite{Bacchetta:2008xw}.
Developing such a unified description—incorporating EOM consistency,
rapidity-divergence cancellation, and NLP $W{+}Y$ matching is essential
for achieving a quantitatively reliable description of azimuthal and
spin-dependent observables in SIDIS and related processes, particularly
in the precision kinematic regime that will be explored at the
Electron--Ion Collider.

\subsection{TMD global analysis }
   
Recent developments in transverse-momentum dependent (TMD) global analyses are reviewed, covering advances in neural network parametrizations, the integration of lattice QCD data, extensive benchmarking against LHC data, theoretical dependencies, and outstanding challenges. Finally, future directions are discussed.

\paragraph{Status of TMD fits.}

\begin{figure}[tbh]
	\centering
    \includegraphics[width=0.7\textwidth]{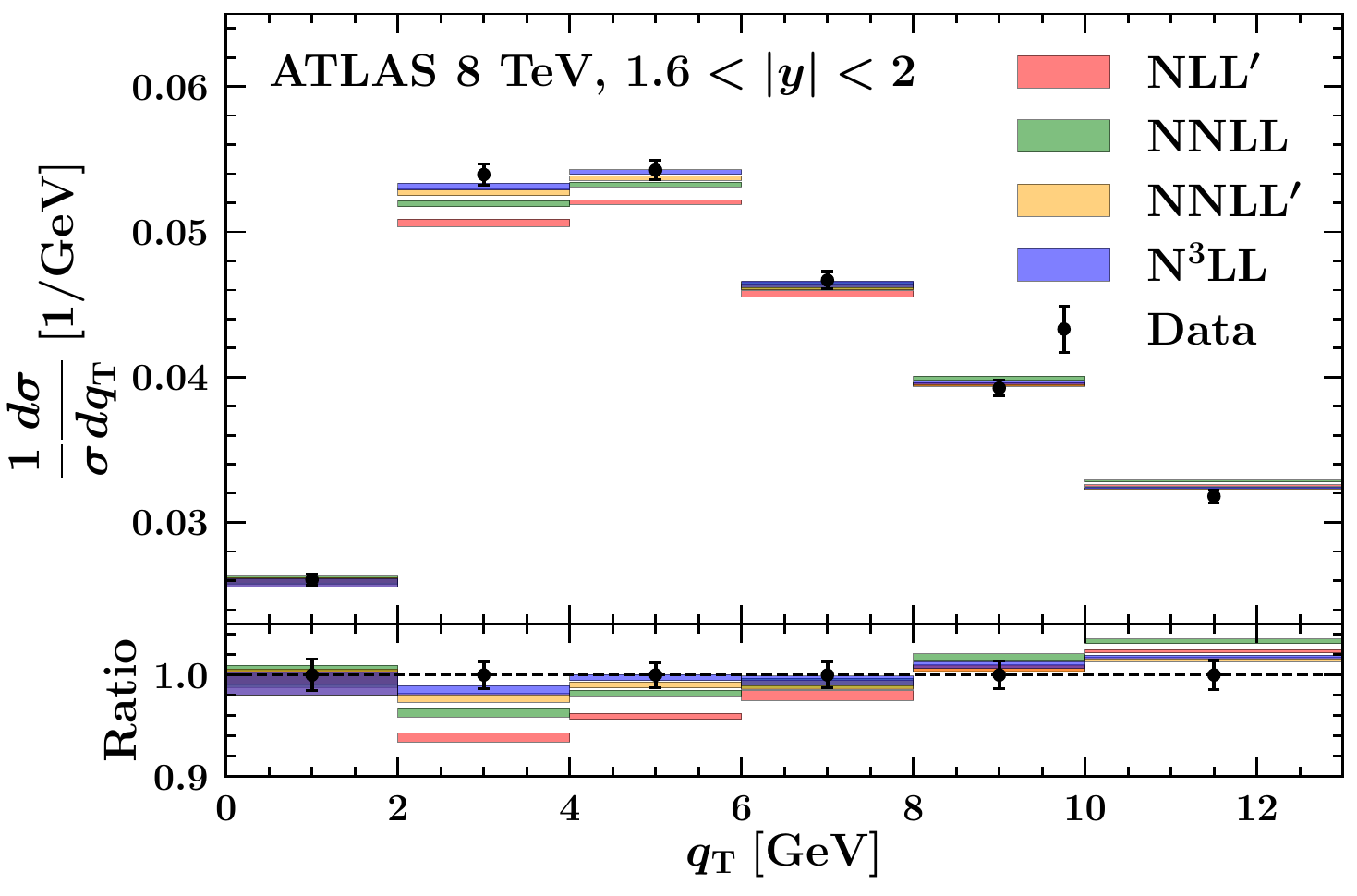}
	\caption{Comparison with experimental data for the ATLAS 8 TeV measurements for 66 GeV $<Q<$ 116 GeV and $1.6 < |y| < 2$. Theoretical predictions were taken from the PV19 fits~\cite{Bacchetta:2019sam} to all perturbative orders considered in that analysis, \textit{i.e.} NLL$'$, NNLL, NNLL$'$, and N$^3$LL. The upper and lower panels show the distributions and their ratios to the experimental central values, respectively. Theoretical uncertainty bands correspond to one-$\sigma$ uncertainties. Error bars on experimental data display uncorrelated uncertainties only. Predictions include systematic shifts.}
	\label{f:PertConv}
\end{figure}

Global TMD fits have reached next-to-next-to-next-to-leading logarithmic (N$^3$LL)~\cite{Barry:2025glq,Bacchetta:2024qre,Bury:2022czx,Bacchetta:2022awv,Scimemi:2019cmh,Bacchetta:2019sam}, and some even next-to-N$^3$LL (N$^4$LL)~\cite{Moos:2025sal,Moos:2023yfa}, accuracy,  combining Drell-Yan (DY) and semi-inclusive deep-inelastic scattering (SIDIS) data. The high theoretical accuracy employed to extract TMDs is not a simple academic exercise. As a matter of fact, it has been shown that next-to-leading logarithmic (NLL) accuracy is insufficient to achieve an accurate description of modern LHC data. However, moving to next-to-next-to-leading logarithmic (NNLL) and N$^3$LL/N$^4$LL dramatically helps improve the description of this data. This is nicely illustrated in Fig.~\ref{f:PertConv}, where theoretical predictions ranging between NLL and N$^3$LL are compared to the precise ATLAS 8 TeV on-peak data in a specific rapidity bin, showing better agreement as the accuracy grows.
Moreover, it has been shown that achieving NNLL and higher accuracies is essential for consistent global fits that simultaneously describe DY and SIDIS experimental data, which in turn range from fixed-target to collider energies.

\begin{figure}[tbh]
	\centering
	\includegraphics[width=0.7\textwidth]{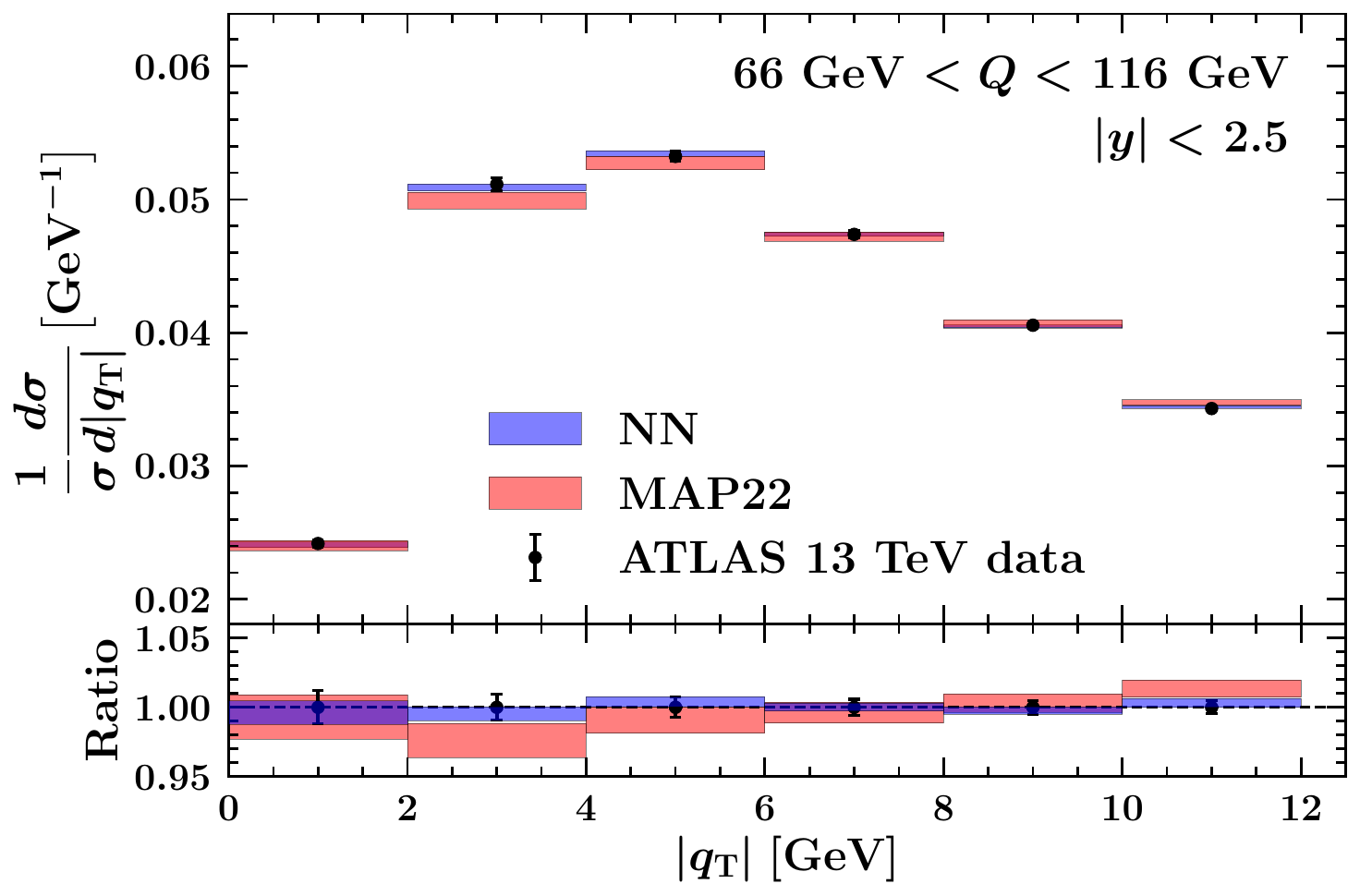}
	\caption{Comparison between experimental data (black dots) and results
		obtained with NN (blue band) and MAP22 (red band) fits. Data corresponds to the ATLAS measurements at 13 TeV
                of the $q_T$-differential cross section normalised by the fiducial one. The layout of the plot is the same as in Fig.~\ref{f:PertConv}}
	\label{f:obs_NNvsMAP22}
\end{figure}

A recent significant innovation is the use of neural networks (NNs) to parametrize the non-perturbative components of TMDs. In a proof-of-concept DY-only extraction, it has been shown that an NN-based parametrization is able to outperform traditional functional forms, yielding a better description of data and providing a more robust uncertainty estimation~\cite{Bacchetta:2025ara}. Fig.~\ref{f:obs_NNvsMAP22} illustrates this achievement for a representative rapidity bin of the ATLAS 13 TeV measurement of the DY transverse momentum distribution~\cite{ATLAS:2019zci}. Specifically, when compared to the traditional MAP22 parametrization, the NN determination reduces correlated systematic shifts and better captures the patterns in high-precision LHC datasets. 

Moreover, preliminary NN-based global TMD fits, including both DY and SIDIS data, feature a better stability with respect to variations of the applied cuts on transverse momentum of SIDIS data. This indicates a reduction of the tension between the description of SIDIS and DY datasets, and better control over TMD fragmentation functions~\cite{Cerutti:GlobalTMDNN}.

\paragraph{Dependence on collinear PDFs.}

Modern global TMD fits rely on external collinear parton distribution functions (PDFs) and fragmentation functions (FFs). Indeed, TMD distributions can be matched onto their collinear counterparts by means of perturbative functions which as of today are known to next-to-next-to-next-to leading order (N$^3$LO) in perturbation QCD. As a consequence, the current uncertainty of collinear PDFs and FFs needs to be propagated to TMDs, which is usually done via Hessian or Monte Carlo methods. Although the choice of the collinear sets can impact the extracted TMDs, it was shown in Refs.~\cite{Bacchetta:2024qre,Bury:2022czx} that the theory/data agreement and the extracted quantities are compatible when different PDF sets are tested. Instead, the dependence on FFs is generally stronger. This is due to the fact that current collinear FF determinations are typically based on significantly reduced data sets as compared to PDFs, which leads to comparably larger uncertainties and broader spreads between different extractions.

\paragraph{SIDIS normalisation.}

Since a decade, it has been observed that theoretical predictions at NLL for SIDIS processes can describe shape and normalization of existing experimental measurements~\cite{Signori:2013mda,Anselmino:2013lza,Echevarria:2014xaa,Bacchetta:2017gcc}. However, when moving to higher accuracies (NNLL or N$^3$LL), it turns out that theoretical predictions tend to severely underestimate data by a seemingly constant (in $q_T$) factor~\cite{Sun:2014dqm,OsvaldoGonzalez-Hernandez:2019iqj,Bacchetta:2022awv}. This issue appears to be driven by perturbative corrections to the SIDIS hard factor, encoding the exclusive high-energy collision of two quarks producing a virtual vector boson, which are particularly large at the typical energy scales of SIDIS measurements and thus compromise the agreement observed at NLL.

In order to fix this issue, some analyses~\cite{Bacchetta:2022awv,Bacchetta:2024qre} introduced \textit{ad hoc} normalisation factors computed by requiring that the integral in $q_T$ of the TMD factorisation formula for SIDIS matches the corresponding collinear cross section. These factors effectively adjust the normalization, allowing for an accurate description of SIDIS data also beyond NLL without, however, introducing any additional parameters to be determined from data. It should also be mentioned that the analysis of Ref.~\cite{Moos:2025sal} does not observe any normalization issue with SIDIS data. This difference between analyses is currently under active investigation.

\paragraph{The Collins-Soper kernel.}

One of the fundamental quantities that enter the computation of TMD distribution is the so-called Collins--Soper kernel (CSK). Indeed, the CSK contributes to 
the evolution of TMDs, which is essential to connect TMDs at different scales. Formally, the CSK has a well-defined operator interpretation in terms of Wilson loops along light-front directions, with a transverse separation given by $\boldsymbol{b}_T$.  When $b\equiv|\boldsymbol{b}_T|$ is small enough, the CSK kernel can be safely computed in perturbation theory. However, as $b$ grows, nonperturbative effects become more and more dominant, making the perturbative calculation unreliable. Since the computation of cross sections in TMD factorization requires accessing the CSK at all possible values of $b$, it is necessary to account for non-perturbative effects. This is typically done by introducing a dedicated function, $g_K(b)$, often parametrized as $g_K(b)=g_2b^2$, with the parameter $g_2$ to be determined from observables sensitive to the CSK.

As of today, several phenomenological TMD analyses have also extracted the CSK from data, typically finding a general mutual agreement. However, the extracted CKS features large uncertainties at large values of $b$, where non-perturbative effects are sizeable. These large uncertainties reflect the fact that experimental data is only indirectly sensitive to the CSK through the scale evolution of TMDs, thus providing limited information on this quantity.

\begin{figure}[tbh]
	\centering
	\includegraphics[width=0.49\textwidth]{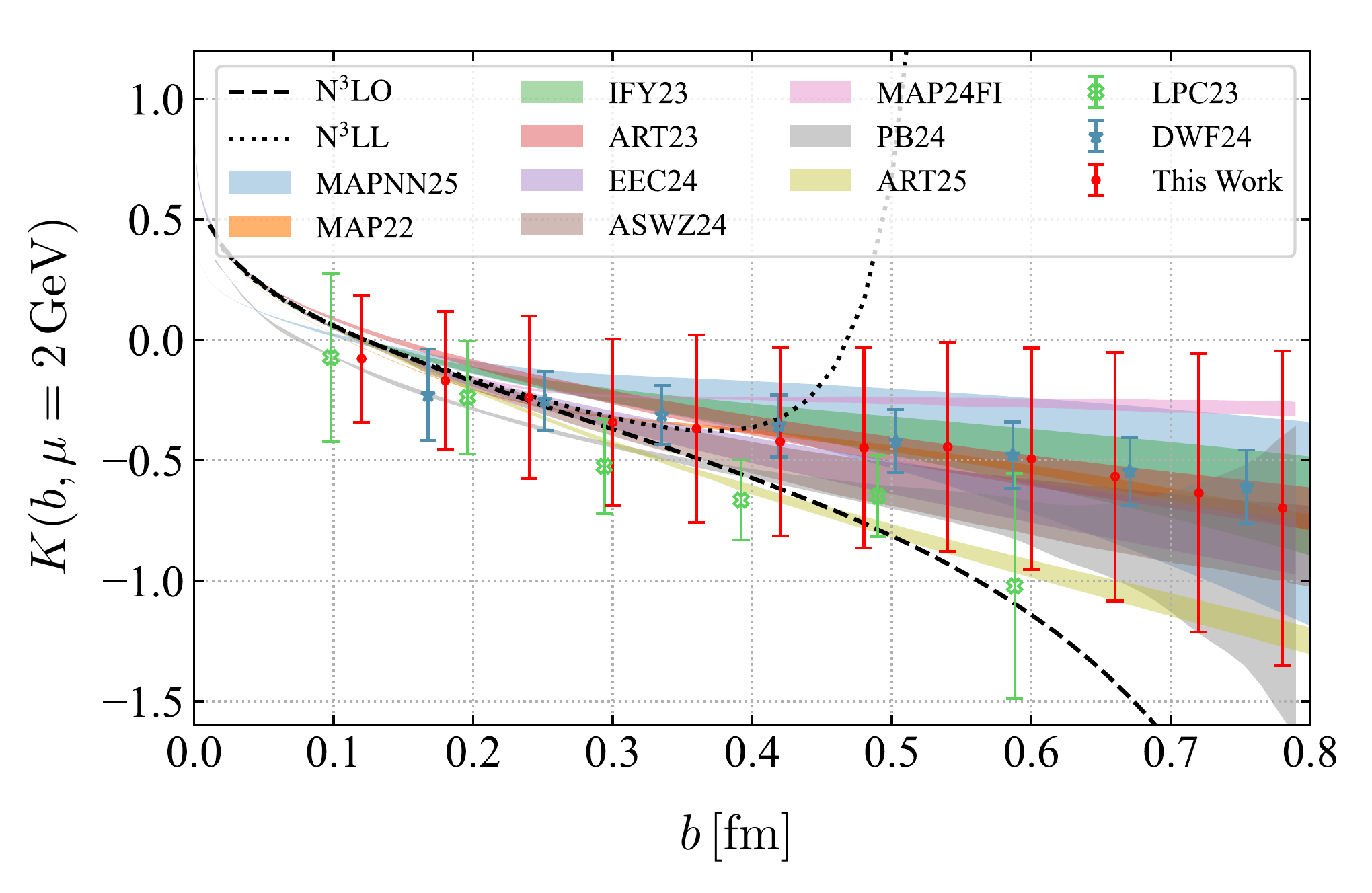}
	\includegraphics[width=0.49\textwidth]{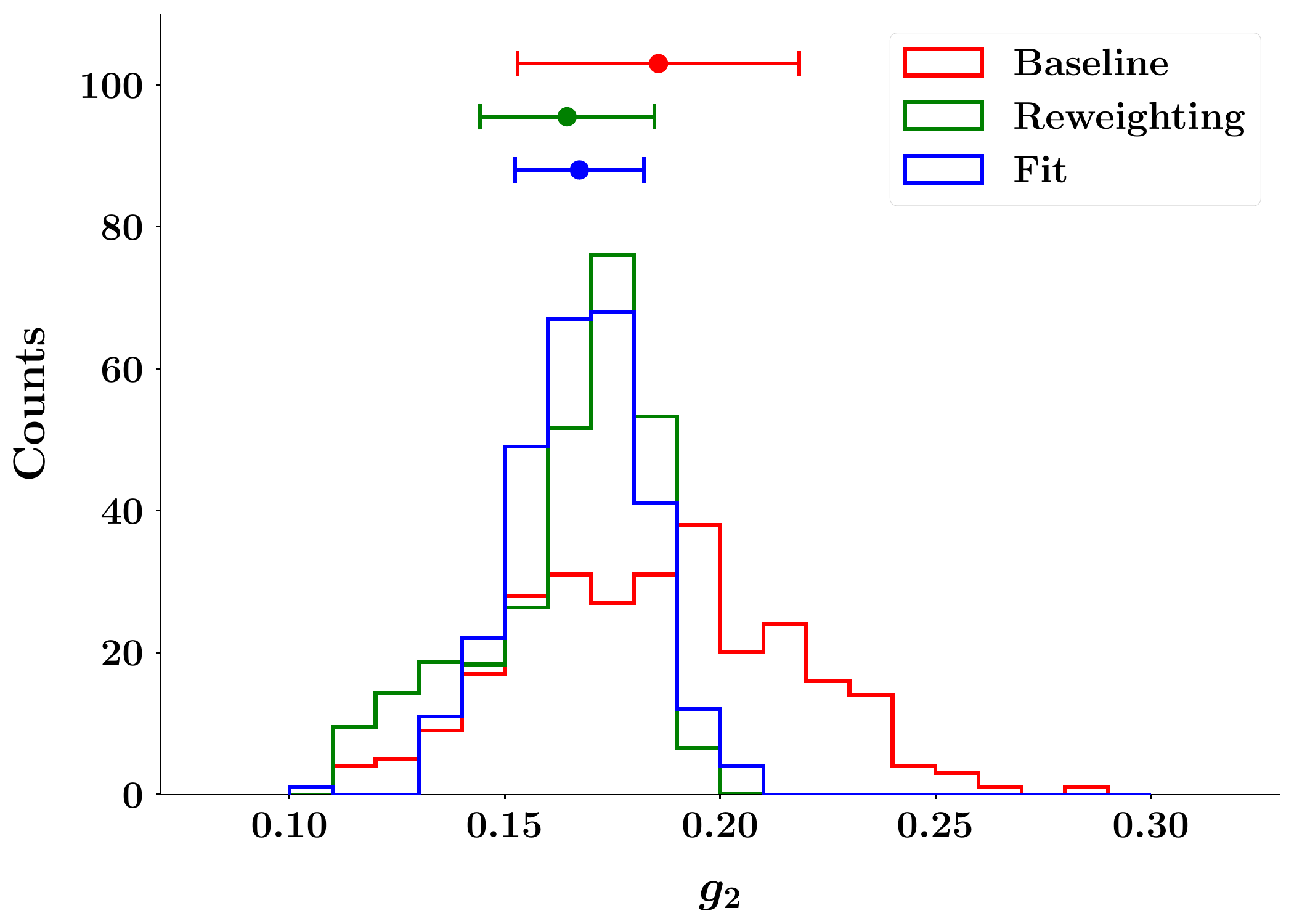}
	\caption{\textit{Left panel:} Comparison between the results for the CS kernel at $\mu = 2$ GeV from a representative set of lattice QCD determinations (colored points) and phenomenological extractions (colored bands). See Ref.~\cite{Bollweg:2025iol} for details. \textit{Right panel:} Monte Carlo replica distributions of the parameter $g_2$ from the baseline fit (red), after reweighting (green), and from the simultaneous fit (blue) of Ref.~\cite{Avkhadiev:2025wps}.
          The points above the distributions display the respective central values and one-$\sigma$ uncertainties.}
	\label{f:CSK_lattice_pheno}
\end{figure}

However, independent determinations of the CSK have also been obtained in lattice QCD. These extractions are generally in good agreement with phenomenological extractions and often present smaller uncertainties in the large-$b$ region. A representative comparison between lattice and phenomenological results is shown in the left panel of Fig.~\ref{f:CSK_lattice_pheno}.
To best exploit the information on the CKS provided by lattice simulations, very recently a joint extraction of the CSK combining DY experimental and lattice QCD data was carried out~\cite{Avkhadiev:2025wps}. This pioneering analysis not only proved the feasibility of including lattice information in TMD analyses, but also highlighted the potential benefits of such extractions by finding a significant reduction of the uncertainty on the CKS when experimental and lattice data are combined. An illustration is given in the right panel of Fig.~\ref{f:CSK_lattice_pheno}, where the parameter $g_2$, parametrizing non-perturbative effects to the CSK, is much better constrained when including lattice data in the analysis of Ref.~\cite{Bacchetta:2025ara}.

\paragraph{Summary and outlook.}

Current TMD global analyses have reached a remarkable theoretical accuracy, combining DY and SIDIS data with sophisticated uncertainty treatments. Key advances include the use of neural networks for flexible non-perturbative parametrizations and a first successful integration of lattice QCD data for the Collins--Soper kernel. 
In spite of these achievements, outstanding challenges still remain which include:
\begin{itemize}
\item Resolving the SIDIS normalisation issue observed beyond NLL accuracy.
\item Systematically exploring the dependence on collinear PDFs/FFs and their uncertainties.
\item Extending the framework to polarized TMDs and $e^+e^-$ annihilation data.
\end{itemize}
Future work is expected to leverage the flexibility of neural networks and the constraining power of lattice QCD to perform simultaneous global fits of TMDs, further refining our understanding of the three-dimensional hadron structure.
    
\subsection{The role of the lattice in global TMD analyses} 

A class of TMD observables that can be evaluated within Lattice QCD, while
avoiding many complications associated with a more general calculation of
individual TMDs that includes their dependence on momentum fraction $x$,
is given by ratios of $x$-integrals of TMDs
\cite{Musch:2011er,Engelhardt:2015xja,Yoon:2017qzo,Engelhardt:2023aem}.
These observables are thus particularly well-suited for achieving a
controlled, quantitative connection to phenomenology. Briefly, one starts
with the fundamental nonlocal correlator
\begin{equation}
\hspace*{-4mm}
\widetilde{\Phi }^{[\Gamma ]}_{\mbox{\scriptsize unsubtr.} } (b,P,S,\ldots )
\equiv \frac{1}{2} \langle P,S | \ \bar{q} (0) \
\Gamma \ {\cal U} [0,\eta v, \eta v+b,b] \ q(b) \ |P,S\rangle
\label{spacecorr}
\end{equation}
evaluated in a hadron state with momentum $P$ and spin $S$. The quark
operator separation $b$ is Fourier conjugate to the quark momentum $k$.
$\Gamma $ denotes an arbitrary Dirac $\gamma $-matrix structure, and
the quark operators are connected by a staple-shaped Wilson line ${\cal U}$ composed of straight-line segments running between the positions specified
in its argument. The direction of the staple legs is given by the unit
vector $v$, whereas their length is parametrized by $\eta $.
The most naive choice of $v$ as a light-cone vector introduces
rapidity divergences, which in the scheme advanced in
\cite{Aybat:2011zv,Collins:2011zzd,Aybat:2011ge} are regulated by
taking $v$ off the light cone
into the space-like region. The Collins-Soper type parameter
$\hat{\zeta } = v \cdot P / (|v|\, |P|)$ can be used to quantify the
deviation of $v$ from the light cone, with the light-cone limit corresponding
to $\hat{\zeta } \rightarrow \infty $. This prescription of employing
space-like $v$ renders the connection to Lattice QCD particularly
straightforward, as detailed below.

An important methodological element of the scheme presented here is the
decomposition of the correlator
$\widetilde{\Phi }^{[\Gamma ]}_{\mbox{\scriptsize unsubtr.} } $
defined in (\ref{spacecorr}) into Lorentz-invariant amplitudes.
This decomposition allows for a simple
translation of results between the Lorentz frame in which TMDs are
defined phenomenologically, and the Lorentz frame suited for a Lattice QCD
calculation, introduced below. At leading twist, the decomposition for
nucleons reads \cite{Musch:2011er}
\begin{eqnarray}
\frac{1}{2P^{+} }
\widetilde{\Phi }^{[\gamma^{+} ]}_{\mbox{\scriptsize unsubtr.} } &=&
\widetilde{A}_{2B} + im_N \epsilon_{ij} b_i S_j \widetilde{A}_{12B}
\nonumber \\
\frac{1}{2P^{+} }
\widetilde{\Phi }^{[\gamma^{+} \gamma^{5} ]}_{\mbox{\scriptsize unsubtr.} }
&=& -\Lambda \widetilde{A}_{6B} +i[(b\cdot P) \Lambda - m_N (b_T \cdot S_T )]
\widetilde{A}_{7B}\nonumber \\
\frac{1}{2P^{+} }
\widetilde{\Phi }^{[i\sigma^{i+} \gamma^{5} ]}_{\mbox{\scriptsize unsubtr.} }
&=& im_N \epsilon_{ij} b_j \widetilde{A}_{4B} -S_i \widetilde{A}_{9B}
-im_N \Lambda b_i \widetilde{A}_{10B} \nonumber \\
& & + m_N [(b\cdot P) \Lambda - m_N (b_T \cdot S_T )] b_i \widetilde{A}_{11B} \ ,
\label{adecomp2}
\end{eqnarray}
where $\Lambda $ denotes the nucleon helicity (i.e.,
$S^{+} =\Lambda P^{+} /m_N $, $S^{-} =-\Lambda m_N /2P^{+} $).

The amplitudes $\widetilde{A}_{iB} $ inherit the divergences associated
with the Wilson line ${\cal U}$ in (\ref{spacecorr}), which must be
compensated by a multiplicative soft factor. Along with additional
wave function renormalization factors attached to the quark operators
at finite separation $b$, these soft factors can be cancelled by forming
appropriate ratios\footnote{The breaking of chiral symmetry implied by
most lattice discretizations induces operator mixing effects
\cite{Constantinou:2019vyb,Shanahan:2019zcq,Green:2020xco}
that can invalidate the cancellation of soft factors in ratios.
To avoid this complication, the results presented below were obtained
using the domain wall discretization, which, though computationally
expensive, preserves chiral symmetry up to negligible residual effects.},
implying a significant simplification of the calculation. Furthermore,
by focusing on momentum fraction $x$-integrated TMDs, one can restrict
the calculation to $b\cdot P=0$; $b\cdot P$ is the variable Fourier
conjugate to $x$. Thus, one arrives at the following TMD ratio observables
that can be evaluated within Lattice QCD in a particularly robust and
controlled fashion (listing only those for which results are presented
below):
\begin{itemize}
\item
The generalized Sivers shift
\begin{equation}
\hspace*{-6mm}
\langle k_y \rangle_{TU} (b_T^2 , \ldots ) =
-m_N \frac{\widetilde{A}_{12B} (-b_T^2 , \ldots )}{\widetilde{A}_{2B}
(-b_T^2 , \ldots )} = m_N \frac{\tilde{f}_{1T}^{\perp [1](1)} (b_T^2 ,
\ldots )}{\tilde{f}_{1}^{[1](0)} (b_T^2 , \ldots )}\,,
\label{sivshift}
\end{equation}
where the expression on the right-hand side establishes the connection
to Fourier-transformed TMD moments, cf.~\cite{Musch:2011er}. In the formal
$b_T \rightarrow 0$ limit, (\ref{sivshift}) represents the average
transverse momentum $k_y $ of unpolarized (``$U$'') quarks orthogonal
to the transverse (``$T$'') spin of the nucleon, normalized to the number
of valence quarks. It should be noted that the $b_T^2 =0$ limit introduces
additional divergences that require adequate treatment compared to the
(``generalized'') finite $b_T $ case.
\item
The generalized $g_{1T} $ worm-gear shift
\begin{equation}
\langle k_x \rangle_{TL} (b_T^2 , \ldots ) =
-m_N \frac{\widetilde{A}_{7B} (-b_T^2 , \ldots )}{\widetilde{A}_{2B}
(-b_T^2 , \ldots )} = m_N \frac{\tilde{g}_{1T}^{[1](1)} (b_T^2 ,
\ldots )}{\tilde{f}_{1}^{[1](0)} (b_T^2 , \ldots )}\,.
\label{wggshift}
\end{equation}
In the formal $b_T \rightarrow 0$ limit, (\ref{wggshift}) represents the
average transverse momentum $k_x $ of longitudinally (``$L$'') polarized
quarks in a nucleon polarized in the same transverse (``$T$'') direction,
i.e., here, the $x$-direction, normalized to the number of valence quarks.
\item
The generalized $h_{1L}^{\perp } $ worm-gear shift
\begin{equation}
\langle k_x \rangle_{LT} (b_T^2 , \ldots ) =
-m_N \frac{\widetilde{A}_{10B} (-b_T^2 , \ldots )}{\widetilde{A}_{2B}
(-b_T^2 , \ldots )} = m_N \frac{\tilde{h}_{1L}^{\perp [1](1)} (b_T^2 ,
\ldots )}{\tilde{f}_{1}^{[1](0)} (b_T^2 , \ldots )}\,.
\label{wgshift}
\end{equation}
In the formal $b_T \rightarrow 0$ limit, (\ref{wgshift}) represents the
average transverse momentum $k_x $ of quarks polarized in the same transverse
(``$T$'') direction, i.e., here, the $x$-direction, in a longitudinally
(``$L$'') polarized nucleon, normalized to the number of valence quarks.
\end{itemize}
As a further simplification, all results presented below pertain to the
isovector $u-d$ quark flavor combination, in which the computationally
demanding disconnected diagrams cancel.

\begin{figure}
\centerline{
\includegraphics[width=11cm]{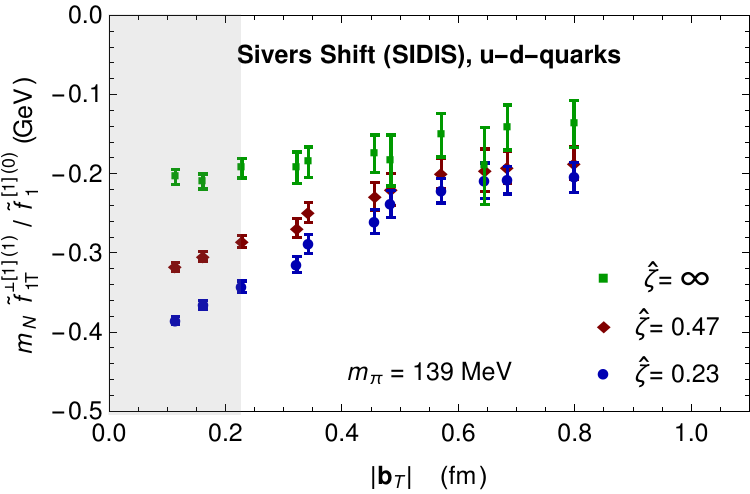}
}
\caption{Generalized Sivers shift as a function of $b_T$, including
extrapolation to $\hat{\zeta } \rightarrow \infty $. The extracted
values compare favorably with a phenomenological determination performed
in \cite{Yoon:2017qzo} on the basis of the global fit provided in
\cite{Echevarria:2014xaa}, namely,
$\langle k_y \rangle_{TU} \, (b_T \sim 0.35\, \mbox{fm} ) = -0.146(49)$.}
\label{siversfig}
\end{figure}

\begin{figure}
\centering
\includegraphics[width=11.7cm]{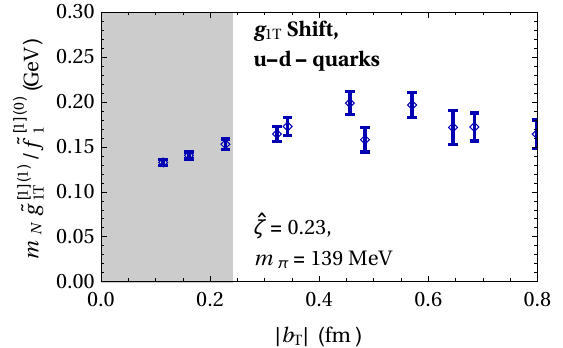} \\
\vspace{1cm}
\includegraphics[width=11cm]{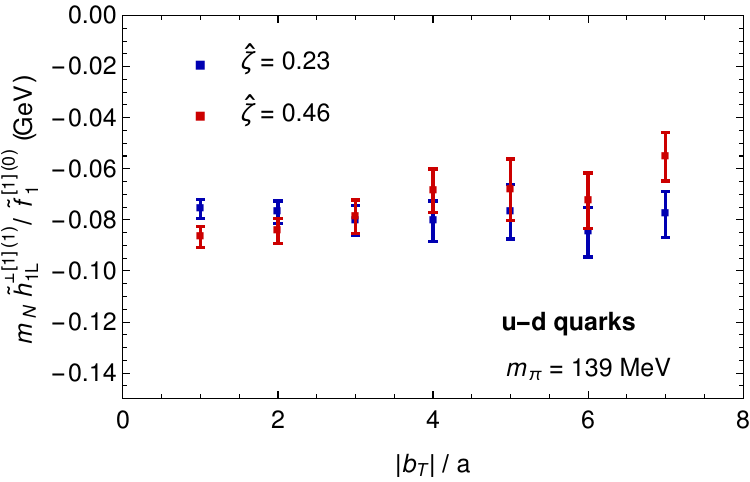}
\caption{$g_{1T} $ and $h_{1L}^{\perp } $ worm-gear shifts as a function
of $b_T$. The lattice spacing $a$ referenced in the lower panel is
$a=0.114\, \mbox{fm}$. Remarkably, these results predict the magnitudes
of the two shifts to differ by around a factor 2, whereas a wide variety
of models of the nucleon (e.g., light-front constituent quark model,
covariant parton model, bag model, light-front quark-diquark model,
light-front version of the chiral quark-soliton model, nonrelativistic
quark model) predict the magnitudes to coincide. Evidently, there are
significant QCD effects not captured by these models, which come to the
fore in a full QCD calculation.}
\label{wormfig}
\end{figure}

In a Lattice QCD calculation, the operator of which one is taking a
hadronic matrix element in (\ref{spacecorr}) has to be located at one
single Euclidean insertion time. On the other hand, the operator as
initially defined above in general extends into the (Minkowski) temporal
direction. However, since all separations in the operator, controlled
by the four-vectors $b$ and $v$, are space-like, the problem can be
boosted to a Lorentz frame in which the entire operator indeed exists
at a single time, and the lattice calculation can be performed in that
frame. This is the point where a formulation employing a space-like
direction $v$ for the purpose of regulating rapidity divergences is
crucial. Furthermore, having decomposed the resulting
$\widetilde{\Phi }^{[\Gamma ]}_{\mbox{\scriptsize unsubtr.} } $
into the invariant amplitudes $\widetilde{A}_{iB} $, the results
for those amplitudes are immediately also applicable in the
Lorentz frame in which (\ref{spacecorr}) was originally defined;
TMD observables of the type (\ref{sivshift})-(\ref{wgshift})
are thus determined.

The data obtained in a lattice calculation are necessarily limited
to a finite staple length $\eta $ and a finite rapidity parameter
$\hat{\zeta } $, and they must be extrapolated to the limits
$\eta \rightarrow \infty $ and $\hat{\zeta } \rightarrow \infty $.
The former extrapolation is typically well-controlled in a substantial
range of the parameter space. The latter extrapolation is more challenging,
since large $\hat{\zeta } $ can only be achieved using large hadron
momentum $P$, and the set of $P$ that can be accessed reliably in a
lattice calculation is fairly limited. Nonetheless, the results
presented for the Sivers shift in Fig.~\ref{siversfig} have been extrapolated
to $\hat{\zeta } \rightarrow \infty $ while retaining a good signal.
The results for the worm gear shifts in Fig.~\ref{wormfig} are presented
at selected fixed $\hat{\zeta } $.

\subsection{TMDs from semi-inclusive channels}
        
Concerning TMD fragmentation functions, we now have measurements for pions, kaons and protons obtained in $e^+e^-$ annihilation at Belle \cite{Belle:2019ywy}, using the event-shape variable thrust. This data has already been used successfully in a fit \cite{Boglione:2023duo,Kang:2020yqw}, although the use of an event shape variable complicates the theoretical interpretation somewhat. From the theoretical point of view, the measurement of nearly back-to-back hadrons would simplify the description in terms of TMD fragmentation functions. Corresponding data have already been analyzed at Belle, but due to the transition from Belle to Belle 2, the publication is significantly delayed and likely requires the inclusion of Belle 2 data before being allowed to proceed. 
        
At the same time, for polarized TMD fragmentation, Collins type asymmetries have been extracted from Belle \cite{Belle:2019nve,Belle:2008fdv}, Babar \cite{BaBar:2015mcn,BaBar:2013jdt}, and BesIII \cite{BESIII:2015fyw} data, also using nearly back-to-back hadron pairs. This was done mostly for charged pions, but also some neutral meson and kaon results are available. At present most of the measurements are projections on typical momentum fractions or transverse momenta, but not yet entirely multidimensional extractions. These are expected to be added eventually. 
The Collins FFs for other particle types, especially for vector mesons which should exhibit an opposite sign 
relative to spinless ones according to the Artru model \cite{Artru:1989zv}, have so far not been measured.   
Polarizing TMD FFs are also accessible for $\Lambda$ baryons, but a better understanding of the flavor contributions is still necessary. 

Collinear fragmentation function measurements, which form the baseline for all TMD FF studies, are available for a large number of final states such as light hadrons, various $D$ mesons and even several strongly or weakly decaying lighter mesons \cite{Belle:2024vua}. 

On the phenomenological side, the theoretical precision of the frameworks used for the extraction of fragmentation functions has improved significantly in recent years, although important differences remain between the unpolarized and polarized cases. Collinear, unpolarized fragmentation functions are typically available at next-to-next-to-leading order (NNLO)~\cite{Borsa:2022vvp,Gao:2025hlm}. Building on this progress, modern extractions of transverse-momentum–dependent (TMD) fragmentation functions combine high-order resummation with fixed-order inputs, achieving up to $\mathrm{N^3LL}$ resummation accuracy matched to NNLO collinear functions in global analyses of Drell–Yan (DY), semi-inclusive deep inelastic scattering (SIDIS), and $e^+e^-$ data~\cite{Bacchetta:2024qre}, and more recently extending to $\mathrm{N^4LL}$ accuracy~\cite{Moos:2025sal}.

By contrast, the extraction of polarized TMD FFs is currently limited to significantly lower perturbative accuracy, typically at the $\mathrm{NLL}$ level, as in the case of the Collins function~\cite{Kang:2015msa}. This disparity reflects both theoretical challenges and the limited availability of precise experimental constraints.

\subsection{TMDs through jets, jet substructure, and energy correlators}
        
Jet measurements across HERA, RHIC, and the LHC have established jets as precision tools for exploring Quantum Chromodynamics (QCD) over a wide range of energies and nuclear environments. Together, these facilities have shaped our understanding of parton dynamics, proton structure, and the properties of hot nuclear matter, while motivating new directions at the future Electron–Ion Collider (EIC).
        
At HERA, the study of jets in electron–proton collisions provided clean tests of perturbative QCD and factorization \cite{H1:2015ubc,Klein_2008}. Inclusive jet and multijet cross sections in deep inelastic scattering (DIS) validated next-to-leading-order calculations and enabled precise extractions of the strong coupling constant, including its scale dependence. Jet measurements also played a critical role in constraining parton distribution functions (PDFs), particularly the gluon density at moderate to high Bjorken-$x$, forming a cornerstone for precision predictions at hadron colliders.
        
At RHIC, jets became probes of strongly interacting nuclear matter. Observations of high-$p_T$ hadron suppression, dijet imbalance, and jet–medium correlations in heavy-ion collisions provided compelling evidence for jet quenching \cite{STAR:2017hhs,STAR:2017ieb} and the formation of a quark–gluon plasma (QGP). These measurements demonstrated that energetic partons lose energy while traversing the medium and allowed constraints on QGP transport properties. Proton–proton jet measurements at RHIC further tested perturbative QCD at intermediate energies\cite{Chang:2021vbq,Robotkova:2024jss,PHENIX:2025pqz}.
        
The LHC extended jet studies to unprecedented energies and precision. In proton–proton collisions, jet and jet-substructure measurements confirmed QCD over many orders of magnitude and enabled high-precision determinations of $\alpha_s$ and PDFs \cite{aaboud2017determination,aaboud2018measurement,acharya2019charged}. In heavy-ion collisions, detailed measurements of groomed jet mass, angular structure, and energy flow revealed how parton showers are modified by a dense medium, significantly advancing quantitative models of jet quenching\cite{Vertesi:2024tdv}.
        
The EIC will open a new frontier by combining clean lepton–hadron collisions with high luminosity and polarized beams, enabling jet studies in a regime inaccessible at previous facilities \cite{Accardi:2012qut,AbdulKhalek:2021gbh}. Forward jets, produced at small angles relative to the proton or ion beam, will probe the low-$x$, gluon-dominated region of hadrons and nuclei. When combined with jet substructure observables, such as groomed momentum sharing, subjet distributions, and angular correlations, these measurements will allow direct access to the early stages of parton shower formation. Unlike hadron colliders, the well-controlled initial state in DIS makes it possible to disentangle perturbative radiation from nonperturbative hadronization effects with unprecedented clarity.
        
Jet substructure in forward kinematics at the EIC will provide sensitive tests of QCD evolution beyond collinear factorization, including transverse-momentum–dependent (TMD) dynamics, small-$x$ evolution, and potential gluon saturation effects. In electron–ion collisions, modifications of jet substructure relative to electron–proton collisions will offer a novel window into cold nuclear matter effects and the spatial distribution of gluons in nuclei. These measurements will bridge precision studies of proton structure at HERA with the emergent collective behavior observed at RHIC and the LHC, completing the jet physics program across QCD regimes.

Building on this foundation, jet production in DIS provides a powerful and systematic framework to access transverse-momentum–dependent parton distributions. In particular, nearly back-to-back configurations of the scattered lepton and a reconstructed jet offer a clean probe of intrinsic partonic transverse momentum. The transverse momentum imbalance between the lepton and the jet directly reflects the underlying TMD parton distribution functions (TMDPDFs), and the corresponding cross section admits a TMD factorization description \cite{Liu:2018trl,Arratia:2020nxw}. Compared to traditional semi-inclusive measurements, the use of jets reduces sensitivity to fragmentation effects and provides a more direct connection to the initial-state transverse dynamics. This makes lepton–jet correlations a particularly clean channel for precision studies of TMD evolution and for extracting both unpolarized and spin-dependent TMDPDFs at the EIC.

Going beyond inclusive jet observables, the internal structure of jets enables direct access to transverse-momentum–dependent fragmentation. Measurements of identified hadrons within jets, differential in their transverse momentum relative to the jet axis, provide sensitivity to TMD fragmentation functions (TMDFFs) in a controlled and factorized framework \cite{Kang:2020xyq,Kang:2021ffh}. In this approach, the jet axis serves as an effective proxy for the fragmenting parton direction, allowing the transverse momentum of hadrons inside the jet to be directly related to fragmentation dynamics. This setup avoids the convolution of initial- and final-state TMDs that appears in traditional semi-inclusive processes, thereby enabling a more direct extraction of TMDFFs. Such measurements establish a natural connection between TMD physics and modern jet substructure techniques.

A particularly promising and theoretically robust class of observables is provided by energy correlators in DIS. Energy–energy correlators (EEC) and their generalizations are infrared- and collinear-safe observables that probe correlations in the energy flow of final-state radiation. In DIS, these observables can be defined with respect to the beam or photon axis and include the one-point energy correlator (OPEC) as a special case \cite{Basham:1977iq}. Recently, energy correlators in DIS have attracted significant attention as versatile probes of transverse-momentum–dependent (TMD) dynamics, nucleon structure, and parton hadronization \cite{Li:2021txc,Kang:2023big,Mi:2025abd,Gao:2025evv,Song:2025bdj,Gao:2025cwy,Zhu:2025qkx,Cao:2025icu}. Importantly, the inclusion of azimuthal angular dependence enables sensitivity to spin-dependent TMDs, providing access to both unpolarized and polarized parton distributions within a unified framework \cite{Kang:2023big}. Unlike hadron-based measurements, energy correlators are less sensitive to hadronization effects and offer a more direct probe of perturbative QCD dynamics, making them particularly well suited for precision studies.

Together, these approaches establish a comprehensive jet-based TMD program at the EIC. Lepton–jet correlations provide clean access to TMDPDFs, hadron-in-jet measurements probe TMDFFs through jet substructure, and energy correlators offer a theoretically controlled and versatile framework that unifies and extends these studies. The combination of these observables in the clean DIS environment of the EIC will enable high-precision extractions of TMDs, stringent tests of their universality, and new insights into the interplay between perturbative evolution, spin dynamics, and nonperturbative transverse structure.
  
\subsection{Gravitational form factors and their connection to GPDs} 
The gravitational form factors (GFFs) are defined through the off-forward matrix element of the QCD energy momentum tensor (EMT) $T^{\mu\nu}$. For a spin-$\frac{1}{2}$ hadron such as the nucleon, they can be parametrized as 
\begin{equation}
\hspace*{-2cm}
\langle p'|T^{\mu\nu}|p\rangle =\bar{u}(p')\left[ \gamma^{(\mu} P^{\nu)} A(t) + \frac{iP^{(\mu}\sigma^{\nu)\rho}\Delta_\rho}{2M}B(t)+\frac{D(t)}{4M}(\Delta^\mu \Delta^\nu -g^{\mu\nu}\Delta^2)\right]u(p). \label{tmunu}
\end{equation} 
Just like the electromagnetic form factors $\langle p'|J^{\mu}_{em}|p\rangle$ represent the interaction of hadrons with the electromagnetic field, GFFs represent their interaction with the gravitational field. However, this analogy is marred by the weakness of the gravitational interaction. Imagine the QED coupling constant $\alpha_{em}=\frac{1}{137}$ is replaced by the gravitational coupling
\begin{equation}
\hspace*{3.7cm}
s G_N  \sim \left(\frac{\sqrt{s}}{M_{\mathrm{Pl}}}\right)^2, \label{GN}
\end{equation}
where $\sqrt{s}$ is the collision energy, $G_N$ is the Newton constant and $M_{\mathrm{Pl}}\sim 10^{19}$ GeV is the Planck energy. 
The enormous suppression  (\ref{GN}) eliminates any realistic prospect of directly measuring GFFs through the gravitational interaction (see however \cite{Hatta:2023fqc}). This is in stark contrast to the electromagnetic form factors that have been measured over the past 70 years, and have recently achieved  percent-level precision  \cite{Xiong:2023zih}. 

An intriguing question that the EIC can address is the following. If a direct measurement is not feasible, how about an {\it indirect} one? The exchange of a spin-two graviton could be mimicked by the exchange of two spin-one particles, such as two photons or two gluons. In fact, this is the basic idea behind the recent attempts to extract GFFs from experiments. The first such attempt \cite{Burkert:2018bqq} exploited the connection between $D(t=0)$ in (\ref{tmunu}) with the dispersion relation \cite{Anikin:2007yh}  between the real and imaginary parts of the Compton amplitudes in Deeply Virtual Compton Scattering (DVCS),
\begin{equation}
\hspace*{-12mm}{\rm Re} {\cal H}_q(\eta,t) 
= \frac{1}{\pi}\int_0^1 dx {\rm P}\left(\frac{1}{\eta-x}-\frac{1}{\eta+x}\right){\rm Im}{\cal H}_q(x,t)+2\int_{-1}^1 dz \frac{D_q(z,t)}{1-z} \,.\label{disd}
\end{equation}
The quark contribution $D_q(t=0)$ to the $D$-term  in (\ref{tmunu}) is given by the second moment of the function $D_{q}(z,t)$ that appears in the subtraction constant:
\begin{equation}
\hspace*{3cm}
\int_{-1}^1 dz z D_q(z,t)= D_q(t). \label{dq}
\end{equation}
However, the problem is that it is not possible to precisely relate the two without the detailed knowledge of $D_q(z,t)$.  This difference boils down to the fact that the two-photon exchange (formed by the incoming virtual photon and the outgoing real photon in DVCS) is not equivalent to the one graviton exchange. In the GPD factorization of DVCS, the two photons can couple to infinitely many twist-two operators with different values of spin, whereas the GFFs are exclusively tied to the EMT operator which is twist-two and spin-two. Isolating the latter is a very challenging inverse problem. The EIC's wide kinematic reach in $Q^2$ certainly helps, but it remains to be seen  what level of accuracy can be achieved in this  extraction, see \cite{Dutrieux:2024bgc} for the current state of the art.   

Meanwhile, an intriguing connection has been suggested between the gluonic part of the GFFs and the near-threshold production of heavy vector mesons such as $J/\psi$ and $\Upsilon$.  The original idea came from holographic QCD \cite{Hatta:2018ina,Mamo:2019mka} which treats the process nonperturbatively. The twist-two, higher spin operators acquire large anomalous dimensions due to strong coupling, and are therefore irrelevant. The scattering amplitude is then dominated by the `protected'  EMT operator, or its matrix elements, the GFFs. More recently, a weak coupling approach based on GPD factorization has been developed. At first sight, this suffers exactly the same problem as in the extraction of the quark GFFs in DVCS, namely, the two gluon exchange is sensitive to infinitely many twist-two gluon operators. However, there are two important differences. First, heavy meson production is primarily sensitive to the gluon GPD $H_g$, instead of the quark GPD $H_q$. Second, the skewness variable $\xi$ becomes order unity near threshold. It turns out that the combined effect of these two features  enhances the sensitivity  to the gluon GFF. Specifically, the following approximation holds \cite{Hatta:2021can,Guo:2021ibg,Guo:2023qgu}:
\begin{equation}
\hspace*{-4mm}
\int_{-1}^1\frac{dx}{x}\left(\frac{1}{\xi-x-i\epsilon}-\frac{1}{\xi+x-i\epsilon}\right)H_g(x,\xi,t)\approx \frac{2}{\xi^2}\frac{5}{4}(A_g(t)+\xi^2D_g)\,, \label{approx}
\end{equation} 
once $\xi$ is moderately large, say $\xi>0.3$. 
The left hand side is the standard leading order DVMP amplitude which features the gluon GPD. Although it   receives contributions from infinitely many twist-two gluon operators, it is nevertheless well approximated by the gluon GFFs on the right hand side. Very recently, the relation (\ref{approx}) has been extended to  next-to-leading order and shown to hold within 10\% errors in both $J/\psi$ photoproduction and $\phi$ electroproduction near threshold \cite{Guo:2025jiz,Hatta:2025vhs}. 
Moreover, the same approximation holds to an even better accuracy in threshold production off the pion in the Sullivan process  \cite{Hatta:2025ryj}.
These developments, while still in their early stages, point toward the possibility of a systematic extraction of both the proton and pion GFFs at the EIC in the future.

\paragraph{Factorization and Sudakov resummation for GPDs and GFFs at large-$t$.} The study of the electromagnetic form factor of the nucleon \(F(t)\) at large momentum transfer \(|t| \gg \Lambda_{\rm QCD}^2\) has a long and winding history \cite{Punjabi:2015bba}. In essence, two mechanisms have been proposed to govern the large-\(t\) behavior. On the one hand, there is the {\it hard-scattering} mechanism, in which all three valence partons participate in the hard interaction with the highly virtual photon. This was contrasted with the so-called {\it Feynman mechanism} (also referred to as the soft or overlap contribution) \cite{Drell:1969km,West:1970av}, where a single ``active'' parton interacts with the virtual photon while the remaining ``spectator'' partons do not.

These two mechanisms represent genuinely distinct contributions to the form factor. While the former can be described within the framework of collinear factorization, this was long thought not to be the case for the latter, which was originally formulated in terms of light-cone wave functions. It is generally assumed that for asymptotically large \(|t|\) the hard-scattering contribution dominates over the Feynman contribution. This expectation is not, however, borne out by data in the accessible range
\(1\,\textrm{GeV}^2 \lesssim |t| \lesssim 25\,\textrm{GeV}^2\)
\cite{Nesterenko:1983ef,Isgur:1984jm,Isgur:1988iw,Radyushkin:1990te,Jakob:1993iw,Diehl:1998kh,Radyushkin:1998rt,Braun:2006hz}.

In \cite{Kivel:2010ns}, it was argued within soft-collinear effective theory (SCET) that the Feynman contribution can be formulated at the operator level and, in fact, exhibits the same power-law behavior as the hard-scattering contribution. Furthermore, the leading contribution to the hard scattering mechanism is suppressed by
\(\sim \alpha_s(\sqrt{|t|})^2\)
compared to the Feynman mechanism. As a result, the Feynman contribution can dominate over the hard-scattering contribution at phenomenologically accessible values of \(t\). This situation is eventually reversed due to Sudakov resummation effects in the Feynman contribution, but only at much larger values of \(|t|\).

As argued in \cite{Hatta:2025xuf}, much of the discussion in \cite{Kivel:2010ns} is directly analogous to the case of generalized parton distributions (GPDs) at large \(t\). In particular, the sole non-perturbative matrix element \(f_q(t)\) entering the quark contribution to the electromagnetic form factor \(F_q(t)\) at leading power is the same as that appearing in the GPD \(H_q(x,\xi,t)\). The ratio between the Feynman contributions \(F_q^{\rm Feyn}\) and \(H_q^{\rm Feyn}\) is a multiplicative hard-scattering kernel, which can be calculated perturbatively and carries the complete \(x\) dependence. Taking for example the first Mellin moment gives
\begin{equation}
\hspace*{-6mm}
\frac{A_q^{\rm Feyn}(t,\mu)}{F_{q}^{\rm Feyn}(t)} =
1 + \frac{\alpha_s(\mu) C_F}{4\pi}
\left(
\frac{8}{3} \ln \frac{-t}{\mu^2}
- \frac{52}{9}
\right)
+ \mathcal O\!\left(\alpha_s^2,\frac{1}{\sqrt{-t}}\right)
\approx 0.85 ,
\end{equation}
where \(A_q^{\rm Feyn}\) denotes the Feynman contribution to the gravitational form factor \(A_q\). This result, however, is subject to potentially large power corrections as well as contamination from the hard-scattering contribution. It will therefore be interesting to see whether this estimate for $A_q/F_q$ remains quantitatively reliable once large-\(t\) data for \(A_q\) become available.

These results have been embedded    into the traditional framework based on light-cone wave functions (the overlap representation of the GPD). This allows one to obtain non-perturbative corrections to the large-\(|t|\) scaling. For example,
\begin{equation}
\hspace*{8mm}
A_q(t) \sim
\underbrace{
U(t)\left(\frac{\Lambda^2}{-t}\right)^{\delta+1}
}_{\mathrm{Feynman}}
+
\underbrace{
\left(\frac{\alpha_s(\sqrt{|t|})}{\pi}\right)^2
\left(\frac{\Lambda^2}{-t}\right)^2
}_{\mathrm{hard\;\, scattering}} ,
\end{equation}
where \(\delta\) is a parameter related to the endpoint behavior of the light-cone wave function and \(U(t)\) is the Sudakov resummation factor. The SCET analysis suggests \(\delta = 1\), although values as large as \(\delta = 2\) have also been proposed \cite{Nesterenko:1983ef,Braun:2006hz,Braun:2001tj}. In any case, at sufficiently large \(|t|\), the Sudakov exponential \(U(t)\) suppresses the Feynman contribution relative to the hard-scattering term, ultimately overcoming the \(\alpha_s^2(\sqrt{|t|})/\pi^2\) suppression. This mirrors precisely the situation encountered earlier for the electromagnetic form factor.

\paragraph {Constraints on GFFs from the averaged null energy condition.} The form factors of the graviton-proton vertex determine the Wigner
transform of the QCD energy-momentum tensor in the proton which may be
used to investigate its mechanical properties such as
the distributions of pressure and shear
forces~\cite{Polyakov:2002yz,Polyakov:2018zvc,Lorce:2018egm} within the
hadron. It is important, however, to check the classification of the EMT
according to
Hawking-Ellis~\cite{Hawking:1973uf}.  The EMT in
the proton is expected to be of ordinary type~I in the radial tails
beyond a few Compton wavelengths where the standard point-wise energy
conditions (such as NEC, WEC, SEC, and DEC) are
satisfied~\cite{Dumitru:2025gzc}. In the interior region, however, the
proton EMT may be of type~IV~\cite{Dumitru:2025gzc},
lacking any causal eigenvectors, and
violating all the standard point-wise energy conditions. This
unconventional behavior originates from the
$T^{0i}$ component of the EMT for definite polarization of the
proton. The exotic properties potentially displayed by the
proton EMT, such as the fact that there may be regions within the
proton where the energy density can vanish or become negative, call
for some caution with definitions of the proton's ``mechanical''
properties.

The averaged null energy condition (ANEC)
represents a constraint that holds in any unitary, Lorentz invariant QFT.
It states that $\int\mathrm{d}\zeta \, \ell^\mu \ell^\nu
\, T_{\mu\nu} = \int\mathrm{d}\zeta\, \left( T^{00} - 2 n^i T^{0i} + n^i n^j
T^{ij} \right) \geq 0$, where the integral is taken over
a complete null geodesic with affine parameter $\zeta \in
(-\infty,\infty)$ and tangent vector $\ell^\mu =
\mathrm{d}x^\mu/\mathrm{d}\zeta$. In
Minkowski spacetime, one has
$x^\mu(\zeta) = \vec{x}_0 + \zeta L \ell^\mu$
with $\ell^\mu = (1,n^i)$ and
$\vec{n}^2 = 1$; $L>0$ represents an arbitrary length scale. 
In the Breit frame, ANEC results in
the following {\it model-independent constraint}
on the $A$ and $J$ GFFs of the proton~\cite{Dumitru:2025gzc}:
\begin{equation}
\hspace*{-4mm}  \int\frac{\mathrm{d}^2\Delta_\perp}{(2\pi)^2}\,
  e^{-i \vec{\Delta}_\perp\cdot\vec{x}_{0\perp}}
  \left[ m A(-\Delta_\perp^2) +
    \frac{\Delta_\perp^2}{4m}\left(A(-\Delta_\perp^2)-2 J(-\Delta_\perp^2)\right)
 \right] \geq 0\, .
 \label{newconstraint}
\end{equation}
The integral in~(\ref{newconstraint}) is finite because $A -
2J \sim 1/t^3$ asymptotically \cite{Tong:2022zax}, but the term
$\Delta_\perp^2 (A-2J)$
is not necessarily positive definite~\cite{Hackett:2023rif}.
This constraint could help improve the precision of form factor models
and parametrizations.

We may compare to ANEC for a spin-0 pion where the asymptotic $A(t)$
GFF takes a monopole form~\cite{Tanaka:2018wea,Tong:2021ctu,Gao:2022vyh}. We
have~\cite{Dumitru:2025gzc}
\begin{equation}
\hspace*{-7mm}
\frac{[\Delta^2 A(-\Delta^2)]_\infty}{4\pi \, |\vec{x}_{0\perp}|}
 +
 \int\frac{\mathrm{d}^2\Delta_\perp}{(2\pi)^2}\,
 e^{-i \vec{\Delta}_\perp\cdot\vec{x}_{0\perp}}
\left[E A(-\Delta_\perp^2) - 
\frac{[\Delta^2 A(-\Delta^2)]_\infty}{2\Delta_\perp}\right] \ge 0
~.
\label{eq:ANEC_pi}
\end{equation}
Here, $[\Delta^2 A(-\Delta^2)]_\infty =
\lim_{\Delta\to\infty}\, \Delta^2 A(-\Delta^2)$ with $\Delta =
|\vec\Delta|$.  The first term in eq.~(\ref{eq:ANEC_pi}) is
positive and ensures that ANEC$_\pi \ge 0$ is
satisfied in the limit $|\vec{x}_{0\perp}| \to 0$.  
However, for finite $|\vec{x}_{0\perp}|$,
Eq.~(\ref{eq:ANEC_pi}) represents a non-trivial constraint on the
$A(t)$ GFF of the pion. For example, if this GFF were described by a
single monopole mass $\Lambda$ then $m_\pi/\Lambda > 0.21$ would
be required by ANEC$_\pi \ge 0$.
\\

\subsection{GPD global analysis} 
       
The high luminosity and versatile polarization capabilities of the EIC offer unprecedented opportunities to explore GPDs~\cite{Ji:1994av, Ji:1996ek, Muller:1994ses} through hard exclusive processes such as deeply virtual Compton scattering (DVCS)~\cite{Ji:1996nm} and deeply virtual meson production (DVMP)~\cite{Radyushkin:1996ru, Collins:1996fb}, among others~\cite{Berger:2001xd, CLAS:2021lky, Belitsky:2002tf, Guidal:2002kt, Deja:2023ahc, Pedrak:2017cpp, Duplancic:2023kwe, Qiu:2022bpq, Qiu:2022pla, Qiu:2023mrm, Grocholski:2022rqj, Nabeebaccus:2023rzr, Qiu:2024mny}. Compared with fixed-target experiments at Jefferson Lab (JLab), the collider kinematics of the EIC are particularly advantageous for DVMP, where suppressing higher-twist contamination from transverse photons is essential for ensuring the validity of collinear factorization and accessing the small-$x_B$ regions. In particular, hard exclusive production of heavy quarkonia such as $J/\psi$ and $\Upsilon$, which are predominantly sensitive to gluon GPDs, constitutes a key focus of the EIC physics program~\cite{Ivanov:2004vd,Chen:2019uit,Flett:2021ghh,Goloskokov:2024egn,Flett:2024htj}.

\begin{figure}[t]
\centering
\raisebox{-0.5\height}{\includegraphics[width=0.4\columnwidth]{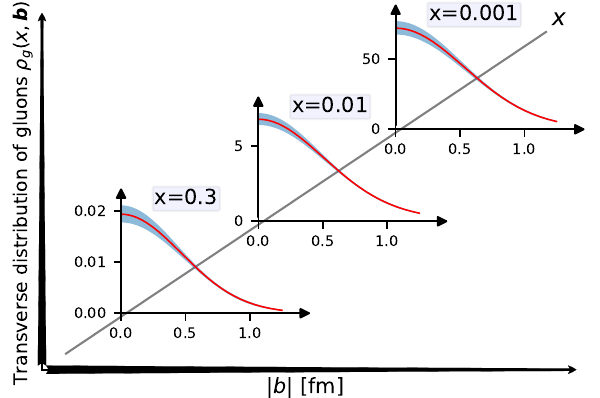}}
\raisebox{-0.5\height}{\includegraphics[width=0.45\columnwidth]{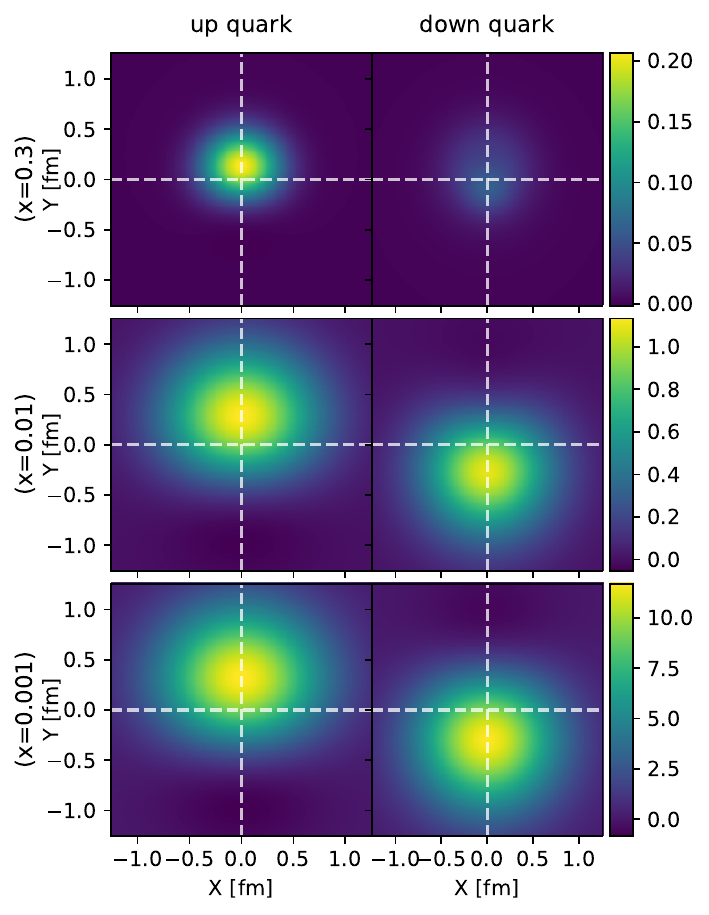}}
\caption
{Left: The transverse-space gluon distribution $\rho_{g}(x,\boldsymbol b)$ in an unpolarized proton with GUMP1.0 GPDs at $\mu=2$ GeV. Bands correspond to $90$\% confidence intervals obtained via a standard Monte Carlo propagation of Hessian errors, neglecting error correlations and other systematic errors. Right: The intrinsic transverse-space distributions $\rho_{q,\rm{In}}^{X}(x,\boldsymbol b)$ of up and down quarks in a proton polarized in the $X$ direction for $x=0.3$, $0.01$, and $0.001$ with GUMP1.0 GPDs at $\mu=2$ GeV.}
\label{fig:gumpgpd}
\end{figure}
  
Substantial progress has been achieved in extracting GPDs from exclusive processes using parametrized models~\cite{Kumericki:2007sa, Kumericki:2009uq, Cuic:2023mki}, neural-network-based reconstructions~\cite{Moutarde:2019tqa, Dutrieux:2021wll}, and analyses of Compton form factors (CFFs)~\cite{Almaeen:2024guo}. The recent {\it GPDs through Universal Moment Parametrization (GUMP)} framework enables global analyses of all four leading-twist GPDs by integrating experimental data with lattice QCD simulations~\cite{Guo:2022upw, Guo:2023ahv, Guo:2025muf}. A central difficulty in GPD extraction is the inverse problem: exclusive data alone do not uniquely determine the underlying GPDs. This ambiguity has also been clearly illustrated in the language of shadow GPDs~\cite{Bertone:2021yyz}, a type of non-vanishing GPDs that do not change the PDF nor contribute to exclusive processes. Incorporating lattice inputs mitigates the inverse problem inherent in GPD extractions, as exclusive measurements generally do not constrain the $x$ dependence directly, whereas lattice calculations within, e.g., the {\it Large Momentum Effective Theory (LaMET)}, do~\cite{Ji:2013dva, Ji:2020ect, Alexandrou:2020zbe, Lin:2021brq, Good:2025daz,Chen:2025xww}. In Fig. \ref{fig:gumpgpd}, the proton tomography based on the GUMP1.0 GPD extraction is presented for both the unpolarized and transversely polarized proton~\cite{Guo:2025muf}. 
        
Despite these advances, several theoretical and phenomenological challenges persist for future EIC studies. It is well understood that disentangling different GPD species and amplitudes demands measurements with diverse polarization configurations~\cite{Shiells:2021xqo}. Measurements involving transverse polarization remain experimentally challenging, making systematic cross-checks among different approaches essential for a consistent global determination. Also, perturbative corrections are expected to be enhanced at small $x_B$, and current calculations are limited to next-to-leading order (NLO), though higher-order effects may be significant, particularly for $J/\psi$ production~\cite{Chen:2019uit, Flett:2021ghh, Goloskokov:2024egn, Flett:2024htj}. For heavy quarkonia where $Q \sim M_{\mathrm{meson}}$, finite-mass corrections can also be substantial and must be carefully treated~\cite{Guo:2024wxy}. Addressing these effects will require continued theoretical developments to achieve the precision demanded by upcoming high-precision measurements at EIC and corresponding phenomenological analyses.
        
Looking ahead, twist-three GPDs have drawn growing interest for their direct connection to parton orbital motion and angular momentum within the nucleon. Beyond the complexities of establishing factorization at subleading twist, twist-three GPDs exhibit intricate intrinsic behavior and pose significant experimental challenges both at JLab and the future EIC.
    
\subsection{GPDs at JLAB and the EIC} 
        
JLab has a broad program for exclusive measurements of GPDs in the valence region. It spans many channels measured with detectors in different experimental halls, providing important cross checks. Common to all is that JLab provides the high luminosity required for precision measurements at high $x$, where cross sections are small, and highly polarized electron beam. Some of the data are taken with polarized and nuclear targets. The measurements are mostly carried out using the CLAS12 large-acceptance detector in Hall B, and focusing spectrometers in conjunction with standalone calorimeters, originally in Hall A but now in Hall C. The former provides a broader kinematic coverage while the latter gives precision data for a smaller number of selected points. In terms of nominal “PAC days,” the program currently approved for CLAS12 is indicative of what we can expect, and includes 239 days for unpolarized protons, 120 for longitudinally polarized protons, 110 for transversely polarized protons, 132 days with deuterium, and 207 days with other nuclei. About half of the unpolarized and longitudinally polarized data have been collected, and the same is true for deuterium and many other nuclear targets, including light nuclei where the recoiling nucleus is measured using the ALERT detector. However, while the amounts of unpolarized and linearly polarized proton data are comparable, the polarized target dilution means that the difference in the figure of merit is significant. Integration of transversely polarized targets is significantly more challenging, and no such proton data have been taken to date. When they become available, these data will be very important for many measurements related to GPDs and TMDs, but additional luminosity limitations mean that JLab data coming out of the 12 GeV program will be rather unbalanced in terms of target polarization. While it has not been utilized for GPD-related measurements to date, Hall C at JLab has $^3$He target providing a high level of polarization, and other similar targets may eventually be used in the other halls. The GlueX detector in Hall D is using a polarized, tagged real-photon beam. As such, it is less focused on GPD-studies, although it has measured near-threshold $J/\psi$ production and could measure timelike Compton scattering. The other halls measure a wide range of channels, including DVCS and different meson channels. Due to the energy dependence of the cross section, pseudoscalars are particularly well suited to the lower c.m. energies of JLab, and can impact our understanding of, for instance, transversity GPDs. Mid-life upgrades of existing detectors will further enhance the 12 GeV capabilities. In addition, there are also longer-term plans for more extensive upgrades, but it is not clear which (if any) of them will be implemented. In summary, JLab 12 GeV has an extensive GPD program covering a wide range of final states and targets, and will be able to carry out precision measurements in the valence region using different complementary detectors, providing luminosities much higher than at other fixed-target facilities.\\
        
The EIC will be the world’s first polarized $ep$ collider, and as such it will provide unique data for GPD-related measurements. Once standard operations have been established after a period of early running, all proton data will be obtained for longitudinal or transverse polarization -- in equal amounts. While the EIC will provide a lower luminosity than JLab for high-$x$ measurements, the rise in the cross sections at higher energies for vector meson production and various Compton processes, in combination with a near-hermetic acceptance that is easier to achieve in a collider than in a fixed-target setting, means that the event rates at the EIC will be quite high, and the polarized figure of merit will be exceptional. In comparison with HERA, one should also note the very significant impact of the fact that the ePIC detector and interaction region were designed with exclusive measurements in mind. Thus, while the c.m. energy of the EIC is lower than that of HERA, events that can be produced can also be detected across essentially all kinematics. In contrast, the HERA near-beam detectors were poorly suited for covering the full kinematic range, and even for the kinematics that were covered the acceptance was not always very high. Thus, the overall coverage for exclusive processes like DVCS is generally comparable for HERA and the EIC, but the EIC will have a much better acceptance, hundred times higher luminosity, and polarized protons and light nuclei (as well as unpolarized heavier nuclei). For GPD-related measurements, the EIC data will thus be a vast improvement over HERA. One should also note that the ePIC central and forward detector have been designed and optimized not only to detect and identify all the produced particles, but also to unambiguously ensure exclusivity and provide adequate resolutions for all the kinematic variables.
        
Any accelerator facility greatly benefits from having more than one detector for cross checks and confirmation of results. In addition, with two detectors it may be possible to reduce the overall systematic uncertainties (\textit{cf.} ZEUS and H1), which would be more even consequential at the EIC than HERA given its higher luminosity. A second EIC detector could, however, also provide additional and complementary capabilities creating new science opportunities, some of which would affect the GPD program. In particular, the introduction of a novel interaction region design incorporating a second focus would greatly expand the far-forward acceptance for the detection of the target protons, light nuclei, and nuclear fragments. The second focus makes it possible to detect particles down to $p_T = 0$. It provides a general improvement for protons -- in particular at smaller values of $x$. For coherent exclusive processes on light ions, where the cross section is already more prominent at lower values of $t$, the forward acceptance in $t$ falls off rapidly with mass. While ePIC can provide a reasonable $t$-coverage for the lightest ions (d and $^3$He), with a second focus one could also extend this to $^4$He and Li (both $^6$Li and $^7$Li may eventually be polarized), which would significantly expand the nuclear GPD program. Being able to detect the nuclear fragments, the second detector would also be able to ensure coherence and exclusivity for GPD-measurements in medium-mass nuclei (such as $^{12}$C, $^{16}$O, $^{40}$Ca) by vetoing breakup with very high efficiency. Since ePIC can only detect fragments from lighter nuclei, such vetoing is more challenging. The ability to detect fragments also makes it possible to measure bound-nucleon structure, for which is it essential to measure the spectator A-1 system. In ePIC, this is also limited to lighter nuclei, whereas a second detector could measure the A-1 spectator system for any nucleus.
        
In addition to an enhancement of the far-forward detection other aspects of complementarity could also have an impact on the EIC GPD program. For instance, a greater emphasis on high-purity muon identification would be beneficial for exclusive production of charmonium, timelike Compton scattering, and may enable measurements of double DVCS at high energy (where the cross sections are larger). A second detector would also benefit from around five years of additional detector R\&D. An example of this is enhanced hadron identification in the barrel (mid-rapidity), where it may be possible to improve the 3$\sigma$ $\pi/K$ separation from around 6 GeV/$c$ to up to 10 GeV/$c$. This would affect exclusive meson production, as well as other non-GPD measurements (e.g. TMD-PDFs).

%% file: outlook.tex
We look back to a wonderful and exciting five-week program. 
The  program’s objective was twofold: (1) to identify and address the most pressing theoretical challenges in preparation for EIC data, and (2) to forge collaborations and define strategies for precision physics, computational innovation, and experimental synergy. We believe that these goals have been met with great success. The program significantly advanced theoretical and computational readiness for the EIC era. It also served as a springboard for further collaboration. We felt that the discussions at the program were of very high quality and most exciting. The participants came from many different and diverse backgrounds; participant interactions were excellent throughout the program. It was also great to see the large number of young postdocs and students at the program and their full commitment in the discussions and activities. 
The organizers express their sincere gratitude to all participants for making this program a truly excellent and memorable event.

This document has summarized the current status of the various topics addressed during the program and the open issues that were identified. It also has given recommendations for the field that we developed through our discussions. We hope that the document will become a most valuable resource for the field in the years leading up to the Electron-Ion Collider era.

%% file: main.bbl
\begin{thebibliography}{100}

\bibitem{Deshpande:2005wd}
Abhay Deshpande, Richard Milner, Raju Venugopalan, and Werner Vogelsang.
\newblock {Study of the fundamental structure of matter with an electron-ion
  collider}.
\newblock {\em Ann. Rev. Nucl. Part. Sci.}, 55:165--228, 2005.

\bibitem{Boer:2011fh}
Daniel Boer et~al.
\newblock {Gluons and the quark sea at high energies: Distributions,
  polarization, tomography. INT report INT-PUB-11-034.}
\newblock arXiv:1108.1713.

\bibitem{Accardi:2012qut}
A.~Accardi et~al.
\newblock {Electron Ion Collider: The Next QCD Frontier}: {Understanding the
  glue that binds us all}.
\newblock {\em Eur. Phys. J. A}, 52(9):268, 2016.

\bibitem{AbdulKhalek:2021gbh}
R.~Abdul~Khalek et~al.
\newblock {Science Requirements and Detector Concepts for the Electron-Ion
  Collider}: {EIC Yellow Report}.
\newblock {\em Nucl. Phys. A}, 1026:122447, 2022.

\bibitem{AbdulKhalek:2022hcn}
R.~Abdul~Khalek et~al.
\newblock {Snowmass 2021 White Paper: Electron Ion Collider for High Energy
  Physics}.
\newblock arXiv:2203.13199.

\bibitem{Abir:2023fpo}
Raktim Abir et~al.
\newblock {The case for an EIC Theory Alliance: Theoretical Challenges of the
  EIC}.
\newblock arXiv:2305.14572.

\bibitem{Vermaseren:2005qc}
J.~A.~M. Vermaseren, A.~Vogt, and S.~Moch.
\newblock {The Third-order QCD corrections to deep-inelastic scattering by
  photon exchange}.
\newblock {\em Nucl. Phys. B}, 724:3--182, 2005.

\bibitem{Ablinger:2025awb}
J.~Ablinger, A.~Behring, J.~Bl{\"u}mlein, A.~De~Freitas, A.~von Manteuffel,
  C.~Schneider, and K.~Sch{\"o}nwald.
\newblock {The three-loop single-mass heavy-flavor corrections to the structure
  functions $F_2(x,Q^2)$ and $g_1(x,Q^2)$}.
\newblock 9 2025.

\bibitem{Falcioni:2023luc}
G.~Falcioni, F.~Herzog, S.~Moch, and A.~Vogt.
\newblock {Four-loop splitting functions in QCD {\textendash} The quark-quark
  case}.
\newblock {\em Phys. Lett. B}, 842:137944, 2023.

\bibitem{Falcioni:2023vqq}
G.~Falcioni, F.~Herzog, S.~Moch, and A.~Vogt.
\newblock {Four-loop splitting functions in QCD {\textendash} The
  gluon-to-quark case}.
\newblock {\em Phys. Lett. B}, 846:138215, 2023.

\bibitem{Falcioni:2024xyt}
G.~Falcioni, F.~Herzog, S.~Moch, A.~Pelloni, and A.~Vogt.
\newblock {Four-loop splitting functions in QCD {\textendash} The
  quark-to-gluon case}.
\newblock {\em Phys. Lett. B}, 856:138906, 2024.

\bibitem{Falcioni:2024qpd}
G.~Falcioni, F.~Herzog, S.~Moch, A.~Pelloni, and A.~Vogt.
\newblock {Four-loop splitting functions in QCD {\textendash} the gluon-gluon
  case {\textendash}}.
\newblock {\em Phys. Lett. B}, 860:139194, 2025.

\bibitem{Gehrmann:2023cqm}
T.~Gehrmann, A.~von Manteuffel, V.~Sotnikov, and T.-Z. Yang.
\newblock {Complete $ {N}_f^2 $ contributions to four-loop pure-singlet
  splitting functions}.
\newblock {\em JHEP}, 01:029, 2024.

\bibitem{Gehrmann:2023iah}
T.~Gehrmann, A.~von Manteuffel, V.~Sotnikov, and T.-Z. Yang.
\newblock {The NfCF3 contribution to the non-singlet splitting function at
  four-loop order}.
\newblock {\em Phys. Lett. B}, 849:138427, 2024.

\bibitem{Zijlstra:1993sh}
E.~B. Zijlstra and W.~L. van Neerven.
\newblock {Order-$\alpha_s^2$ corrections to the polarized structure function
  $g_1 (x,Q^2)$}.
\newblock {\em Nucl. Phys. B}, 417:61--100, 1994.
\newblock [Erratum: Nucl.Phys.B 426, 245 (1994), Erratum: Nucl.Phys.B 773,
  105--106 (2007), Erratum: Nucl.Phys.B 501, 599--599 (1997)].

\bibitem{Moch:2014sna}
S.~Moch, J.~A.~M. Vermaseren, and A.~Vogt.
\newblock {The Three-Loop Splitting Functions in QCD: The Helicity-Dependent
  Case}.
\newblock {\em Nucl. Phys. B}, 889:351--400, 2014.

\bibitem{Blumlein:2021enk}
J.~Bl{\"u}mlein, P.~Marquard, C.~Schneider, and K.~Sch{\"o}nwald.
\newblock {The three-loop unpolarized and polarized non-singlet anomalous
  dimensions from off shell operator matrix elements}.
\newblock {\em Nucl. Phys. B}, 971:115542, 2021.

\bibitem{Blumlein:2022gpp}
J.~Bl{\"u}mlein, P.~Marquard, C.~Schneider, and K.~Sch{\"o}nwald.
\newblock {The massless three-loop Wilson coefficients for the deep-inelastic
  structure functions F$_{2}$, F$_{L}$, xF$_{3}$ and g$_{1}$}.
\newblock {\em JHEP}, 11:156, 2022.

\bibitem{Goyal:2023zdi}
Saurav Goyal, Sven-Olaf Moch, Vaibhav Pathak, Narayan Rana, and V.~Ravindran.
\newblock {Next-to-Next-to-Leading Order QCD Corrections to Semi-Inclusive
  Deep-Inelastic Scattering}.
\newblock {\em Phys. Rev. Lett.}, 132(25):251902, 2024.

\bibitem{Bonino:2024qbh}
Leonardo Bonino, Thomas Gehrmann, and Giovanni Stagnitto.
\newblock {Semi-Inclusive Deep-Inelastic Scattering at Next-to-Next-to-Leading
  Order in QCD}.
\newblock {\em Phys. Rev. Lett.}, 132(25):251901, 2024.

\bibitem{Goyal:2024tmo}
Saurav Goyal, Roman~N. Lee, Sven-Olaf Moch, Vaibhav Pathak, Narayan Rana, and
  V.~Ravindran.
\newblock {Next-to-Next-to-Leading Order QCD Corrections to Polarized
  Semi-Inclusive Deep-Inelastic Scattering}.
\newblock {\em Phys. Rev. Lett.}, 133:211905, 2024.

\bibitem{Bonino:2024wgg}
Leonardo Bonino, Thomas Gehrmann, Markus L{\"o}chner, Kay Sch{\"o}nwald, and
  Giovanni Stagnitto.
\newblock {Polarized Semi-Inclusive Deep-Inelastic Scattering at
  Next-to-Next-to-Leading Order in QCD}.
\newblock {\em Phys. Rev. Lett.}, 133(21):211904, 2024.

\bibitem{Bonino:2025qta}
Leonardo Bonino, Thomas Gehrmann, Markus L{\"o}chner, Kay Sch{\"o}nwald, and
  Giovanni Stagnitto.
\newblock {Neutral and charged current semi-inclusive deep-inelastic scattering
  at NNLO QCD}.
\newblock {\em JHEP}, 10:016, 2025.

\bibitem{Bonino:2025bqa}
Leonardo Bonino, Thomas Gehrmann, Markus L{\"o}chner, Kay Sch{\"o}nwald, and
  Giovanni Stagnitto.
\newblock {Polarized Neutral and Charged Current Semi-Inclusive Deep-Inelastic
  Scattering at NNLO in QCD}.
\newblock 9 2025.

\bibitem{Goyal:2025qyu}
Saurav Goyal, Roman~N. Lee, Sven-Olaf Moch, Vaibhav Pathak, and V.~Ravindran.
\newblock {NNLO QCD$\otimes$QED corrections to unpolarized and polarized
  SIDIS}.
\newblock arXiv:2510.18872.

\bibitem{Haug:2025ava}
Juliane Haug and Fabian Wunder.
\newblock {Single-valued representation of unpolarized and polarized
  semi-inclusive deep inelastic scattering at next-to-next-to-leading order}.
\newblock {\em Phys. Rev. D}, 112(11):114036, 2025.

\bibitem{Cacciari:2001cw}
Matteo Cacciari and Stefano Catani.
\newblock {Soft gluon resummation for the fragmentation of light and heavy
  quarks at large x}.
\newblock {\em Nucl. Phys. B}, 617:253--290, 2001.

\bibitem{Anderle:2012rq}
Daniele~P. Anderle, Felix Ringer, and Werner Vogelsang.
\newblock {QCD resummation for semi-inclusive hadron production processes}.
\newblock {\em Phys. Rev. D}, 87(3):034014, 2013.

\bibitem{Anderle:2013lka}
Daniele~P. Anderle, Felix Ringer, and Werner Vogelsang.
\newblock {Threshold resummation for polarized (semi-)inclusive deep inelastic
  scattering}.
\newblock {\em Phys. Rev. D}, 87:094021, 2013.

\bibitem{Abele:2021nyo}
Maurizio Abele, Daniel de~Florian, and Werner Vogelsang.
\newblock {Approximate NNLO QCD corrections to semi-inclusive DIS}.
\newblock {\em Phys. Rev. D}, 104(9):094046, 2021.

\bibitem{Abele:2022wuy}
Maurizio Abele, Daniel de~Florian, and Werner Vogelsang.
\newblock {Threshold resummation at NLL3 accuracy and approximate N3LO
  corrections to semi-inclusive DIS}.
\newblock {\em Phys. Rev. D}, 106(1):014015, 2022.

\bibitem{Borsa:2024mss}
Ignacio Borsa, Marco Stratmann, Werner Vogelsang, Daniel de~Florian, and
  Rodolfo Sassot.
\newblock {Next-to-Next-to-Leading Order Global Analysis of Polarized Parton
  Distribution Functions}.
\newblock {\em Phys. Rev. Lett.}, 133(15):151901, 2024.

\bibitem{Bertone:2024taw}
Valerio Bertone, Amedeo Chiefa, and Emanuele~R. Nocera.
\newblock {Helicity-dependent parton distribution functions at
  next-to-next-to-leading order accuracy from inclusive and semi-inclusive
  deep-inelastic scattering data}.
\newblock {\em Phys. Lett. B}, 865:139497, 2025.

\bibitem{Goyal:2025bzf}
Saurav Goyal, Sven-Olaf Moch, Vaibhav Pathak, Narayan Rana, and V.~Ravindran.
\newblock {Soft and virtual corrections to semi-inclusive DIS up to four loops
  in QCD}.
\newblock {\em Phys. Rev. D}, 113(3):034004, 2026.

\bibitem{Catani:1996vz}
S.~Catani and M.~H. Seymour.
\newblock {A General algorithm for calculating jet cross-sections in NLO QCD}.
\newblock {\em Nucl. Phys. B}, 485:291--419, 1997.
\newblock [Erratum: Nucl.Phys.B 510, 503--504 (1998)].

\bibitem{Frixione:1995ms}
S.~Frixione, Z.~Kunszt, and A.~Signer.
\newblock {Three jet cross-sections to next-to-leading order}.
\newblock {\em Nucl. Phys. B}, 467:399--442, 1996.

\bibitem{Anastasiou:2003gr}
Charalampos Anastasiou, Kirill Melnikov, and Frank Petriello.
\newblock {A new method for real radiation at NNLO}.
\newblock {\em Phys. Rev. D}, 69:076010, 2004.

\bibitem{Binoth:2004jv}
T.~Binoth and G.~Heinrich.
\newblock {Numerical evaluation of phase space integrals by sector
  decomposition}.
\newblock {\em Nucl. Phys. B}, 693:134--148, 2004.

\bibitem{Gehrmann-DeRidder:2005btv}
A.~Gehrmann-De~Ridder, T.~Gehrmann, and E.~W.~Nigel Glover.
\newblock {Antenna subtraction at NNLO}.
\newblock {\em JHEP}, 09:056, 2005.

\bibitem{Catani:2007vq}
Stefano Catani and Massimiliano Grazzini.
\newblock {An NNLO subtraction formalism in hadron collisions and its
  application to Higgs boson production at the LHC}.
\newblock {\em Phys. Rev. Lett.}, 98:222002, 2007.

\bibitem{Czakon:2011ve}
M.~Czakon.
\newblock {Double-real radiation in hadronic top quark pair production as a
  proof of a certain concept}.
\newblock {\em Nucl. Phys. B}, 849:250--295, 2011.

\bibitem{Caola:2017dug}
Fabrizio Caola, Kirill Melnikov, and Raoul R{\"o}ntsch.
\newblock {Nested soft-collinear subtractions in NNLO QCD computations}.
\newblock {\em Eur. Phys. J. C}, 77(4):248, 2017.

\bibitem{DelDuca:2016ily}
Vittorio Del~Duca, Claude Duhr, Adam Kardos, G{\'a}bor Somogyi, Zolt{\'a}n
  Sz{\H{o}}r, Zolt{\'a}n Tr{\'o}cs{\'a}nyi, and Zolt{\'a}n Tulip{\'a}nt.
\newblock {Jet production in the CoLoRFulNNLO method: event shapes in
  electron-positron collisions}.
\newblock {\em Phys. Rev. D}, 94(7):074019, 2016.

\bibitem{Bertolotti:2022aih}
Gloria Bertolotti, Lorenzo Magnea, Giovanni Pelliccioli, Alessandro Ratti,
  Chiara Signorile-Signorile, Paolo Torrielli, and Sandro Uccirati.
\newblock {NNLO subtraction for any massless final state: a complete analytic
  expression}.
\newblock {\em JHEP}, 07:140, 2023.
\newblock [Erratum: JHEP 05, 019 (2024)].

\bibitem{Klasen:1996it}
M.~Klasen and G.~Kramer.
\newblock {Inclusive two jet production at HERA: Direct and resolved
  cross-sections in next-to-leading order QCD}.
\newblock {\em Z. Phys. C}, 76:67--74, 1997.

\bibitem{Frixione:1997ks}
Stefano Frixione and Giovanni Ridolfi.
\newblock {Jet photoproduction at HERA}.
\newblock {\em Nucl. Phys. B}, 507:315--333, 1997.

\bibitem{Nagy:2001xb}
Zoltan Nagy and Zoltan Trocsanyi.
\newblock {Multijet cross-sections in deep inelastic scattering at
  next-to-leading order}.
\newblock {\em Phys. Rev. Lett.}, 87:082001, 2001.

\bibitem{Laenen:1992zk}
Eric Laenen, S.~Riemersma, J.~Smith, and W.~L. van Neerven.
\newblock {Complete O (alpha-s) corrections to heavy flavor structure functions
  in electroproduction}.
\newblock {\em Nucl. Phys. B}, 392:162--228, 1993.

\bibitem{Frixione:1995qc}
Stefano Frixione, Paolo Nason, and Giovanni Ridolfi.
\newblock {Differential distributions for heavy flavor production at HERA}.
\newblock {\em Nucl. Phys. B}, 454:3--24, 1995.

\bibitem{Cacciari:2001td}
Matteo Cacciari, Stefano Frixione, and Paolo Nason.
\newblock {The p(T) spectrum in heavy flavor photoproduction}.
\newblock {\em JHEP}, 03:006, 2001.

\bibitem{Alwall:2014hca}
J.~Alwall, R.~Frederix, S.~Frixione, V.~Hirschi, F.~Maltoni, O.~Mattelaer,
  H.~S. Shao, T.~Stelzer, P.~Torrielli, and M.~Zaro.
\newblock {The automated computation of tree-level and next-to-leading order
  differential cross sections, and their matching to parton shower
  simulations}.
\newblock {\em JHEP}, 07:079, 2014.

\bibitem{Sherpa:2024mfk}
Enrico Bothmann et~al.
\newblock {Event generation with Sherpa 3}.
\newblock {\em JHEP}, 12:156, 2024.

\bibitem{Frixione:2007vw}
Stefano Frixione, Paolo Nason, and Carlo Oleari.
\newblock {Matching NLO QCD computations with Parton Shower simulations: the
  POWHEG method}.
\newblock {\em JHEP}, 11:070, 2007.

\bibitem{Bierlich:2022pfr}
Christian Bierlich et~al.
\newblock {A comprehensive guide to the physics and usage of PYTHIA 8.3}.
\newblock {\em SciPost Phys. Codeb.}, 2022:8, 2022.

\bibitem{Bewick:2023tfi}
Gavin Bewick et~al.
\newblock {Herwig 7.3 release note}.
\newblock {\em Eur. Phys. J. C}, 84(10):1053, 2024.

\bibitem{Helenius:2024rth}
Ilkka Helenius, Peter Meinzinger, Simon Pl{\"a}tzer, and Peter Richardson.
\newblock {Photoproduction in general-purpose event generators}.
\newblock arXiv:2406.08026.

\bibitem{Hoeche:2023gme}
Stefan Hoeche, Frank Krauss, and Peter Meinzinger.
\newblock {Resolved photons in Sherpa}.
\newblock {\em Eur. Phys. J. C}, 84(2):178, 2024.

\bibitem{Meinzinger:2023xuf}
Peter Meinzinger and Frank Krauss.
\newblock {Hadron-level NLO predictions for QCD observables in photo-production
  at the Electron-Ion Collider}.
\newblock {\em Phys. Rev. D}, 109(3):034037, 2024.

\bibitem{Helenius:2019gbd}
Ilkka Helenius and Christine~O. Rasmussen.
\newblock {Hard diffraction in photoproduction with Pythia 8}.
\newblock {\em Eur. Phys. J. C}, 79(5):413, 2019.

\bibitem{Banfi:2023mhz}
Andrea Banfi, Silvia Ferrario~Ravasio, Barbara J{\"a}ger, Alexander Karlberg,
  Felix Reichenbach, and Giulia Zanderighi.
\newblock {A POWHEG generator for deep inelastic scattering}.
\newblock {\em JHEP}, 02:023, 2024.

\bibitem{Meinzinger:2025pam}
Peter Meinzinger, Daniel Reichelt, and Federico Silvetti.
\newblock {Event generation at MEPS@NLO accuracy in neutral and charged current
  DIS at the EIC}.
\newblock {\em Phys. Rev. D}, 112(7):074039, 2025.

\bibitem{Borsa:2020yxh}
Ignacio Borsa, Daniel de~Florian, and Iv{\'a}n Pedron.
\newblock {Inclusive-jet and dijet production in polarized deep inelastic
  scattering}.
\newblock {\em Phys. Rev. D}, 103(1):014008, 2021.

\bibitem{Heinrich:2020ybq}
Gudrun Heinrich.
\newblock {Collider Physics at the Precision Frontier}.
\newblock {\em Phys. Rept.}, 922:1--69, 2021.

\bibitem{Currie:2017tpe}
James Currie, Thomas Gehrmann, Alexander Huss, and Jan Niehues.
\newblock {NNLO QCD corrections to jet production in deep inelastic
  scattering}.
\newblock {\em JHEP}, 07:018, 2017.
\newblock [Erratum: JHEP 12, 042 (2020)].

\bibitem{NNLOJET:2025rno}
A.~Huss et~al.
\newblock {NNLOJET: a parton-level event generator for jet cross sections at
  NNLO QCD accuracy}.
\newblock arXiv:2503.22804.

\bibitem{Gehrmann:2025xab}
Thomas Gehrmann and Markus L{\"o}chner.
\newblock {The unresolved behaviour of polarized scattering matrix elements at
  NNLO in QCD}.
\newblock {\em JHEP}, 02:097, 2026.

\bibitem{Boussarie:2023izj}
Renaud Boussarie et~al.
\newblock {TMD Handbook}.
\newblock arXiv:2304.03302.

\bibitem{Sun:2014dqm}
Peng Sun, Joshua Isaacson, C.~P. Yuan, and Feng Yuan.
\newblock {Nonperturbative functions for SIDIS and Drell\textendash{}Yan
  processes}.
\newblock {\em Int. J. Mod. Phys. A}, 33(11):1841006, 2018.

\bibitem{Gonzalez-Hernandez:2018ipj}
J.~O. Gonzalez-Hernandez, T.~C. Rogers, N.~Sato, and B.~Wang.
\newblock {Challenges with Large Transverse Momentum in Semi-Inclusive Deeply
  Inelastic Scattering}.
\newblock {\em Phys. Rev. D}, 98(11):114005, 2018.

\bibitem{Bacchetta:2019tcu}
Alessandro Bacchetta, Giuseppe Bozzi, Martin Lambertsen, Fulvio Piacenza,
  Julius Steiglechner, and Werner Vogelsang.
\newblock {Difficulties in the description of Drell-Yan processes at moderate
  invariant mass and high transverse momentum}.
\newblock {\em Phys. Rev. D}, 100(1):014018, 2019.

\bibitem{Gauld:2021pkr}
R.~Gauld, A.~Gehrmann-De~Ridder, T.~Gehrmann, E.~W.~N. Glover, A.~Huss,
  I.~Majer, and A.~Rodriguez~Garcia.
\newblock {Transverse momentum distributions in low-mass Drell-Yan lepton pair
  production at NNLO QCD}.
\newblock {\em Phys. Lett. B}, 829:137111, 2022.

\bibitem{Dong:2026eas}
Liang Dong, Shen Fang, Jun Gao, Hai~Tao Li, Ding~Yu Shao, and Yu~Jiao Zhu.
\newblock {NNLO QCD corrections to hadron production in DIS at finite
  transverse momentum}.
\newblock arXiv:2602.22972.

\bibitem{Gao:2026tnd}
Jun Gao, Hai~Tao Li, Hua~Xing Zhu, and Yu~Jiao Zhu.
\newblock {Transverse-Momentum Subtraction for Semi-Inclusive Deep-Inelastic
  Scattering}.
\newblock arXiv:2602.06364.

\bibitem{Eskola:2002yc}
K.~J. Eskola, H.~Honkanen, V.~J. Kolhinen, Jian-wei Qiu, and C.~A. Salgado.
\newblock {Nonlinear corrections to the DGLAP equations in view of the hera
  data}.
\newblock {\em Nucl. Phys. B}, 660:211--224, 2003.

\bibitem{Qiu:2002mh}
Jian-wei Qiu.
\newblock {Structure functions and parton distributions}.
\newblock {\em Nucl. Phys. A}, 715:309--318, 2003.

\bibitem{Qiu:2003vd}
Jian-wei Qiu and Ivan Vitev.
\newblock {Resummed QCD power corrections to nuclear shadowing}.
\newblock {\em Phys. Rev. Lett.}, 93:262301, 2004.

\bibitem{Qiu:2001hj}
Jian-wei Qiu and George~F. Sterman.
\newblock {QCD and rescattering in nuclear targets}.
\newblock {\em Int. J. Mod. Phys. E}, 12:149, 2003.

\bibitem{Luo:1994np}
M.~Luo, Jian-wei Qiu, and George~F. Sterman.
\newblock {Anomalous nuclear enhancement in deeply inelastic scattering and
  photoproduction}.
\newblock {\em Phys. Rev. D}, 50:1951--1971, 1994.

\bibitem{Guo:2000nz}
Xiao-feng Guo and Xin-Nian Wang.
\newblock {Multiple scattering, parton energy loss and modified fragmentation
  functions in deeply inelastic e A scattering}.
\newblock {\em Phys. Rev. Lett.}, 85:3591--3594, 2000.

\bibitem{Wang:2001ifa}
Xin-Nian Wang and Xiao-feng Guo.
\newblock {Multiple parton scattering in nuclei: Parton energy loss}.
\newblock {\em Nucl. Phys. A}, 696:788--832, 2001.

\bibitem{HERMES:2007plz}
A.~Airapetian et~al.
\newblock {Hadronization in semi-inclusive deep-inelastic scattering on
  nuclei}.
\newblock {\em Nucl. Phys. B}, 780:1--27, 2007.

\bibitem{HERMES:2005mar}
A.~Airapetian et~al.
\newblock {Double hadron leptoproduction in the nuclear medium}.
\newblock {\em Phys. Rev. Lett.}, 96:162301, 2006.

\bibitem{Catani:2003zt}
Stefano Catani, Daniel de~Florian, Massimiliano Grazzini, and Paolo Nason.
\newblock {Soft gluon resummation for Higgs boson production at hadron
  colliders}.
\newblock {\em JHEP}, 07:028, 2003.

\bibitem{Ahrens:2009cxz}
Valentin Ahrens, Thomas Becher, Matthias Neubert, and Li~Lin Yang.
\newblock {Renormalization-Group Improved Prediction for Higgs Production at
  Hadron Colliders}.
\newblock {\em Eur. Phys. J. C}, 62:333--353, 2009.

\bibitem{Bonvini:2014joa}
Marco Bonvini and Simone Marzani.
\newblock {Resummed Higgs cross section at N$^{3}$LL}.
\newblock {\em JHEP}, 09:007, 2014.

\bibitem{Catani:2014uta}
Stefano Catani, Leandro Cieri, Daniel de~Florian, Giancarlo Ferrera, and
  Massimiliano Grazzini.
\newblock {Threshold resummation at N$^3$LL accuracy and soft-virtual cross
  sections at N$^3$LO}.
\newblock {\em Nucl. Phys. B}, 888:75--91, 2014.

\bibitem{Kidonakis:1997gm}
Nikolaos Kidonakis and George~F. Sterman.
\newblock {Resummation for QCD hard scattering}.
\newblock {\em Nucl. Phys. B}, 505:321--348, 1997.

\bibitem{Bonciani:2003nt}
Roberto Bonciani, Stefano Catani, Michelangelo~L. Mangano, and Paolo Nason.
\newblock {Sudakov resummation of multiparton QCD cross-sections}.
\newblock {\em Phys. Lett. B}, 575:268--278, 2003.

\bibitem{Catani:2013vaa}
Stefano Catani, Massimiliano Grazzini, and Alessandro Torre.
\newblock {Soft-gluon resummation for single-particle inclusive hadroproduction
  at high transverse momentum}.
\newblock {\em Nucl. Phys. B}, 874:720--745, 2013.

\bibitem{Beneke:2011mq}
M.~Beneke, P.~Falgari, S.~Klein, and C.~Schwinn.
\newblock {Hadronic top-quark pair production with NNLL threshold resummation}.
\newblock {\em Nucl. Phys. B}, 855:695--741, 2012.

\bibitem{Czakon:2009zw}
Michal Czakon, Alexander Mitov, and George~F. Sterman.
\newblock {Threshold Resummation for Top-Pair Hadroproduction to
  Next-to-Next-to-Leading Log}.
\newblock {\em Phys. Rev. D}, 80:074017, 2009.

\bibitem{Yang:2014hya}
Li~Lin Yang, Chong~Sheng Li, Jun Gao, and Jian Wang.
\newblock {NNLL momentum-space threshold resummation in direct top quark
  production at the LHC}.
\newblock {\em JHEP}, 12:123, 2014.

\bibitem{Kidonakis:2010tc}
Nikolaos Kidonakis.
\newblock {NNLL resummation for s-channel single top quark production}.
\newblock {\em Phys. Rev. D}, 81:054028, 2010.

\bibitem{Becher:2009th}
Thomas Becher and Matthew~D. Schwartz.
\newblock {Direct photon production with effective field theory}.
\newblock {\em JHEP}, 02:040, 2010.

\bibitem{Ferroglia:2013awa}
Andrea Ferroglia, Simone Marzani, Ben~D. Pecjak, and Li~Lin Yang.
\newblock {Boosted top production: factorization and resummation for
  single-particle inclusive distributions}.
\newblock {\em JHEP}, 01:028, 2014.

\bibitem{Hinderer:2018nkb}
Patriz Hinderer, Felix Ringer, George Sterman, and Werner Vogelsang.
\newblock {Threshold Resummation at NNLL for Single-particle Production in
  Hadronic Collisions}.
\newblock {\em Phys. Rev. D}, 99(5):054019, 2019.

\bibitem{Beneke:2010da}
Martin Beneke, Pietro Falgari, and Christian Schwinn.
\newblock {Threshold resummation for pair production of coloured heavy
  (s)particles at hadron colliders}.
\newblock {\em Nucl. Phys. B}, 842:414--474, 2011.

\bibitem{Beenakker:2013mva}
Wim Beenakker, Tim Janssen, Susanne Lepoeter, Michael Kr{\"a}mer, Anna Kulesza,
  Eric Laenen, Irene Niessen, Silja Thewes, and Tom Van~Daal.
\newblock {Towards NNLL resummation: hard matching coefficients for squark and
  gluino hadroproduction}.
\newblock {\em JHEP}, 10:120, 2013.

\bibitem{Sterman:2006hu}
George~F. Sterman and Werner Vogelsang.
\newblock {Crossed Threshold Resummation}.
\newblock {\em Phys. Rev. D}, 74:114002, 2006.

\bibitem{Catani:1989ne}
S.~Catani and L.~Trentadue.
\newblock {Resummation of the QCD Perturbative Series for Hard Processes}.
\newblock {\em Nucl. Phys. B}, 327:323--352, 1989.

\bibitem{Lustermans:2019cau}
Gillian Lustermans, Johannes K.~L. Michel, and Frank~J. Tackmann.
\newblock {Generalized Threshold Factorization with Full Collinear Dynamics}.
\newblock arXiv:1908.00985.

\bibitem{Mistlberger:2025lee}
Bernhard Mistlberger and Gherardo Vita.
\newblock {Collinear Approximations for LHC Cross Sections: Factorization and
  Resummation}.
\newblock arXiv:2502.17548.

\bibitem{DeRos:2026bcv}
Lorenzo De~Ros, Stefano Forte, Giovanni Ridolfi, and Davide~Maria Tagliabue.
\newblock {Threshold resummation of rapidity distributions at fixed partonic
  rapidity}.
\newblock arXiv:2601.04309.

\bibitem{Forte:2026tva}
Stefano Forte, Giovanni Ridolfi, and Francesco Ventola.
\newblock {Threshold resummation of Semi-Inclusive Deep-Inelastic Scattering}.
\newblock arXiv:2601.06245.

\bibitem{Borsa:2022vvp}
Ignacio Borsa, Rodolfo Sassot, Daniel de~Florian, Marco Stratmann, and Werner
  Vogelsang.
\newblock {Towards a Global QCD Analysis of Fragmentation Functions at
  Next-to-Next-to-Leading Order Accuracy}.
\newblock {\em Phys. Rev. Lett.}, 129(1):012002, 2022.

\bibitem{Goyal:2024emo}
Saurav Goyal, Roman~N. Lee, Sven-Olaf Moch, Vaibhav Pathak, Narayan Rana, and
  V.~Ravindran.
\newblock {NNLO QCD corrections to unpolarized and polarized SIDIS}.
\newblock {\em Phys. Rev. D}, 111(9):094007, 2025.

\bibitem{Bonocore:2015esa}
D.~Bonocore, E.~Laenen, L.~Magnea, S.~Melville, L.~Vernazza, and C.~D. White.
\newblock {A factorization approach to next-to-leading-power threshold
  logarithms}.
\newblock {\em JHEP}, 06:008, 2015.

\bibitem{vanBijleveld:2023vck}
Robin van Bijleveld, Eric Laenen, Leonardo Vernazza, and Guoxing Wang.
\newblock {Next-to-leading power resummed rapidity distributions near threshold
  for Drell-Yan and diphoton production}.
\newblock {\em JHEP}, 10:126, 2023.

\bibitem{Ebert:2017uel}
Markus~A. Ebert, Johannes K.~L. Michel, and Frank~J. Tackmann.
\newblock {Resummation Improved Rapidity Spectrum for Gluon Fusion Higgs
  Production}.
\newblock {\em JHEP}, 05:088, 2017.

\bibitem{deFlorian:2013taa}
Daniel de~Florian, Melanie Pfeuffer, Andreas Sch{\"a}fer, and Werner Vogelsang.
\newblock {Soft-gluon Resummation for High-pT Inclusive-Hadron Production at
  COMPASS}.
\newblock {\em Phys. Rev. D}, 88(1):014024, 2013.

\bibitem{Uebler:2015ria}
Claudia Uebler, Andreas Sch{\"a}fer, and Werner Vogelsang.
\newblock {Threshold resummation for polarized high-$p_T$ hadron production at
  COMPASS}.
\newblock {\em Phys. Rev. D}, 92:094029, 2015.

\bibitem{LHCHiggsCrossSectionWorkingGroup:2016ypw}
D.~de~Florian et~al.
\newblock {Handbook of LHC Higgs Cross Sections: 4. Deciphering the Nature of
  the Higgs Sector}.
\newblock {\em CERN Yellow Rep. Monogr.}, 2:1--869, 2017.

\bibitem{ParticleDataGroup:2016lqr}
C.~Patrignani et~al.
\newblock {Review of Particle Physics}.
\newblock {\em Chin. Phys. C}, 40(10):100001, 2016.

\bibitem{ParticleDataGroup:2024cfk}
S.~Navas et~al.
\newblock {Review of particle physics}.
\newblock {\em Phys. Rev. D}, 110(3):030001, 2024.

\bibitem{Bethke:2000ai}
S.~Bethke.
\newblock {Determination of the QCD coupling $\alpha_s$}.
\newblock {\em J. Phys. G}, 26:R27, 2000.

\bibitem{Bethke:2006ac}
Siegfried Bethke.
\newblock {Experimental tests of asymptotic freedom}.
\newblock {\em Prog. Part. Nucl. Phys.}, 58:351--386, 2007.

\bibitem{Bethke:2009jm}
Siegfried Bethke.
\newblock {The 2009 World Average of alpha(s)}.
\newblock {\em Eur. Phys. J. C}, 64:689--703, 2009.

\bibitem{ParticleDataGroup:2006fqo}
W.~M. Yao et~al.
\newblock {Review of Particle Physics}.
\newblock {\em J. Phys. G}, 33:1--1232, 2006.

\bibitem{ParticleDataGroup:2010dbb}
K.~Nakamura et~al.
\newblock {Review of particle physics}.
\newblock {\em J. Phys. G}, 37:075021, 2010.

\bibitem{Altarelli:2013bpa}
Guido Altarelli.
\newblock {The QCD Running Coupling and its Measurement}.
\newblock {\em PoS}, Corfu2012:002, 2013.

\bibitem{dEnterria:2022hzv}
D.~d'Enterria et~al.
\newblock {The strong coupling constant: state of the art and the decade
  ahead}.
\newblock {\em J. Phys. G}, 51(9):090501, 2024.

\bibitem{Whitlow:1990gk}
L.~W. Whitlow, Stephen Rock, A.~Bodek, E.~M. Riordan, and S.~Dasu.
\newblock {A Precise extraction of R = sigma-L / sigma-T from a global analysis
  of the SLAC deep inelastic e p and e d scattering cross-sections}.
\newblock {\em Phys. Lett. B}, 250:193--198, 1990.

\bibitem{BCDMS:1989eab}
A.~C. Benvenuti et~al.
\newblock {Test of {QCD} and a Measurement of $\Lambda$ From Scaling Violations
  in the Proton Structure Function F(2) (X, $Q^2$) at High $Q^2$}.
\newblock {\em Phys. Lett. B}, 223:490--496, 1989.

\bibitem{NewMuon:1996fwh}
M.~Arneodo et~al.
\newblock {Measurement of the proton and deuteron structure functions, F2(p)
  and F2(d), and of the ratio sigma-L / sigma-T}.
\newblock {\em Nucl. Phys. B}, 483:3--43, 1997.

\bibitem{H1:2015ubc}
H.~Abramowicz et~al.
\newblock {Combination of measurements of inclusive deep inelastic ${e^{\pm
  }p}$ scattering cross sections and QCD analysis of HERA data}.
\newblock {\em Eur. Phys. J. C}, 75(12):580, 2015.

\bibitem{H1:2017bml}
V.~Andreev et~al.
\newblock {Determination of the strong coupling constant $\alpha_s(m_Z)$ in
  next-to-next-to-leading order QCD using H1 jet cross section measurements}.
\newblock {\em Eur. Phys. J. C}, 77(11):791, 2017.
\newblock [Erratum: Eur.Phys.J.C 81, 738 (2021)].

\bibitem{Alekhin:2024bhs}
S.~Alekhin, M.~V. Garzelli, S.~O. Moch, and O.~Zenaiev.
\newblock {NNLO PDFs driven by top-quark data}.
\newblock {\em Eur. Phys. J. C}, 85(2):162, 2025.

\bibitem{Alekhin:2025qdj}
S.~Alekhin, M.~V. Garzelli, S.~Moch, and O.~Zenaiev.
\newblock {Determination of the strong coupling from high-energy data}.
\newblock {\em Eur. Phys. J. C}, 86(2):161, 2026.

\bibitem{Ablat:2025gbp}
Alim Ablat, Sayipjamal Dulat, Marco Guzzi, Joey Huston, Kirtimaan Mohan, Pavel
  Nadolsky, Dan Stump, and C.~P. Yuan.
\newblock {Strong Coupling Constant Determination from the new CTEQ-TEA Global
  QCD Analysis}.
\newblock arXiv:2512.23792.

\bibitem{Alekhin:2018pai}
S.~Alekhin, J.~Bl{\"u}mlein, and S.~Moch.
\newblock {NLO PDFs from the ABMP16 fit}.
\newblock {\em Eur. Phys. J. C}, 78(6):477, 2018.

\bibitem{Alekhin:2017kpj}
S.~Alekhin, J.~Bl{\"u}mlein, S.~Moch, and R.~Placakyte.
\newblock {Parton distribution functions, $\alpha_s$, and heavy-quark masses
  for LHC Run II}.
\newblock {\em Phys. Rev. D}, 96(1):014011, 2017.

\bibitem{Cridge:2021qfd}
T.~Cridge, L.~A. Harland-Lang, A.~D. Martin, and R.~S. Thorne.
\newblock {An investigation of the $\alpha _S$ and heavy quark mass dependence
  in the MSHT20 global PDF analysis}.
\newblock {\em Eur. Phys. J. C}, 81(8):744, 2021.

\bibitem{Ball:2025xgq}
Richard~D. Ball, Andrea Barontini, Juan Cruz-Martinez, Stefano Forte, Felix
  Hekhorn, Emanuele~R. Nocera, Juan Rojo, and Roy Stegeman.
\newblock {A determination of $\alpha _s(m_Z)$ at
  ${{\textrm{aN}}}^3{{\textrm{LO}}}_{{\textrm{QCD}}}\otimes
  {{\textrm{NLO}}}_{{\textrm{QED}}}$ accuracy from a global PDF analysis}.
\newblock {\em Eur. Phys. J. C}, 85(9):1001, 2025.

\bibitem{Cridge:2024exf}
T.~Cridge, L.~A. Harland-Lang, and R.~S. Thorne.
\newblock {A first determination of the strong coupling $\alpha _S$ at
  approximate $\textrm{N}^3$LO order in a global PDF fit}.
\newblock {\em Eur. Phys. J. C}, 84(10):1009, 2024.

\bibitem{FlavourLatticeAveragingGroupFLAG:2024oxs}
Y.~Aoki et~al.
\newblock {FLAG review 2024}.
\newblock {\em Phys. Rev. D}, 113(1):014508, 2026.

\bibitem{Mo:1968cg}
Luke~W. Mo and Yung-Su Tsai.
\newblock {Radiative Corrections to Elastic and Inelastic e p and mu p
  Scattering}.
\newblock {\em Rev. Mod. Phys.}, 41:205--235, 1969.

\bibitem{Bardin:1988by}
D.~Yu. Bardin, C.~Burdik, P.~Ch. Khristova, and T.~Riemann.
\newblock {Electroweak radiative corrections to deep inelastic scattering at
  HERA: neutral current scattering}.
\newblock {\em Z. Phys. C}, 42:679, 1989.

\bibitem{Bardin:1989vz}
D.~Yu. Bardin, K.~C. Burdik, P.~Kh. Khristova, and T.~Riemann.
\newblock {Electroweak radiative corrections to deep inelastic scattering at
  HERA: charged current scattering}.
\newblock {\em Z. Phys. C}, 44:149, 1989.

\bibitem{Kripfganz:1987yu}
J.~Kripfganz and H.~J. Mohring.
\newblock {Electromagnetic corrections to deep inelastic scattering at HERA}.
\newblock {\em Z. Phys. C}, 38:653--658, 1988.

\bibitem{Cammarota:2025jyr}
Justin Cammarota, Jian-Wei Qiu, Kazuhiro Watanabe, and Jia-Yue Zhang.
\newblock {Factorized QED and QCD contribution to deeply inelastic scattering}.
\newblock {\em Phys. Rev. D}, 112(5):056007, 2025.

\bibitem{Ball:2018iqk}
Richard~D. Ball, Stefano Carrazza, Luigi Del~Debbio, Stefano Forte, Zahari
  Kassabov, Juan Rojo, Emma Slade, and Maria Ubiali.
\newblock {Precision determination of the strong coupling constant within a
  global PDF analysis}.
\newblock {\em Eur. Phys. J. C}, 78(5):408, 2018.

\bibitem{Forte:2020pyp}
Stefano Forte and Zahari Kassabov.
\newblock {Why $\alpha _s$ cannot be determined from hadronic processes without
  simultaneously determining the parton distributions}.
\newblock {\em Eur. Phys. J. C}, 80(3):182, 2020.

\bibitem{Forte:2025pvf}
Stefano Forte, Juan Rojo, and Roy Stegeman.
\newblock {Extractions of the strong coupling from collider data without PDF
  refitting are biased}.
\newblock In {\em {2025 European Physical Society Conference on High Energy
  Physics}}, 11 2025.

\bibitem{DAgostini:1993arp}
G.~D'Agostini.
\newblock {On the use of the covariance matrix to fit correlated data}.
\newblock {\em Nucl. Instrum. Meth. A}, 346:306--311, 1994.

\bibitem{Ball:2009qv}
Richard~D. Ball, Luigi Del~Debbio, Stefano Forte, Alberto Guffanti, Jose~I.
  Latorre, Juan Rojo, and Maria Ubiali.
\newblock {Fitting Parton Distribution Data with Multiplicative Normalization
  Uncertainties}.
\newblock {\em JHEP}, 05:075, 2010.

\bibitem{Ball:2012wy}
Richard~D. Ball et~al.
\newblock {Parton Distribution Benchmarking with LHC Data}.
\newblock {\em JHEP}, 04:125, 2013.

\bibitem{Gao:2013xoa}
Jun Gao, Marco Guzzi, Joey Huston, Hung-Liang Lai, Zhao Li, Pavel Nadolsky, Jon
  Pumplin, Daniel Stump, and C.~P. Yuan.
\newblock {CT10 next-to-next-to-leading order global analysis of QCD}.
\newblock {\em Phys. Rev. D}, 89(3):033009, 2014.

\bibitem{Ball:2011us}
Richard~D. Ball, Valerio Bertone, Luigi Del~Debbio, Stefano Forte, Alberto
  Guffanti, Jose~I. Latorre, Simone Lionetti, Juan Rojo, and Maria Ubiali.
\newblock {Precision NNLO determination of $\alpha_s(M_Z)$ using an unbiased
  global parton set}.
\newblock {\em Phys. Lett. B}, 707:66--71, 2012.

\bibitem{Feynman:356451}
Richard~P. Feynman.
\newblock {\em {P}hoton-hadron interactions}.
\newblock Frontiers in physics. Benjamin, Reading, 1972.
\newblock Includes unchanged reprints with later publication date.

\bibitem{Metz:2016swz}
Andreas Metz and Anselm Vossen.
\newblock {Parton Fragmentation Functions}.
\newblock {\em Prog. Part. Nucl. Phys.}, 91:136--202, 2016.

\bibitem{Aschenauer:2019kzf}
Elke~C. Aschenauer, Ignacio Borsa, Rodolfo Sassot, and Charlotte Van~Hulse.
\newblock {Semi-inclusive Deep-Inelastic Scattering, Parton Distributions and
  Fragmentation Functions at a Future Electron-Ion Collider}.
\newblock {\em Phys. Rev. D}, 99(9):094004, 2019.

\bibitem{Borsa:2020lsz}
Ignacio Borsa, Gonzalo Lucero, Rodolfo Sassot, Elke~C. Aschenauer, and Ana~S.
  Nunes.
\newblock {Revisiting helicity parton distributions at a future electron-ion
  collider}.
\newblock {\em Phys. Rev. D}, 102(9):094018, 2020.

\bibitem{Anderle:2015lqa}
Daniele~P. Anderle, Felix Ringer, and Marco Stratmann.
\newblock {Fragmentation Functions at Next-to-Next-to-Leading Order Accuracy}.
\newblock {\em Phys. Rev. D}, 92(11):114017, 2015.

\bibitem{AbdulKhalek:2022laj}
Rabah Abdul~Khalek, Valerio Bertone, Alice Khoudli, and Emanuele~R. Nocera.
\newblock {Pion and kaon fragmentation functions at next-to-next-to-leading
  order}.
\newblock {\em Phys. Lett. B}, 834:137456, 2022.

\bibitem{Gao:2025hlm}
Jun Gao, XiaoMin Shen, Hongxi Xing, Yuxiang Zhao, and Bin Zhou.
\newblock {Fragmentation Functions of Charged Hadrons at
  Next-to-Next-to-Leading Order and Constraints on the Proton Parton
  Distribution Functions}.
\newblock {\em Phys. Rev. Lett.}, 135(4):041902, 2025.

\bibitem{Czakon:2025yti}
Micha{\l} Czakon, Terry Generet, Alexander Mitov, and Rene Poncelet.
\newblock {Identified Hadron Production at Hadron Colliders in
  Next-to-Next-to-Leading-Order QCD}.
\newblock {\em Phys. Rev. Lett.}, 135(17):17, 2025.

\bibitem{Caletti:2024xaw}
Simone Caletti, Aude Gehrmann-De~Ridder, Alexander Huss, Adrian~Rodriguez
  Garcia, and Giovanni Stagnitto.
\newblock {QCD predictions for vector boson plus hadron production at the LHC}.
\newblock {\em JHEP}, 10:027, 2024.

\bibitem{Gao:2024nkz}
Jun Gao, ChongYang Liu, XiaoMin Shen, Hongxi Xing, and Yuxiang Zhao.
\newblock {Simultaneous Determination of Fragmentation Functions and Test on
  Momentum Sum Rule}.
\newblock {\em Phys. Rev. Lett.}, 132(26):261903, 2024.

\bibitem{Bonino:2026dvr}
Leonardo Bonino, Aude Gehrmann-De~Ridder, Thomas Gehrmann, Alexander Huss,
  Francesco Merlotti, and Giovanni Stagnitto.
\newblock {Precise QCD Predictions for Hadron-in-jet Production in $e^+e^-$
  Collisions}.
\newblock arXiv:2602.12344.

\bibitem{Generet:2025bqx}
Terry Generet, Rene Poncelet, and Miha Mu{\v{s}}kinja.
\newblock {Associated production of a W-boson and a charm meson at NNLO in
  QCD}.
\newblock {\em JHEP}, 02:023, 2026.

\bibitem{dEnterria:2013sgr}
David d'Enterria, Kari~J. Eskola, Ilkka Helenius, and Hannu Paukkunen.
\newblock {Confronting current NLO parton fragmentation functions with
  inclusive charged-particle spectra at hadron colliders}.
\newblock {\em Nucl. Phys. B}, 883:615--628, 2014.

\bibitem{Borsa:2021ran}
Ignacio Borsa, Daniel de~Florian, Rodolfo Sassot, and Marco Stratmann.
\newblock {Pion fragmentation functions at high energy colliders}.
\newblock {\em Phys. Rev. D}, 105(3):L031502, 2022.

\bibitem{Kripfganz:1990vm}
J.~Kripfganz, H.~J. Mohring, and H.~Spiesberger.
\newblock {Higher order leading logarithmic QED corrections to deep inelastic e
  p scattering at very high-energies}.
\newblock {\em Z. Phys. C}, 49:501--510, 1991.

\bibitem{Bohm:1986na}
M.~Bohm and H.~Spiesberger.
\newblock {Radiative corrections to neutral current deep inelastic lepton
  nucleon scattering at HERA energies}.
\newblock {\em Nucl. Phys. B}, 294:1081--1110, 1987.

\bibitem{Kwiatkowski:1990es}
A.~Kwiatkowski, H.~Spiesberger, and H.~J. Mohring.
\newblock {Heracles: An Event Generator for $e p$ Interactions at {HERA}
  Energies Including Radiative Processes: Version 1.0}.
\newblock {\em Comput. Phys. Commun.}, 69:155--172, 1992.

\bibitem{Charchula:1994kf}
K.~Charchula, G.~A. Schuler, and H.~Spiesberger.
\newblock {Combined QED and QCD radiative effects in deep inelastic lepton -
  proton scattering: The Monte Carlo generator DJANGO6}.
\newblock {\em Comput. Phys. Commun.}, 81:381--402, 1994.

\bibitem{Liu:2020rvc}
Tianbo Liu, W.~Melnitchouk, Jian-Wei Qiu, and N.~Sato.
\newblock {Factorized approach to radiative corrections for inelastic
  lepton-hadron collisions}.
\newblock {\em Phys. Rev. D}, 104(9):094033, 2021.

\bibitem{Liu:2021jfp}
Tianbo Liu, W.~Melnitchouk, Jian-Wei Qiu, and N.~Sato.
\newblock {A new approach to semi-inclusive deep-inelastic scattering with QED
  and QCD factorization}.
\newblock {\em JHEP}, 11:157, 2021.

\bibitem{ZEUS:2023zie}
I.~Abt et~al.
\newblock {Measurement of jet production in deep inelastic scattering and NNLO
  determination of the strong coupling at ZEUS}.
\newblock {\em Eur. Phys. J. C}, 83(11):1082, 2023.

\bibitem{Lorkowski:2024hwx}
Florian Lorkowski.
\newblock {Treatment of QED corrections in jet production in deep inelastic
  scattering at ZEUS}.
\newblock In {\em {58th Rencontres de Moriond on QCD and High Energy
  Interactions}}, 5 2024.

\bibitem{Schonherr:2017qcj}
Marek Sch{\"o}nherr.
\newblock {An automated subtraction of NLO EW infrared divergences}.
\newblock {\em Eur. Phys. J. C}, 78(2):119, 2018.

\bibitem{Frederix:2018nkq}
R.~Frederix, S.~Frixione, V.~Hirschi, D.~Pagani, H.~S. Shao, and M.~Zaro.
\newblock {The automation of next-to-leading order electroweak calculations}.
\newblock {\em JHEP}, 07:185, 2018.
\newblock [Erratum: JHEP 11, 085 (2021)].

\bibitem{Zhang:2014gcy}
Yu~Zhang, Wen-Gan Ma, Ren-You Zhang, Chong Chen, and Lei Guo.
\newblock {QCD NLO and EW NLO corrections to $t\bar{t}H$ production with top
  quark decays at hadron collider}.
\newblock {\em Phys. Lett. B}, 738:1--5, 2014.

\bibitem{Kallweit:2015dum}
Stefan Kallweit, Jonas~M. Lindert, Philipp Maierhofer, Stefano Pozzorini, and
  Marek Sch{\"o}nherr.
\newblock {NLO QCD+EW predictions for V + jets including off-shell vector-boson
  decays and multijet merging}.
\newblock {\em JHEP}, 04:021, 2016.

\bibitem{Biedermann:2017yoi}
Benedikt Biedermann, Stephan Br{\"a}uer, Ansgar Denner, Mathieu Pellen, Steffen
  Schumann, and Jennifer~M. Thompson.
\newblock {Automation of NLO QCD and EW corrections with Sherpa and Recola}.
\newblock {\em Eur. Phys. J. C}, 77:492, 2017.

\bibitem{Kallweit:2017khh}
S.~Kallweit, J.~M. Lindert, S.~Pozzorini, and M.~Sch{\"o}nherr.
\newblock {NLO QCD+EW predictions for $2\ell2\nu$ diboson signatures at the
  LHC}.
\newblock {\em JHEP}, 11:120, 2017.

\bibitem{Czakon:2017wor}
Michal Czakon, David Heymes, Alexander Mitov, Davide Pagani, Ioannis Tsinikos,
  and Marco Zaro.
\newblock {Top-pair production at the LHC through NNLO QCD and NLO EW}.
\newblock {\em JHEP}, 10:186, 2017.

\bibitem{Grazzini:2019jkl}
Massimiliano Grazzini, S.~Kallweit, Jonas~M. Lindert, Stefano Pozzorini, and
  Marius Wiesemann.
\newblock {NNLO QCD + NLO EW with Matrix+OpenLoops: precise predictions for
  vector-boson pair production}.
\newblock {\em JHEP}, 02:087, 2020.

\bibitem{Lindert:2022qdd}
Jonas~M. Lindert, Daniele Lombardi, Marius Wiesemann, Giulia Zanderighi, and
  Silvia Zanoli.
\newblock {W$^{±}$Z production at NNLO QCD and NLO EW matched to parton
  showers with MiNNLO$_{PS}$}.
\newblock {\em JHEP}, 11:036, 2022.

\bibitem{Flower:2026byh}
Lois Flower and Marek Sch{\"o}nherr.
\newblock {$e^+e^- \to ZH$ at NLO EW matched to a QED parton shower}.
\newblock arXiv:2603.05585.

\bibitem{Yennie:1961ad}
D.~R. Yennie, Steven~C. Frautschi, and H.~Suura.
\newblock {The infrared divergence phenomena and high-energy processes}.
\newblock {\em Annals Phys.}, 13:379--452, 1961.

\bibitem{Lin:2017snn}
Huey-Wen Lin et~al.
\newblock {Parton distributions and lattice QCD calculations: a community white
  paper}.
\newblock {\em Prog. Part. Nucl. Phys.}, 100:107--160, 2018.

\bibitem{Alexandrou:2024ozj}
C.~Alexandrou, S.~Bacchio, J.~Finkenrath, C.~Iona, G.~Koutsou, Y.~Li, and
  G.~Spanoudes.
\newblock {Nucleon charges and {\ensuremath{\sigma}}-terms in lattice QCD}.
\newblock {\em Phys. Rev. D}, 111(5):054505, 2025.

\bibitem{Park:2025rxi}
Sungwoo Park, Rajan Gupta, Tanmoy Bhattacharya, Fangcheng He, Santanu Mondal,
  Huey-Wen Lin, and Boram Yoon.
\newblock {Flavor diagonal nucleon charges using clover fermions on MILC HISQ
  ensembles}.
\newblock {\em Phys. Rev. D}, 112(5):054508, 2025.

\bibitem{Djukanovic:2023beb}
Dalibor Djukanovic, Georg von Hippel, Harvey~B. Meyer, Konstantin Ottnad,
  Miguel Salg, and Hartmut Wittig.
\newblock {Electromagnetic form factors of the nucleon from Nf=2+1 lattice
  QCD}.
\newblock {\em Phys. Rev. D}, 109(9):094510, 2024.

\bibitem{Djukanovic:2023jag}
Dalibor Djukanovic, Georg von Hippel, Harvey~B. Meyer, Konstantin Ottnad,
  Miguel Salg, and Hartmut Wittig.
\newblock {Precision Calculation of the Electromagnetic Radii of the Proton and
  Neutron from Lattice QCD}.
\newblock {\em Phys. Rev. Lett.}, 132(21):211901, 2024.

\bibitem{Alexandrou:2025vto}
Constantia Alexandrou, Simone Bacchio, Giannis Koutsou, Bhavna Prasad, and
  Gregoris Spanoudes.
\newblock {Proton and neutron electromagnetic form factors from lattice QCD in
  the continuum limit}.
\newblock arXiv:2507.20910.

\bibitem{Jang:2023zts}
Yong-Chull Jang, Rajan Gupta, Tanmoy Bhattacharya, Boram Yoon, and Huey-Wen
  Lin.
\newblock {Nucleon isovector axial form factors}.
\newblock {\em Phys. Rev. D}, 109(1):014503, 2024.

\bibitem{Alexandrou:2023qbg}
Constantia Alexandrou, Simone Bacchio, Martha Constantinou, Jacob Finkenrath,
  Roberto Frezzotti, Bartosz Kostrzewa, Giannis Koutsou, Gregoris Spanoudes,
  and Carsten Urbach.
\newblock {Nucleon axial and pseudoscalar form factors using twisted-mass
  fermion ensembles at the physical point}.
\newblock {\em Phys. Rev. D}, 109(3):034503, 2024.

\bibitem{Djukanovic:2022wru}
Dalibor Djukanovic, Georg von Hippel, Jonna Koponen, Harvey~B. Meyer,
  Konstantin Ottnad, Tobias Schulz, and Hartmut Wittig.
\newblock {Isovector axial form factor of the nucleon from lattice QCD}.
\newblock {\em Phys. Rev. D}, 106(7):074503, 2022.

\bibitem{Alexandrou:2020sml}
C.~Alexandrou, S.~Bacchio, M.~Constantinou, J.~Finkenrath, K.~Hadjiyiannakou,
  K.~Jansen, G.~Koutsou, H.~Panagopoulos, and G.~Spanoudes.
\newblock {Complete flavor decomposition of the spin and momentum fraction of
  the proton using lattice QCD simulations at physical pion mass}.
\newblock {\em Phys. Rev. D}, 101(9):094513, 2020.

\bibitem{Hackett:2023rif}
Daniel~C. Hackett, Dimitra~A. Pefkou, and Phiala~E. Shanahan.
\newblock {Gravitational Form Factors of the Proton from Lattice QCD}.
\newblock {\em Phys. Rev. Lett.}, 132(25):251904, 2024.

\bibitem{Fan:2022qve}
Zhouyou Fan, Huey-Wen Lin, and Matthew Zeilbeck.
\newblock {Nonperturbatively renormalized nucleon gluon momentum fraction in
  the continuum limit of Nf=2+1+1 lattice QCD}.
\newblock {\em Phys. Rev. D}, 107(3):034505, 2023.

\bibitem{Lin:2020hdm}
Huey-Wen Lin et~al.
\newblock {Parton distributions and lattice-QCD calculations: Toward 3D
  structure}.
\newblock {\em Prog. Part. Nucl. Phys.}, 121:103908, 2021.

\bibitem{Alexandrou2026}
C.~Alexandrou.
\newblock {Hadron Structure from lattice QCD in the context of the Electron-Ion
  Collider}.
\newblock {\em PoS}, LATTICE2025:001, 2026.

\bibitem{Good:2023ecp}
William Good, Kinza Hasan, Allison Chevis, and Huey-Wen Lin.
\newblock {Gluon moment and parton distribution function of the pion from
  Nf=2+1+1 lattice QCD}.
\newblock {\em Phys. Rev. D}, 109(11):114509, 2024.

\bibitem{ExtendedTwistedMass:2024kjf}
Constantia Alexandrou et~al.
\newblock {Quark and Gluon Momentum Fractions in the Pion and in the Kaon}.
\newblock {\em Phys. Rev. Lett.}, 134(13):131902, 2025.

\bibitem{NieMiera:2025inn}
Alex NieMiera, William Good, and Huey-Wen Lin.
\newblock {Kaon gluon parton distribution and momentum fraction from 2+1+1
  lattice QCD with high statistics}.
\newblock {\em Phys. Rev. D}, 112(7):074504, 2025.

\bibitem{Alexandrou:2021mmi}
Constantia Alexandrou, Simone Bacchio, Ian Clo{\"e}t, Martha Constantinou,
  Kyriakos Hadjiyiannakou, Giannis Koutsou, and Colin Lauer.
\newblock {Pion and kaon {\ensuremath{\langle}}x3{\ensuremath{\rangle}} from
  lattice QCD and PDF reconstruction from Mellin moments}.
\newblock {\em Phys. Rev. D}, 104(5):054504, 2021.

\bibitem{Alexandrou:2020gxs}
Constantia Alexandrou, Simone Bacchio, Ian Cloet, Martha Constantinou, Kyriakos
  Hadjiyiannakou, Giannis Koutsou, and Colin Lauer.
\newblock {Mellin moments $\langle x \rangle$ and $\langle x^2 \rangle$ for the
  pion and kaon from lattice QCD}.
\newblock {\em Phys. Rev. D}, 103(1):014508, 2021.

\bibitem{Shindler:2023xpd}
Andrea Shindler.
\newblock {Moments of parton distribution functions of any order from lattice
  QCD}.
\newblock {\em Phys. Rev. D}, 110(5):L051503, 2024.

\bibitem{Detmold:2021uru}
William Detmold, Anthony~V. Grebe, Issaku Kanamori, C.~J.~David Lin, Robert~J.
  Perry, and Yong Zhao.
\newblock {Parton physics from a heavy-quark operator product expansion:
  Formalism and Wilson coefficients}.
\newblock {\em Phys. Rev. D}, 104(7):074511, 2021.

\bibitem{Francis:2025rya}
Anthony Francis et~al.
\newblock {Gradient flow for parton distribution functions: first application
  to the pion}.
\newblock arXiv:2509.02472.

\bibitem{Francis:2025pgf}
Anthony Francis, Patrick Fritzsch, Rohith Karur, Jangho Kim, Giovanni Pederiva,
  Dimitra~A. Pefkou, Antonio Rago, Andrea Shindler, Andr{\'e} Walker-Loud, and
  Savvas Zafeiropoulos.
\newblock {Moments of parton distributions functions of the pion from lattice
  QCD using gradient flow}.
\newblock arXiv:2510.26738.

\bibitem{Detmold:2025lyb}
William Detmold, Anthony~V. Grebe, Issaku Kanamori, C.~J.~David Lin, Robert~J.
  Perry, and Yong Zhao.
\newblock {Parton physics from a heavy-quark operator product expansion:
  Lattice QCD calculation of the fourth moment of the pion distribution
  amplitude}.
\newblock {\em Phys. Rev. D}, 113(1):014510, 2026.

\bibitem{Lin:2020fsj}
Huey-Wen Lin, Jiunn-Wei Chen, and Rui Zhang.
\newblock {Lattice Nucleon Isovector Unpolarized Parton Distribution in the
  Physical-Continuum Limit}.
\newblock arXiv:2011.14971.

\bibitem{Chen:2018xof}
Jiunn-Wei Chen, Luchang Jin, Huey-Wen Lin, Yu-Sheng Liu, Yi-Bo Yang, Jian-Hui
  Zhang, and Yong Zhao.
\newblock {Lattice Calculation of Parton Distribution Function from LaMET at
  Physical Pion Mass with Large Nucleon Momentum}.
\newblock arXiv:1803.04393.

\bibitem{Alexandrou:2018pbm}
Constantia Alexandrou, Krzysztof Cichy, Martha Constantinou, Karl Jansen,
  Aurora Scapellato, and Fernanda Steffens.
\newblock {Light-Cone Parton Distribution Functions from Lattice QCD}.
\newblock {\em Phys. Rev. Lett.}, 121(11):112001, 2018.

\bibitem{Bhat:2020ktg}
Manjunath Bhat, Krzysztof Cichy, Martha Constantinou, and Aurora Scapellato.
\newblock {Flavor nonsinglet parton distribution functions from lattice QCD at
  physical quark masses via the pseudodistribution approach}.
\newblock {\em Phys. Rev. D}, 103(3):034510, 2021.

\bibitem{Joo:2020spy}
B{\'a}lint Jo{\'o}, Joseph Karpie, Kostas Orginos, Anatoly~V. Radyushkin,
  David~G. Richards, and Savvas Zafeiropoulos.
\newblock {Parton Distribution Functions from Ioffe Time Pseudodistributions
  from Lattice Calculations: Approaching the Physical Point}.
\newblock {\em Phys. Rev. Lett.}, 125(23):232003, 2020.

\bibitem{Chu:2025kew}
Min-Huan Chu, Manuel Cola{\c{c}}o, Shohini Bhattacharya, Krzysztof Cichy,
  Martha Constantinou, Andreas Metz, and Fernanda Steffens.
\newblock {Generalized parton distributions from lattice QCD with asymmetric
  momentum transfer: Unpolarized quarks at nonzero skewness}.
\newblock {\em Phys. Rev. D}, 112(9):094510, 2025.

\bibitem{Bhattacharya:2025yba}
Shohini Bhattacharya, Krzysztof Cichy, Martha Constantinou, Andreas Metz,
  Joshua Miller, Peter Petreczky, and Fernanda Steffens.
\newblock {Generalized parton distributions from lattice QCD with asymmetric
  momentum transfer: Tensor case}.
\newblock {\em Phys. Rev. D}, 112(11):114504, 2025.

\bibitem{Bhattacharya:2023jsc}
Shohini Bhattacharya et~al.
\newblock {Generalized parton distributions from lattice QCD with asymmetric
  momentum transfer: Axial-vector case}.
\newblock {\em Phys. Rev. D}, 109(3):034508, 2024.

\bibitem{Bhattacharya:2022aob}
Shohini Bhattacharya, Krzysztof Cichy, Martha Constantinou, Jack Dodson, Xiang
  Gao, Andreas Metz, Swagato Mukherjee, Aurora Scapellato, Fernanda Steffens,
  and Yong Zhao.
\newblock {Generalized parton distributions from lattice QCD with asymmetric
  momentum transfer: Unpolarized quarks}.
\newblock {\em Phys. Rev. D}, 106(11):114512, 2022.

\bibitem{Gao:2021hxl}
Xiang Gao, Kyle Lee, Swagato Mukherjee, Charles Shugert, and Yong Zhao.
\newblock {Origin and resummation of threshold logarithms in the lattice QCD
  calculations of PDFs}.
\newblock {\em Phys. Rev. D}, 103(9):094504, 2021.

\bibitem{Su:2022fiu}
Yushan Su, Jack Holligan, Xiangdong Ji, Fei Yao, Jian-Hui Zhang, and Rui Zhang.
\newblock {Resumming quark's longitudinal momentum logarithms in LaMET
  expansion of lattice PDFs}.
\newblock {\em Nucl. Phys. B}, 991:116201, 2023.

\bibitem{Holligan:2023rex}
Jack Holligan, Xiangdong Ji, Huey-Wen Lin, Yushan Su, and Rui Zhang.
\newblock {Precision control in lattice calculation of x-dependent pion
  distribution amplitude}.
\newblock {\em Nucl. Phys. B}, 993:116282, 2023.

\bibitem{Zhang:2023bxs}
Rui Zhang, Jack Holligan, Xiangdong Ji, and Yushan Su.
\newblock {Leading power accuracy in lattice calculations of parton
  distributions}.
\newblock {\em Phys. Lett. B}, 844:138081, 2023.

\bibitem{Ji:2023pba}
Xiangdong Ji, Yizhuang Liu, and Yushan Su.
\newblock {Threshold resummation for computing large-x parton distribution
  through large-momentum effective theory}.
\newblock {\em JHEP}, 08:037, 2023.

\bibitem{Ji:2024hit}
Xiangdong Ji, Yizhuang Liu, Yushan Su, and Rui Zhang.
\newblock {Effects of threshold resummation for large-x PDF in large momentum
  effective theory}.
\newblock {\em JHEP}, 03:045, 2025.

\bibitem{Holligan:2025baj}
Jack Holligan, Huey-Wen Lin, Rui Zhang, and Yong Zhao.
\newblock {Resummation for lattice QCD calculation of generalized parton
  distributions at nonzero skewness}.
\newblock {\em JHEP}, 207:241, 2025.

\bibitem{Lin:2025hka}
Huey-Wen Lin.
\newblock {Mapping parton distributions of hadrons with lattice QCD}.
\newblock {\em Prog. Part. Nucl. Phys.}, 144:104177, 2025.

\bibitem{Ji:2020ect}
Xiangdong Ji, Yu-Sheng Liu, Yizhuang Liu, Jian-Hui Zhang, and Yong Zhao.
\newblock {Large-momentum effective theory}.
\newblock {\em Rev. Mod. Phys.}, 93(3):035005, 2021.

\bibitem{Burkert:2022hjz}
V.~D. Burkert et~al.
\newblock {Precision studies of QCD in the low energy domain of the EIC}.
\newblock {\em Prog. Part. Nucl. Phys.}, 131:104032, 2023.

\bibitem{Ji:2013dva}
Xiangdong Ji.
\newblock {Parton Physics on a Euclidean Lattice}.
\newblock {\em Phys. Rev. Lett.}, 110:262002, 2013.

\bibitem{Ji:2014gla}
Xiangdong Ji.
\newblock {Parton Physics from Large-Momentum Effective Field Theory}.
\newblock {\em Sci. China Phys. Mech. Astron.}, 57:1407--1412, 2014.

\bibitem{Radyushkin:2017cyf}
A.~V. Radyushkin.
\newblock {Quasi-parton distribution functions, momentum distributions, and
  pseudo-parton distribution functions}.
\newblock {\em Phys. Rev. D}, 96(3):034025, 2017.

\bibitem{Fan:2020cpa}
Zhouyou Fan, Rui Zhang, and Huey-Wen Lin.
\newblock {Nucleon gluon distribution function from 2 + 1 + 1-flavor lattice
  QCD}.
\newblock {\em Int. J. Mod. Phys. A}, 36(13):2150080, 2021.

\bibitem{Fan:2021bcr}
Zhouyou Fan and Huey-Wen Lin.
\newblock {Gluon parton distribution of the pion from lattice QCD}.
\newblock {\em Phys. Lett. B}, 823:136778, 2021.

\bibitem{Fan:2022kcb}
Zhouyou Fan, William Good, and Huey-Wen Lin.
\newblock {Gluon parton distribution of the nucleon from (2+1+1)-flavor lattice
  QCD in the physical-continuum limit}.
\newblock {\em Phys. Rev. D}, 108(1):014508, 2023.

\bibitem{Salas-Chavira:2021wui}
Alejandro Salas-Chavira, Zhouyou Fan, and Huey-Wen Lin.
\newblock {First glimpse into the kaon gluon parton distribution using lattice
  QCD}.
\newblock {\em Phys. Rev. D}, 106(9):094510, 2022.

\bibitem{HadStruc:2021wmh}
Tanjib Khan et~al.
\newblock {Unpolarized gluon distribution in the nucleon from lattice quantum
  chromodynamics}.
\newblock {\em Phys. Rev. D}, 104(9):094516, 2021.

\bibitem{Delmar:2023agv}
Joseph Delmar, Constantia Alexandrou, Krzysztof Cichy, Martha Constantinou, and
  Kyriakos Hadjiyiannakou.
\newblock {Gluon PDF of the proton using twisted mass fermions}.
\newblock {\em Phys. Rev. D}, 108(9):094515, 2023.

\bibitem{Good:2024iur}
William Good, Kinza Hasan, and Huey-Wen Lin.
\newblock {Toward the first gluon parton distribution from the LaMET}.
\newblock {\em J. Phys. G}, 52(3):035105, 2025.

\bibitem{NieMiera:2025mwj}
Alex NieMiera, William Good, Huey-Wen Lin, and Fei Yao.
\newblock {First Self-Renormalized Gluon PDF of Nucleon from Large-Momentum
  Effective Theory in the Continuum Limit}.
\newblock arXiv:2510.17758.

\bibitem{Chen:2025xww}
Chen Chen, Hongxin Dong, Liuming Liu, Peng Sun, Xiaonu Xiong, Yi-Bo Yang, Fei
  Yao, Jian-Hui Zhang, Chunhua Zeng, and Shiyi Zhong.
\newblock {Unpolarized gluon PDF of the nucleon from lattice QCD in the
  continuum limit}.
\newblock arXiv:2510.26425.

\bibitem{NieMiera:2025vcx}
Alex NieMiera, William Good, Huey-Wen Lin, and Fei Yao.
\newblock {Systematic Study of the Self-Renormalized Nucleon Gluon PDF in
  Large-Momentum Effective Theory}.
\newblock arXiv:2511.14708.

\bibitem{Good:2025nny}
William Good, Patrick~C. Barry, Huey-Wen Lin, W.~Melnitchouk, Alex NieMiera,
  and Nobuo Sato.
\newblock {Pionic gluons from global QCD analysis of experimental and lattice
  data}.
\newblock arXiv:2507.22730.

\bibitem{Balitsky:2021cwr}
Ian Balitsky, Wayne Morris, and Anatoly Radyushkin.
\newblock {Polarized gluon pseudodistributions at short distances}.
\newblock {\em JHEP}, 02:193, 2022.

\bibitem{HadStruc:2022yaw}
Colin Egerer et~al.
\newblock {Toward the determination of the gluon helicity distribution in the
  nucleon from lattice quantum chromodynamics}.
\newblock {\em Phys. Rev. D}, 106(9):094511, 2022.

\bibitem{Khan:2022vot}
Tanjib Khan, Tianbo Liu, and Raza~Sabbir Sufian.
\newblock {Gluon helicity in the nucleon from lattice QCD and machine
  learning}.
\newblock {\em Phys. Rev. D}, 108(7):074502, 2023.

\bibitem{Chowdhury:2024ymm}
Talal~Ahmed Chowdhury, Taku Izubuchi, Methun Kamruzzaman, Nikhil Karthik,
  Tanjib Khan, Tianbo Liu, Arpon Paul, Jakob Schoenleber, and Raza~Sabbir
  Sufian.
\newblock {Polarized and unpolarized gluon PDFs: Generative machine learning
  applications for lattice QCD matrix elements at short distance and large
  momentum}.
\newblock {\em Phys. Rev. D}, 111(7):074509, 2025.

\bibitem{Yang:2016plb}
Yi-Bo Yang, Raza~Sabbir Sufian, Andrei Alexandru, Terrence Draper, Michael~J.
  Glatzmaier, Keh-Fei Liu, and Yong Zhao.
\newblock {Glue Spin and Helicity in the Proton from Lattice QCD}.
\newblock {\em Phys. Rev. Lett.}, 118(10):102001, 2017.

\bibitem{Ji:2013fga}
Xiangdong Ji, Jian-Hui Zhang, and Yong Zhao.
\newblock {Physics of the Gluon-Helicity Contribution to Proton Spin}.
\newblock {\em Phys. Rev. Lett.}, 111:112002, 2013.

\bibitem{deFlorian:2024utd}
Daniel de~Florian, Stefano Forte, and Werner Vogelsang.
\newblock {Higgs production at RHIC and the positivity of the gluon helicity
  distribution}.
\newblock {\em Phys. Rev. D}, 109(7):074007, 2024.

\bibitem{Schoenleber:2024auy}
Jakob Schoenleber, Raza~Sabbir Sufian, Taku Izubuchi, and Yi-Bo Yang.
\newblock {Gluon unpolarized, polarized, and transversity GPDs from lattice
  QCD: Lorentz-covariant parametrization}.
\newblock {\em Phys. Rev. D}, 111(9):094510, 2025.

\bibitem{Ethier:2017zbq}
J.~J. Ethier, N.~Sato, and W.~Melnitchouk.
\newblock {First simultaneous extraction of spin-dependent parton distributions
  and fragmentation functions from a global QCD analysis}.
\newblock {\em Phys. Rev. Lett.}, 119(13):132001, 2017.

\bibitem{Ji:2020ena}
Xiangdong Ji, Feng Yuan, and Yong Zhao.
\newblock {What we know and what we don{\textquoteright}t know about the proton
  spin after 30 years}.
\newblock {\em Nature Rev. Phys.}, 3(1):27--38, 2021.

\bibitem{Cruz-Martinez:2025ahf}
Juan Cruz-Martinez, Toon Hasenack, Felix Hekhorn, Giacomo Magni, Emanuele~R.
  Nocera, Tanjona~R. Rabemananjara, Juan Rojo, Tanishq Sharma, and Gijs van
  Seeventer.
\newblock {NNPDFpol2.0: a global determination of polarised PDFs and their
  uncertainties at next-to-next-to-leading order}.
\newblock {\em JHEP}, 07:168, 2025.

\bibitem{Cocuzza:2025qvf}
C.~Cocuzza, N.~T. Hunt-Smith, W.~Melnitchouk, N.~Sato, and A.~W. Thomas.
\newblock {Global QCD analysis of spin PDFs in the proton with high-x and
  lattice constraints}.
\newblock {\em Phys. Rev. D}, 112(11):114017, 2025.

\bibitem{Adamiak:2021ppq}
Daniel Adamiak, Yuri~V. Kovchegov, W.~Melnitchouk, Daniel Pitonyak, Nobuo Sato,
  and Matthew~D. Sievert.
\newblock {First analysis of world polarized DIS data with small-x helicity
  evolution}.
\newblock {\em Phys. Rev. D}, 104(3):L031501, 2021.

\bibitem{Adamiak:2023yhz}
Daniel Adamiak, Nicholas Baldonado, Yuri~V. Kovchegov, W.~Melnitchouk, Daniel
  Pitonyak, Nobuo Sato, Matthew~D. Sievert, Andrey Tarasov, and Yossathorn
  Tawabutr.
\newblock {Global analysis of polarized DIS and SIDIS data with improved
  small-x helicity evolution}.
\newblock {\em Phys. Rev. D}, 108(11):114007, 2023.

\bibitem{Anderle:2021wcy}
Daniele~P. Anderle et~al.
\newblock {Electron-ion collider in China}.
\newblock {\em Front. Phys. (Beijing)}, 16(6):64701, 2021.

\bibitem{Taghavi-Shahri:2016idw}
F.~Taghavi-Shahri, Hamzeh Khanpour, S.~Atashbar~Tehrani, and Z.~Alizadeh~Yazdi.
\newblock {Next-to-next-to-leading order QCD analysis of spin-dependent parton
  distribution functions and their uncertainties: Jacobi polynomials approach}.
\newblock {\em Phys. Rev. D}, 93(11):114024, 2016.

\bibitem{Sato:2016tuz}
Nobuo Sato, W.~Melnitchouk, S.~E. Kuhn, J.~J. Ethier, and A.~Accardi.
\newblock {Iterative Monte Carlo analysis of spin-dependent parton
  distributions}.
\newblock {\em Phys. Rev. D}, 93(7):074005, 2016.

\bibitem{Cocuzza:2022jye}
C.~Cocuzza, W.~Melnitchouk, A.~Metz, and N.~Sato.
\newblock {Polarized antimatter in the proton from a global QCD analysis}.
\newblock {\em Phys. Rev. D}, 106(3):L031502, 2022.

\bibitem{Zhou:2022wzm}
Y.~Zhou, N.~Sato, and W.~Melnitchouk.
\newblock {How well do we know the gluon polarization in the proton?}
\newblock {\em Phys. Rev. D}, 105(7):074022, 2022.

\bibitem{Whitehill:2022mpq}
R.~M. Whitehill, Yiyu Zhou, N.~Sato, and W.~Melnitchouk.
\newblock {Accessing gluon polarization with high-PT hadrons in SIDIS}.
\newblock {\em Phys. Rev. D}, 107(3):034033, 2023.

\bibitem{Hunt-Smith:2024khs}
N.~T. Hunt-Smith, C.~Cocuzza, W.~Melnitchouk, N.~Sato, A.~W. Thomas, and M.~J.
  White.
\newblock {New Data-Driven Constraints on the Sign of Gluon Polarization in the
  Proton}.
\newblock {\em Phys. Rev. Lett.}, 133(16):161901, 2024.

\bibitem{Bringewatt:2020ixn}
J.~Bringewatt, N.~Sato, W.~Melnitchouk, Jian-Wei Qiu, F.~Steffens, and
  M.~Constantinou.
\newblock {Confronting lattice parton distributions with global QCD analysis}.
\newblock {\em Phys. Rev. D}, 103(1):016003, 2021.

\bibitem{Karpie:2023nyg}
J.~Karpie, R.~M. Whitehill, W.~Melnitchouk, C.~Monahan, K.~Orginos, J.~W. Qiu,
  D.~G. Richards, N.~Sato, and S.~Zafeiropoulos.
\newblock {Gluon helicity from global analysis of experimental data and lattice
  QCD Ioffe time distributions}.
\newblock {\em Phys. Rev. D}, 109(3):036031, 2024.

\bibitem{Flores-Mendieta:1998tfv}
Ruben Flores-Mendieta, Elizabeth~Ellen Jenkins, and Aneesh~V. Manohar.
\newblock {SU(3) symmetry breaking in hyperon semileptonic decays}.
\newblock {\em Phys. Rev. D}, 58:094028, 1998.

\bibitem{Altarelli:1998gn}
Guido Altarelli, Stefano Forte, and Giovanni Ridolfi.
\newblock {On positivity of parton distributions}.
\newblock {\em Nucl. Phys. B}, 534:277--296, 1998.

\bibitem{Rojo:2015acz}
Juan Rojo et~al.
\newblock {The PDF4LHC report on PDFs and LHC data: Results from Run I and
  preparation for Run II}.
\newblock {\em J. Phys. G}, 42:103103, 2015.

\bibitem{PDF4LHCWorkingGroup:2022cjn}
Richard~D. Ball et~al.
\newblock {The PDF4LHC21 combination of global PDF fits for the LHC Run III}.
\newblock {\em J. Phys. G}, 49(8):080501, 2022.

\bibitem{Bertone:2013vaa}
Valerio Bertone, Stefano Carrazza, and Juan Rojo.
\newblock {APFEL: A PDF Evolution Library with QED corrections}.
\newblock {\em Comput. Phys. Commun.}, 185:1647--1668, 2014.

\bibitem{Bertone:2017gds}
Valerio Bertone.
\newblock {APFEL++: A new PDF evolution library in C++}.
\newblock {\em PoS}, DIS2017:201, 2018.

\bibitem{Candido:2022tld}
Alessandro Candido, Felix Hekhorn, and Giacomo Magni.
\newblock {EKO: evolution kernel operators}.
\newblock {\em Eur. Phys. J. C}, 82(10):976, 2022.

\bibitem{Candido:2024rkr}
Alessandro Candido, Felix Hekhorn, Giacomo Magni, Tanjona~R. Rabemananjara, and
  Roy Stegeman.
\newblock {Yadism: yet another deep-inelastic scattering module}.
\newblock {\em Eur. Phys. J. C}, 84(7):698, 2024.

\bibitem{Hekhorn:2024tqm}
Felix Hekhorn, Giacomo Magni, Emanuele~R. Nocera, Tanjona~R. Rabemananjara,
  Juan Rojo, Adrianne Schaus, and Roy Stegeman.
\newblock {Heavy quarks in polarised deep-inelastic scattering at the
  electron-ion collider}.
\newblock {\em Eur. Phys. J. C}, 84(2):189, 2024.

\bibitem{Carrazza:2020gss}
S.~Carrazza, E.~R. Nocera, C.~Schwan, and M.~Zaro.
\newblock {PineAPPL: combining EW and QCD corrections for fast evaluation of
  LHC processes}.
\newblock {\em JHEP}, 12:108, 2020.

\bibitem{Buckley:2014ana}
Andy Buckley, James Ferrando, Stephen Lloyd, Karl Nordstr{\"o}m, Ben Page,
  Martin R{\"u}fenacht, Marek Sch{\"o}nherr, and Graeme Watt.
\newblock {LHAPDF6: parton density access in the LHC precision era}.
\newblock {\em Eur. Phys. J. C}, 75:132, 2015.

\bibitem{valerio_bertone_2024_10933177}
Valerio Bertone and Amedeo Chiefa, MapCollaboration/Denali: Sheldon Chalet,
  April 2024. See \url{https://github.com/MapCollaboration/Denali}.

\bibitem{NNPDF:2021uiq}
Richard~D. Ball et~al.
\newblock {An open-source machine learning framework for global analyses of
  parton distributions}.
\newblock {\em Eur. Phys. J. C}, 81(10):958, 2021.

\bibitem{Zhou:2021llj}
Y.~Zhou, C.~Cocuzza, F.~Delcarro, W.~Melnitchouk, A.~Metz, and N.~Sato.
\newblock {Revisiting quark and gluon polarization in the proton at the EIC}.
\newblock {\em Phys. Rev. D}, 104(3):034028, 2021.

\bibitem{Amoroso:2022eow}
S.~Amoroso et~al.
\newblock {Snowmass 2021 Whitepaper: Proton Structure at the Precision
  Frontier}.
\newblock {\em Acta Phys. Polon. B}, 53(12):12--A1, 2022.

\bibitem{Hou:2019efy}
Tie-Jiun Hou et~al.
\newblock {New CTEQ global analysis of quantum chromodynamics with
  high-precision data from the LHC}.
\newblock {\em Phys. Rev. D}, 103(1):014013, 2021.

\bibitem{Bailey:2020ooq}
S.~Bailey, T.~Cridge, L.~A. Harland-Lang, A.~D. Martin, and R.~S. Thorne.
\newblock {Parton distributions from LHC, HERA, Tevatron and fixed target data:
  MSHT20 PDFs}.
\newblock {\em Eur. Phys. J. C}, 81(4):341, 2021.

\bibitem{NNPDF:2021njg}
Richard~D. Ball et~al.
\newblock {The path to proton structure at 1{\%} accuracy}.
\newblock {\em Eur. Phys. J. C}, 82(5):428, 2022.

\bibitem{Moffat:2021dji}
Eric Moffat, Wally Melnitchouk, T.~C. Rogers, and Nobuo Sato.
\newblock {Simultaneous Monte~Carlo analysis of parton densities and
  fragmentation functions}.
\newblock {\em Phys. Rev. D}, 104(1):016015, 2021.

\bibitem{ATLAS:2021vod}
Georges Aad et~al.
\newblock {Determination of the parton distribution functions of the proton
  using diverse ATLAS data from $pp$ collisions at $\sqrt{s} = 7$, 8 and
  13~TeV}.
\newblock {\em Eur. Phys. J. C}, 82(5):438, 2022.

\bibitem{McGowan:2022nag}
J.~McGowan, T.~Cridge, L.~A. Harland-Lang, and R.~S. Thorne.
\newblock {Approximate N$^{3}$LO parton distribution functions with theoretical
  uncertainties: MSHT20aN$^3$LO PDFs}.
\newblock {\em Eur. Phys. J. C}, 83(3):185, 2023.
\newblock [Erratum: Eur.Phys.J.C 83, 302 (2023)].

\bibitem{NNPDF:2024nan}
Richard~D. Ball et~al.
\newblock {The path to $\hbox {N}^3\hbox {LO}$ parton distributions}.
\newblock {\em Eur. Phys. J. C}, 84(7):659, 2024.

\bibitem{NNPDF:2024dpb}
Richard~D. Ball et~al.
\newblock {Determination of the theory uncertainties from missing higher orders
  on NNLO parton distributions with percent accuracy}.
\newblock {\em Eur. Phys. J. C}, 84(5):517, 2024.

\bibitem{Cridge:2024icl}
Thomas Cridge et~al.
\newblock {Combination of aN$^3$LO PDFs and implications for Higgs production
  cross-sections at the LHC}.
\newblock {\em J. Phys. G}, 52:6, 2025.

\bibitem{Xie:2021ajm}
Keping Xie, Timothy Hobbs, Tie-Jiun Hou, Carl Schmidt, Mengshi Yan, and
  Chien-Peng Yuan.
\newblock {The photon content of the proton in the CT18 global analysis}.
\newblock {\em SciPost Phys. Proc.}, 8:074, 2022.

\bibitem{Cridge:2021pxm}
T.~Cridge, L.~A. Harland-Lang, A.~D. Martin, and R.~S. Thorne.
\newblock {QED parton distribution functions in the MSHT20 fit}.
\newblock {\em Eur. Phys. J. C}, 82(1):90, 2022.

\bibitem{NNPDF:2024djq}
Richard~D. Ball et~al.
\newblock {Photons in the proton: implications for the LHC}.
\newblock {\em Eur. Phys. J. C}, 84(5):540, 2024.

\bibitem{Ball:2022qtp}
Richard~D. Ball, Alessandro Candido, Stefano Forte, Felix Hekhorn, Emanuele~R.
  Nocera, Juan Rojo, and Christopher Schwan.
\newblock {Parton distributions and new physics searches: the
  Drell\textendash{}Yan forward\textendash{}backward asymmetry as a case
  study}.
\newblock {\em Eur. Phys. J. C}, 82(12):1160, 2022.

\bibitem{xFitter:2022zjb}
H.~Abdolmaleki et~al.
\newblock {xFitter: An Open Source QCD Analysis Framework. A resource and
  reference document for the Snowmass study}.
\newblock arXiv:2206.12465.

\bibitem{Kotz:2025mcj}
Lucas Kotz, Aurore Courtoy, Pavel Nadolsky, Fredrick Olness, and Maximiliano
  Ponce-Chavez.
\newblock {Fant{\^o}mas: An analysis of parton distributions in a pion with
  B{\'e}zier parametrizations}.
\newblock {\em PoS}, DIS2025:032, 2025.

\bibitem{Costantini:2025wxp}
Mark~N. Costantini, Luca Mantani, James~M. Moore, and Maria Ubiali.
\newblock {A linear PDF model for Bayesian inference}.
\newblock arXiv:2507.16913.

\bibitem{Costantini:2026mxm}
Mark~N. Costantini, Luca Mantani, James~M. Moore, Valentina~Sch{\"u}tze
  S{\'a}nchez, and Maria Ubiali.
\newblock {Colibri: A new tool for fast-flying PDF fits}.
\newblock {\em Eur. Phys. J. C}, 86(1):22, 2026.

\bibitem{Cruz-Martinez:2026rct}
J.~M. Cruz-Martinez, T.~Giani, and L.~A. Harland-Lang.
\newblock {Assessing the Impact of Fitting Methodology at aN$^3$LO with FPPDF:
  an Open Source Tool for Extracting Parton Distribution Functions in the
  Hessian Approach}.
\newblock arXiv:2602.07118.

\bibitem{Wang:2018heo}
Bo-Ting Wang, T.~J. Hobbs, Sean Doyle, Jun Gao, Tie-Jiun Hou, Pavel~M.
  Nadolsky, and Fredrick~I. Olness.
\newblock {Mapping the sensitivity of hadronic experiments to nucleon
  structure}.
\newblock {\em Phys. Rev. D}, 98(9):094030, 2018.

\bibitem{Jing:2023isu}
Xiaoxian Jing et~al.
\newblock {Quantifying the interplay of experimental constraints in analyses of
  parton distributions}.
\newblock {\em Phys. Rev. D}, 108(3):034029, 2023.

\bibitem{Khalek:2021ulf}
Rabah~Abdul Khalek, Jacob~J. Ethier, Emanuele~R. Nocera, and Juan Rojo.
\newblock {Self-consistent determination of proton and nuclear PDFs at the
  Electron Ion Collider}.
\newblock {\em Phys. Rev. D}, 103(9):096005, 2021.

\bibitem{Hobbs:2019gob}
T.~J. Hobbs, Bo-Ting Wang, Pavel~M. Nadolsky, and Fredrick~I. Olness.
\newblock {Charting the coming synergy between lattice QCD and high-energy
  phenomenology}.
\newblock {\em Phys. Rev. D}, 100(9):094040, 2019.

\bibitem{PDFSenseWebsite}
\url{https://metapdf.hepforge.org/PDFSense/}.

\bibitem{Carrazza:2019sec}
Stefano Carrazza, Celine Degrande, Shayan Iranipour, Juan Rojo, and Maria
  Ubiali.
\newblock {Can New Physics hide inside the proton?}
\newblock {\em Phys. Rev. Lett.}, 123(13):132001, 2019.

\bibitem{Greljo:2021kvv}
Admir Greljo, Shayan Iranipour, Zahari Kassabov, Maeve Madigan, James Moore,
  Juan Rojo, Maria Ubiali, and Cameron Voisey.
\newblock {Parton distributions in the SMEFT from high-energy Drell-Yan tails}.
\newblock {\em JHEP}, 07:122, 2021.

\bibitem{Iranipour:2022iak}
Shayan Iranipour and Maria Ubiali.
\newblock {A new generation of simultaneous fits to LHC data using deep
  learning}.
\newblock {\em JHEP}, 05:032, 2022.

\bibitem{Hammou:2023heg}
Elie Hammou, Zahari Kassabov, Maeve Madigan, Michelangelo~L. Mangano, Luca
  Mantani, James Moore, Manuel~Morales Alvarado, and Maria Ubiali.
\newblock {Hide and seek: how PDFs can conceal new physics}.
\newblock {\em JHEP}, 11:090, 2023.

\bibitem{Kassabov:2023hbm}
Zahari Kassabov, Maeve Madigan, Luca Mantani, James Moore, Manuel
  Morales~Alvarado, Juan Rojo, and Maria Ubiali.
\newblock {The top quark legacy of the LHC Run II for PDF and SMEFT analyses}.
\newblock {\em JHEP}, 05:205, 2023.

\bibitem{Gao:2022srd}
Jun Gao, MeiSen Gao, T.~J. Hobbs, DianYu Liu, and XiaoMin Shen.
\newblock {Simultaneous CTEQ-TEA extraction of PDFs and SMEFT parameters from
  jet and $ t\overline{t} $ data}.
\newblock {\em JHEP}, 05:003, 2023.

\bibitem{Costantini:2024xae}
Mark~N. Costantini, Elie Hammou, Zahari Kassabov, Maeve Madigan, Luca Mantani,
  Manuel Morales~Alvarado, James~M. Moore, and Maria Ubiali.
\newblock {SIMUnet: an open-source tool for simultaneous global fits of EFT
  Wilson coefficients and PDFs}.
\newblock {\em Eur. Phys. J. C}, 84(8):805, 2024.

\bibitem{Hammou:2024xuj}
Elie Hammou and Maria Ubiali.
\newblock {Unravelling new physics signals at the HL-LHC with EIC and FPF
  constraints}.
\newblock {\em Phys. Rev. D}, 111(9):095028, 2025.

\bibitem{Cole:2026eex}
Ella Cole, Mark~N. Costantini, Elie Hammou, Luca Mantani, Francesco Merlotti,
  Manuel Morales-Alvarado, and Maria Ubiali.
\newblock {Tailored PDFs for New Physics searches}.
\newblock arXiv:2602.20235.

\bibitem{Farina:2016rws}
Marco Farina, Giuliano Panico, Duccio Pappadopulo, Joshua~T. Ruderman, Riccardo
  Torre, and Andrea Wulzer.
\newblock {Energy helps accuracy: electroweak precision tests at hadron
  colliders}.
\newblock {\em Phys. Lett. B}, 772:210--215, 2017.

\bibitem{Cruz-Martinez:2023sdv}
Juan~M. Cruz-Martinez, Max Fieg, Tommaso Giani, Peter Krack, Toni
  M{\"a}kel{\"a}, Tanjona~R. Rabemananjara, and Juan Rojo.
\newblock {The LHC as a Neutrino-Ion Collider}.
\newblock {\em Eur. Phys. J. C}, 84(4):369, 2024.

\bibitem{Bissolotti:2023vdw}
Chiara Bissolotti, Radja Boughezal, and Kaan Simsek.
\newblock {SMEFT probes in future precision DIS experiments}.
\newblock {\em Phys. Rev. D}, 108(7):075007, 2023.

\bibitem{Cichy:2018mum}
Krzysztof Cichy and Martha Constantinou.
\newblock {A guide to light-cone PDFs from Lattice QCD: an overview of
  approaches, techniques and results}.
\newblock {\em Adv. High Energy Phys.}, 2019:3036904, 2019.

\bibitem{Alexandrou:2020zbe}
Constantia Alexandrou, Krzysztof Cichy, Martha Constantinou, Kyriakos
  Hadjiyiannakou, Karl Jansen, Aurora Scapellato, and Fernanda Steffens.
\newblock {Unpolarized and helicity generalized parton distributions of the
  proton within lattice QCD}.
\newblock {\em Phys. Rev. Lett.}, 125(26):262001, 2020.

\bibitem{Kotz:2023pbu}
Lucas Kotz, Aurore Courtoy, Pavel Nadolsky, Fredrick Olness, and Maximiliano
  Ponce-Chavez.
\newblock {Analysis of parton distributions in a pion with B{\'e}zier
  parametrizations}.
\newblock {\em Phys. Rev. D}, 109(7):074027, 2024.

\bibitem{Gao:2021dbh}
Xiang Gao, Andrew~D. Hanlon, Swagato Mukherjee, Peter Petreczky, Philipp Scior,
  Sergey Syritsyn, and Yong Zhao.
\newblock {Lattice QCD Determination of the Bjorken-x Dependence of Parton
  Distribution Functions at Next-to-Next-to-Leading Order}.
\newblock {\em Phys. Rev. Lett.}, 128(14):142003, 2022.

\bibitem{Gao:2022iex}
Xiang Gao, Andrew~D. Hanlon, Nikhil Karthik, Swagato Mukherjee, Peter
  Petreczky, Philipp Scior, Shuzhe Shi, Sergey Syritsyn, Yong Zhao, and Kai
  Zhou.
\newblock {Continuum-extrapolated NNLO valence PDF of the pion at the physical
  point}.
\newblock {\em Phys. Rev. D}, 106(11):114510, 2022.

\bibitem{Kotz:2025lio}
Lucas Kotz, Aurore Courtoy, Pavel Nadolsky, and Maximiliano Ponce-Chavez.
\newblock {Epistemic and nuclear uncertainties for the parton distributions of
  the pion}.
\newblock {\em Phys. Rev. D}, 112(7):L071502, 2025.

\bibitem{Barry:2021osv}
P.~C. Barry, Chueng-Ryong Ji, N.~Sato, and W.~Melnitchouk.
\newblock {Global QCD Analysis of Pion Parton Distributions with Threshold
  Resummation}.
\newblock {\em Phys. Rev. Lett.}, 127(23):232001, 2021.

\bibitem{Novikov:2020snp}
Ivan Novikov et~al.
\newblock {Parton Distribution Functions of the Charged Pion Within The xFitter
  Framework}.
\newblock {\em Phys. Rev. D}, 102(1):014040, 2020.

\bibitem{JeffersonLabAngularMomentumJAM:2022aix}
P.~C. Barry et~al.
\newblock {Complementarity of experimental and lattice QCD data on pion parton
  distributions}.
\newblock {\em Phys. Rev. D}, 105(11):114051, 2022.

\bibitem{Chiefa:2025loi}
Amedeo Chiefa, Mark~N. Costantini, Juan Cruz-Martinez, Emanuele~R. Nocera,
  Tanjona~R. Rabemananjara, Juan Rojo, Tanishq Sharma, Roy Stegeman, and Maria
  Ubiali.
\newblock {Parton distributions confront LHC Run II data: a quantitative
  appraisal}.
\newblock {\em JHEP}, 07:067, 2025.

\bibitem{ATLAS:2023lsr}
Georges Aad et~al.
\newblock {A precise measurement of the Z-boson double-differential transverse
  momentum and rapidity distributions in the full phase space of the decay
  leptons with the ATLAS experiment at~$\sqrt{s}=8$~TeV}.
\newblock {\em Eur. Phys. J. C}, 84(3):315, 2024.

\bibitem{CMS:2024ony}
Aram Hayrapetyan et~al.
\newblock {Measurement of the Drell{\textendash}Yan forward-backward asymmetry
  and of the effective leptonic weak mixing angle in proton-proton collisions
  at s=13TeV}.
\newblock {\em Phys. Lett. B}, 866:139526, 2025.

\bibitem{ATLAS:2023fsi}
{Improved W boson Mass Measurement using 7 TeV Proton-Proton Collisions with
  the ATLAS Detector}.
\newblock 2023.

\bibitem{CMS:2024lrd}
Vladimir Chekhovsky et~al.
\newblock {High-precision measurement of the W boson mass with the CMS
  experiment at the LHC}.
\newblock arXiv:2412.13872.

\bibitem{DelDebbio:2021whr}
Luigi Del~Debbio, Tommaso Giani, and Michael Wilson.
\newblock {Bayesian approach to inverse problems: an application to NNPDF
  closure testing}.
\newblock {\em Eur. Phys. J. C}, 82(4):330, 2022.

\bibitem{Barontini:2025lnl}
Andrea Barontini, Mark~N. Costantini, Giovanni De~Crescenzo, Stefano Forte, and
  Maria Ubiali.
\newblock {Evaluating the faithfulness of PDF uncertainties in the presence of
  inconsistent data}.
\newblock arXiv:2503.17447.

\bibitem{Harland-Lang:2024kvt}
L.~A. Harland-Lang, T.~Cridge, and R.~S. Thorne.
\newblock {A stress test of global PDF fits: closure testing the MSHT PDFs and
  a first direct comparison to the neural net approach}.
\newblock {\em Eur. Phys. J. C}, 85(3):316, 2025.

\bibitem{Cruz-Martinez:2021rgy}
Juan Cruz-Martinez, Stefano Forte, and Emanuele~R. Nocera.
\newblock {Future tests of parton distributions}.
\newblock {\em Acta Phys. Polon. B}, 52:243, 2021.

\bibitem{Carli:2010rw}
Tancredi Carli, Dan Clements, Amanda Cooper-Sarkar, Claire Gwenlan, Gavin~P.
  Salam, Frank Siegert, Pavel Starovoitov, and Mark Sutton.
\newblock {A posteriori inclusion of parton density functions in NLO QCD
  final-state calculations at hadron colliders: The APPLGRID Project}.
\newblock {\em Eur. Phys. J. C}, 66:503--524, 2010.

\bibitem{Kluge:2006xs}
T.~Kluge, K.~Rabbertz, and M.~Wobisch.
\newblock {FastNLO: Fast pQCD calculations for PDF fits}.
\newblock In {\em {14th International Workshop on Deep Inelastic Scattering}},
  pages 483--486, 9 2006.

\bibitem{Devoto:2025cuf}
S.~Devoto, T.~Jezo, S.~Kallweit, and C.~Schwan.
\newblock {MATRIX HAWAII: PineAPPL interpolation grids with MATRIX}.
\newblock 6 2025.

\bibitem{Courtoy:2022ocu}
Aurore Courtoy, Joey Huston, Pavel Nadolsky, Keping Xie, Mengshi Yan, and C.~P.
  Yuan.
\newblock {Parton distributions need representative sampling}.
\newblock {\em Phys. Rev. D}, 107(3):034008, 2023.

\bibitem{Courtoy:2025ppd}
Aurore Courtoy and Arturo Ibsen.
\newblock {Information criteria for selecting parton distribution function
  solutions}.
\newblock {\em Eur. Phys. J. C}, 86(1):86, 2026.

\bibitem{Klasen:2023uqj}
M.~Klasen and H.~Paukkunen.
\newblock {Nuclear PDFs After the First Decade of LHC Data}.
\newblock {\em Ann. Rev. Nucl. Part. Sci.}, 74:49--87, 2024.

\bibitem{Duwentaster:2022kpv}
P.~Duwent{\"a}ster, T.~Je{\v{z}}o, M.~Klasen, K.~Kova{\v{r}}{\'\i}k, A.~Kusina,
  K.~F. Muzakka, F.~I. Olness, R.~Ruiz, I.~Schienbein, and J.~Y. Yu.
\newblock {Impact of heavy quark and quarkonium data on nuclear gluon PDFs}.
\newblock {\em Phys. Rev. D}, 105(11):114043, 2022.

\bibitem{Eskola:2021nhw}
Kari~J. Eskola, Petja Paakkinen, Hannu Paukkunen, and Carlos~A. Salgado.
\newblock {EPPS21: a global QCD analysis of nuclear PDFs}.
\newblock {\em Eur. Phys. J. C}, 82(5):413, 2022.

\bibitem{AbdulKhalek:2022fyi}
Rabah Abdul~Khalek, Rhorry Gauld, Tommaso Giani, Emanuele~R. Nocera, Tanjona~R.
  Rabemananjara, and Juan Rojo.
\newblock {nNNPDF3.0: evidence for a modified partonic structure in heavy
  nuclei}.
\newblock {\em Eur. Phys. J. C}, 82(6):507, 2022.

\bibitem{Helenius:2021tof}
Ilkka Helenius, Marina Walt, and Werner Vogelsang.
\newblock {NNLO nuclear parton distribution functions with electroweak-boson
  production data from the LHC}.
\newblock {\em Phys. Rev. D}, 105(9):094031, 2022.

\bibitem{Khanpour:2020zyu}
Hamzeh Khanpour, Maryam Soleymaninia, S.~Atashbar~Tehrani, Hubert Spiesberger,
  and Vadim Guzey.
\newblock {Nuclear parton distribution functions with uncertainties in a
  general mass variable flavor number scheme}.
\newblock {\em Phys. Rev. D}, 104(3):034010, 2021.

\bibitem{Arleo:2025oos}
F.~Arleo et~al.
\newblock {Nuclear Cold QCD: Review and Future Strategy}.
\newblock arXiv:2506.17454.

\bibitem{Klasen:2017kwb}
M.~Klasen, K.~Kovarik, and J.~Potthoff.
\newblock {Nuclear parton density functions from jet production in DIS at an
  EIC}.
\newblock {\em Phys. Rev. D}, 95(9):094013, 2017.

\bibitem{Armesto:2023hnw}
N{\'e}stor Armesto, Thomas Cridge, Francesco Giuli, Lucian Harland-Lang, Paul
  Newman, Barak Schmookler, Robert Thorne, and Katarzyna Wichmann.
\newblock {Impact of inclusive electron ion collider data on collinear parton
  distributions}.
\newblock {\em Phys. Rev. D}, 109(5):054019, 2024.

\bibitem{Ruiz:2023ozv}
R.~Ruiz et~al.
\newblock {Target mass corrections in lepton{\textendash}nucleus DIS: Theory
  and applications to nuclear PDFs}.
\newblock {\em Prog. Part. Nucl. Phys.}, 136:104096, 2024.

\bibitem{nCTEQ:2023cpo}
A.~W. Denniston et~al.
\newblock {Modification of Quark-Gluon Distributions in Nuclei by Correlated
  Nucleon Pairs}.
\newblock {\em Phys. Rev. Lett.}, 133(15):152502, 2024.

\bibitem{Winter:2017bfs}
Frank Winter, William Detmold, Arjun~S. Gambhir, Kostas Orginos, Martin~J.
  Savage, Phiala~E. Shanahan, and Michael~L. Wagman.
\newblock {First lattice QCD study of the gluonic structure of light nuclei}.
\newblock {\em Phys. Rev. D}, 96(9):094512, 2017.

\bibitem{Detmold:2020snb}
William Detmold, Marc Illa, David~J. Murphy, Patrick Oare, Kostas Orginos,
  Phiala~E. Shanahan, Michael~L. Wagman, and Frank Winter.
\newblock {Lattice QCD Constraints on the Parton Distribution Functions of
  $^3$He}.
\newblock {\em Phys. Rev. Lett.}, 126(20):202001, 2021.

\bibitem{Sullivan:1971kd}
J.~D. Sullivan.
\newblock {One pion exchange and deep inelastic electron - nucleon scattering}.
\newblock {\em Phys. Rev. D}, 5:1732--1737, 1972.

\bibitem{Aicher:2010cb}
Matthias Aicher, Andreas Schafer, and Werner Vogelsang.
\newblock {Soft-gluon resummation and the valence parton distribution function
  of the pion}.
\newblock {\em Phys. Rev. Lett.}, 105:252003, 2010.

\bibitem{Barry:2018ort}
P.~C. Barry, N.~Sato, W.~Melnitchouk, and Chueng-Ryong Ji.
\newblock {First Monte Carlo Global QCD Analysis of Pion Parton Distributions}.
\newblock {\em Phys. Rev. Lett.}, 121(15):152001, 2018.

\bibitem{Karpie:2018zaz}
Joseph Karpie, Kostas Orginos, and Savvas Zafeiropoulos.
\newblock {Moments of Ioffe time parton distribution functions from non-local
  matrix elements}.
\newblock {\em JHEP}, 11:178, 2018.

\bibitem{Joo:2019bzr}
B{\'a}lint Jo{\'o}, Joseph Karpie, Kostas Orginos, Anatoly~V. Radyushkin,
  David~G. Richards, Raza~Sabbir Sufian, and Savvas Zafeiropoulos.
\newblock {Pion valence structure from Ioffe-time parton pseudodistribution
  functions}.
\newblock {\em Phys. Rev. D}, 100(11):114512, 2019.

\bibitem{Sufian:2020vzb}
Raza~Sabbir Sufian, Colin Egerer, Joseph Karpie, Robert~G. Edwards, B{\'a}lint
  Jo{\'o}, Yan-Qing Ma, Kostas Orginos, Jian-Wei Qiu, and David~G. Richards.
\newblock {Pion Valence Quark Distribution from Current-Current Correlation in
  Lattice QCD}.
\newblock {\em Phys. Rev. D}, 102(5):054508, 2020.

\bibitem{Cao:2021aci}
N.~Y. Cao, P.~C. Barry, N.~Sato, and W.~Melnitchouk.
\newblock {Towards the three-dimensional parton structure of the pion:
  Integrating transverse momentum data into global QCD analysis}.
\newblock {\em Phys. Rev. D}, 103(11):114014, 2021.

\bibitem{Barry:2025wjx}
P.~C. Barry, Chueng-Ryong Ji, W.~Melnitchouk, N.~Sato, and Fernanda Steffens.
\newblock {First simultaneous global QCD analysis of kaon and pion parton
  distributions with lattice QCD constraints}.
\newblock arXiv:2510.11979.

\bibitem{Adams:2018pwt}
B.~Adams et~al.
\newblock {Letter of Intent: A New QCD facility at the M2 beam line of the CERN
  SPS (COMPASS++/AMBER)}.
\newblock arXiv:1808.00848.

\bibitem{Klasen:2002xb}
Michael Klasen.
\newblock {Theory of hard photoproduction}.
\newblock {\em Rev. Mod. Phys.}, 74:1221--1282, 2002.

\bibitem{Nisius:1999cv}
Richard Nisius.
\newblock {The Photon structure from deep inelastic electron photon
  scattering}.
\newblock {\em Phys. Rept.}, 332:165--317, 2000.

\bibitem{Moch:2001im}
S.~Moch, J.~A.~M. Vermaseren, and A.~Vogt.
\newblock {Next-to-next-to leading order QCD corrections to the photon's parton
  structure}.
\newblock {\em Nucl. Phys. B}, 621:413--458, 2002.

\bibitem{Klasen:2013cba}
M.~Klasen, G.~Kramer, and M.~Michael.
\newblock {Next-to-next-to-leading order contributions to jet photoproduction
  and determination of $\alpha_s$}.
\newblock {\em Phys. Rev. D}, 89(7):074032, 2014.

\bibitem{Feike:2025plq}
Alexander Feike, Tom{\'a}{\v{s}} Je{\v{z}}o, and Michael Klasen.
\newblock {Dijet Photoproduction in POWHEG BOX}.
\newblock In {\em {2nd International Workshop on the Physics of Ultra
  Peripheral Collisions}}, 9 2025.

\bibitem{Cornet:2004nb}
F.~Cornet, P.~Jankowski, and M.~Krawczyk.
\newblock {A New 5 flavor NLO analysis and parametrizations of parton
  distributions of the real photon}.
\newblock {\em Phys. Rev. D}, 70:093004, 2004.

\bibitem{Slominski:2005bw}
W.~Slominski, H.~Abramowicz, and A.~Levy.
\newblock {NLO photon parton parametrization using ee and ep data}.
\newblock {\em Eur. Phys. J. C}, 45:633--641, 2006.

\bibitem{Gluck:1991jc}
M.~Gluck, E.~Reya, and A.~Vogt.
\newblock {Photonic parton distributions}.
\newblock {\em Phys. Rev. D}, 46:1973--1979, 1992.

\bibitem{Klasen:2001sg}
M.~Klasen and G.~Kramer.
\newblock {Photoproduction of jets on a virtual pion target in next-to-leading
  order QCD}.
\newblock {\em Phys. Lett. B}, 508:259--268, 2001.

\bibitem{ARRINGTON2022103985}
J.~Arrington, M.~Battaglieri, A.~Boehnlein, S.A. Bogacz, W.K. Brooks,
  E.~Chudakov, I.~Cloët, R.~Ent, H.~Gao, J.~Grames, L.~Harwood, X.~Ji,
  C.~Keppel, G.~Krafft, R.D. McKeown, J.~Napolitano, J.W. Qiu, P.~Rossi,
  M.~Schram, S.~Stepanyan, J.~Stevens, A.P. Szczepaniak, N.~Toro, and X.~Zheng.
\newblock Physics with cebaf at 12 gev and future opportunities.
\newblock {\em Progress in Particle and Nuclear Physics}, 127:103985, 2022.

\bibitem{Andre:2022xeh}
K.~D.~J. Andr{\'e} et~al.
\newblock {An experiment for electron-hadron scattering at the LHC}.
\newblock {\em Eur. Phys. J. C}, 82(1):40, 2022.

\bibitem{FPF:LOI}
Luis~A. Anchordoqui and et~al.
\newblock {LETTER OF INTENT: THE FORWARD PHYSICS FACILITY}.

\bibitem{ANCHORDOQUI:20221}
Anchordoqui et~al.
\newblock The forward physics facility: Sites, experiments, and physics
  potential.
\newblock {\em Physics Reports}, 968:1--50, 2022.

\bibitem{Arrington_2023}
J~Arrington, J~Benesch, A~Camsonne, J~Caylor, J-P Chen, S~Covrig~Dusa,
  A~Emmert, G~Evans, H~Gao, J-O Hansen, G~M Huber, S~Joosten, V~Khachatryan,
  N~Liyanage, Z-E Meziani, M~Nycz, C~Peng, M~Paolone, W~Seay, P~A Souder,
  N~Sparveris, H~Spiesberger, Y~Tian, E~Voutier, J~Xie, W~Xiong, Z-Y Ye, Z~Ye,
  J~Zhang, Z-W Zhao, X~Zheng, and For the Jefferson Lab SoLID~Collaboration.
\newblock The solenoidal large intensity device (solid) for jlab 12 gev.
\newblock {\em Journal of Physics G: Nuclear and Particle Physics},
  50(11):110501, sep 2023.

\bibitem{Ullrich:2020}
Thomas Ullrich.
\newblock Private communication.

\bibitem{PhysRevD.93.114017}
A.~Accardi, L.~T. Brady, W.~Melnitchouk, J.~F. Owens, and N.~Sato.
\newblock Constraints on large-$x$ parton distributions from new weak boson
  production and deep-inelastic scattering data.
\newblock {\em Phys. Rev. D}, 93:114017, Jun 2016.

\bibitem{PhysRevD.109.054019}
N\'estor Armesto, Thomas Cridge, Francesco Giuli, Lucian Harland-Lang, Paul
  Newman, Barak Schmookler, Robert Thorne, and Katarzyna Wichmann.
\newblock Impact of inclusive electron ion collider data on collinear parton
  distributions.
\newblock {\em Phys. Rev. D}, 109:054019, Mar 2024.

\bibitem{Gribov:1983ivg}
L.~V. Gribov, E.~M. Levin, and M.~G. Ryskin.
\newblock {Semihard Processes in QCD}.
\newblock {\em Phys. Rept.}, 100:1--150, 1983.

\bibitem{Mueller:1985wy}
Alfred~H. Mueller and Jian-wei Qiu.
\newblock {Gluon Recombination and Shadowing at Small Values of x}.
\newblock {\em Nucl. Phys. B}, 268:427--452, 1986.

\bibitem{McLerran:1993ni}
Larry~D. McLerran and Raju Venugopalan.
\newblock {Computing quark and gluon distribution functions for very large
  nuclei}.
\newblock {\em Phys. Rev. D}, 49:2233--2241, 1994.

\bibitem{McLerran:1993ka}
Larry~D. McLerran and Raju Venugopalan.
\newblock {Gluon distribution functions for very large nuclei at small
  transverse momentum}.
\newblock {\em Phys. Rev. D}, 49:3352--3355, 1994.

\bibitem{McLerran:1994vd}
Larry~D. McLerran and Raju Venugopalan.
\newblock {Green's functions in the color field of a large nucleus}.
\newblock {\em Phys. Rev. D}, 50:2225--2233, 1994.

\bibitem{Iancu:2003xm}
Edmond Iancu and Raju Venugopalan.
\newblock {\em {The Color glass condensate and high-energy scattering in QCD}},
  pages 249--3363.
\newblock 3 2003.

\bibitem{Gelis:2010nm}
Francois Gelis, Edmond Iancu, Jamal Jalilian-Marian, and Raju Venugopalan.
\newblock {The Color Glass Condensate}.
\newblock {\em Ann. Rev. Nucl. Part. Sci.}, 60:463--489, 2010.

\bibitem{Kowalski:2007rw}
H.~Kowalski, T.~Lappi, and R.~Venugopalan.
\newblock {Nuclear enhancement of universal dynamics of high parton densities}.
\newblock {\em Phys. Rev. Lett.}, 100:022303, 2008.

\bibitem{Balitsky:1995ub}
I.~Balitsky.
\newblock {Operator expansion for high-energy scattering}.
\newblock {\em Nucl. Phys. B}, 463:99--160, 1996.

\bibitem{Ayala:1995kg}
Alejandro Ayala, Jamal Jalilian-Marian, Larry~D. McLerran, and Raju
  Venugopalan.
\newblock {The Gluon propagator in nonAbelian Weizsacker-Williams fields}.
\newblock {\em Phys. Rev. D}, 52:2935--2943, 1995.

\bibitem{Albacete:2004gw}
J.~L. Albacete, N.~Armesto, J.~G. Milhano, C.~A. Salgado, and U.~A. Wiedemann.
\newblock {Numerical analysis of the Balitsky-Kovchegov equation with running
  coupling: Dependence of the saturation scale on nuclear size and rapidity}.
\newblock {\em Phys. Rev. D}, 71:014003, 2005.

\bibitem{Kovchegov:2006vj}
Yuri~V. Kovchegov and Heribert Weigert.
\newblock {Triumvirate of Running Couplings in Small-x Evolution}.
\newblock {\em Nucl. Phys. A}, 784:188--226, 2007.

\bibitem{Balitsky:2006wa}
Ian Balitsky.
\newblock {Quark contribution to the small-x evolution of color dipole}.
\newblock {\em Phys. Rev. D}, 75:014001, 2007.

\bibitem{Lappi:2012vw}
T.~Lappi and H.~M{\"a}ntysaari.
\newblock {On the running coupling in the JIMWLK equation}.
\newblock {\em Eur. Phys. J. C}, 73(2):2307, 2013.

\bibitem{Albacete:2009fh}
Javier~L. Albacete, Nestor Armesto, Jose~Guilherme Milhano, and Carlos~A.
  Salgado.
\newblock {Non-linear QCD meets data: A Global analysis of lepton-proton
  scattering with running coupling BK evolution}.
\newblock {\em Phys. Rev. D}, 80:034031, 2009.

\bibitem{Albacete:2010sy}
Javier~L. Albacete, Nestor Armesto, Jose~Guilherme Milhano, Paloma
  Quiroga-Arias, and Carlos~A. Salgado.
\newblock {AAMQS: A non-linear QCD analysis of new HERA data at small-x
  including heavy quarks}.
\newblock {\em Eur. Phys. J. C}, 71:1705, 2011.

\bibitem{Lappi:2013zma}
T.~Lappi and H.~M{\"a}ntysaari.
\newblock {Single inclusive particle production at high energy from HERA data
  to proton-nucleus collisions}.
\newblock {\em Phys. Rev. D}, 88:114020, 2013.

\bibitem{Mantysaari:2018zdd}
Heikki M\"antysaari and Bj\"orn Schenke.
\newblock {Confronting impact parameter dependent JIMWLK evolution with HERA
  data}.
\newblock {\em Phys. Rev. D}, 98(3):034013, 2018.

\bibitem{Balitsky:2008zza}
Ian Balitsky and Giovanni~A. Chirilli.
\newblock {Next-to-leading order evolution of color dipoles}.
\newblock {\em Phys. Rev. D}, 77:014019, 2008.

\bibitem{Balitsky:2009xg}
Ian Balitsky and Giovanni~A. Chirilli.
\newblock {NLO evolution of color dipoles in N=4 SYM}.
\newblock {\em Nucl. Phys. B}, 822:45--87, 2009.

\bibitem{Balitsky:2009yp}
Ian Balitsky and Giovanni~A. Chirilli.
\newblock {High-energy amplitudes in N=4 SYM in the next-to-leading order}.
\newblock {\em Phys. Lett. B}, 687:204--213, 2010.

\bibitem{Lappi:2015fma}
T.~Lappi and H.~M{\"a}ntysaari.
\newblock {Direct numerical solution of the coordinate space Balitsky-Kovchegov
  equation at next to leading order}.
\newblock {\em Phys. Rev. D}, 91(7):074016, 2015.

\bibitem{Kwiecinski:1997ee}
J.~Kwiecinski, Alan~D. Martin, and A.~M. Stasto.
\newblock {A Unified BFKL and GLAP description of F2 data}.
\newblock {\em Phys. Rev. D}, 56:3991--4006, 1997.

\bibitem{Salam:1998tj}
G.~P. Salam.
\newblock {A Resummation of large subleading corrections at small x}.
\newblock {\em JHEP}, 07:019, 1998.

\bibitem{Ciafaloni:1999yw}
M.~Ciafaloni, D.~Colferai, and G.~P. Salam.
\newblock {Renormalization group improved small x equation}.
\newblock {\em Phys. Rev. D}, 60:114036, 1999.

\bibitem{Altarelli:1999vw}
Guido Altarelli, Richard~D. Ball, and Stefano Forte.
\newblock {Resummation of singlet parton evolution at small x}.
\newblock {\em Nucl. Phys. B}, 575:313--329, 2000.

\bibitem{Ciafaloni:2003rd}
M.~Ciafaloni, D.~Colferai, G.~P. Salam, and A.~M. Stasto.
\newblock {Renormalization group improved small x Green's function}.
\newblock {\em Phys. Rev. D}, 68:114003, 2003.

\bibitem{SabioVera:2005tiv}
Agustin Sabio~Vera.
\newblock {An 'All-poles' approximation to collinear resummations in the Regge
  limit of perturbative QCD}.
\newblock {\em Nucl. Phys. B}, 722:65--80, 2005.

\bibitem{Iancu:2015vea}
E.~Iancu, J.~D. Madrigal, A.~H. Mueller, G.~Soyez, and D.~N.
  Triantafyllopoulos.
\newblock {Resumming double logarithms in the QCD evolution of color dipoles}.
\newblock {\em Phys. Lett. B}, 744:293--302, 2015.

\bibitem{Ducloue:2019ezk}
B.~Duclou{\'e}, E.~Iancu, A.~H. Mueller, G.~Soyez, and D.~N.
  Triantafyllopoulos.
\newblock {Non-linear evolution in QCD at high-energy beyond leading order}.
\newblock {\em JHEP}, 04:081, 2019.

\bibitem{Ducloue:2019jmy}
B.~Duclou\'e, E.~Iancu, G.~Soyez, and D.~N. Triantafyllopoulos.
\newblock {HERA data and collinearly-improved BK dynamics}.
\newblock {\em Phys. Lett. B}, 803:135305, 2020.

\bibitem{Lappi:2016fmu}
T.~Lappi and H.~Mäntysaari.
\newblock {Next-to-leading order Balitsky-Kovchegov equation with resummation}.
\newblock {\em Phys. Rev. D}, 93(9):094004, 2016.

\bibitem{Balitsky:2013fea}
Ian Balitsky and Giovanni~A. Chirilli.
\newblock {Rapidity evolution of Wilson lines at the next-to-leading order}.
\newblock {\em Phys. Rev. D}, 88:111501, 2013.

\bibitem{Kovner:2013ona}
Alex Kovner, Michael Lublinsky, and Yair Mulian.
\newblock {Jalilian-Marian, Iancu, McLerran, Weigert, Leonidov, Kovner
  evolution at next to leading order}.
\newblock {\em Phys. Rev. D}, 89(6):061704, 2014.

\bibitem{Kovner:2014lca}
Alex Kovner, Michael Lublinsky, and Yair Mulian.
\newblock {NLO JIMWLK evolution unabridged}.
\newblock {\em JHEP}, 08:114, 2014.

\bibitem{Lublinsky:2016meo}
Michael Lublinsky and Yair Mulian.
\newblock {High Energy QCD at NLO: from light-cone wave function to JIMWLK
  evolution}.
\newblock {\em JHEP}, 05:097, 2017.

\bibitem{Hatta:2016ujq}
Yoshitaka Hatta and Edmond Iancu.
\newblock {Collinearly improved JIMWLK evolution in Langevin form}.
\newblock {\em JHEP}, 08:083, 2016.

\bibitem{Balitsky:2010ze}
Ian Balitsky and Giovanni~A. Chirilli.
\newblock {Photon impact factor in the next-to-leading order}.
\newblock {\em Phys. Rev. D}, 83:031502, 2011.

\bibitem{Beuf:2017bpd}
Guillaume Beuf.
\newblock {Dipole factorization for DIS at NLO: Combining the $q\bar{q}$ and
  $q\bar{q}g$ contributions}.
\newblock {\em Phys. Rev. D}, 96(7):074033, 2017.

\bibitem{Hanninen:2017ddy}
H.~H\"anninen, T.~Lappi, and R.~Paatelainen.
\newblock {One-loop corrections to light cone wave functions: the dipole
  picture DIS cross section}.
\newblock {\em Annals Phys.}, 393:358--412, 2018.

\bibitem{Ducloue:2017ftk}
B.~Duclou\'e, H.~H\"anninen, T.~Lappi, and Y.~Zhu.
\newblock {Deep inelastic scattering in the dipole picture at next-to-leading
  order}.
\newblock {\em Phys. Rev. D}, 96(9):094017, 2017.

\bibitem{Beuf:2020dxl}
G.~Beuf, H.~H{\"a}nninen, T.~Lappi, and H.~M{\"a}ntysaari.
\newblock {Color Glass Condensate at next-to-leading order meets HERA data}.
\newblock {\em Phys. Rev. D}, 102:074028, 2020.

\bibitem{Boussarie:2016bkq}
R.~Boussarie, A.~V. Grabovsky, D.~{\relax Yu}. Ivanov, L.~Szymanowski, and
  S.~Wallon.
\newblock {Next-to-Leading Order Computation of Exclusive Diffractive Light
  Vector Meson Production in a Saturation Framework}.
\newblock {\em Phys. Rev. Lett.}, 119(7):072002, 2017.

\bibitem{Boussarie:2016ogo}
R.~Boussarie, A.~V. Grabovsky, L.~Szymanowski, and S.~Wallon.
\newblock {On the one loop $ {\gamma}^{\left(\ast \right)}\to q\overline{q} $
  impact factor and the exclusive diffractive cross sections for the production
  of two or three jets}.
\newblock {\em JHEP}, 11:149, 2016.

\bibitem{Fucilla:2022wcg}
Michael Fucilla, Andrey~V. Grabovsky, Emilie Li, Lech Szymanowski, and Samuel
  Wallon.
\newblock {NLO computation of diffractive di-hadron production in a saturation
  framework}.
\newblock {\em JHEP}, 03:159, 2023.

\bibitem{Beuf:2024msh}
Guillaume Beuf, Tuomas Lappi, Heikki M{\"a}ntysaari, Risto Paatelainen, and
  Jani Penttala.
\newblock {Diffractive deep inelastic scattering at NLO in the dipole picture}.
\newblock {\em JHEP}, 05:024, 2024.

\bibitem{Mantysaari:2021ryb}
Heikki M{\"a}ntysaari and Jani Penttala.
\newblock {Exclusive heavy vector meson production at next-to-leading order in
  the dipole picture}.
\newblock {\em Phys. Lett. B}, 823:136723, 2021.

\bibitem{Chirilli:2011km}
Giovanni~A. Chirilli, Bo-Wen Xiao, and Feng Yuan.
\newblock {One-loop Factorization for Inclusive Hadron Production in $pA$
  Collisions in the Saturation Formalism}.
\newblock {\em Phys. Rev. Lett.}, 108:122301, 2012.

\bibitem{Chirilli:2012jd}
Giovanni~A. Chirilli, Bo-Wen Xiao, and Feng Yuan.
\newblock {Inclusive Hadron Productions in pA Collisions}.
\newblock {\em Phys. Rev. D}, 86:054005, 2012.

\bibitem{Caucal:2024cdq}
Paul Caucal, Elouan Ferrand, and Farid Salazar.
\newblock {Semi-inclusive single-jet production in DIS at next-to-leading order
  in the Color Glass Condensate}.
\newblock {\em JHEP}, 05:110, 2024.

\bibitem{Altinoluk:2024vgg}
Tolga Altinoluk, Jamal Jalilian-Marian, and Cyrille Marquet.
\newblock {Sudakov double logs in single-inclusive hadron production in DIS at
  small x from the color glass condensate formalism}.
\newblock {\em Phys. Rev. D}, 110(9):094056, 2024.

\bibitem{Bergabo:2022zhe}
Filip Bergabo and Jamal Jalilian-Marian.
\newblock {Single inclusive hadron production in DIS at small x: next to
  leading order corrections}.
\newblock {\em JHEP}, 01:095, 2023.

\bibitem{Altinoluk:2025dwd}
Tolga Altinoluk, Filip Bergabo, Jamal Jalilian-Marian, Cyrille Marquet, and
  Yu~Shi.
\newblock {SIDIS at small x at next-to-leading order: Transverse photon}.
\newblock {\em Phys. Rev. D}, 112(5):054020, 2025.

\bibitem{Caucal:2021ent}
Paul Caucal, Farid Salazar, and Raju Venugopalan.
\newblock {Dijet impact factor in DIS at next-to-leading order in the Color
  Glass Condensate}.
\newblock {\em JHEP}, 11:222, 2021.

\bibitem{Bergabo:2022tcu}
Filip Bergabo and Jamal Jalilian-Marian.
\newblock {One-loop corrections to dihadron production in DIS at small x}.
\newblock {\em Phys. Rev. D}, 106(5):054035, 2022.

\bibitem{Bergabo:2023wed}
Filip Bergabo and Jamal Jalilian-Marian.
\newblock {Dihadron production in DIS at small x at next-to-leading order:
  Transverse photons}.
\newblock {\em Phys. Rev. D}, 107(5):054036, 2023.

\bibitem{Roy:2019hwr}
Kaushik Roy and Raju Venugopalan.
\newblock {NLO impact factor for inclusive photon$+$dijet production in $e+A$
  DIS at small $x$}.
\newblock {\em Phys. Rev. D}, 101(3):034028, 2020.

\bibitem{Stasto:2013cha}
Anna~M. Stasto, Bo-Wen Xiao, and David Zaslavsky.
\newblock {Towards the Test of Saturation Physics Beyond Leading Logarithm}.
\newblock {\em Phys. Rev. Lett.}, 112(1):012302, 2014.

\bibitem{Stewart:2023lwz}
Iain~W. Stewart and Varun Vaidya.
\newblock {Power counting to saturation}.
\newblock {\em Phys. Rev. D}, 110(1):L011504, 2024.

\bibitem{Neill:2023jcd}
Duff Neill, Aditya Pathak, and Iain~W. Stewart.
\newblock {Small-x factorization from effective field theory}.
\newblock {\em JHEP}, 09:089, 2023.

\bibitem{Altinoluk:2018byz}
Tolga Altinoluk, Renaud Boussarie, Cyrille Marquet, and Pieter Taels.
\newblock {TMD factorization for dijets + photon production from the
  dilute-dense CGC framework}.
\newblock {\em JHEP}, 07:079, 2019.

\bibitem{Altinoluk:2020qet}
Tolga Altinoluk, Renaud Boussarie, Cyrille Marquet, and Pieter Taels.
\newblock {Photoproduction of three jets in the CGC: gluon TMDs and dilute
  limit}.
\newblock {\em JHEP}, 07:143, 2020.

\bibitem{Tong:2022zwp}
Xuan-Bo Tong, Bo-Wen Xiao, and Yuan-Yuan Zhang.
\newblock {Harmonics of Parton Saturation in Lepton-Jet Correlations at the
  Electron-Ion Collider}.
\newblock {\em Phys. Rev. Lett.}, 130(15):151902, 2023.

\bibitem{Tong:2023bus}
Xuan-Bo Tong, Bo-Wen Xiao, and Yuan-Yuan Zhang.
\newblock {Harmonics of lepton-jet correlations in inclusive and diffractive
  scatterings}.
\newblock {\em Phys. Rev. D}, 109(5):054004, 2024.

\bibitem{Iancu:2021rup}
E.~Iancu, A.~H. Mueller, and D.~N. Triantafyllopoulos.
\newblock {Probing Parton Saturation and the Gluon Dipole via Diffractive Jet
  Production at the Electron-Ion Collider}.
\newblock {\em Phys. Rev. Lett.}, 128(20):202001, 2022.

\bibitem{Iancu:2022lcw}
E.~Iancu, A.~H. Mueller, D.~N. Triantafyllopoulos, and S.~Y. Wei.
\newblock {Gluon dipole factorisation for diffractive dijets}.
\newblock {\em JHEP}, 10:103, 2022.

\bibitem{Iancu:2023lel}
E.~Iancu, A.~H. Mueller, D.~N. Triantafyllopoulos, and S.~Y. Wei.
\newblock {Probing gluon saturation via diffractive jets in ultra-peripheral
  nucleus-nucleus collisions}.
\newblock {\em Eur. Phys. J. C}, 83(11):1078, 2023.

\bibitem{Hatta:2022lzj}
Yoshitaka Hatta, Bo-Wen Xiao, and Feng Yuan.
\newblock {Semi-inclusive diffractive deep inelastic scattering at small x}.
\newblock {\em Phys. Rev. D}, 106(9):094015, 2022.

\bibitem{Hauksson:2024bvv}
S.~Hauksson, E.~Iancu, A.~H. Mueller, D.~N. Triantafyllopoulos, and S.~Y. Wei.
\newblock {TMD factorisation for diffractive jets in photon-nucleus
  interactions}.
\newblock {\em JHEP}, 06:180, 2024.

\bibitem{Caucal:2025qjg}
Paul Caucal and Farid Salazar.
\newblock {Small-x Factorization in the Target Fragmentation Region}.
\newblock {\em Phys. Rev. Lett.}, 136(8):081901, 2026.

\bibitem{Altinoluk:2024tyx}
Tolga Altinoluk, Guillaume Beuf, Etienne Blanco, and Swaleha Mulani.
\newblock {Quark TMDs from back-to-back dijet production at forward rapidities
  in pA collisions beyond eikonal accuracy in the CGC}.
\newblock {\em JHEP}, 06:097, 2025.

\bibitem{Altinoluk:2024zom}
Tolga Altinoluk, Guillaume Beuf, Alina Czajka, and Cyrille Marquet.
\newblock {Back-to-back dijet production in DIS at next-to-eikonal accuracy and
  twist-3 gluon TMDs}.
\newblock {\em Phys. Rev. D}, 111(1):014010, 2025.

\bibitem{Agostini:2024xqs}
Pedro Agostini, Tolga Altinoluk, and N{\'e}stor Armesto.
\newblock {Next-to-eikonal corrections to dijet production in Deep Inelastic
  Scattering in the dilute limit of the Color Glass Condensate}.
\newblock {\em JHEP}, 07:137, 2024.

\bibitem{Altinoluk:2023qfr}
Tolga Altinoluk, Nestor Armesto, and Guillaume Beuf.
\newblock {Probing quark transverse momentum distributions in the color glass
  condensate: Quark-gluon dijets in deep inelastic scattering at
  next-to-eikonal accuracy}.
\newblock {\em Phys. Rev. D}, 108(7):074023, 2023.

\bibitem{Altinoluk:2022jkk}
Tolga Altinoluk, Guillaume Beuf, Alina Czajka, and Arantxa Tymowska.
\newblock {DIS dijet production at next-to-eikonal accuracy in the CGC}.
\newblock {\em Phys. Rev. D}, 107(7):074016, 2023.

\bibitem{Altinoluk:2019fui}
Tolga Altinoluk, Renaud Boussarie, and Piotr Kotko.
\newblock {Interplay of the CGC and TMD frameworks to all orders in kinematic
  twist}.
\newblock {\em JHEP}, 05:156, 2019.

\bibitem{Altinoluk:2019wyu}
Tolga Altinoluk and Renaud Boussarie.
\newblock {Low $x$ physics as an infinite twist (G)TMD framework: unravelling
  the origins of saturation}.
\newblock {\em JHEP}, 10:208, 2019.

\bibitem{Mantysaari:2019hkq}
Heikki M\"antysaari, Niklas Mueller, Farid Salazar, and Bj\"orn Schenke.
\newblock {Multigluon Correlations and Evidence of Saturation from Dijet
  Measurements at an Electron-Ion Collider}.
\newblock {\em Phys. Rev. Lett.}, 124(11):112301, 2020.

\bibitem{Boussarie:2021ybe}
Renaud Boussarie, Heikki M\"antysaari, Farid Salazar, and Bj\"orn Schenke.
\newblock {The importance of kinematic twists and genuine saturation effects in
  dijet production at the Electron-Ion Collider}.
\newblock {\em JHEP}, 09:178, 2021.

\bibitem{Mueller:2012uf}
A.~H. Mueller, Bo-Wen Xiao, and Feng Yuan.
\newblock {Sudakov Resummation in Small-$x$ Saturation Formalism}.
\newblock {\em Phys. Rev. Lett.}, 110(8):082301, 2013.

\bibitem{Mueller:2013wwa}
A.~H. Mueller, Bo-Wen Xiao, and Feng Yuan.
\newblock {Sudakov double logarithms resummation in hard processes in the
  small-x saturation formalism}.
\newblock {\em Phys. Rev. D}, 88(11):114010, 2013.

\bibitem{Kovchegov:2015zha}
Yuri~V. Kovchegov and Matthew~D. Sievert.
\newblock {Calculating TMDs of a Large Nucleus: Quasi-Classical Approximation
  and Quantum Evolution}.
\newblock {\em Nucl. Phys. B}, 903:164--203, 2016.

\bibitem{Taels:2022tza}
Pieter Taels, Tolga Altinoluk, Guillaume Beuf, and Cyrille Marquet.
\newblock {Dijet photoproduction at low x at next-to-leading order and its
  back-to-back limit}.
\newblock {\em JHEP}, 10:184, 2022.

\bibitem{Caucal:2022ulg}
Paul Caucal, Farid Salazar, Bj\"orn Schenke, and Raju Venugopalan.
\newblock {Back-to-back inclusive dijets in DIS at small x: Sudakov suppression
  and gluon saturation at NLO}.
\newblock {\em JHEP}, 11:169, 2022.

\bibitem{Caucal:2024bae}
Paul Caucal and Edmond Iancu.
\newblock {Evolution of the transverse-momentum dependent gluon distribution at
  small x}.
\newblock {\em Phys. Rev. D}, 111(7):074008, 2025.

\bibitem{Caucal:2024vbv}
Paul Caucal, Edmond Iancu, A.~H. Mueller, and Feng Yuan.
\newblock {Jet Definition and Transverse-Momentum{\textendash}Dependent
  Factorization in Semi-inclusive Deep-Inelastic Scattering}.
\newblock {\em Phys. Rev. Lett.}, 134(6):061903, 2025.

\bibitem{Duan:2024nlr}
Haowu Duan, Alex Kovner, and Michael Lublinsky.
\newblock {Collins-Soper-Sterman Hamiltonian: High energy evolution of rapidity
  dependent observables}.
\newblock {\em Phys. Rev. D}, 111(5):054022, 2025.

\bibitem{Duan:2024qev}
Haowu Duan, Alex Kovner, and Michael Lublinsky.
\newblock {Born-Oppenheimer renormalization group for high energy scattering:
  CSS, DGLAP and all that}.
\newblock {\em JHEP}, 08:137, 2025.

\bibitem{Mukherjee:2023snp}
Swagato Mukherjee, Vladimir~V. Skokov, Andrey Tarasov, and Shaswat Tiwari.
\newblock {Unified description of DGLAP, CSS, and BFKL evolution: TMD
  factorization bridging large and small x}.
\newblock {\em Phys. Rev. D}, 109(3):034035, 2024.

\bibitem{Kovchegov:1999yj}
Yuri~V. Kovchegov.
\newblock {Small x F(2) structure function of a nucleus including multiple
  pomeron exchanges}.
\newblock {\em Phys. Rev. D}, 60:034008, 1999.

\bibitem{JalilianMarian:1997jx}
Jamal Jalilian-Marian, Alex Kovner, Andrei Leonidov, and Heribert Weigert.
\newblock {The BFKL equation from the Wilson renormalization group}.
\newblock {\em Nucl. Phys. B}, 504:415--431, 1997.

\bibitem{JalilianMarian:1997gr}
Jamal Jalilian-Marian, Alex Kovner, Andrei Leonidov, and Heribert Weigert.
\newblock {The Wilson renormalization group for low x physics: Towards the high
  density regime}.
\newblock {\em Phys. Rev. D}, 59:014014, 1998.

\bibitem{Kovner:2000pt}
Alex Kovner, J.~Guilherme Milhano, and Heribert Weigert.
\newblock {Relating different approaches to nonlinear QCD evolution at finite
  gluon density}.
\newblock {\em Phys. Rev. D}, 62:114005, 2000.

\bibitem{Weigert:2000gi}
Heribert Weigert.
\newblock {Unitarity at small Bjorken x}.
\newblock {\em Nucl. Phys. A}, 703:823--860, 2002.

\bibitem{Iancu:2000hn}
Edmond Iancu, Andrei Leonidov, and Larry~D. McLerran.
\newblock {Nonlinear gluon evolution in the color glass condensate. 1.}
\newblock {\em Nucl. Phys. A}, 692:583--645, 2001.

\bibitem{Iancu:2001ad}
Edmond Iancu, Andrei Leonidov, and Larry~D. McLerran.
\newblock {The Renormalization group equation for the color glass condensate}.
\newblock {\em Phys. Lett. B}, 510:133--144, 2001.

\bibitem{Ferreiro:2001qy}
Elena Ferreiro, Edmond Iancu, Andrei Leonidov, and Larry McLerran.
\newblock {Nonlinear gluon evolution in the color glass condensate. 2.}
\newblock {\em Nucl. Phys. A}, 703:489--538, 2002.

\bibitem{Collins:1981uk}
John~C. Collins and Davison~E. Soper.
\newblock {Back-To-Back Jets in QCD}.
\newblock {\em Nucl. Phys. B}, 193:381, 1981.
\newblock [Erratum: Nucl.Phys.B 213, 545 (1983)].

\bibitem{Collins:1981uw}
John~C. Collins and Davison~E. Soper.
\newblock {Parton Distribution and Decay Functions}.
\newblock {\em Nucl. Phys. B}, 194:445--492, 1982.

\bibitem{Collins:1984kg}
John~C. Collins, Davison~E. Soper, and George~F. Sterman.
\newblock {Transverse Momentum Distribution in Drell-Yan Pair and W and Z Boson
  Production}.
\newblock {\em Nucl. Phys.}, B250:199--224, 1985.

\bibitem{Collins:2011zzd}
John Collins.
\newblock {\em {Foundations of Perturbative QCD}}, volume~32.
\newblock Cambridge University Press, 2011.

\bibitem{Gribov:1972ri}
V.~N. Gribov and L.~N. Lipatov.
\newblock {Deep inelastic e p scattering in perturbation theory}.
\newblock {\em Sov. J. Nucl. Phys.}, 15:438--450, 1972.

\bibitem{Altarelli:1977zs}
Guido Altarelli and G.~Parisi.
\newblock {Asymptotic Freedom in Parton Language}.
\newblock {\em Nucl. Phys. B}, 126:298--318, 1977.

\bibitem{Dokshitzer:1977sg}
Yuri~L. Dokshitzer.
\newblock {Calculation of the Structure Functions for Deep Inelastic Scattering
  and e+ e- Annihilation by Perturbation Theory in Quantum Chromodynamics.}
\newblock {\em Sov. Phys. JETP}, 46:641--653, 1977.

\bibitem{Ram}
Ramkumar Radhakrishnan.
\newblock {Single inclusive gluon production at central rapidity at NLO}.
\newblock Nuclear theroy group seminar, NCSU, 2025.

\bibitem{Morreale:2021pnn}
Astrid Morreale and Farid Salazar.
\newblock {Mining for Gluon Saturation at Colliders}.
\newblock {\em Universe}, 7(8):312, 2021.

\bibitem{Altinoluk:2014oxa}
Tolga Altinoluk, N{\'e}stor Armesto, Guillaume Beuf, Mauricio Mart{\'\i}nez,
  and Carlos~A. Salgado.
\newblock {Next-to-eikonal corrections in the CGC: gluon production and spin
  asymmetries in pA collisions}.
\newblock {\em JHEP}, 07:068, 2014.

\bibitem{Altinoluk:2015gia}
Tolga Altinoluk, N{\'e}stor Armesto, Guillaume Beuf, and Alexis Moscoso.
\newblock {Next-to-next-to-eikonal corrections in the CGC}.
\newblock {\em JHEP}, 01:114, 2016.

\bibitem{Altinoluk:2015xuy}
Tolga Altinoluk and Adrian Dumitru.
\newblock {Particle production in high-energy collisions beyond the shockwave
  limit}.
\newblock {\em Phys. Rev. D}, 94(7):074032, 2016.

\bibitem{Agostini:2019avp}
Pedro Agostini, Tolga Altinoluk, and N{\'e}stor Armesto.
\newblock {Non-eikonal corrections to multi-particle production in the Color
  Glass Condensate}.
\newblock {\em Eur. Phys. J. C}, 79(7):600, 2019.

\bibitem{Agostini:2019hkj}
Pedro Agostini, Tolga Altinoluk, and N{\'e}stor Armesto.
\newblock {Effect of non-eikonal corrections on azimuthal asymmetries in the
  Color Glass Condensate}.
\newblock {\em Eur. Phys. J. C}, 79(9):790, 2019.

\bibitem{Altinoluk:2020oyd}
Tolga Altinoluk, Guillaume Beuf, Alina Czajka, and Arantxa Tymowska.
\newblock {Quarks at next-to-eikonal accuracy in the CGC: Forward quark-nucleus
  scattering}.
\newblock {\em Phys. Rev. D}, 104(1):014019, 2021.

\bibitem{Altinoluk:2021lvu}
Tolga Altinoluk and Guillaume Beuf.
\newblock {Quark and scalar propagators at next-to-eikonal accuracy in the CGC
  through a dynamical background gluon field}.
\newblock {\em Phys. Rev. D}, 105(7):074026, 2022.

\bibitem{Agostini:2022ctk}
Pedro Agostini, Tolga Altinoluk, N{\'e}stor Armesto, Fabio Dominguez, and
  Jos{\'e}~Guilherme Milhano.
\newblock {Multiparticle production in proton{\textendash}nucleus collisions
  beyond eikonal accuracy}.
\newblock {\em Eur. Phys. J. C}, 82(11):1001, 2022.

\bibitem{Agostini:2022oge}
Pedro Agostini, Tolga Altinoluk, and N{\'e}stor Armesto.
\newblock {Finite width effects on the azimuthal asymmetry in proton-nucleus
  collisions in the Color Glass Condensate}.
\newblock {\em Phys. Lett. B}, 840:137892, 2023.

\bibitem{Agostini:2023cvc}
Pedro Agostini.
\newblock {Scalar propagator in a background gluon field beyond the eikonal
  approximation}.
\newblock {\em JHEP}, 11:099, 2023.

\bibitem{Altinoluk:2024dba}
Tolga Altinoluk, Guillaume Beuf, and Swaleha Mulani.
\newblock {Forward parton-nucleus scattering at next-to-eikonal accuracy in the
  color glass condensate}.
\newblock {\em Phys. Rev. D}, 111(3):034028, 2025.

\bibitem{Altinoluk:2025ang}
Tolga Altinoluk, Guillaume Beuf, and Swaleha Mulani.
\newblock {Parton model contributions as next-to-eikonal corrections to the
  dipole factorization of DIS and SIDIS at low xBj}.
\newblock {\em Phys. Rev. D}, 113(3):034011, 2026.

\bibitem{Altinoluk:2025ivn}
Tolga Altinoluk, Guillaume Beuf, Jules Favrel, and Michael Fucilla.
\newblock {Next-to-Leading Order corrections to the Next-to-Eikonal DIS
  structure functions}.
\newblock arXiv:2512.16788.

\bibitem{Kovchegov:2015pbl}
Yuri~V. Kovchegov, Daniel Pitonyak, and Matthew~D. Sievert.
\newblock {Helicity Evolution at Small-x}.
\newblock {\em JHEP}, 01:072, 2016.
\newblock [Erratum: JHEP 10, 148 (2016)].

\bibitem{Kovchegov:2016zex}
Yuri~V. Kovchegov, Daniel Pitonyak, and Matthew~D. Sievert.
\newblock {Helicity Evolution at Small $x$: Flavor Singlet and Non-Singlet
  Observables}.
\newblock {\em Phys. Rev. D}, 95(1):014033, 2017.

\bibitem{Kovchegov:2016weo}
Yuri~V. Kovchegov, Daniel Pitonyak, and Matthew~D. Sievert.
\newblock {Small-$x$ asymptotics of the quark helicity distribution}.
\newblock {\em Phys. Rev. Lett.}, 118(5):052001, 2017.

\bibitem{Kovchegov:2017jxc}
Yuri~V. Kovchegov, Daniel Pitonyak, and Matthew~D. Sievert.
\newblock {Small-$x$ Asymptotics of the Quark Helicity Distribution: Analytic
  Results}.
\newblock {\em Phys. Lett. B}, 772:136--140, 2017.

\bibitem{Kovchegov:2017lsr}
Yuri~V. Kovchegov, Daniel Pitonyak, and Matthew~D. Sievert.
\newblock {Small-$x$ Asymptotics of the Gluon Helicity Distribution}.
\newblock {\em JHEP}, 10:198, 2017.

\bibitem{Kovchegov:2018znm}
Yuri~V. Kovchegov and Matthew~D. Sievert.
\newblock {Small-$x$ Helicity Evolution: an Operator Treatment}.
\newblock {\em Phys. Rev. D}, 99(5):054032, 2019.

\bibitem{Kovchegov:2018zeq}
Yuri~V. Kovchegov and Matthew~D. Sievert.
\newblock {Valence Quark Transversity at Small $x$}.
\newblock {\em Phys. Rev. D}, 99(5):054033, 2019.

\bibitem{Kovchegov:2020kxg}
Yuri~V. Kovchegov and M.~Gabriel Santiago.
\newblock {Lensing mechanism meets small- $x$ physics: Single transverse spin
  asymmetry in $p^{\uparrow}+p$ and $p^{\uparrow}+A$ collisions}.
\newblock {\em Phys. Rev. D}, 102(1):014022, 2020.

\bibitem{Kovchegov:2020hgb}
Yuri~V. Kovchegov and Yossathorn Tawabutr.
\newblock {Helicity at Small $x$: Oscillations Generated by Bringing Back the
  Quarks}.
\newblock {\em JHEP}, 08:014, 2020.

\bibitem{Kovchegov:2021lvz}
Yuri~V. Kovchegov, Andrey Tarasov, and Yossathorn Tawabutr.
\newblock {Helicity evolution at small x: the single-logarithmic contribution}.
\newblock {\em JHEP}, 03:184, 2022.

\bibitem{Kovchegov:2021iyc}
Yuri~V. Kovchegov and M.~Gabriel Santiago.
\newblock {Quark sivers function at small $x$: spin-dependent odderon and the
  sub-eikonal evolution}.
\newblock {\em JHEP}, 11:200, 2021.
\newblock [Erratum: JHEP 09, 186 (2022)].

\bibitem{Cougoulic:2022gbk}
Florian Cougoulic, Yuri~V. Kovchegov, Andrey Tarasov, and Yossathorn Tawabutr.
\newblock {Quark and gluon helicity evolution at small x: revised and updated}.
\newblock {\em JHEP}, 07:095, 2022.
\newblock [Erratum: JHEP 09, 052 (2024)].

\bibitem{Kovchegov:2022kyy}
Yuri~V. Kovchegov and M.~Gabriel Santiago.
\newblock {T-odd leading-twist quark TMDs at small x}.
\newblock {\em JHEP}, 11:098, 2022.

\bibitem{Borden:2023ugd}
Jeremy Borden and Yuri~V. Kovchegov.
\newblock {Analytic solution for the revised helicity evolution at small x and
  large Nc: New resummed gluon-gluon polarized anomalous dimension and
  intercept}.
\newblock {\em Phys. Rev. D}, 108(1):014001, 2023.

\bibitem{Kovchegov:2024aus}
Yuri~V. Kovchegov and Ming Li.
\newblock {Gluon double-spin asymmetry in the longitudinally polarized p + p
  collisions}.
\newblock {\em JHEP}, 05:177, 2024.

\bibitem{Borden:2024bxa}
Jeremy Borden, Yuri~V. Kovchegov, and Ming Li.
\newblock {Helicity evolution at small x: quark to gluon and gluon to quark
  transition operators}.
\newblock {\em JHEP}, 09:037, 2024.

\bibitem{Balitsky:2015qba}
I.~Balitsky and A.~Tarasov.
\newblock {Rapidity evolution of gluon TMD from low to moderate x}.
\newblock {\em JHEP}, 10:017, 2015.

\bibitem{Balitsky:2016dgz}
I.~Balitsky and A.~Tarasov.
\newblock {Gluon TMD in particle production from low to moderate x}.
\newblock {\em JHEP}, 06:164, 2016.

\bibitem{Balitsky:2017flc}
I.~Balitsky and A.~Tarasov.
\newblock {Higher-twist corrections to gluon TMD factorization}.
\newblock {\em JHEP}, 07:095, 2017.

\bibitem{Hatta:2016aoc}
Yoshitaka Hatta, Yuya Nakagawa, Feng Yuan, Yong Zhao, and Bowen Xiao.
\newblock {Gluon orbital angular momentum at small-$x$}.
\newblock {\em Phys. Rev. D}, 95(11):114032, 2017.

\bibitem{Kovchegov:2019rrz}
Yuri~V. Kovchegov.
\newblock {Orbital Angular Momentum at Small $x$}.
\newblock {\em JHEP}, 03:174, 2019.

\bibitem{Boussarie:2019icw}
Renaud Boussarie, Yoshitaka Hatta, and Feng Yuan.
\newblock {Proton Spin Structure at Small-$x$}.
\newblock {\em Phys. Lett. B}, 797:134817, 2019.

\bibitem{Kovchegov:2023yzd}
Yuri~V. Kovchegov and Brandon Manley.
\newblock {Orbital angular momentum at small x revisited}.
\newblock {\em JHEP}, 02:060, 2024.
\newblock [Erratum: JHEP 08, 140 (2024)].

\bibitem{Kovchegov:2024wjs}
Yuri~V. Kovchegov and Brandon Manley.
\newblock {Elastic dijet production in electron scattering on a longitudinally
  polarized proton at small x: A portal to orbital angular momentum
  distributions}.
\newblock {\em Phys. Rev. D}, 111(5):054017, 2025.

\bibitem{Cougoulic:2020tbc}
Florian Cougoulic and Yuri~V. Kovchegov.
\newblock {Helicity-dependent extension of the McLerran{\textendash}Venugopalan
  model}.
\newblock {\em Nucl. Phys. A}, 1004:122051, 2020.

\bibitem{Li:2024fdb}
Ming Li.
\newblock {Quasiclassical Gluon Fields and Low{\textquoteright}s Soft Theorem
  at Small Momentum-Fraction x}.
\newblock {\em Phys. Rev. Lett.}, 133(2):021902, 2024.

\bibitem{Agostini:2025vvx}
Pedro Agostini, Tolga Altinoluk, N{\'e}stor Armesto, Guillaume Beuf, Florian
  Cougoulic, and Swaleha Mulani.
\newblock {Dijet production in DIS off a large nucleus at next-to-eikonal
  accuracy in a Gaussian model within the CGC framework}.
\newblock {\em Phys. Rev. D}, 113(5):054035, 2026.

\bibitem{Mukherjee:2026cte}
Swagato Mukherjee, Vladimir~V. Skokov, Andrey Tarasov, Shaswat Tiwari, and Fei
  Yao.
\newblock {Back-to-back dijet production in DIS at arbitrary Bjorken-x: TMD
  gluon distributions to twist-3 accuracy}.
\newblock arXiv:2602.15137.

\bibitem{Dominguez:2011wm}
Fabio Dominguez, Cyrille Marquet, Bo-Wen Xiao, and Feng Yuan.
\newblock {Universality of Unintegrated Gluon Distributions at small x}.
\newblock {\em Phys. Rev. D}, 83:105005, 2011.

\bibitem{Metz:2011wb}
Andreas Metz and Jian Zhou.
\newblock {Distribution of linearly polarized gluons inside a large nucleus}.
\newblock {\em Phys. Rev. D}, 84:051503, 2011.

\bibitem{Dumitru:2015gaa}
Adrian Dumitru, Tuomas Lappi, and Vladimir Skokov.
\newblock {Distribution of Linearly Polarized Gluons and Elliptic Azimuthal
  Anisotropy in Deep Inelastic Scattering Dijet Production at High Energy}.
\newblock {\em Phys. Rev. Lett.}, 115(25):252301, 2015.

\bibitem{Marquet:2016cgx}
C.~Marquet, E.~Petreska, and C.~Roiesnel.
\newblock {Transverse-momentum-dependent gluon distributions from JIMWLK
  evolution}.
\newblock {\em JHEP}, 10:065, 2016.

\bibitem{KovnerAndCo}
Alexander Kovner and et~al.
\newblock {MV model for the sub-eikonal corrections in the dense regime}.
\newblock in preparation, 2026.

\bibitem{Armesto:2019mna}
Nestor Armesto, Fabio Dominguez, Alex Kovner, Michael Lublinsky, and Vladimir
  Skokov.
\newblock {The Color Glass Condensate density matrix: Lindblad evolution,
  entanglement entropy and Wigner functional}.
\newblock {\em JHEP}, 05:025, 2019.

\bibitem{Li:2020bys}
Ming Li and Alex Kovner.
\newblock {JIMWLK Evolution, Lindblad Equation and Quantum-Classical
  Correspondence}.
\newblock {\em JHEP}, 05:036, 2020.

\bibitem{Lidar:2019qog}
Daniel~A. Lidar.
\newblock {Lecture Notes on the Theory of Open Quantum Systems}.
\newblock arXiv:1902.00967.

\bibitem{Agrawal:2026kis}
Anjali~A. Agrawal, Evan Budd, Alexander~F. Kemper, Vladimir~V. Skokov, Andrey
  Tarasov, and Shaswat Tiwari.
\newblock {JIMWLK on a quantum computer}.
\newblock arXiv:2603.02516.

\bibitem{Raychowdhury:2019iki}
Indrakshi Raychowdhury and Jesse~R. Stryker.
\newblock {Loop, string, and hadron dynamics in SU(2) Hamiltonian lattice gauge
  theories}.
\newblock {\em Phys. Rev. D}, 101(11):114502, 2020.

\bibitem{Klco:2019evd}
Natalie Klco, Jesse~R. Stryker, and Martin~J. Savage.
\newblock {SU(2) non-Abelian gauge field theory in one dimension on digital
  quantum computers}.
\newblock {\em Phys. Rev. D}, 101(7):074512, 2020.

\bibitem{Ciavarella:2024fzw}
Anthony~N. Ciavarella and Christian~W. Bauer.
\newblock {Quantum Simulation of SU(3) Lattice Yang-Mills Theory at Leading
  Order in Large-Nc Expansion}.
\newblock {\em Phys. Rev. Lett.}, 133(11):111901, 2024.

\bibitem{Byrnes:2005qx}
Tim Byrnes and Yoshihisa Yamamoto.
\newblock {Simulating lattice gauge theories on a quantum computer}.
\newblock {\em Phys. Rev. A}, 73:022328, 2006.

\bibitem{Schlimgen:2022aji}
Anthony~W. Schlimgen, Kade Head-Marsden, LeeAnn~M. Sager, Prineha Narang, and
  David~A. Mazziotti.
\newblock {Quantum simulation of the Lindblad equation~using a unitary
  decomposition of operators}.
\newblock {\em Phys. Rev. Res.}, 4(2):023216, 2022.

\bibitem{Schlimgen:2021hxs}
Anthony~W. Schlimgen, Kade Head-Marsden, LeeAnn~M. Sager, Prineha Narang, and
  David~A. Mazziotti.
\newblock {Quantum Simulation of Open Quantum Systems Using a Unitary
  Decomposition of Operators}.
\newblock {\em Phys. Rev. Lett.}, 127:270503, 2021.

\bibitem{STAR:2006dgg}
J.~Adams et~al.
\newblock {Forward neutral pion production in p+p and d+Au collisions at
  s(NN)**(1/2) = 200-GeV}.
\newblock {\em Phys. Rev. Lett.}, 97:152302, 2006.

\bibitem{PHENIX:2011puq}
A.~Adare et~al.
\newblock {Suppression of back-to-back hadron pairs at forward rapidity in
  $d+$Au Collisions at $\sqrt{s_{NN}}=200$ GeV}.
\newblock {\em Phys. Rev. Lett.}, 107:172301, 2011.

\bibitem{STAR:2021fgw}
M.~S. Abdallah et~al.
\newblock {Evidence for Nonlinear Gluon Effects in QCD and Their Mass Number
  Dependence at STAR}.
\newblock {\em Phys. Rev. Lett.}, 129(9):092501, 2022.

\bibitem{LHCb:2015coe}
Roel Aaij et~al.
\newblock {Measurements of long-range near-side angular correlations in
  $\sqrt{s_{\mathrm{NN}}}=5$TeV proton-lead collisions in the forward region}.
\newblock {\em Phys. Lett. B}, 762:473--483, 2016.

\bibitem{ATLAS:2019jgo}
Morad Aaboud et~al.
\newblock {Dijet azimuthal correlations and conditional yields in pp and p+Pb
  collisions at sNN=5.02TeV with the ATLAS detector}.
\newblock {\em Phys. Rev. C}, 100(3):034903, 2019.

\bibitem{Adare:2011sc}
A.~Adare et~al.
\newblock {Suppression of back-to-back hadron pairs at forward rapidity in
  $d+$Au Collisions at $\sqrt{s_{NN}}=200$ GeV}.
\newblock {\em Phys. Rev. Lett.}, 107:172301, 2011.

\bibitem{Cassar:2025vdp}
Kiera Cassar, Zhen Wang, Xiaoxuan Chu, and Elke-Caroline Aschenauer.
\newblock {Investigating the broadening phenomenon in two-particle correlations
  induced by gluon saturation}.
\newblock {\em Phys. Rev. D}, 112(3):034034, 2025.

\bibitem{Zheng:2014vka}
L.~Zheng, E.~C. Aschenauer, J.~H. Lee, and Bo-Wen Xiao.
\newblock {Probing Gluon Saturation through Dihadron Correlations at an
  Electron-Ion Collider}.
\newblock {\em Phys. Rev. D}, 89(7):074037, 2014.

\bibitem{Caucal:2023fsf}
Paul Caucal, Farid Salazar, Bj{\"o}rn Schenke, Tomasz Stebel, and Raju
  Venugopalan.
\newblock {Back-to-Back Inclusive Dijets in Deep Inelastic Scattering at Small
  x: Complete NLO Results and Predictions}.
\newblock {\em Phys. Rev. Lett.}, 132(8):081902, 2024.

\bibitem{Kharzeev:2004bw}
Dmitri Kharzeev, Eugene Levin, and Larry McLerran.
\newblock {Jet azimuthal correlations and parton saturation in the color glass
  condensate}.
\newblock {\em Nucl. Phys. A}, 748:627--640, 2005.

\bibitem{Stasto:2011ru}
Anna Stasto, Bo-Wen Xiao, and Feng Yuan.
\newblock {Back-to-Back Correlations of Di-hadrons in dAu Collisions at RHIC}.
\newblock {\em Phys. Lett. B}, 716:430--434, 2012.

\bibitem{Stasto:2018rci}
Anna Stasto, Shu-Yi Wei, Bo-Wen Xiao, and Feng Yuan.
\newblock {On the Dihadron Angular Correlations in Forward $pA$ collisions}.
\newblock {\em Phys. Lett. B}, 784:301--306, 2018.

\bibitem{Lappi:2012nh}
T.~Lappi and H.~Mantysaari.
\newblock {Forward dihadron correlations in deuteron-gold collisions with the
  Gaussian approximation of JIMWLK}.
\newblock {\em Nucl. Phys. A}, 908:51--72, 2013.

\bibitem{Albacete:2010pg}
Javier~L. Albacete and Cyrille Marquet.
\newblock {Azimuthal correlations of forward di-hadrons in d+Au collisions at
  RHIC in the Color Glass Condensate}.
\newblock {\em Phys. Rev. Lett.}, 105:162301, 2010.

\bibitem{Albacete:2018ruq}
Javier~L. Albacete, Giuliano Giacalone, Cyrille Marquet, and Marek Matas.
\newblock {Forward dihadron back-to-back correlations in $pA$ collisions}.
\newblock {\em Phys. Rev. D}, 99(1):014002, 2019.

\bibitem{Caucal:2025zkl}
Paul Caucal, Zhong-Bo Kang, Piotr Korcyl, Farid Salazar, Bj{\"o}rn Schenke,
  Tomasz Stebel, Raju Venugopalan, and Wenbin Zhao.
\newblock {Probing gluon saturation with forward di-hadron correlations in
  proton-nucleus collisions}.
\newblock arXiv:2512.21466.

\bibitem{EuropeanMuon:1987isl}
J.~Ashman et~al.
\newblock A measurement of the spin asymmetry and determination of the
  structure function g1 in deep inelastic muon-proton scattering.
\newblock {\em Physics Letters B}, 206(2):364--370, 1988.

\bibitem{Jaffe:1989jz}
R.~L. Jaffe and Aneesh Manohar.
\newblock {The G(1) Problem: Fact and Fantasy on the Spin of the Proton}.
\newblock {\em Nucl. Phys.}, B337:509--546, 1990.

\bibitem{Ji:1996ek}
Xiang-Dong Ji.
\newblock {Gauge-Invariant Decomposition of Nucleon Spin}.
\newblock {\em Phys. Rev. Lett.}, 78:610--613, 1997.

\bibitem{Aidala:2012mv}
Christine~A. Aidala, Steven~D. Bass, Delia Hasch, and Gerhard~K. Mallot.
\newblock {The Spin Structure of the Nucleon}.
\newblock {\em Rev. Mod. Phys.}, 85:655--691, 2013.

\bibitem{Leader:2013jra}
E.~Leader and C.~Lorcé.
\newblock {The angular momentum controversy: What's it all about and does it
  matter?}
\newblock {\em Phys. Rept.}, 541(3):163--248, 2014.

\bibitem{Aschenauer:2013woa}
E.~C. Aschenauer et~al.
\newblock {The RHIC Spin Program: Achievements and Future Opportunities}.
\newblock arXiv:1304.0079.

\bibitem{Aschenauer:2015eha}
Elke-Caroline Aschenauer et~al.
\newblock {The RHIC SPIN Program: Achievements and Future Opportunities}.
\newblock arXiv:1501.01220.

\bibitem{Proceedings:2020eah}
Alexei Prokudin, Yoshitaka Hatta, Yuri Kovchegov, and Cyrille Marquet, editors.
\newblock {\em {Proceedings, Probing Nucleons and Nuclei in High Energy
  Collisions:} {Dedicated to the Physics of the Electron Ion Collider}:
  {Seattle (WA), United States, Oct. 2018}}. WSP, 2020.

\bibitem{Bartels:1995iu}
Jochen Bartels, B.I. Ermolaev, and M.G. Ryskin.
\newblock {Nonsinglet contributions to the structure function g1 at small x}.
\newblock {\em Z.Phys.}, C70:273--280, 1996.

\bibitem{Bartels:1996wc}
Jochen Bartels, B.~I. Ermolaev, and M.~G. Ryskin.
\newblock {Flavor singlet contribution to the structure function G(1) at small
  x}.
\newblock {\em Z. Phys. C}, 72:627--635, 1996.

\bibitem{Gorshkov:1966ht}
V.~G. Gorshkov, V.~N. Gribov, L.~N. Lipatov, and G.~V. Frolov.
\newblock {Doubly logarithmic asymptotic behavior in quantum electrodynamics}.
\newblock {\em Sov. J. Nucl. Phys.}, 6:95, 1968.
\newblock [Yad. Fiz.6,129(1967)].

\bibitem{Kirschner:1983di}
R.~Kirschner and L.n. Lipatov.
\newblock {Double Logarithmic Asymptotics and Regge Singularities of Quark
  Amplitudes with Flavor Exchange}.
\newblock {\em Nucl.Phys.}, B213:122--148, 1983.

\bibitem{Kirschner:1994rq}
R.~Kirschner.
\newblock {Reggeon interactions in perturbative QCD}.
\newblock {\em Z.Phys.}, C65:505--510, 1995.

\bibitem{Kirschner:1994vc}
R.~Kirschner.
\newblock {Regge asymptotics of scattering with flavor exchange in QCD}.
\newblock {\em Z.Phys.}, C67:459--466, 1995.

\bibitem{Blumlein:1996hb}
J.~Bl{\"u}mlein and A.~Vogt.
\newblock {The Singlet contribution to the structure function g1 (x, Q**2) at
  small x}.
\newblock {\em Phys. Lett. B}, 386:350--358, 1996.

\bibitem{Griffiths:1999dj}
S.~Griffiths and D.A. Ross.
\newblock {Studying the perturbative Reggeon}.
\newblock {\em Eur.Phys.J.}, C12:277--286, 2000.

\bibitem{Blumlein:1995jp}
J.~Blumlein and A.~Vogt.
\newblock {On the behavior of nonsinglet structure functions at small x}.
\newblock {\em Phys. Lett. B}, 370:149--155, 1996.

\bibitem{Ermolaev:1999jx}
B.~I. Ermolaev, Mario Greco, and S.~I. Troian.
\newblock {QCD running coupling effects for the nonsinglet structure function
  at small $x$}.
\newblock {\em Nucl. Phys.}, B571:137--150, 2000.

\bibitem{Ermolaev:2000sg}
B.~I. Ermolaev, Mario Greco, and S.~I. Troyan.
\newblock {Intercepts of the nonsinglet structure functions}.
\newblock {\em Nucl. Phys.}, B594:71--88, 2001.

\bibitem{Ermolaev:2003zx}
B.~I. Ermolaev, Mario Greco, and S.~I. Troyan.
\newblock {Running coupling effects for the singlet structure function $g_1$ at
  small $x$}.
\newblock {\em Phys. Lett.}, B579:321--330, 2004.

\bibitem{Ermolaev:2009cq}
B.~I. Ermolaev, M.~Greco, and S.~I. Troyan.
\newblock {Overview of the spin structure function $g_1$ at arbitrary $x$ and
  $Q^2$}.
\newblock {\em Riv. Nuovo Cim.}, 33:57--122, 2010.

\bibitem{Cougoulic:2019aja}
Florian Cougoulic and Yuri~V. Kovchegov.
\newblock {Helicity-dependent generalization of the JIMWLK evolution}.
\newblock {\em Phys. Rev.}, D100(11):114020, 2019.

\bibitem{Chirilli:2021lif}
Giovanni~Antonio Chirilli.
\newblock {High-energy operator product expansion at sub-eikonal level}.
\newblock {\em JHEP}, 06:096, 2021.

\bibitem{Adamiak:2023okq}
Daniel Adamiak, Yuri~V. Kovchegov, and Yossathorn Tawabutr.
\newblock {Helicity evolution at small x: Revised asymptotic results at large
  Nc and Nf}.
\newblock {\em Phys. Rev. D}, 108(5):054005, 2023.

\bibitem{Borden:2025ehe}
Jeremy Borden and Yuri~V. Kovchegov.
\newblock {Analytic solution for the helicity evolution equations at small x
  and large Nc and Nf}.
\newblock {\em Phys. Rev. D}, 113(3):034002, 2026.

\bibitem{Mueller:1994rr}
Alfred~H. Mueller.
\newblock Soft gluons in the infinite momentum wave function and the {BFKL}
  pomeron.
\newblock {\em Nucl. Phys.}, B415:373--385, 1994.

\bibitem{Mueller:1994jq}
Alfred~H. Mueller and Bimal Patel.
\newblock Single and double {BFKL} pomeron exchange and a dipole picture of
  high-energy hard processes.
\newblock {\em Nucl. Phys.}, B425:471--488, 1994.

\bibitem{Mueller:1995gb}
Alfred~H. Mueller.
\newblock Unitarity and the {BFKL} pomeron.
\newblock {\em Nucl. Phys.}, B437:107--126, 1995.

\bibitem{Balitsky:1998ya}
Ian Balitsky.
\newblock Factorization and high-energy effective action.
\newblock {\em Phys. Rev.}, D60:014020, 1999.

\bibitem{Kovchegov:1999ua}
Yuri~V. Kovchegov.
\newblock Unitarization of the {BFKL} pomeron on a nucleus.
\newblock {\em Phys. Rev.}, D61:074018, 2000.

\bibitem{Jalilian-Marian:1997dw}
Jamal Jalilian-Marian, Alex Kovner, and Heribert Weigert.
\newblock The {Wilson} renormalization group for low x physics: Gluon evolution
  at finite parton density.
\newblock {\em Phys. Rev.}, D59:014015, 1998.

\bibitem{Jalilian-Marian:1997gr}
Jamal Jalilian-Marian, Alex Kovner, Andrei Leonidov, and Heribert Weigert.
\newblock The {Wilson} renormalization group for low x physics: Towards the
  high density regime.
\newblock {\em Phys. Rev.}, D59:014014, 1998.

\bibitem{Chirilli:2018kkw}
Giovanni~Antonio Chirilli.
\newblock {Sub-eikonal corrections to scattering amplitudes at high energy}.
\newblock {\em JHEP}, 01:118, 2019.

\bibitem{Jalilian-Marian:2018iui}
Jamal Jalilian-Marian.
\newblock {Quark jets scattering from a gluon field: from saturation to high
  $p_t$}.
\newblock {\em Phys. Rev.}, D99(1):014043, 2019.

\bibitem{Jalilian-Marian:2019kaf}
Jamal Jalilian-Marian.
\newblock {Rapidity loss, spin, and angular asymmetries in the scattering of a
  quark from the color field of a proton or nucleus}.
\newblock {\em Phys. Rev. D}, 102(1):014008, 2020.

\bibitem{Boussarie:2020vzf}
Renaud Boussarie and Yacine Mehtar-Tani.
\newblock {Gauge invariance of transverse momentum dependent distributions at
  small $x$}.
\newblock {\em Phys. Rev. D}, 103(9):094012, 2021.

\bibitem{Boussarie:2020fpb}
Renaud Boussarie and Yacine Mehtar-Tani.
\newblock {A novel formulation of the unintegrated gluon distribution for DIS}.
\newblock {\em Phys. Lett. B}, 831:137125, 2022.

\bibitem{Altinoluk:2023dww}
Tolga Altinoluk, Guillaume Beuf, and Jamal Jalilian-Marian.
\newblock {Renormalization of the gluon distribution function in the background
  field formalism}.
\newblock {\em Phys. Rev. D}, 112(3):034021, 2025.

\bibitem{Li:2023tlw}
Ming Li.
\newblock {Small x physics beyond eikonal approximation: an effective
  Hamiltonian approach}.
\newblock {\em JHEP}, 07:158, 2023.

\bibitem{Altinoluk:2025ewj}
Tolga Altinoluk, Guillaume Beuf, and Jamal Jalilian-Marian.
\newblock {One-loop renormalization of quark TMD in the light-cone gauge: CSS
  evolution}.
\newblock arXiv:2505.20467.

\bibitem{Mertig:1995ny}
R.~Mertig and W.~L. van Neerven.
\newblock {The Calculation of the two loop spin splitting functions
  P(ij)(1)(x)}.
\newblock {\em Z. Phys. C}, 70:637--654, 1996.

\bibitem{Moch:1999eb}
S.~Moch and J.~A.~M. Vermaseren.
\newblock {Deep inelastic structure functions at two loops}.
\newblock {\em Nucl. Phys. B}, 573:853--907, 2000.

\bibitem{vanNeerven:2000uj}
W.~L. van Neerven and A.~Vogt.
\newblock {NNLO evolution of deep inelastic structure functions: The Singlet
  case}.
\newblock {\em Nucl. Phys. B}, 588:345--373, 2000.

\bibitem{Blumlein:2021ryt}
J.~Bl\"umlein, P.~Marquard, C.~Schneider, and K.~Sch\"onwald.
\newblock {The three-loop polarized singlet anomalous dimensions from off-shell
  operator matrix elements}.
\newblock {\em JHEP}, 01:193, 2022.

\bibitem{Blumlein:2021lmf}
J.~Bl\"umlein and M.~Saragnese.
\newblock {The N3LO scheme-invariant QCD evolution of the non-singlet structure
  functions F2NS(x,Q2) and g1NS(x,Q2)}.
\newblock {\em Phys. Lett. B}, 820:136589, 2021.

\bibitem{Davies:2022ofz}
J.~Davies, C.~H. Kom, S.~Moch, and A.~Vogt.
\newblock {Resummation of small-x double logarithms in QCD: inclusive
  deep-inelastic scattering}.
\newblock {\em JHEP}, 08:135, 2022.

\bibitem{Adamiak:2025mdy}
Daniel Adamiak, Heikki M{\"a}ntysaari, and Yossathorn Tawabutr.
\newblock {Proton spin from small-x with constraints from the valence quark
  model}.
\newblock {\em Phys. Lett. B}, 870:139911, 2025.

\bibitem{Dumitru:2024pcv}
Adrian Dumitru, Heikki M{\"a}ntysaari, and Yossathorn Tawabutr.
\newblock {Polarized dipole scattering amplitudes meet the valence quark
  model}.
\newblock {\em Phys. Rev. D}, 110(5):054030, 2024.

\bibitem{Adamiak:2025dpw}
Daniel Adamiak, Nicholas Baldonado, Yuri~V. Kovchegov, Ming Li, W.~Melnitchouk,
  Daniel Pitonyak, Nobuo Sato, Matthew~D. Sievert, Andrey Tarasov, and
  Yossathorn Tawabutr.
\newblock {First study of polarized proton-proton scattering with small-x
  helicity evolution}.
\newblock {\em Phys. Rev. D}, 112(9):094032, 2025.

\bibitem{Balitsky:1987bk}
I.~I. Balitsky and Vladimir~M. Braun.
\newblock {Evolution Equations for QCD String Operators}.
\newblock {\em Nucl. Phys. B}, 311:541--584, 1989.

\bibitem{Kuraev:1998ht}
E.~V. Kuraev, Nikolai~N. Nikolaev, and B.~G. Zakharov.
\newblock {Diffractive vector mesons beyond the s channel helicity
  conservation}.
\newblock {\em JETP Lett.}, 68:696--703, 1998.

\bibitem{Ivanov:2004ax}
I.~P. Ivanov, N.~N. Nikolaev, and A.~A. Savin.
\newblock {Diffractive vector meson production at HERA: From soft to hard QCD}.
\newblock {\em Phys. Part. Nucl.}, 37:1--85, 2006.

\bibitem{Boussarie:2019vmk}
Renaud Boussarie, Yoshitaka Hatta, Lech Szymanowski, and Samuel Wallon.
\newblock {Probing the Gluon Sivers Function with an Unpolarized Target: GTMD
  Distributions and the Odderons}.
\newblock {\em Phys. Rev. Lett.}, 124(17):172501, 2020.

\bibitem{Hagiwara:2020mqb}
Yoshikazu Hagiwara, Yoshitaka Hatta, Roman Pasechnik, and Jian Zhou.
\newblock {Spin-dependent Pomeron and Odderon in elastic proton-proton
  scattering}.
\newblock {\em Eur. Phys. J. C}, 80(5):427, 2020.

\bibitem{Koempel:2011rc}
John Koempel, Peter Kroll, Andreas Metz, and Jian Zhou.
\newblock {Exclusive production of quarkonia as a probe of the GPD E for
  gluons}.
\newblock {\em Phys. Rev. D}, 85:051502, 2012.

\bibitem{Hatta:2022bxn}
Yoshitaka Hatta and Jian Zhou.
\newblock {Small-$x$ evolution of the gluon GPD $E_g$}.
\newblock {\em Phys. Rev. Lett.}, 129(25):252002, 2022.

\bibitem{Benic:2025ral}
Sanjin Beni{\'c} and Adrian Dumitru.
\newblock {Off forward non-s-channel helicity conserving contributions to
  exclusive vector quarkonium production from the spin dependent BFKL Pomeron}.
\newblock {\em Phys. Rev. D}, 112(3):034025, 2025.

\bibitem{Mantysaari:2025mht}
Heikki M{\"a}ntysaari, Yossathorn Tawabutr, and Xuan-Bo Tong.
\newblock {Nucleon energy correlators for the odderon}.
\newblock {\em Phys. Rev. D}, 112(11):114027, 2025.

\bibitem{Ma:2003py}
J.~P. Ma.
\newblock {Diffractive photoproduction of eta(c)}.
\newblock {\em Nucl. Phys. A}, 727:333--352, 2003.

\bibitem{Boer:2015pni}
Dani\"el Boer, Miguel~G. Echevarria, Piet Mulders, and Jian Zhou.
\newblock {Single spin asymmetries from a single Wilson loop}.
\newblock {\em Phys. Rev. Lett.}, 116(12):122001, 2016.

\bibitem{Benic:2024fbf}
Sanjin Benic, Adrian Dumitru, Leszek Motyka, and Tomasz Stebel.
\newblock {Gluon Sivers function from forward exclusive \ensuremath{\chi}c1
  photoproduction on unpolarized protons}.
\newblock {\em Phys. Rev. D}, 111(5):054008, 2025.

\bibitem{Lappi:2010ek}
T.~Lappi.
\newblock {Small x physics and RHIC data}.
\newblock {\em Int. J. Mod. Phys.}, E20(1):1--43, 2011.

\bibitem{Albacete:2014fwa}
Javier~L. Albacete and Cyrille Marquet.
\newblock {Gluon saturation and initial conditions for relativistic heavy ion
  collisions}.
\newblock {\em Prog.Part.Nucl.Phys.}, 76:1--42, 2014.

\bibitem{Stasto:2000er}
A.~M. Stasto, K.~Golec-Biernat, and J.~Kwiecinski.
\newblock Geometric scaling for the total $\gamma^* p$ cross-section in the low
  x region.
\newblock {\em Phys. Rev. Lett.}, 86:596--599, 2001.

\bibitem{Iancu:2002xk}
Edmond Iancu, Andrei Leonidov, and Larry McLerran.
\newblock The colour glass condensate: An introduction.
\newblock hep-ph/0202270.

\bibitem{Marquet:2007vb}
Cyrille Marquet.
\newblock {Forward inclusive dijet production and azimuthal correlations in pA
  collisions}.
\newblock {\em Nucl. Phys.}, A796:41--60, 2007.

\bibitem{ALICE:2015ppz}
Jaroslav Adam et~al.
\newblock {Measurement of dijet $k_T$ in p{\textendash}Pb collisions at
  $\sqrt{s}_{NN}$=5.02 TeV}.
\newblock {\em Phys. Lett. B}, 746:385--395, 2015.

\bibitem{McLerran:1998nk}
Larry~D. McLerran and Raju Venugopalan.
\newblock {Fock space distributions, structure functions, higher twists and
  small x}.
\newblock {\em Phys. Rev.}, D59:094002, 1999.

\bibitem{Mueller:1999wm}
Alfred~H. Mueller.
\newblock Parton saturation at small x and in large nuclei.
\newblock {\em Nucl. Phys.}, B558:285--303, 1999.

\bibitem{Marquet:2009ca}
Cyrille Marquet, Bo-Wen Xiao, and Feng Yuan.
\newblock {Semi-inclusive Deep Inelastic Scattering at small x}.
\newblock {\em Phys. Lett.}, B682:207--211, 2009.

\bibitem{Kowalski:2008sa}
H.~Kowalski, T.~Lappi, C.~Marquet, and R.~Venugopalan.
\newblock {Nuclear enhancement and suppression of diffractive structure
  functions at high energies}.
\newblock {\em Phys. Rev. C}, 78:045201, 2008.

\bibitem{Golec-Biernat:1998zce}
Krzysztof~J. Golec-Biernat and M.~Wusthoff.
\newblock {Saturation effects in deep inelastic scattering at low Q**2 and its
  implications on diffraction}.
\newblock {\em Phys. Rev. D}, 59:014017, 1998.

\bibitem{Golec-Biernat:1999qor}
Krzysztof~J. Golec-Biernat and M.~Wusthoff.
\newblock {Saturation in diffractive deep inelastic scattering}.
\newblock {\em Phys. Rev. D}, 60:114023, 1999.

\bibitem{Dominguez:2010xd}
Fabio Dominguez, Bo-Wen Xiao, and Feng Yuan.
\newblock {$k_t$-factorization for Hard Processes in Nuclei}.
\newblock {\em Phys.Rev.Lett.}, 106:022301, 2011.

\bibitem{Dominguez:2011br}
Fabio Dominguez, Jian-Wei Qiu, Bo-Wen Xiao, and Feng Yuan.
\newblock {On the linearly polarized gluon distributions in the color dipole
  model}.
\newblock {\em Phys.Rev.}, D85:045003, 2012.

\bibitem{Xiao:2017yya}
Bo-Wen Xiao, Feng Yuan, and Jian Zhou.
\newblock {Transverse Momentum Dependent Parton Distributions at Small-x}.
\newblock {\em Nucl. Phys. B}, 921:104--126, 2017.

\bibitem{Marquet:2017xwy}
Cyrille Marquet, Claude Roiesnel, and Pieter Taels.
\newblock {Linearly polarized small-$x$ gluons in forward heavy-quark pair
  production}.
\newblock {\em Phys. Rev.}, D97(1):014004, 2018.

\bibitem{Caucal:2025xxh}
Paul Caucal, Marcos~Guerrero Morales, Edmond Iancu, Farid Salazar, and Feng
  Yuan.
\newblock {Unveiling the sea: universality of the transverse momentum dependent
  quark distributions at small x}.
\newblock {\em Phys. Lett. B}, 874:140271, 2026.

\bibitem{Sterman:1993hfp}
George~F. Sterman.
\newblock {\em {An Introduction to quantum field theory}}.
\newblock Cambridge University Press, 8 1993.

\bibitem{Bhattacharya:2025fnz}
Shohini Bhattacharya, Chuan-Qi He, Zhong-Bo Kang, Diego Padilla, and Jani
  Penttala.
\newblock {Parton distributions in the shockwave formalism}.
\newblock arXiv:2510.02254.

\bibitem{Kovchegov:2025yyl}
Yuri~V. Kovchegov, M.~Gabriel Santiago, and Huachen Sun.
\newblock {Unpolarized GPDs at small $x$ and non-zero skewness}.
\newblock arXiv:2512.10086.

\bibitem{Benic:2026idy}
Sanjin Beni{\'c}, Yoshikazu Hagiwara, Boris {\v{S}}ari{\'c}, and Eric~Andreas
  Vivoda.
\newblock {Generalized transverse momentum distributions at small-$x$}.
\newblock arXiv:2603.06092.

\bibitem{Ayala:2016lhd}
A.~Ayala, M.~Hentschinski, J.~Jalilian-Marian, and M.~E. Tejeda-Yeomans.
\newblock {Polarized 3 parton production in inclusive DIS at small x}.
\newblock {\em Phys. Lett. B}, 761:229--233, 2016.

\bibitem{Ayala:2017rmh}
Alejandro Ayala, Martin Hentschinski, Jamal Jalilian-Marian, and Maria~Elena
  Tejeda-Yeomans.
\newblock {Spinor helicity methods in high-energy factorization: efficient
  momentum-space calculations in the Color Glass Condensate formalism}.
\newblock {\em Nucl. Phys. B}, 920:232--255, 2017.

\bibitem{Iancu:2020mos}
Edmond Iancu and Yair Mulian.
\newblock {Forward dijets in proton-nucleus collisions at next-to-leading
  order: the real corrections}.
\newblock {\em JHEP}, 03:005, 2021.

\bibitem{Iancu:2022gpw}
Edmond Iancu and Yair Mulian.
\newblock {Dihadron production in DIS at NLO: the real corrections}.
\newblock {\em JHEP}, 07:121, 2023.

\bibitem{Fucilla:2023mkl}
Michael Fucilla, Andrey Grabovsky, Emilie Li, Lech Szymanowski, and Samuel
  Wallon.
\newblock {Diffractive single hadron production in a saturation framework at
  the NLO}.
\newblock {\em JHEP}, 02:165, 2024.

\bibitem{Caucal:2025mth}
Paul Caucal, Edmond Iancu, Farid Salazar, and Feng Yuan.
\newblock {Gluon splitting at small $x$: a unified derivation for the JIMWLK,
  DGLAP and CSS equations}.
\newblock arXiv:2510.08454.

\bibitem{vanHameren:2016ftb}
A.~van Hameren, P.~Kotko, K.~Kutak, C.~Marquet, E.~Petreska, and S.~Sapeta.
\newblock {Forward di-jet production in p+Pb collisions in the small-x improved
  TMD factorization framework}.
\newblock {\em JHEP}, 12:034, 2016.
\newblock [Erratum: JHEP 02, 158 (2019)].

\bibitem{vanHameren:2019ysa}
Andreas van Hameren, Piotr Kotko, Krzysztof Kutak, and Sebastian Sapeta.
\newblock {Broadening and saturation effects in dijet azimuthal correlations in
  p-p and p-Pb collisions at $\mathbf{\sqrt{s}} = $ 5.02 TeV}.
\newblock {\em Phys. Lett. B}, 795:511--515, 2019.

\bibitem{Al-Mashad:2022zbq}
M.~Abdullah Al-Mashad, A.~van Hameren, H.~Kakkad, P.~Kotko, K.~Kutak, P.~van
  Mechelen, and S.~Sapeta.
\newblock {Dijet azimuthal correlations in p-p and p-Pb collisions at forward
  LHC calorimeters}.
\newblock {\em JHEP}, 12:131, 2022.

\bibitem{vanHameren:2023oiq}
A.~van Hameren, H.~Kakkad, P.~Kotko, K.~Kutak, and S.~Sapeta.
\newblock {Searching for saturation in forward dijet production at the LHC}.
\newblock {\em Eur. Phys. J. C}, 83(10):947, 2023.

\bibitem{Iancu:2025jsu}
E.~Iancu, D.~N. Triantafyllopoulos, S.~Y. Wei, and F.~Yuan.
\newblock {The quantum evolutions of the diffractive transverse-momentum
  dependent gluon distribution}.
\newblock arXiv:2512.11730.

\bibitem{Braidot:2010zh}
Ermes Braidot.
\newblock {Suppression of Forward Pion Correlations in d+Au Interactions at
  STAR}.
\newblock hep-ph/1005.2378.

\bibitem{Ewerz:2003xi}
Carlo Ewerz.
\newblock {The Odderon in quantum chromodynamics}.
\newblock hep-ph/0306137.

\bibitem{Kovchegov:2003dm}
Yuri~V. Kovchegov, Lech Szymanowski, and Samuel Wallon.
\newblock {Perturbative odderon in the dipole model}.
\newblock {\em Phys. Lett. B}, 586:267--281, 2004.

\bibitem{Hatta:2005as}
Y.~Hatta, E.~Iancu, K.~Itakura, and L.~McLerran.
\newblock {Odderon in the color glass condensate}.
\newblock {\em Nucl. Phys. A}, 760:172--207, 2005.

\bibitem{Lappi:2016gqe}
T.~Lappi, A.~Ramnath, K.~Rummukainen, and H.~Weigert.
\newblock {JIMWLK evolution of the odderon}.
\newblock {\em Phys. Rev. D}, 94(5):054014, 2016.

\bibitem{Yao:2018vcg}
Xiaojun Yao, Yoshikazu Hagiwara, and Yoshitaka Hatta.
\newblock {Computing the gluon Sivers function at small-$x$}.
\newblock {\em Phys. Lett. B}, 790:361--366, 2019.

\bibitem{Benic:2024pqe}
Sanjin Beni{\'c}, Adrian Dumitru, Abhiram Kaushik, Leszek Motyka, and Tomasz
  Stebel.
\newblock {Photon-odderon interference in exclusive {\ensuremath{\chi}}c
  charmonium production at the Electron-Ion Collider}.
\newblock {\em Phys. Rev. D}, 110(1):014025, 2024.

\bibitem{Dumitru:2019qec}
Adrian Dumitru and Tomasz Stebel.
\newblock {Multiquark matrix elements in the proton and three gluon exchange
  for exclusive $\eta_c$ production in photon-proton diffractive scattering}.
\newblock {\em Phys. Rev. D}, 99(9):094038, 2019.

\bibitem{Dumitru:2022ooz}
Adrian Dumitru, Heikki M{\"a}ntysaari, and Risto Paatelainen.
\newblock {Stronger C-odd color charge correlations in the proton at higher
  energy}.
\newblock {\em Phys. Rev. D}, 107(1):L011501, 2023.
\newblock [Erratum: Phys.Rev.D 111, 059902 (2025)].

\bibitem{Siddikov:2025orq}
M.~Siddikov, I.~Zemlyakov, and M.~Roa.
\newblock {Exclusive photoproduction of
  {\ensuremath{\chi}}c{\ensuremath{\gamma}} pairs in the small-x kinematics}.
\newblock {\em Phys. Rev. D}, 113(3):034024, 2026.

\bibitem{Bautista:2016xnp}
I.~Bautista, A.~Fernandez~Tellez, and Martin Hentschinski.
\newblock {BFKL evolution and the growth with energy of exclusive $J/\psi$ and
  $\Upsilon$ photoproduction cross sections}.
\newblock {\em Phys. Rev. D}, 94(5):054002, 2016.

\bibitem{ArroyoGarcia:2019cfl}
A.~Arroyo~Garcia, M.~Hentschinski, and K.~Kutak.
\newblock {QCD evolution based evidence for the onset of gluon saturation in
  exclusive photo-production of vector mesons}.
\newblock {\em Phys. Lett. B}, 795:569--575, 2019.

\bibitem{Hentschinski:2020yfm}
Martin Hentschinski and Emilio Padr{\'o}n~Molina.
\newblock {Exclusive $J/\Psi$ and $\Psi(2s)$ photo-production as a probe of QCD
  low $x$ evolution equations}.
\newblock {\em Phys. Rev. D}, 103(7):074008, 2021.

\bibitem{Peredo:2023oym}
Marco~Alcazar Peredo and Martin Hentschinski.
\newblock {Ratio of J/{\ensuremath{\Psi}} and {\ensuremath{\Psi}}(2s) exclusive
  photoproduction cross sections as an indicator for the presence of nonlinear
  QCD evolution}.
\newblock {\em Phys. Rev. D}, 109(1):014032, 2024.

\bibitem{Mantysaari:2018nng}
Heikki M{\"a}ntysaari and Pia Zurita.
\newblock {In depth analysis of the combined HERA data in the dipole models
  with and without saturation}.
\newblock {\em Phys. Rev. D}, 98:036002, 2018.

\bibitem{Penttala:2024hvp}
Jani Penttala and Christophe Royon.
\newblock {Gluon saturation effects in exclusive heavy vector meson
  photoproduction}.
\newblock {\em Phys. Lett. B}, 864:139394, 2025.

\bibitem{Mantysaari:2022sux}
Heikki M{\"a}ntysaari, Farid Salazar, and Bj{\"o}rn Schenke.
\newblock {Nuclear geometry at high energy from exclusive vector meson
  production}.
\newblock {\em Phys. Rev. D}, 106(7):074019, 2022.

\bibitem{Mantysaari:2023xcu}
Heikki M{\"a}ntysaari, Farid Salazar, and Bj{\"o}rn Schenke.
\newblock {Energy dependent nuclear suppression from gluon saturation in
  exclusive vector meson production}.
\newblock {\em Phys. Rev. D}, 109(7):L071504, 2024.

\bibitem{Mantysaari:2024zxq}
Heikki M{\"a}ntysaari, Jani Penttala, Farid Salazar, and Bj{\"o}rn Schenke.
\newblock {Finite-size effects on small-x evolution and saturation in proton
  and nuclear targets}.
\newblock {\em Phys. Rev. D}, 111(5):054033, 2025.

\bibitem{Mantysaari:2025ltq}
Heikki M{\"a}ntysaari, Hendrik Roch, Farid Salazar, Bj{\"o}rn Schenke, Chun
  Shen, and Wenbin Zhao.
\newblock {Global Bayesian analysis of J/{\ensuremath{\psi}} photoproduction on
  proton and lead targets}.
\newblock {\em Phys. Rev. D}, 113(1):014038, 2026.

\bibitem{Mantysaari:2022kdm}
Heikki M{\"a}ntysaari and Jani Penttala.
\newblock {Complete calculation of exclusive heavy vector meson production at
  next-to-leading order in the dipole picture}.
\newblock {\em JHEP}, 08:247, 2022.

\bibitem{Lappi:2020ufv}
Tuomas Lappi, Heikki M{\"a}ntysaari, and Jani Penttala.
\newblock {Relativistic corrections to the vector meson light front wave
  function}.
\newblock {\em Phys. Rev. D}, 102(5):054020, 2020.

\bibitem{Bodwin:1994jh}
Geoffrey~T. Bodwin, Eric Braaten, and G.~Peter Lepage.
\newblock {Rigorous QCD analysis of inclusive annihilation and production of
  heavy quarkonium}.
\newblock {\em Phys. Rev. D}, 51:1125--1171, 1995.
\newblock [Erratum: Phys.Rev.D 55, 5853 (1997)].

\bibitem{Gremm:1997dq}
Martin Gremm and Anton Kapustin.
\newblock {Annihilation of S wave quarkonia and the measurement of alpha-s}.
\newblock {\em Phys. Lett. B}, 407:323--330, 1997.

\bibitem{Bodwin:2006dn}
Geoffrey~T. Bodwin, Daekyoung Kang, and Jungil Lee.
\newblock {Potential-model calculation of an order-v(2) NRQCD matrix element}.
\newblock {\em Phys. Rev. D}, 74:014014, 2006.

\bibitem{Bodwin:2007fz}
Geoffrey~T. Bodwin, Hee~Sok Chung, Daekyoung Kang, Jungil Lee, and Chaehyun Yu.
\newblock {Improved determination of color-singlet nonrelativistic QCD matrix
  elements for S-wave charmonium}.
\newblock {\em Phys. Rev. D}, 77:094017, 2008.

\bibitem{Li:2019ncs}
Rong Li, Yu~Feng, and Yan-Qing Ma.
\newblock {Exclusive quarkonium production or decay in soft gluon
  factorization}.
\newblock {\em JHEP}, 05:009, 2020.

\bibitem{Babiarz:2025agk}
Izabela Babiarz, Chris~A. Flett, Melih~A. Ozcelik, Wolfgang Sch{\"a}fer, and
  Antoni Szczurek.
\newblock {Transition form-factor for $\eta _Q$ at NNLO in the strong coupling
  $\alpha _s$ and with all-order $v^2$ resummation}.
\newblock {\em Eur. Phys. J. C}, 85(12):1474, 2025.

\bibitem{Echevarria:2019ynx}
Miguel~G. Echevarria.
\newblock {Proper TMD factorization for quarkonia production: $pp\to\eta_{c,b}$
  as a study case}.
\newblock {\em JHEP}, 10:144, 2019.

\bibitem{Fleming:2019pzj}
Sean Fleming, Yiannis Makris, and Thomas Mehen.
\newblock {An effective field theory approach to quarkonium at small transverse
  momentum}.
\newblock {\em JHEP}, 04:122, 2020.

\bibitem{Boer:2023zit}
Dani\"el Boer, Jelle Bor, Luca Maxia, Cristian Pisano, and Feng Yuan.
\newblock {Transverse momentum dependent shape function for J/\ensuremath{\psi}
  production in SIDIS}.
\newblock {\em JHEP}, 08:105, 2023.

\bibitem{Echevarria:2025oab}
Miguel~G. Echevarria, Raj Kishore, and Samuel~F. Romera.
\newblock {Modeling the TMD shape function in $J/\psi$ electroproduction}.
\newblock arXiv:2510.11809.

\bibitem{Basham:1978bw}
C.~Louis Basham, Lowell~S. Brown, Stephen~D. Ellis, and Sherwin~T. Love.
\newblock {Energy Correlations in electron - Positron Annihilation: Testing
  QCD}.
\newblock {\em Phys. Rev. Lett.}, 41:1585, 1978.

\bibitem{Basham:1978zq}
C.~L. Basham, L.~S. Brown, S.~D. Ellis, and S.~T. Love.
\newblock {Energy Correlations in electron-Positron Annihilation in Quantum
  Chromodynamics: Asymptotically Free Perturbation Theory}.
\newblock {\em Phys. Rev. D}, 19:2018, 1979.

\bibitem{Moult:2025nhu}
Ian Moult and Hua~Xing Zhu.
\newblock {Energy Correlators: A Journey From Theory to Experiment}.
\newblock arXiv:2506.09119.

\bibitem{Kang:2023oqj}
Zhong-Bo Kang, Jani Penttala, Fanyi Zhao, and Yiyu Zhou.
\newblock {Transverse energy-energy correlators in the color-glass condensate
  at the electron-ion collider}.
\newblock {\em Phys. Rev. D}, 109(9):094012, 2024.

\bibitem{Kang:2025vjk}
Zhong-Bo Kang, Robert Kao, Meijian Li, and Jani Penttala.
\newblock {Transverse energy-energy correlators at small x for photon-hadron
  production}.
\newblock {\em Phys. Rev. D}, 112(7):076006, 2025.

\bibitem{Bhattacharya:2025bqa}
Shohini Bhattacharya, Zhong-Bo Kang, Diego Padilla, and Jani Penttala.
\newblock {Probing the Sivers Asymmetry with Transverse Energy-Energy
  Correlators in the Small-$x$ Regime}.
\newblock arXiv:2504.10475.

\bibitem{Ganguli:2025aqa}
Ishita Ganguli and Piotr Kotko.
\newblock {Transverse energy-energy and azimuthal correlations in forward
  {\ensuremath{\gamma}}-hadron production in proton-proton and proton-lead
  collisions at the LHC}.
\newblock {\em Phys. Rev. D}, 112(11):114020, 2025.

\bibitem{Kang:2026hig}
Zhong-Bo Kang, Robert Kao, Meijian Li, and Jani Penttala.
\newblock {One-point energy correlator for deep inelastic scattering at small
  $x$}.
\newblock arXiv:2603.02300.

\bibitem{Liu:2023aqb}
Hao-Yu Liu, Xiaohui Liu, Ji-Chen Pan, Feng Yuan, and Hua~Xing Zhu.
\newblock {Nucleon Energy Correlators for the Color Glass Condensate}.
\newblock {\em Phys. Rev. Lett.}, 130(18):181901, 2023.

\bibitem{Almaeen:2021}
Manal Almaeen, Yasir Alanazi, Nobuo Sato, W.~Melnitchouk, Michelle~P. Kuchera,
  and Yaohang Li.
\newblock {Variational Autoencoder Inverse Mapper: An End-to-End Deep Learning
  Framework for Inverse Problems}.
\newblock {\em Proc. Int. Joint Conf. Neural Netw. (IJCNN)}, pages 1--8, 2021.

\bibitem{Almaeen:2024guo}
Manal Almaeen, Tareq Alghamdi, Brandon Kriesten, Douglas Adams, Yaohang Li,
  Huey-Wen Lin, and Simonetta Liuti.
\newblock {VAIM-CFF: a variational autoencoder inverse mapper solution to
  Compton form factor extraction from deeply virtual exclusive reactions}.
\newblock {\em Eur. Phys. J. C}, 85(5):499, 2025.

\bibitem{Hossen:2024qwo}
Fayaz Hossen et~al.
\newblock {Variational autoencoder inverse mapper for extraction of Compton
  form factors: Benchmarks and conditional learning}.
\newblock arXiv:2408.11681.

\bibitem{Kriesten:2021sqc}
Brandon Kriesten, Philip Velie, Emma Yeats, Fernanda~Yepez Lopez, and Simonetta
  Liuti.
\newblock {Parametrization of quark and gluon generalized parton distributions
  in a dynamical framework}.
\newblock {\em Phys. Rev. D}, 105(5):056022, 2022.

\bibitem{Kriesten:2019jep}
Brandon Kriesten, Simonetta Liuti, Liliet Calero-Diaz, Dustin Keller, Andrew
  Meyer, Gary~R. Goldstein, and J.~Osvaldo Gonzalez-Hernandez.
\newblock {Extraction of generalized parton distribution observables from
  deeply virtual electron proton scattering experiments}.
\newblock {\em Phys. Rev. D}, 101(5):054021, 2020.

\bibitem{Kriesten:2020wcx}
Brandon Kriesten and Simonetta Liuti.
\newblock {Theory of deeply virtual Compton scattering off the unpolarized
  proton}.
\newblock {\em Phys. Rev. D}, 105(1):016015, 2022.

\bibitem{Almaeen:2022imx}
Manal Almaeen, Jake Grigsby, Joshua Hoskins, Brandon Kriesten, Yaohang Li,
  Huey-Wen Lin, and Simonetta Liuti.
\newblock {Benchmarks for a Global Extraction of Information from Deeply
  Virtual Exclusive Scattering}.
\newblock arXiv:2207.10766.

\bibitem{Karpie:2019eiq}
Joseph Karpie, Kostas Orginos, Alexander Rothkopf, and Savvas Zafeiropoulos.
\newblock {Reconstructing parton distribution functions from Ioffe time data:
  from Bayesian methods to Neural Networks}.
\newblock {\em JHEP}, 04:057, 2019.

\bibitem{DelDebbio:2020rgv}
Luigi Del~Debbio, Tommaso Giani, Joseph Karpie, Kostas Orginos, Anatoly
  Radyushkin, and Savvas Zafeiropoulos.
\newblock {Neural-network analysis of Parton Distribution Functions from
  Ioffe-time pseudodistributions}.
\newblock {\em JHEP}, 02:138, 2021.

\bibitem{Dutrieux:2024rem}
Herv{\'e} Dutrieux, Joseph Karpie, Kostas Orginos, and Savvas Zafeiropoulos.
\newblock {Simple nonparametric reconstruction of parton distributions from
  limited Fourier information}.
\newblock {\em Phys. Rev. D}, 111(3):034515, 2025.

\bibitem{Zhang:2020gaj}
Rui Zhang, Carson Honkala, Huey-Wen Lin, and Jiunn-Wei Chen.
\newblock {Pion and kaon distribution amplitudes in the continuum limit}.
\newblock {\em Phys. Rev. D}, 102(9):094519, 2020.

\bibitem{Boyda:2022nmh}
Denis Boyda et~al.
\newblock {Applications of Machine Learning to Lattice Quantum Field Theory}.
\newblock In {\em {Snowmass 2021}}, 2 2022.

\bibitem{Balitsky:2021bds}
Ian Balitsky, Wayne Morris, and Anatoly Radyushkin.
\newblock {Gluon pseudo-distributions at short distances}.
\newblock {\em SciPost Phys. Proc.}, 8:161, 2022.

\bibitem{Kriesten:2025gti}
Brandon Kriesten, Alex NieMiera, William Good, T.~J. Hobbs, and Huey-Wen Lin.
\newblock {Decoding the proton{\textquoteright}s gluonic density with lattice
  QCD-informed machine learning}.
\newblock {\em Phys. Lett. B}, 874:140132, 2026.

\bibitem{Kriesten:2023uoi}
Brandon Kriesten and T.~J. Hobbs.
\newblock {Learning PDFs through interpretable latent representations in Mellin
  space}.
\newblock {\em Phys. Rev. D}, 111(1):014028, 2025.

\bibitem{Kriesten:2024are}
Brandon Kriesten, Jonathan Gomprecht, and T.~J. Hobbs.
\newblock {Explainable AI classification for parton density theory}.
\newblock {\em JHEP}, 11:007, 2024.

\bibitem{Alanazi:ijcai2021p588}
Y.~Alanazi et~al.
\newblock A survey of machine learning-based physics event generation.
\newblock In {\em Proceedings of IJCAI2021}, pages 4286--4293, August 2021.
\newblock Survey Track.

\bibitem{Torbunov:2024iki}
Dmitrii Torbunov, Yi~Huang, Meifeng Lin, Yihui Ren, Yeonju Go, Timothy Rinn,
  Haiwang Yu, Brett Viren, and Jin Huang.
\newblock {Effectiveness of denoising diffusion probabilistic models for fast
  and high-fidelity whole-event simulation in high-energy heavy-ion
  experiments}.
\newblock {\em Phys. Rev. C}, 110(3):034912, 2024.

\bibitem{Devlin:2023jzp}
Peter Devlin, Jian-Wei Qiu, Felix Ringer, and Nobuo Sato.
\newblock {Diffusion model approach to simulating electron-proton scattering
  events}.
\newblock {\em Phys. Rev. D}, 110(1):016030, 2024.

\bibitem{Go:2025exm}
Yeonju Go et~al.
\newblock {Robust and Generalizable Background Subtraction on Images of
  Calorimeter Jets using Unsupervised Generative Learning}.
\newblock arXiv:2510.23717.

\bibitem{Park:2025ebs}
David Park et~al.
\newblock {FM4NPP: A Scaling Foundation Model for Nuclear and Particle
  Physics}.
\newblock 8 2025.

\bibitem{Giroux:2025elr}
James Giroux and Cristiano Fanelli.
\newblock {Towards foundation models for experimental readout systems combining
  discrete and continuous data}.
\newblock {\em Mach. Learn. Sci. Tech.}, 7(1):015031, 2026.

\bibitem{Huang:2021ymz}
Yi~Huang, Yihui Ren, Shinjae Yoo, and Jin Huang.
\newblock {Efficient Data Compression for 3D Sparse TPC via Bicephalous
  Convolutional Autoencoder}.
\newblock arXiv:2111.05423.

\bibitem{Huang:2023ype}
Yi~Huang, Yihui Ren, Shinjae Yoo, and Jin Huang.
\newblock {Fast 2D Bicephalous Convolutional Autoencoder for Compressing 3D
  Time Projection Chamber Data}.
\newblock In {\em {International Conference on High Performance Computing,
  Network, Storage, and Analysis}}, arXiv:2310.15026.

\bibitem{Huang:2024tir}
Yi~Huang et~al.
\newblock {Variable Rate Neural Compression for Sparse Detector Data}.
\newblock arXiv:2411.11942.

\bibitem{Qu:2019gqs}
Huilin Qu and Loukas Gouskos.
\newblock {ParticleNet: Jet Tagging via Particle Clouds}.
\newblock {\em Phys. Rev. D}, 101(5):056019, 2020.

\bibitem{Qu:2022mxj}
Huilin Qu, Congqiao Li, and Sitian Qian.
\newblock {Particle Transformer for Jet Tagging}.
\newblock arXiv:2202.03772.

\bibitem{Stein:2022nvf}
Annika Stein, Xavier Coubez, Spandan Mondal, Andrzej Novak, and Alexander
  Schmidt.
\newblock {Improving Robustness of Jet Tagging Algorithms with Adversarial
  Training}.
\newblock {\em Comput. Softw. Big Sci.}, 6(1):15, 2022.

\bibitem{CMS-DP-2024-066}
{CMS Collaboration}.
\newblock {A unified approach for jet tagging in Run 3 at $\sqrt{s}$=13.6 TeV
  in CMS}.
\newblock CMS Detector Performance Summary CMS-DP-2024-066, 2024.

\bibitem{6795963}
Sepp Hochreiter and Jürgen Schmidhuber.
\newblock Long short-term memory.
\newblock {\em Neural Computation}, 9(8):1735--1780, 1997.

\bibitem{Li:2023koj}
Xuan Li.
\newblock {Exploration of hadronization through heavy flavor production at the
  future Electron-Ion Collider}.
\newblock In {\em {30th International Workshop on Deep-Inelastic Scattering and
  Related Subjects}}, volume 296, page 16001, 2024.

\bibitem{Li:2025lxr}
Xuan Li.
\newblock {Recent open heavy flavor studies for the Electron-Ion Collider}.
\newblock {\em PoS}, QNP2024:113, 2025.

\bibitem{ulmer2023priorposteriornetworkssurvey}
Dennis Ulmer, Christian Hardmeier, and Jes Frellsen.
\newblock Prior and posterior networks: A survey on evidential deep learning
  methods for uncertainty estimation, 2023.

\bibitem{Kriesten:2024ist}
Brandon Kriesten and T.~J. Hobbs.
\newblock {Anomalous electroweak physics unraveled via evidential deep
  learning}.
\newblock {\em Eur. Phys. J. C}, 85(8):883, 2025.

\bibitem{Metodiev:2017vrx}
Eric~M. Metodiev, Benjamin Nachman, and Jesse Thaler.
\newblock {Classification without labels: Learning from mixed samples in high
  energy physics}.
\newblock {\em JHEP}, 10:174, 2017.

\bibitem{Cerri:2018anq}
Olmo Cerri, Thong~Q. Nguyen, Maurizio Pierini, Maria Spiropulu, and Jean-Roch
  Vlimant.
\newblock {Variational Autoencoders for New Physics Mining at the Large Hadron
  Collider}.
\newblock {\em JHEP}, 05:036, 2019.

\bibitem{Blance:2019ibf}
Andrew Blance, Michael Spannowsky, and Philip Waite.
\newblock {Adversarially-trained autoencoders for robust unsupervised new
  physics searches}.
\newblock {\em JHEP}, 10:047, 2019.

\bibitem{Dillon:2021nxw}
Barry~M. Dillon, Tilman Plehn, Christof Sauer, and Peter Sorrenson.
\newblock {Better Latent Spaces for Better Autoencoders}.
\newblock {\em SciPost Phys.}, 11:061, 2021.

\bibitem{Hallin:2022eoq}
Anna Hallin, Gregor Kasieczka, Tobias Quadfasel, David Shih, and Manuel
  Sommerhalder.
\newblock {Resonant anomaly detection without background sculpting}.
\newblock {\em Phys. Rev. D}, 107(11):114012, 2023.

\bibitem{Buhmann:2023acn}
Erik Buhmann, Cedric Ewen, Gregor Kasieczka, Vinicius Mikuni, Benjamin Nachman,
  and David Shih.
\newblock {Full phase space resonant anomaly detection}.
\newblock {\em Phys. Rev. D}, 109(5):055015, 2024.

\bibitem{Araz:2025oax}
Jack~Y. Araz, Dimitrios Athanasakos, Mateusz Ploskon, and Felix Ringer.
\newblock {Graph theory inspired anomaly detection at the LHC}.
\newblock {\em JHEP}, 02:254, 2026.

\bibitem{Kasieczka:2021xcg}
Gregor Kasieczka et~al.
\newblock {The LHC Olympics 2020 a community challenge for anomaly detection in
  high energy physics}.
\newblock {\em Rept. Prog. Phys.}, 84(12):124201, 2021.

\bibitem{GEANT4:2002zbu}
S.~Agostinelli et~al.
\newblock {GEANT4 - A Simulation Toolkit}.
\newblock {\em Nucl. Instrum. Meth. A}, 506:250--303, 2003.

\bibitem{Lafferty:1994cj}
G.~D. Lafferty and T.~R. Wyatt.
\newblock {Where to stick your data points: The treatment of measurements
  within wide bins}.
\newblock {\em Nucl. Instrum. Meth. A}, 355:541--547, 1995.

\bibitem{H1:2009pze}
F.~D. Aaron et~al.
\newblock {Combined Measurement and QCD Analysis of the Inclusive e+- p
  Scattering Cross Sections at HERA}.
\newblock {\em JHEP}, 01:109, 2010.

\bibitem{Buckley:2019xhk}
Andy Buckley, Philip Ilten, Dmitri Konstantinov, Leif L{\"o}nnblad, James Monk,
  Witold Pokorski, Tomasz Przedzinski, and Andrii Verbytskyi.
\newblock {The HepMC3 event record library for Monte Carlo event generators}.
\newblock {\em Comput. Phys. Commun.}, 260:107310, 2021.

\bibitem{Ji:2004wu}
Xiang-dong Ji, Jian-ping Ma, and Feng Yuan.
\newblock {QCD factorization for semi-inclusive deep-inelastic scattering at
  low transverse momentum}.
\newblock {\em Phys. Rev. D}, 71:034005, 2005.

\bibitem{Ji:2004xq}
Xiang-dong Ji, Jian-Ping Ma, and Feng Yuan.
\newblock Qcd factorization for spin-dependent cross sections in dis and
  drell-yan processes at low transverse momentum.
\newblock {\em Phys. Lett.}, B597:299--308, 2004.

\bibitem{Aybat:2011zv}
S.~Mert Aybat and Ted~C. Rogers.
\newblock {TMD Parton Distribution and Fragmentation Functions with QCD
  Evolution}.
\newblock {\em Phys. Rev.}, D83:114042, 2011.

\bibitem{Aybat:2011ge}
S.~Mert Aybat, John~C. Collins, Jian-Wei Qiu, and Ted~C. Rogers.
\newblock {The QCD Evolution of the Sivers Function}.
\newblock {\em Phys.Rev.}, D85:034043, 2012.

\bibitem{Echevarria:2011epo}
Miguel~G. Echevarria, Ahmad Idilbi, and Ignazio Scimemi.
\newblock {Factorization Theorem For Drell-Yan At Low $q_T$ And Transverse
  Momentum Distributions On-The-Light-Cone}.
\newblock {\em JHEP}, 07:002, 2012.

\bibitem{Bain:2016rrv}
Reggie Bain, Yiannis Makris, and Thomas Mehen.
\newblock {Transverse Momentum Dependent Fragmenting Jet Functions with
  Applications to Quarkonium Production}.
\newblock {\em JHEP}, 11:144, 2016.

\bibitem{Makris:2017arq}
Yiannis Makris, Duff Neill, and Varun Vaidya.
\newblock {Probing Transverse-Momentum Dependent Evolution With Groomed Jets}.
\newblock {\em JHEP}, 07:167, 2018.

\bibitem{Kang:2017glf}
Zhong-Bo Kang, Xiaohui Liu, Felix Ringer, and Hongxi Xing.
\newblock {The transverse momentum distribution of hadrons within jets}.
\newblock {\em JHEP}, 11:068, 2017.

\bibitem{Kang:2019ahe}
Zhong-Bo Kang, Kyle Lee, John Terry, and Hongxi Xing.
\newblock {Jet fragmentation functions for $Z$-tagged jets}.
\newblock {\em Phys. Lett. B}, 798:134978, 2019.

\bibitem{Kang:2020xyq}
Zhong-Bo Kang, Kyle Lee, and Fanyi Zhao.
\newblock {Polarized jet fragmentation functions}.
\newblock {\em Phys. Lett. B}, 809:135756, 2020.

\bibitem{Kang:2021ffh}
Zhong-Bo Kang, Kyle Lee, Ding~Yu Shao, and Fanyi Zhao.
\newblock {Spin asymmetries in electron-jet production at the future electron
  ion collider}.
\newblock {\em JHEP}, 11:005, 2021.

\bibitem{Musch:2011er}
B.U. Musch, Ph. Hagler, M.~Engelhardt, J.W. Negele, and A.~Schafer.
\newblock {Sivers and Boer-Mulders observables from lattice QCD}.
\newblock {\em Phys.Rev.}, D85:094510, 2012.

\bibitem{Engelhardt:2015xja}
M.~Engelhardt, P.~H{\"a}gler, B.~Musch, J.~Negele, and A.~Sch{\"a}fer.
\newblock {Lattice QCD study of the Boer-Mulders effect in a pion}.
\newblock {\em Phys. Rev. D}, 93(5):054501, 2016.

\bibitem{Yoon:2017qzo}
Boram Yoon, Michael Engelhardt, Rajan Gupta, Tanmoy Bhattacharya, Jeremy~R.
  Green, Bernhard~U. Musch, John~W. Negele, Andrew~V. Pochinsky, Andreas
  Sch{\"a}fer, and Sergey~N. Syritsyn.
\newblock {Nucleon Transverse Momentum-dependent Parton Distributions in
  Lattice QCD: Renormalization Patterns and Discretization Effects}.
\newblock {\em Phys. Rev. D}, 96(9):094508, 2017.

\bibitem{Engelhardt:2023aem}
Michael Engelhardt, Nesreen Hasan, Taku Izubuchi, Christos Kallidonis, Stefan
  Krieg, Stefan Meinel, John Negele, Andrew Pochinsky, Giorgio Silvi, and
  Sergey Syritsyn.
\newblock {Transverse momentum-dependent parton distributions for
  longitudinally polarized nucleons from domain wall fermion calculations at
  the physical pion mass}.
\newblock {\em PoS}, LATTICE2022:103, 2023.

\bibitem{Ebert:2020gxr}
Markus~A. Ebert, Stella~T. Schindler, Iain~W. Stewart, and Yong Zhao.
\newblock {One-loop Matching for Spin-Dependent Quasi-TMDs}.
\newblock {\em JHEP}, 09:099, 2020.

\bibitem{Ebert:2022fmh}
Markus~A. Ebert, Stella~T. Schindler, Iain~W. Stewart, and Yong Zhao.
\newblock {Factorization connecting continuum \& lattice TMDs}.
\newblock {\em JHEP}, 04:178, 2022.

\bibitem{Collins:2014jpa}
John Collins and Ted Rogers.
\newblock {Understanding the large-distance behavior of
  transverse-momentum-dependent parton densities and the Collins-Soper
  evolution kernel}.
\newblock {\em Phys.Rev.}, D91(7):074020, 2015.

\bibitem{Collins:2016hqq}
J.~Collins, L.~Gamberg, A.~Prokudin, T.~C. Rogers, N.~Sato, and B.~Wang.
\newblock {Relating Transverse Momentum Dependent and Collinear Factorization
  Theorems in a Generalized Formalism}.
\newblock {\em Phys. Rev.}, D94(3):034014, 2016.

\bibitem{Balazs:1995nz}
Csaba Balazs, Jian-wei Qiu, and C.~P. Yuan.
\newblock {Effects of QCD resummation on distributions of leptons from the
  decay of electroweak vector bosons}.
\newblock {\em Phys. Lett. B}, 355:548--554, 1995.

\bibitem{Balazs:1997xd}
C.~Balazs and C.~P. Yuan.
\newblock {Soft gluon effects on lepton pairs at hadron colliders}.
\newblock {\em Phys. Rev. D}, 56:5558--5583, 1997.

\bibitem{Ellis:1997sc}
R.~Keith Ellis, D.~A. Ross, and Sinisa Veseli.
\newblock {Vector boson production in hadronic collisions}.
\newblock {\em Nucl. Phys. B}, 503:309--338, 1997.

\bibitem{Bozzi:2010xn}
Giuseppe Bozzi, Stefano Catani, Giancarlo Ferrera, Daniel de~Florian, and
  Massimiliano Grazzini.
\newblock {Production of Drell-Yan lepton pairs in hadron collisions:
  Transverse-momentum resummation at next-to-next-to-leading logarithmic
  accuracy}.
\newblock {\em Phys. Lett. B}, 696:207--213, 2011.

\bibitem{Banfi:2012du}
Andrea Banfi, Mrinal Dasgupta, Simone Marzani, and Lee Tomlinson.
\newblock {Predictions for Drell-Yan $\phi^*$ and $Q_T$ observables at the
  LHC}.
\newblock {\em Phys. Lett. B}, 715:152--156, 2012.

\bibitem{Echevarria:2015uaa}
Miguel~G. Echevarria, Tomas Kasemets, Piet~J. Mulders, and Cristian Pisano.
\newblock {QCD evolution of (un)polarized gluon TMDPDFs and the Higgs
  $q_T$-distribution}.
\newblock {\em JHEP}, 07:158, 2015.
\newblock [Erratum: JHEP 05, 073 (2017)].

\bibitem{Catani:2015vma}
Stefano Catani, Daniel de~Florian, Giancarlo Ferrera, and Massimiliano
  Grazzini.
\newblock {Vector boson production at hadron colliders: transverse-momentum
  resummation and leptonic decay}.
\newblock {\em JHEP}, 12:047, 2015.

\bibitem{Monni:2016ktx}
Pier~Francesco Monni, Emanuele Re, and Paolo Torrielli.
\newblock {Higgs Transverse-Momentum Resummation in Direct Space}.
\newblock {\em Phys. Rev. Lett.}, 116(24):242001, 2016.

\bibitem{Ebert:2016gcn}
Markus~A. Ebert and Frank~J. Tackmann.
\newblock {Resummation of Transverse Momentum Distributions in Distribution
  Space}.
\newblock {\em JHEP}, 02:110, 2017.

\bibitem{Kang:2017cjk}
Daekyoung Kang, Christopher Lee, and Varun Vaidya.
\newblock {A fast and accurate method for perturbative resummation of
  transverse momentum-dependent observables}.
\newblock {\em JHEP}, 04:149, 2018.

\bibitem{Chen:2016hgw}
A.~P. Chen and J.~P. Ma.
\newblock {Light-Cone Singularities and Transverse-Momentum-Dependent
  Factorization at Twist-3}.
\newblock {\em Phys. Lett. B}, 768:380--386, 2017.

\bibitem{Chen:2018pzu}
Xuan Chen, Thomas Gehrmann, E.~W.~Nigel Glover, Alexander Huss, Ye~Li, Duff
  Neill, Markus Schulze, Iain~W. Stewart, and Hua~Xing Zhu.
\newblock {Precise QCD Description of the Higgs Boson Transverse Momentum
  Spectrum}.
\newblock {\em Phys. Lett.}, B788:425--430, 2019.

\bibitem{Becher:2019bnm}
Thomas Becher and Monika Hager.
\newblock {Event-Based Transverse Momentum Resummation}.
\newblock {\em Eur. Phys. J. C}, 79(8):665, 2019.

\bibitem{Scimemi:2019cmh}
Ignazio Scimemi and Alexey Vladimirov.
\newblock {Non-perturbative structure of semi-inclusive deep-inelastic and
  Drell-Yan scattering at small transverse momentum}.
\newblock {\em JHEP}, 06:137, 2020.

\bibitem{Bacchetta:2019sam}
Alessandro Bacchetta, Valerio Bertone, Chiara Bissolotti, Giuseppe Bozzi,
  Filippo Delcarro, Fulvio Piacenza, and Marco Radici.
\newblock {Transverse-momentum-dependent parton distributions up to N$^{3}$LL
  from Drell-Yan data}.
\newblock {\em JHEP}, 07:117, 2020.

\bibitem{Ebert:2020dfc}
Markus~A. Ebert, Johannes K.~L. Michel, Iain~W. Stewart, and Frank~J. Tackmann.
\newblock {Drell-Yan $q_{T}$ resummation of fiducial power corrections at
  N$^{3}$LL}.
\newblock {\em JHEP}, 04:102, 2021.

\bibitem{Scimemi:2017etj}
Ignazio Scimemi and Alexey Vladimirov.
\newblock {Analysis of vector boson production within TMD factorization}.
\newblock {\em Eur. Phys. J. C}, 78(2):89, 2018.

\bibitem{Bacchetta:2022awv}
Alessandro Bacchetta, Valerio Bertone, Chiara Bissolotti, Giuseppe Bozzi,
  Matteo Cerutti, Fulvio Piacenza, Marco Radici, and Andrea Signori.
\newblock {Unpolarized transverse momentum distributions from a global fit of
  Drell-Yan and semi-inclusive deep-inelastic scattering data}.
\newblock {\em JHEP}, 10:127, 2022.

\bibitem{Echevarria:2020hpy}
Miguel~G. Echevarria, Zhong-Bo Kang, and John Terry.
\newblock {Global analysis of the Sivers functions at NLO+NNLL in QCD}.
\newblock {\em JHEP}, 01:126, 2021.

\bibitem{Bacchetta:2020gko}
Alessandro Bacchetta, Filippo Delcarro, Cristian Pisano, and Marco Radici.
\newblock {The 3-dimensional distribution of quarks in momentum space}.
\newblock {\em Phys. Lett. B}, 827:136961, 2022.

\bibitem{Bury:2020vhj}
Marcin Bury, Alexei Prokudin, and Alexey Vladimirov.
\newblock {Extraction of the Sivers Function from SIDIS, Drell-Yan, and
  $W^{\pm}/Z$ Data at Next-to-Next-to-Next-to Leading Order}.
\newblock {\em Phys. Rev. Lett.}, 126(11):112002, 2021.

\bibitem{Bury:2021sue}
Marcin Bury, Alexei Prokudin, and Alexey Vladimirov.
\newblock {Extraction of the Sivers function from SIDIS, Drell-Yan, and
  $W^\pm/Z$ boson production data with TMD evolution}.
\newblock {\em JHEP}, 05:151, 2021.

\bibitem{Bury:2022czx}
Marcin Bury, Francesco Hautmann, Sergio Leal-Gomez, Ignazio Scimemi, Alexey
  Vladimirov, and Pia Zurita.
\newblock {PDF bias and flavor dependence in TMD distributions}.
\newblock {\em JHEP}, 10:118, 2022.

\bibitem{Moos:2023yfa}
Valentin Moos, Ignazio Scimemi, Alexey Vladimirov, and Pia Zurita.
\newblock {Extraction of unpolarized transverse momentum distributions from the
  fit of Drell-Yan data at N$^{4}$LL}.
\newblock {\em JHEP}, 05:036, 2024.

\bibitem{Bacchetta:2024qre}
Alessandro Bacchetta, Valerio Bertone, Chiara Bissolotti, Giuseppe Bozzi,
  Matteo Cerutti, Filippo Delcarro, Marco Radici, Lorenzo Rossi, and Andrea
  Signori.
\newblock {Flavor dependence of unpolarized quark transverse momentum
  distributions from a global fit}.
\newblock {\em JHEP}, 08:232, 2024.

\bibitem{Bacchetta:2025ara}
Alessandro Bacchetta, Valerio Bertone, Chiara Bissolotti, Matteo Cerutti, Marco
  Radici, Simone Rodini, and Lorenzo Rossi.
\newblock {Neural-Network Extraction of Unpolarized
  Transverse-Momentum-Dependent Distributions}.
\newblock {\em Phys. Rev. Lett.}, 135(2):021904, 2025.

\bibitem{Barry:2025glq}
P.~C. Barry et~al.
\newblock {First simultaneous analysis of transverse momentum dependent and
  collinear parton distributions in the proton}.
\newblock arXiv:2510.13771.

\bibitem{Collins:1981va}
John~C. Collins and Davison~E. Soper.
\newblock {Back-To-Back Jets: Fourier Transform from B to K-Transverse}.
\newblock {\em Nucl. Phys. B}, 197:446--476, 1982.

\bibitem{Meng:1991da}
Rui-bin Meng, Fredrick~I. Olness, and Davison~E. Soper.
\newblock {Semiinclusive deeply inelastic scattering at electron - proton
  colliders}.
\newblock {\em Nucl. Phys. B}, 371:79--110, 1992.

\bibitem{Meng:1995yn}
Ruibin Meng, Fredrick~I. Olness, and Davison~E. Soper.
\newblock {Semiinclusive deeply inelastic scattering at small q(T)}.
\newblock {\em Phys. Rev. D}, 54:1919--1935, 1996.

\bibitem{Idilbi:2004vb}
Ahmad Idilbi, Xiang-dong Ji, Jian-Ping Ma, and Feng Yuan.
\newblock {Collins-Soper equation for the energy evolution of
  transverse-momentum and spin dependent parton distributions}.
\newblock {\em Phys. Rev. D}, 70:074021, 2004.

\bibitem{Yuan:2007nd}
Feng Yuan.
\newblock {Azimuthal asymmetric distribution of hadrons inside a jet at hadron
  collider}.
\newblock {\em Phys. Rev. Lett.}, 100:032003, 2008.

\bibitem{Sivers:1989cc}
Dennis~W. Sivers.
\newblock {Single Spin Production Asymmetries from the Hard Scattering of
  Point-Like Constituents}.
\newblock {\em Phys. Rev. D}, 41:83, 1990.

\bibitem{Boer:1997nt}
Daniel Boer and P.J. Mulders.
\newblock {Time reversal odd distribution functions in leptoproduction}.
\newblock {\em Phys.Rev.}, D57:5780--5786, 1998.

\bibitem{Tangerman:1994eh}
R.~D. Tangerman and P.~J. Mulders.
\newblock {Intrinsic transverse momentum and the polarized Drell-Yan process}.
\newblock {\em Phys. Rev. D}, 51:3357--3372, 1995.

\bibitem{Kotzinian:1995cz}
A.~M. Kotzinian and P.~J. Mulders.
\newblock {Longitudinal quark polarization in transversely polarized nucleons}.
\newblock {\em Phys. Rev. D}, 54:1229--1232, 1996.

\bibitem{Mulders:1995dh}
P.~J. Mulders and R.~D. Tangerman.
\newblock {The Complete tree level result up to order 1/Q for polarized deep
  inelastic leptoproduction}.
\newblock {\em Nucl. Phys.}, B461:197--237, 1996.
\newblock [Erratum: Nucl. Phys.B484,538(1997)].

\bibitem{Boer:1999si}
Daniel Boer and P.~J. Mulders.
\newblock {Color gauge invariance in the Drell-Yan process}.
\newblock {\em Nucl. Phys. B}, 569:505--526, 2000.

\bibitem{Belitsky:2002sm}
Andrei~V. Belitsky, X.~Ji, and F.~Yuan.
\newblock {Final state interactions and gauge invariant parton distributions}.
\newblock {\em Nucl.Phys.}, B656:165--198, 2003.

\bibitem{Bomhof:2006dp}
C.~J. Bomhof, P.~J. Mulders, and F.~Pijlman.
\newblock {The Construction of gauge-links in arbitrary hard processes}.
\newblock {\em Eur. Phys. J.}, C47:147--162, 2006.

\bibitem{Boer:1997mf}
Daniel Boer, R.~Jakob, and P.J. Mulders.
\newblock {Asymmetries in polarized hadron production in e+ e- annihilation up
  to order 1/Q}.
\newblock {\em Nucl. Phys. B}, 504:345--380, 1997.

\bibitem{Goldstein:2002vv}
Gary~R. Goldstein and Leonard Gamberg.
\newblock {Transversity and meson photoproduction}.
\newblock In {\em {31st International Conference on High Energy Physics}},
  pages 452--454, 9 2002.

\bibitem{Brodsky:2002cx}
Stanley~J. Brodsky, Dae~Sung Hwang, and Ivan Schmidt.
\newblock {Final state interactions and single spin asymmetries in
  semiinclusive deep inelastic scattering}.
\newblock {\em Phys.Lett.}, B530:99--107, 2002.

\bibitem{Brodsky:2002rv}
Stanley~J. Brodsky, Dae~Sung Hwang, and Ivan Schmidt.
\newblock {Initial state interactions and single spin asymmetries in Drell-Yan
  processes}.
\newblock {\em Nucl.Phys.}, B642:344--356, 2002.

\bibitem{Collins:2002kn}
John~C. Collins.
\newblock {Leading twist single transverse-spin asymmetries: Drell-Yan and deep
  inelastic scattering}.
\newblock {\em Phys.Lett.}, B536:43--48, 2002.

\bibitem{Georgi:1977tv}
Howard Georgi and H.~David Politzer.
\newblock {Clean Tests of QCD in mu p Scattering}.
\newblock {\em Phys. Rev. Lett.}, 40:3, 1978.

\bibitem{Cahn:1978se}
Robert~N. Cahn.
\newblock {Azimuthal Dependence in Leptoproduction: A Simple Parton Model
  Calculation}.
\newblock {\em Phys. Lett. B}, 78:269--273, 1978.

\bibitem{Cahn:1989yf}
R.N. Cahn.
\newblock {Critique of Parton Model Calculations of Azimuthal Dependence in
  Leptoproduction}.
\newblock {\em Phys. Rev. D}, 40:3107--3110, 1989.

\bibitem{Ravndal:1973kt}
F.~Ravndal.
\newblock {On the azimuthal dependence of semiinclusive, deep inelastic
  electroproduction cross-sections}.
\newblock {\em Phys. Lett. B}, 43:301--303, 1973.

\bibitem{Anselmino:2005nn}
M.~Anselmino, M.~Boglione, U.~D'Alesio, A.~Kotzinian, F.~Murgia, et~al.
\newblock {The Role of Cahn and sivers effects in deep inelastic scattering}.
\newblock {\em Phys.Rev.}, D71:074006, 2005.

\bibitem{Kotzinian:1994dv}
Aram Kotzinian.
\newblock {New quark distributions and semiinclusive electroproduction on the
  polarized nucleons}.
\newblock {\em Nucl. Phys.}, B441:234, 1995.

\bibitem{Levelt:1993ac}
J.~Levelt and P.~J. Mulders.
\newblock {Quark correlation functions in deep inelastic semiinclusive
  processes}.
\newblock {\em Phys. Rev. D}, 49:96--113, 1994.

\bibitem{Boer:2003cm}
Daniel Boer, P.~J. Mulders, and F.~Pijlman.
\newblock {Universality of T odd effects in single spin and azimuthal
  asymmetries}.
\newblock {\em Nucl. Phys. B}, 667:201--241, 2003.

\bibitem{Bacchetta:2004zf}
Alessandro Bacchetta, Piet~J. Mulders, and Fetze Pijlman.
\newblock New observables in longitudinal single-spin asymmetries in
  semi-inclusive dis.
\newblock {\em Phys. Lett.}, B595:309--317, 2004.

\bibitem{Bacchetta:2006tn}
Alessandro Bacchetta, Markus Diehl, Klaus Goeke, Andreas Metz, Piet~J. Mulders,
  and Marc Schlegel.
\newblock {Semi-inclusive deep inelastic scattering at small transverse
  momentum}.
\newblock {\em JHEP}, 02:093, 2007.

\bibitem{Lu:2011th}
Zhun Lu and Ivan Schmidt.
\newblock {Transverse momentum dependent twist-three result for polarized
  Drell-Yan processes}.
\newblock {\em Phys. Rev. D}, 84:114004, 2011.

\bibitem{Burkardt:2008ps}
Matthias Burkardt.
\newblock {Transverse Force on Quarks in DIS}.
\newblock 2008.

\bibitem{EuropeanMuon:1983tsy}
J.~J. Aubert et~al.
\newblock {Measurement of Hadronic Azimuthal Distributions in Deep Inelastic
  Muon Proton Scattering}.
\newblock {\em Phys. Lett. B}, 130:118--122, 1983.

\bibitem{EuropeanMuon:1986ulc}
M.~Arneodo et~al.
\newblock {Measurement of Hadron Azimuthal Distributions in Deep Inelastic Muon
  Proton Scattering}.
\newblock {\em Z. Phys. C}, 34:277, 1987.

\bibitem{Adams:1993hs}
M.~R. Adams et~al.
\newblock {Perturbative QCD effects observed in 490 GeV deep inelastic muon
  scattering}.
\newblock {\em Phys. Rev.}, D48:5057--5066, 1993.

\bibitem{Breitweg:2000qh}
J.~Breitweg et~al.
\newblock {Measurement of azimuthal asymmetries in deep inelastic scattering}.
\newblock {\em Phys. Lett.}, B481:199--212, 2000.

\bibitem{Chekanov:2002sz}
S.~Chekanov et~al.
\newblock {Study of the azimuthal asymmetry of jets in neutral current deep
  inelastic scattering at HERA}.
\newblock {\em Phys. Lett.}, B551:226--240, 2003.

\bibitem{Mkrtchyan:2007sr}
H.~Mkrtchyan et~al.
\newblock {Transverse momentum dependence of semi-inclusive pion production}.
\newblock {\em Phys. Lett. B}, 665:20--25, 2008.

\bibitem{CLAS:2008nzy}
M.~Osipenko et~al.
\newblock {Measurement of unpolarized semi-inclusive pi+ electroproduction off
  the proton}.
\newblock {\em Phys. Rev. D}, 80:032004, 2009.

\bibitem{Airapetian:2013bim}
A.~Airapetian et~al.
\newblock {Transverse target single-spin asymmetry in inclusive
  electroproduction of charged pions and kaons}.
\newblock {\em Phys.~Lett.}, B728:183--190, 2014.

\bibitem{Adolph:2014zba}
C.~Adolph et~al.
\newblock {Collins and Sivers asymmetries in muonproduction of pions and kaons
  off transversely polarised protons}.
\newblock {\em Phys. Lett.}, B744:250--259, 2015.

\bibitem{Airapetian:1999tv}
A.~Airapetian et~al.
\newblock Observation of a single-spin azimuthal asymmetry in semi- inclusive
  pion electro-production.
\newblock {\em Phys. Rev. Lett.}, 84:4047--4051, 2000.

\bibitem{Airapetian:2001eg}
A.~Airapetian et~al.
\newblock Single-spin azimuthal asymmetries in electroproduction of neutral
  pions in semi-inclusive deep-inelastic scattering.
\newblock {\em Phys. Rev.}, D64:097101, 2001.

\bibitem{Gamberg:2006ru}
L.~P. Gamberg, D.~S. Hwang, A.~Metz, and M.~Schlegel.
\newblock Light-cone divergence in twist-3 correlation functions.
\newblock {\em Phys. Lett.}, B639:508--512, 2006.

\bibitem{Boer:2006eq}
Daniel Boer and Werner Vogelsang.
\newblock {Drell-Yan lepton angular distribution at small transverse momentum}.
\newblock {\em Phys. Rev. D}, 74:014004, 2006.

\bibitem{Lam:1978pu}
C.~S. Lam and Wu-Ki Tung.
\newblock {A Systematic Approach to Inclusive Lepton Pair Production in
  Hadronic Collisions}.
\newblock {\em Phys. Rev. D}, 18:2447, 1978.

\bibitem{Koike:2006fn}
Yuji Koike, Junji Nagashima, and Werner Vogelsang.
\newblock {Resummation for polarized semi-inclusive deep-inelastic scattering
  at small transverse momentum}.
\newblock {\em Nucl. Phys.}, B744:59--79, 2006.

\bibitem{Bacchetta:2008xw}
Alessandro Bacchetta, Daniel Boer, Markus Diehl, and Piet~J. Mulders.
\newblock {Matches and mismatches in the descriptions of semi-inclusive
  processes at low and high transverse momentum}.
\newblock {\em JHEP}, 08:023, 2008.

\bibitem{Bacchetta:2019qkv}
Alessandro Bacchetta, Giuseppe Bozzi, Miguel~G. Echevarria, Cristian Pisano,
  Alexey Prokudin, and Marco Radici.
\newblock {Azimuthal asymmetries in unpolarized SIDIS and Drell-Yan processes:
  a case study towards TMD factorization at subleading twist}.
\newblock {\em Phys. Lett. B}, 797:134850, 2019.

\bibitem{Vladimirov:2021hdn}
Alexey Vladimirov, Valentin Moos, and Ignazio Scimemi.
\newblock {Transverse momentum dependent operator expansion at next-to-leading
  power}.
\newblock {\em JHEP}, 01:110, 2022.

\bibitem{Rodini:2022wki}
Simone Rodini and Alexey Vladimirov.
\newblock {Definition and evolution of transverse momentum dependent
  distribution of twist-three}.
\newblock {\em JHEP}, 08:031, 2022.

\bibitem{Ebert:2021jhy}
Markus~A. Ebert, Anjie Gao, and Iain~W. Stewart.
\newblock {Factorization for azimuthal asymmetries in SIDIS at next-to-leading
  power}.
\newblock {\em JHEP}, 06:007, 2022.

\bibitem{Gamberg:2022lju}
Leonard Gamberg, Zhong-Bo Kang, Ding~Yu Shao, John Terry, and Fanyi Zhao.
\newblock {Transverse-momentum-dependent factorization at next-to-leading
  power}.
\newblock arXiv:2211.13209.

\bibitem{Rodini:2023plb}
Simone Rodini and Alexey Vladimirov.
\newblock {Transverse momentum dependent factorization for SIDIS at
  next-to-leading power}.
\newblock {\em Phys. Rev. D}, 110(3):034009, 2024.

\bibitem{Diehl:2005pc}
M.~Diehl and S.~Sapeta.
\newblock {On the analysis of lepton scattering on longitudinally or
  transversely polarized protons}.
\newblock {\em Eur. Phys. J.}, C41:515, 2005.

\bibitem{Chiu:2012ir}
Jui-Yu Chiu, Ambar Jain, Duff Neill, and Ira~Z. Rothstein.
\newblock {A Formalism for the Systematic Treatment of Rapidity Logarithms in
  Quantum Field Theory}.
\newblock {\em JHEP}, 05:084, 2012.

\bibitem{Balitsky:2023hmh}
Ian Balitsky.
\newblock {Rapidity-only TMD factorization at one loop}.
\newblock {\em JHEP}, 03:029, 2023.

\bibitem{Chay:1991jc}
June-gone Chay, Stephen~D. Ellis, and W.~James Stirling.
\newblock {Azimuthal asymmetry in lepton - photon scattering at high-energies}.
\newblock {\em Phys. Lett. B}, 269:175--182, 1991.

\bibitem{Oganesian:1997jq}
K.~A. Oganesian, H.~R. Avakian, N.~Bianchi, and P.~Di~Nezza.
\newblock {Investigations of azimuthal asymmetry in semiinclusive
  leptoproduction}.
\newblock {\em Eur. Phys. J. C}, 5:681--685, 1998.

\bibitem{Piloneta:2025jjb}
Sara Piloneta and Alexey Vladimirov.
\newblock {Kinematic power corrections for TMD factorization theorem of
  semi-inclusive deep-inelastic scattering}.
\newblock {\em JHEP}, 03:049, 2026.

\bibitem{Balitsky:2026nux}
Ian Balitsky and Alexei Prokudin.
\newblock {Next-to-next-to-leading power corrections to unpolarized
  Semi-Inclusive Deep Inelastic Scattering}.
\newblock arXiv:2601.18882.

\bibitem{Moos:2025sal}
Valentin Moos, Ignazio Scimemi, Alexey Vladimirov, and Pia Zurita.
\newblock {Determination of unpolarized TMD distributions from the fit of
  Drell-Yan and SIDIS data at N$^{4}$LL}.
\newblock {\em JHEP}, 11:134, 2025.

\bibitem{ATLAS:2019zci}
Georges Aad et~al.
\newblock {Measurement of the transverse momentum distribution of
  Drell{\textendash}Yan lepton pairs in proton{\textendash}proton collisions at
  $\sqrt{s}=13$ TeV with the ATLAS detector}.
\newblock {\em Eur. Phys. J. C}, 80(7):616, 2020.

\bibitem{Cerutti:GlobalTMDNN}
Matteo Cerutti.
\newblock A global fit of drell--yan and semi-inclusive deep inelastic
  scattering data with neural networks.
\newblock Manuscript in preparation.

\bibitem{Signori:2013mda}
Andrea Signori, Alessandro Bacchetta, Marco Radici, and Gunar Schnell.
\newblock {Investigations into the flavor dependence of partonic transverse
  momentum}.
\newblock {\em JHEP}, 11:194, 2013.

\bibitem{Anselmino:2013lza}
M.~Anselmino, M.~Boglione, J.~O. Gonzalez~Hernandez, S.~Melis, and A.~Prokudin.
\newblock {Unpolarised Transverse Momentum Dependent Distribution and
  Fragmentation Functions from SIDIS Multiplicities}.
\newblock {\em JHEP}, 04:005, 2014.

\bibitem{Echevarria:2014xaa}
Miguel~G. Echevarria, Ahmad Idilbi, Zhong-Bo Kang, and Ivan Vitev.
\newblock {QCD Evolution of the Sivers Asymmetry}.
\newblock {\em Phys. Rev. D}, 89:074013, 2014.

\bibitem{Bacchetta:2017gcc}
Alessandro Bacchetta, Filippo Delcarro, Cristian Pisano, Marco Radici, and
  Andrea Signori.
\newblock {Extraction of partonic transverse momentum distributions from
  semi-inclusive deep-inelastic scattering, Drell-Yan and Z-boson production}.
\newblock {\em JHEP}, 06:081, 2017.
\newblock [Erratum: JHEP 06, 051 (2019)].

\bibitem{OsvaldoGonzalez-Hernandez:2019iqj}
J.~Osvaldo Gonzalez-Hernandez.
\newblock {Comments on the perturbative and non-perturbative contributions in
  unpolarized SIDIS}.
\newblock {\em PoS}, DIS2019:176, 2019.

\bibitem{Bollweg:2025iol}
Dennis Bollweg, Xiang Gao, Jinchen He, Swagato Mukherjee, and Yong Zhao.
\newblock {Transverse-momentum-dependent pion structures from lattice QCD:
  Collins-Soper kernel, soft factor, TMDWF, and TMDPDF}.
\newblock {\em Phys. Rev. D}, 112(3):034501, 2025.

\bibitem{Avkhadiev:2025wps}
Artur Avkhadiev, Valerio Bertone, Chiara Bissolotti, Matteo Cerutti, Yang Fu,
  Simone Rodini, Phiala Shanahan, Michael Wagman, and Yong Zhao.
\newblock {An extraction of the Collins-Soper kernel from a joint analysis of
  experimental and lattice data}.
\newblock arXiv:2510.26489.

\bibitem{Constantinou:2019vyb}
Martha Constantinou, Haralambos Panagopoulos, and Gregoris Spanoudes.
\newblock {One-loop renormalization of staple-shaped operators in continuum and
  lattice regularizations}.
\newblock {\em Phys. Rev. D}, 99(7):074508, 2019.

\bibitem{Shanahan:2019zcq}
Phiala Shanahan, Michael~L. Wagman, and Yong Zhao.
\newblock {Nonperturbative renormalization of staple-shaped Wilson line
  operators in lattice QCD}.
\newblock {\em Phys. Rev. D}, 101(7):074505, 2020.

\bibitem{Green:2020xco}
Jeremy~R. Green, Karl Jansen, and Fernanda Steffens.
\newblock {Improvement, generalization, and scheme conversion of Wilson-line
  operators on the lattice in the auxiliary field approach}.
\newblock {\em Phys. Rev. D}, 101(7):074509, 2020.

\bibitem{Belle:2019ywy}
R.~Seidl et~al.
\newblock {Transverse momentum dependent production cross sections of charged
  pions, kaons and protons produced in inclusive $e^+e^-$ annihilation at
  $\sqrt{s}=$ 10.58 GeV}.
\newblock {\em Phys. Rev. D}, 99(11):112006, 2019.

\bibitem{Boglione:2023duo}
M.~Boglione and A.~Simonelli.
\newblock {Full treatment of the thrust distribution in single inclusive
  $^+e^-\to h X$ processes}.
\newblock {\em JHEP}, 09:006, 2023.

\bibitem{Kang:2020yqw}
Zhong-Bo Kang, Ding~Yu Shao, and Fanyi Zhao.
\newblock {QCD resummation on single hadron transverse momentum distribution
  with the thrust axis}.
\newblock {\em JHEP}, 12:127, 2020.

\bibitem{Belle:2019nve}
H.~Li et~al.
\newblock {Azimuthal asymmetries of back-to-back $\pi^\pm-(\pi^0,\eta,\pi^\pm)$
  pairs in $e^+e^-$ annihilation}.
\newblock {\em Phys. Rev. D}, 100(9):092008, 2019.

\bibitem{Belle:2008fdv}
R.~Seidl et~al.
\newblock {Measurement of Azimuthal Asymmetries in Inclusive Production of
  Hadron Pairs in e+e- Annihilation at s**(1/2) = 10.58-GeV}.
\newblock {\em Phys. Rev. D}, 78:032011, 2008.
\newblock [Erratum: Phys.Rev.D 86, 039905 (2012)].

\bibitem{BaBar:2015mcn}
J.~P. Lees et~al.
\newblock {Collins asymmetries in inclusive charged $KK$ and $K\pi$ pairs
  produced in $e^+e^-$ annihilation}.
\newblock {\em Phys. Rev. D}, 92(11):111101, 2015.

\bibitem{BaBar:2013jdt}
J.~P. Lees et~al.
\newblock {Measurement of Collins asymmetries in inclusive production of
  charged pion pairs in $e^+e^-$ annihilation at BABAR}.
\newblock {\em Phys. Rev. D}, 90(5):052003, 2014.

\bibitem{BESIII:2015fyw}
M.~Ablikim et~al.
\newblock {Measurement of azimuthal asymmetries in inclusive charged dipion
  production in $e^+e^-$ annihilations at $\sqrt{s}$ = 3.65 GeV}.
\newblock {\em Phys. Rev. Lett.}, 116(4):042001, 2016.

\bibitem{Artru:1989zv}
Xavier Artru and Mustapha Mekhfi.
\newblock {Transversely Polarized Parton Densities, their Evolution and their
  Measurement}.
\newblock {\em Z. Phys. C}, 45:669, 1990.

\bibitem{Belle:2024vua}
R.~Seidl et~al.
\newblock {Production cross sections of light and charmed mesons in e+e-
  annihilation near 10.58~GeV}.
\newblock {\em Phys. Rev. D}, 111(5):052003, 2025.

\bibitem{Kang:2015msa}
Zhong-Bo Kang, Alexei Prokudin, Peng Sun, and Feng Yuan.
\newblock {Extraction of Quark Transversity Distribution and Collins
  Fragmentation Functions with QCD Evolution}.
\newblock {\em Phys. Rev. D}, 93(1):014009, 2016.

\bibitem{Klein_2008}
M.~Klein and R.~Yoshida.
\newblock Collider physics at hera.
\newblock {\em Progress in Particle and Nuclear Physics}, 61(2):343–393,
  October 2008.

\bibitem{STAR:2017hhs}
L.~Adamczyk et~al.
\newblock {Measurements of jet quenching with semi-inclusive hadron+jet
  distributions in Au+Au collisions at $\sqrt{s_{NN}}$ = 200 GeV}.
\newblock {\em Phys. Rev. C}, 96(2):024905, 2017.

\bibitem{STAR:2017ieb}
L.~Adamczyk et~al.
\newblock {Beam Energy Dependence of Jet-Quenching Effects in Au+Au Collisions
  at $\sqrt{s_{_{ \mathrm{NN}}}}$ = 7.7, 11.5, 14.5, 19.6, 27, 39, and 62.4
  GeV}.
\newblock {\em Phys. Rev. Lett.}, 121(3):032301, 2018.

\bibitem{Chang:2021vbq}
Zilong Chang.
\newblock {Inclusive Jet Cross-section Measurements in $pp$ Collisions at
  $\sqrt{s} =$ 200 and 510 GeV with STAR}.
\newblock {\em PoS}, PANIC2021:377, 2022.

\bibitem{Robotkova:2024jss}
Monika Robotkov{\'a}.
\newblock {Systematic exploration of multi-scale jet substructure in $p$ + $p$
  collisions at $\sqrt{s}=200$ GeV by the STAR experiment}.
\newblock {\em PoS}, HardProbes2023:177, 2024.

\bibitem{PHENIX:2025pqz}
N.~J. Abdulameer et~al.
\newblock {Measurement of inclusive jet cross section and substructure in p+p
  collisions at s=200{\,}{\,}GeV}.
\newblock {\em Phys. Rev. D}, 111(11):112008, 2025.

\bibitem{aaboud2017determination}
Morad Aaboud, Georges Aad, Brad Abbott, Jalal Abdallah, Ovsat Abdinov, Baptiste
  Abeloos, Syed~Haider Abidi, Ossama~Sherif AbouZeid, NL~Abraham, Halina
  Abramowicz, et~al.
\newblock Determination of the strong coupling constant $\alpha$ s from
  transverse energy--energy correlations in multijet events at s= 8 tev using
  the atlas detector.
\newblock {\em The European Physical Journal C}, 77(12):872, 2017.

\bibitem{aaboud2018measurement}
Morad Aaboud, Georges Aad, Brad Abbott, B~Abeloos, SH~Abidi, OS~AbouZeid,
  NL~Abraham, H~Abramowicz, H~Abreu, R~Abreu, et~al.
\newblock Measurement of inclusive jet and dijet cross-sections in
  proton-proton collisions at $\sqrt{s}= 13$ tev with the atlas detector.
\newblock {\em Journal of High Energy Physics}, 2018(5):195, 2018.

\bibitem{acharya2019charged}
Shreyasi Acharya, Dagmar Adamov{\'a}, A~Adler, Jonatan Adolfsson, Madan~M
  Aggarwal, Gianluca Aglieri~Rinella, Michelangelo Agnello, Nikita Agrawal,
  Zubayer Ahammed, Sang~Un Ahn, et~al.
\newblock Charged jet cross section and fragmentation in proton-proton
  collisions at s= 7 tev.
\newblock {\em Physical Review D}, 99(1):012016, 2019.

\bibitem{Vertesi:2024tdv}
Robert Vertesi.
\newblock {Jet substructure measurements in heavy-ion collisions}.
\newblock {\em Int. J. Mod. Phys. A}, 40(09):2444002, 2025.

\bibitem{Liu:2018trl}
Xiaohui Liu, Felix Ringer, Werner Vogelsang, and Feng Yuan.
\newblock {Lepton-jet Correlations in Deep Inelastic Scattering at the
  Electron-Ion Collider}.
\newblock {\em Phys. Rev. Lett.}, 122(19):192003, 2019.

\bibitem{Arratia:2020nxw}
Miguel Arratia, Zhong-Bo Kang, Alexei Prokudin, and Felix Ringer.
\newblock {Jet-based measurements of Sivers and Collins asymmetries at the
  future electron-ion collider}.
\newblock {\em Phys. Rev. D}, 102(7):074015, 2020.

\bibitem{Basham:1977iq}
C.~Louis Basham, Lowell~S. Brown, S.~D. Ellis, and S.~T. Love.
\newblock {Electron - Positron Annihilation Energy Pattern in Quantum
  Chromodynamics: Asymptotically Free Perturbation Theory}.
\newblock {\em Phys. Rev. D}, 17:2298, 1978.

\bibitem{Li:2021txc}
Hai~Tao Li, Yiannis Makris, and Ivan Vitev.
\newblock {Energy-energy correlators in Deep Inelastic Scattering}.
\newblock {\em Phys. Rev. D}, 103(9):094005, 2021.

\bibitem{Kang:2023big}
Zhong-Bo Kang, Kyle Lee, Ding~Yu Shao, and Fanyi Zhao.
\newblock {Probing transverse momentum dependent structures with azimuthal
  dependence of energy correlators}.
\newblock {\em JHEP}, 03:153, 2024.

\bibitem{Mi:2025abd}
Zihao Mi and Zhan Wang.
\newblock {One-point energy correlator inside jets}.
\newblock {\em JHEP}, 11:090, 2025.

\bibitem{Gao:2025evv}
Mei-Sen Gao, Zhong-Bo Kang, Wanchen Li, and Ding~Yu Shao.
\newblock {Accessing nucleon transversity with one-point energy correlators}.
\newblock arXiv:2509.15809.

\bibitem{Song:2025bdj}
Yu-Kun Song, Shu-Yi Wei, Lei Yang, and Jian Zhou.
\newblock {Gluon Polarimetry with Energy-Energy Correlators}.
\newblock arXiv:2509.14960.

\bibitem{Gao:2025cwy}
Jun Gao, Hai~Tao Li, and Yu~Jiao Zhu.
\newblock {Energy correlators resolving proton spin}.
\newblock {\em Phys. Rev. D}, 113(3):034028, 2026.

\bibitem{Zhu:2025qkx}
Yu~Jiao Zhu.
\newblock {Energy correlators in semi-inclusive electron-positron
  annihilation}.
\newblock {\em Phys. Rev. D}, 113(1):014025, 2026.

\bibitem{Cao:2025icu}
Qing-Hong Cao, Zhite Yu, C.~P. Yuan, Shutao Zhang, and Hua~Xing Zhu.
\newblock {Collins-type fragmentation energy correlator in semi-inclusive deep
  inelastic lepton-hadron scattering}.
\newblock {\em JHEP}, 02:244, 2026.

\bibitem{Hatta:2023fqc}
Yoshitaka Hatta.
\newblock {Accessing the gravitational form factors of the nucleon and nuclei
  through a massive graviton}.
\newblock {\em Phys. Rev. D}, 109(5):L051502, 2024.

\bibitem{Xiong:2023zih}
Weizhi Xiong and Chao Peng.
\newblock {Proton Electric Charge Radius from Lepton Scattering}.
\newblock {\em Universe}, 9(4):182, 2023.

\bibitem{Burkert:2018bqq}
V.~D. Burkert, L.~Elouadrhiri, and F.~X. Girod.
\newblock {The pressure distribution inside the proton}.
\newblock {\em Nature}, 557(7705):396--399, 2018.

\bibitem{Anikin:2007yh}
I.~V. Anikin and O.~V. Teryaev.
\newblock {Dispersion relations and subtractions in hard exclusive processes}.
\newblock {\em Phys. Rev. D}, 76:056007, 2007.

\bibitem{Dutrieux:2024bgc}
Herv{\'e} Dutrieux, Thibaud Meisgny, C{\'e}dric Mezrag, and Herv{\'e} Moutarde.
\newblock {Proton internal pressure from deeply virtual Compton scattering on
  collider kinematics}.
\newblock {\em Eur. Phys. J. C}, 85(1):105, 2025.

\bibitem{Hatta:2018ina}
Yoshitaka Hatta and Di-Lun Yang.
\newblock {Holographic $J/\psi$ production near threshold and the proton mass
  problem}.
\newblock {\em Phys. Rev. D}, 98(7):074003, 2018.

\bibitem{Mamo:2019mka}
Kiminad~A. Mamo and Ismail Zahed.
\newblock {Diffractive photoproduction of $J/\psi$ and $\Upsilon$ using
  holographic QCD: gravitational form factors and GPD of gluons in the proton}.
\newblock {\em Phys. Rev. D}, 101(8):086003, 2020.

\bibitem{Hatta:2021can}
Yoshitaka Hatta and Mark Strikman.
\newblock {$\phi$-meson lepto-production near threshold and the strangeness
  $D$-term}.
\newblock {\em Phys. Lett. B}, 817:136295, 2021.

\bibitem{Guo:2021ibg}
Yuxun Guo, Xiangdong Ji, and Yizhuang Liu.
\newblock {QCD Analysis of Near-Threshold Photon-Proton Production of Heavy
  Quarkonium}.
\newblock {\em Phys. Rev. D}, 103(9):096010, 2021.

\bibitem{Guo:2023qgu}
Yuxun Guo, Xiangdong Ji, and Feng Yuan.
\newblock {Proton{\textquoteright}s gluon GPDs at large skewness and
  gravitational form factors from near threshold heavy quarkonium
  photoproduction}.
\newblock {\em Phys. Rev. D}, 109(1):014014, 2024.

\bibitem{Guo:2025jiz}
Yuxun Guo, Feng Yuan, and Wenbin Zhao.
\newblock {Bayesian Inferring Nucleon Gravitational Form Factors via
  Near-Threshold J/{\ensuremath{\psi}} Photoproduction}.
\newblock {\em Phys. Rev. Lett.}, 135(11):111902, 2025.

\bibitem{Hatta:2025vhs}
Yoshitaka Hatta, Henry~T. Klest, Kornelija Passek-K., and Jakob Schoenleber.
\newblock {Deeply virtual $\phi$-meson production near threshold}.
\newblock arXiv:2501.12343.

\bibitem{Hatta:2025ryj}
Yoshitaka Hatta and Jakob Schoenleber.
\newblock {Sullivan Process near Threshold and the Pion Gravitational Form
  Factors}.
\newblock {\em Phys. Rev. Lett.}, 134(25):251901, 2025.

\bibitem{Punjabi:2015bba}
V.~Punjabi, C.~F. Perdrisat, M.~K. Jones, E.~J. Brash, and C.~E. Carlson.
\newblock {The Structure of the Nucleon: Elastic Electromagnetic Form Factors}.
\newblock {\em Eur. Phys. J. A}, 51:79, 2015.

\bibitem{Drell:1969km}
S.~D. Drell and Tung-Mow Yan.
\newblock {Connection of Elastic Electromagnetic Nucleon Form-Factors at Large
  Q**2 and Deep Inelastic Structure Functions Near Threshold}.
\newblock {\em Phys. Rev. Lett.}, 24:181--185, 1970.

\bibitem{West:1970av}
Geoffrey~B. West.
\newblock {Phenomenological model for the electromagnetic structure of the
  proton}.
\newblock {\em Phys. Rev. Lett.}, 24:1206--1209, 1970.

\bibitem{Nesterenko:1983ef}
V.~A. Nesterenko and A.~V. Radyushkin.
\newblock {Local Quark - Hadron Duality and Nucleon Form-factors in {QCD}}.
\newblock {\em Phys. Lett. B}, 128:439--444, 1983.

\bibitem{Isgur:1984jm}
Nathan Isgur and C.~H. Llewellyn~Smith.
\newblock {Asymptopia in High q**2 Exclusive Processes in QCD}.
\newblock {\em Phys. Rev. Lett.}, 52:1080, 1984.

\bibitem{Isgur:1988iw}
Nathan Isgur and C.~H. Llewellyn~Smith.
\newblock {The Applicability of Perturbative QCD to Exclusive Processes}.
\newblock {\em Nucl. Phys. B}, 317:526--572, 1989.

\bibitem{Radyushkin:1990te}
A.~V. Radyushkin.
\newblock {Hadronic form-factors: Perturbative QCD versus QCD sum rules}.
\newblock {\em Nucl. Phys. A}, 532:141--154, 1991.

\bibitem{Jakob:1993iw}
R.~Jakob and P.~Kroll.
\newblock {The Pion form-factor: Sudakov suppressions and intrinsic transverse
  momentum}.
\newblock {\em Phys. Lett. B}, 315:463--470, 1993.
\newblock [Erratum: Phys.Lett.B 319, 545 (1993)].

\bibitem{Diehl:1998kh}
M.~Diehl, T.~Feldmann, R.~Jakob, and P.~Kroll.
\newblock {Linking parton distributions to form-factors and Compton
  scattering}.
\newblock {\em Eur. Phys. J. C}, 8:409--434, 1999.

\bibitem{Radyushkin:1998rt}
A.~V. Radyushkin.
\newblock {Nonforward parton densities and soft mechanism for form-factors and
  wide angle Compton scattering in QCD}.
\newblock {\em Phys. Rev. D}, 58:114008, 1998.

\bibitem{Braun:2006hz}
V.~M. Braun, A.~Lenz, and M.~Wittmann.
\newblock {Nucleon Form Factors in QCD}.
\newblock {\em Phys. Rev. D}, 73:094019, 2006.

\bibitem{Kivel:2010ns}
Nikolai Kivel and Marc Vanderhaeghen.
\newblock {Soft spectator scattering in the nucleon form factors at large $Q^2$
  within the SCET approach}.
\newblock {\em Phys. Rev. D}, 83:093005, 2011.

\bibitem{Hatta:2025xuf}
Yoshitaka Hatta and Jakob Schoenleber.
\newblock {Generalized parton distributions and gravitational form factors at
  large momentum transfer}.
\newblock {\em JHEP}, 12:058, 2025.

\bibitem{Braun:2001tj}
Vladimir~M. Braun, A.~Lenz, N.~Mahnke, and E.~Stein.
\newblock {Light cone sum rules for the nucleon form-factors}.
\newblock {\em Phys. Rev. D}, 65:074011, 2002.

\bibitem{Polyakov:2002yz}
M.~V. Polyakov.
\newblock {Generalized parton distributions and strong forces inside nucleons
  and nuclei}.
\newblock {\em Phys. Lett. B}, 555:57--62, 2003.

\bibitem{Polyakov:2018zvc}
Maxim~V. Polyakov and Peter Schweitzer.
\newblock {Forces inside hadrons: pressure, surface tension, mechanical radius,
  and all that}.
\newblock {\em Int. J. Mod. Phys. A}, 33(26):1830025, 2018.

\bibitem{Lorce:2018egm}
C\'edric Lorc\'e, Herv\'e Moutarde, and Arkadiusz~P. Trawi\'nski.
\newblock {Revisiting the mechanical properties of the nucleon}.
\newblock {\em Eur. Phys. J. C}, 79(1):89, 2019.

\bibitem{Hawking:1973uf}
Stephen~W. Hawking and George F.~R. Ellis.
\newblock {\em {The Large Scale Structure of Space-Time}}.
\newblock Cambridge Monographs on Mathematical Physics. Cambridge University
  Press, 2 2023.

\bibitem{Dumitru:2025gzc}
Adrian Dumitru and Jorge Noronha.
\newblock {Violation of energy conditions and the gravitational radius of the
  proton}.
\newblock {\em Phys. Rev. D}, 112(9):094057, 2025.

\bibitem{Tong:2022zax}
Xuan-Bo Tong, Jian-Ping Ma, and Feng Yuan.
\newblock {Perturbative calculations of gravitational form factors at large
  momentum transfer}.
\newblock {\em JHEP}, 10:046, 2022.

\bibitem{Tanaka:2018wea}
Kazuhiro Tanaka.
\newblock {Operator relations for gravitational form factors of a spin-0
  hadron}.
\newblock {\em Phys. Rev. D}, 98(3):034009, 2018.

\bibitem{Tong:2021ctu}
Xuan-Bo Tong, Jian-Ping Ma, and Feng Yuan.
\newblock {Gluon gravitational form factors at large momentum transfer}.
\newblock {\em Phys. Lett. B}, 823:136751, 2021.

\bibitem{Gao:2022vyh}
Xiang Gao, Andrew~D. Hanlon, Nikhil Karthik, Swagato Mukherjee, Peter
  Petreczky, Philipp Scior, Sergey Syritsyn, and Yong Zhao.
\newblock {Pion distribution amplitude at the physical point using the
  leading-twist expansion of the quasi-distribution-amplitude matrix element}.
\newblock {\em Phys. Rev. D}, 106(7):074505, 2022.

\bibitem{Ji:1994av}
Xiang-Dong Ji.
\newblock {A QCD analysis of the mass structure of the nucleon}.
\newblock {\em Phys. Rev. Lett.}, 74:1071--1074, 1995.

\bibitem{Muller:1994ses}
Dieter M{\"u}ller, D.~Robaschik, B.~Geyer, F.~M. Dittes, and
  J.~Ho{\v{r}}ej{\v{s}}i.
\newblock {Wave functions, evolution equations and evolution kernels from light
  ray operators of QCD}.
\newblock {\em Fortsch. Phys.}, 42:101--141, 1994.

\bibitem{Ji:1996nm}
Xiang-Dong Ji.
\newblock {Deeply virtual Compton scattering}.
\newblock {\em Phys. Rev. D}, 55:7114--7125, 1997.

\bibitem{Radyushkin:1996ru}
A.~V. Radyushkin.
\newblock {Asymmetric gluon distributions and hard diffractive
  electroproduction}.
\newblock {\em Phys. Lett. B}, 385:333--342, 1996.

\bibitem{Collins:1996fb}
John~C. Collins, Leonid Frankfurt, and Mark Strikman.
\newblock {Factorization for hard exclusive electroproduction of mesons in
  QCD}.
\newblock {\em Phys. Rev. D}, 56:2982--3006, 1997.

\bibitem{Berger:2001xd}
Edgar~R. Berger, M.~Diehl, and B.~Pire.
\newblock {Time - like Compton scattering: Exclusive photoproduction of lepton
  pairs}.
\newblock {\em Eur. Phys. J. C}, 23:675--689, 2002.

\bibitem{CLAS:2021lky}
P.~Chatagnon et~al.
\newblock {First Measurement of Timelike Compton Scattering}.
\newblock {\em Phys. Rev. Lett.}, 127(26):262501, 2021.

\bibitem{Belitsky:2002tf}
Andrei~V. Belitsky and Dieter Mueller.
\newblock {Exclusive electroproduction of lepton pairs as a probe of nucleon
  structure}.
\newblock {\em Phys. Rev. Lett.}, 90:022001, 2003.

\bibitem{Guidal:2002kt}
M.~Guidal and M.~Vanderhaeghen.
\newblock {Double deeply virtual Compton scattering off the nucleon}.
\newblock {\em Phys. Rev. Lett.}, 90:012001, 2003.

\bibitem{Deja:2023ahc}
K.~Deja, V.~Martinez-Fernandez, B.~Pire, P.~Sznajder, and J.~Wagner.
\newblock {Phenomenology of double deeply virtual Compton scattering in the era
  of new experiments}.
\newblock {\em Phys. Rev. D}, 107(9):094035, 2023.

\bibitem{Pedrak:2017cpp}
A.~Pedrak, B.~Pire, L.~Szymanowski, and J.~Wagner.
\newblock {Hard photoproduction of a diphoton with a large invariant mass}.
\newblock {\em Phys. Rev. D}, 96(7):074008, 2017.
\newblock [Erratum: Phys.Rev.D 100, 039901 (2019)].

\bibitem{Duplancic:2023kwe}
Goran Duplan{\v{c}}i{\'c}, Saad Nabeebaccus, Kornelija Passek-Kumeri{\v{c}}ki,
  Bernard Pire, Lech Szymanowski, and Samuel Wallon.
\newblock {Probing chiral-even and chiral-odd leading twist quark generalized
  parton distributions through the exclusive photoproduction of a
  {\ensuremath{\gamma}}{\ensuremath{\rho}} pair}.
\newblock {\em Phys. Rev. D}, 107(9):094023, 2023.

\bibitem{Qiu:2022bpq}
Jian-Wei Qiu and Zhite Yu.
\newblock {Exclusive production of a pair of high transverse momentum photons
  in pion-nucleon collisions for extracting generalized parton distributions}.
\newblock {\em JHEP}, 08:103, 2022.

\bibitem{Qiu:2022pla}
Jian-Wei Qiu and Zhite Yu.
\newblock {Single diffractive hard exclusive processes for the study of
  generalized parton distributions}.
\newblock {\em Phys. Rev. D}, 107(1):014007, 2023.

\bibitem{Qiu:2023mrm}
Jian-Wei Qiu and Zhite Yu.
\newblock {Extraction of the Parton Momentum-Fraction Dependence of Generalized
  Parton Distributions from Exclusive Photoproduction}.
\newblock {\em Phys. Rev. Lett.}, 131(16):161902, 2023.

\bibitem{Grocholski:2022rqj}
Oskar Grocholski, Bernard Pire, Pawe{\l} Sznajder, Lech Szymanowski, and Jakub
  Wagner.
\newblock {Phenomenology of diphoton photoproduction at next-to-leading order}.
\newblock {\em Phys. Rev. D}, 105(9):094025, 2022.

\bibitem{Nabeebaccus:2023rzr}
Saad Nabeebaccus, Jakob Schoenleber, Lech Szymanowski, and Samuel Wallon.
\newblock {Breakdown of collinear factorization in the exclusive
  photoproduction of a {\ensuremath{\pi}}0{\ensuremath{\gamma}} pair with large
  invariant mass}.
\newblock {\em Phys. Rev. D}, 111(3):034040, 2025.

\bibitem{Qiu:2024mny}
Jian-Wei Qiu and Zhite Yu.
\newblock {Extracting transition generalized parton distributions from hard
  exclusive pion-nucleon scattering}.
\newblock {\em Phys. Rev. D}, 109(7):074023, 2024.

\bibitem{Ivanov:2004vd}
D.~Yu. Ivanov, A.~Schafer, L.~Szymanowski, and G.~Krasnikov.
\newblock {Exclusive photoproduction of a heavy vector meson in QCD}.
\newblock {\em Eur. Phys. J. C}, 34(3):297--316, 2004.
\newblock [Erratum: Eur.Phys.J.C 75, 75 (2015)].

\bibitem{Chen:2019uit}
Zi-Qiang Chen and Cong-Feng Qiao.
\newblock {NLO QCD corrections to exclusive electroproduction of quarkonium}.
\newblock {\em Phys. Lett. B}, 797:134816, 2019.
\newblock [Erratum: Phys.Lett. B, 135759 (2020)].

\bibitem{Flett:2021ghh}
C.~A. Flett, J.~A. Gracey, S.~P. Jones, and T.~Teubner.
\newblock {Exclusive heavy vector meson electroproduction to NLO in collinear
  factorisation}.
\newblock {\em JHEP}, 08:150, 2021.

\bibitem{Goloskokov:2024egn}
S.~V. Goloskokov, Ya-Ping Xie, and Xurong Chen.
\newblock {Study of gluon GPDs in exclusive J/{\ensuremath{\psi}} production in
  electron-proton scattering}.
\newblock {\em Phys. Rev. D}, 110(7):076029, 2024.

\bibitem{Flett:2024htj}
C.~A. Flett, J.~P. Lansberg, S.~Nabeebaccus, M.~Nefedov, P.~Sznajder, and
  J.~Wagner.
\newblock {Exclusive vector-quarkonium photoproduction at NLO in
  {\ensuremath{\alpha}}s in collinear factorisation with evolution of the
  generalised parton distributions and high-energy resummation}.
\newblock {\em Phys. Lett. B}, 859:139117, 2024.

\bibitem{Kumericki:2007sa}
K.~Kumericki, Dieter Mueller, and K.~Passek-Kumericki.
\newblock {Towards a fitting procedure for deeply virtual Compton scattering at
  next-to-leading order and beyond}.
\newblock {\em Nucl. Phys. B}, 794:244--323, 2008.

\bibitem{Kumericki:2009uq}
Kresimir Kumeri{\v{c}}ki and Dieter Mueller.
\newblock {Deeply virtual Compton scattering at small $x_B$ and the access to
  the GPD H}.
\newblock {\em Nucl. Phys. B}, 841:1--58, 2010.

\bibitem{Cuic:2023mki}
Marija {\v{C}}ui{\'c}, Goran Duplan{\v{c}}i{\'c}, Kre{\v{s}}imir
  Kumeri{\v{c}}ki, and Kornelija Passek-K.
\newblock {NLO corrections to the deeply virtual meson production revisited:
  impact on the extraction of generalized parton distributions}.
\newblock {\em JHEP}, 12:192, 2023.
\newblock [Erratum: JHEP 02, 225 (2024)].

\bibitem{Moutarde:2019tqa}
H.~Moutarde, P.~Sznajder, and J.~Wagner.
\newblock {Unbiased determination of DVCS Compton Form Factors}.
\newblock {\em Eur. Phys. J. C}, 79(7):614, 2019.

\bibitem{Dutrieux:2021wll}
H.~Dutrieux, O.~Grocholski, H.~Moutarde, and P.~Sznajder.
\newblock {Artificial neural network modelling of generalised parton
  distributions}.
\newblock {\em Eur. Phys. J. C}, 82(3):252, 2022.
\newblock [Erratum: Eur.Phys.J.C 82, 389 (2022)].

\bibitem{Guo:2022upw}
Yuxun Guo, Xiangdong Ji, and Kyle Shiells.
\newblock {Generalized parton distributions through universal moment
  parameterization: zero skewness case}.
\newblock {\em JHEP}, 09:215, 2022.

\bibitem{Guo:2023ahv}
Yuxun Guo, Xiangdong Ji, M.~Gabriel Santiago, Kyle Shiells, and Jinghong Yang.
\newblock {Generalized parton distributions through universal moment
  parameterization: non-zero skewness case}.
\newblock {\em JHEP}, 05:150, 2023.

\bibitem{Guo:2025muf}
Yuxun Guo, Fatma~P. Aslan, Xiangdong Ji, and M.~Gabriel Santiago.
\newblock {First Global Extraction of Generalized Parton Distributions from
  Experiment and Lattice Data with Next-to-Leading-Order Accuracy}.
\newblock {\em Phys. Rev. Lett.}, 135(26):261903, 2025.

\bibitem{Bertone:2021yyz}
V.~Bertone, H.~Dutrieux, C.~Mezrag, H.~Moutarde, and P.~Sznajder.
\newblock {Deconvolution problem of deeply virtual Compton scattering}.
\newblock {\em Phys. Rev. D}, 103(11):114019, 2021.

\bibitem{Lin:2021brq}
Huey-Wen Lin.
\newblock {Nucleon helicity generalized parton distribution at physical pion
  mass from lattice QCD}.
\newblock {\em Phys. Lett. B}, 824:136821, 2022.

\bibitem{Good:2025daz}
William Good, Fei Yao, and Huey-Wen Lin.
\newblock {First nucleon gluon PDF from large momentum effective theory}.
\newblock {\em Phys. Lett. B}, 872:140067, 2026.

\bibitem{Shiells:2021xqo}
Kyle Shiells, Yuxun Guo, and Xiangdong Ji.
\newblock {On extraction of twist-two Compton form factors from DVCS
  observables through harmonic analysis}.
\newblock {\em JHEP}, 08:048, 2022.

\bibitem{Guo:2024wxy}
Yuxun Guo, Xiangdong Ji, M.~Gabriel Santiago, Jinghong Yang, and Hao-Cheng
  Zhang.
\newblock {Small-x gluon GPD constrained from deeply virtual
  J/{\ensuremath{\psi}} production and gluon PDF through universal-moment
  parametrization}.
\newblock {\em Phys. Rev. D}, 112(5):054036, 2025.

\end{thebibliography}
